\documentclass[graybox, secnum]{svmult}

\usepackage{mathptmx}       
\usepackage{helvet}         
\usepackage{courier}        
\usepackage{type1cm}        
\usepackage{makeidx}         
\usepackage{graphicx}        
\usepackage{multicol}        
\usepackage[bottom]{footmisc}
\usepackage{hyperref}        
\usepackage{soul}            
\hypersetup{
   colorlinks=true,
   urlcolor=blue,
   citecolor=blue,
   linkcolor=blue
   }
\usepackage{natbib}
\makeindex             

\usepackage[utf8]{inputenc}
\usepackage{indentfirst}
\usepackage{pdflscape}
\usepackage{longtable}
\usepackage{amssymb}
\usepackage{amsmath}
\usepackage{adjustbox}
\usepackage{threeparttable}
\usepackage{tabto}
\usepackage{float}
\usepackage{color}
\usepackage{multirow}
\usepackage{multicol}
\usepackage{geometry}
\usepackage{graphicx}
\usepackage{txfonts}
\usepackage{multicol}
\usepackage[dvipsnames]{xcolor}
\usepackage{setspace}

\newcommand{\Ion}[2]{#1{\,\textsc{#2}}}

\newcommand{\Lsun}{L$_{\odot}$}
\newcommand{\Msun}{M$_{\odot}$}

\newcommand{\bhl}{Bondi-Hoyle-Lyttleton}
\newcommand{\hergap}{Hertzsprung gap}

\newcommand{\pceb}{post-CE binary}
\newcommand{\pcebs}{post-CE binaries}
\newcommand{\msto}{MS stars close to the turn-off point}

\newcommand{\mesa}{{\sc mesa}}

\definecolor{smalt(darkpowderblue)}{rgb}{0.0, 0.2, 0.6}
\definecolor{forestgreen(traditional)}{rgb}{0.0, 0.5, 0.0}

\newcommand{\eprint}[2][]{\href{https://arxiv.org/abs/#2}{arXiv:~\nolinkurl{#2}}}

\newcommand\memsai{Mem. S.A.It.}%
\newcommand\actaa{Acta Astron.} 
\newcommand{\adspr}{AdSpR}			
\newcommand{\araa}{ARA\&A}			
\newcommand{\apj}{ApJ}				
\newcommand{\apjl}{ApJL}			
\newcommand{\apjs}{ApJS}			
\newcommand{\aj}{AJ}				
\newcommand{\apss}{Ap\&SS}          
\newcommand{\aap}{A\&A~}	        
\newcommand{\aapr}{A\&AR}			
\newcommand{\mnras}{MNRAS}			
\newcommand{\pasp}{PASP}			
\newcommand{\physrep}{Phys. Rep.}			
\newcommand{\nat}{Nat.}				
\newcommand{\nar}{New Astron. Rev.}		
\newcommand{\ssr}{Space Sci. Rev.}		

\begin{document}
\title*{Formation and Evolution of Accreting Compact Objects}
\author{
Diogo Belloni  \thanks{corresponding author} and Matthias R. Schreiber
}
\institute{
Diogo Belloni \at Departamento de F\'isica, Universidad T\'ecnica Federico Santa Mar\'ia, Av. España 1680, Valpara\'iso, Chile \email{diogobellonizorzi@gmail.com}
\and 
Matthias R. Schreiber \at Departamento de F\'isica, Universidad T\'ecnica Federico Santa Mar\'ia, Av. España 1680, Valpara\'iso, Chile; Millennium Nucleus for Planet Formation (NPF), Valpara\'iso, Chile \email{Matthias.Schreiber@usm.cl}
}
\maketitle
\abstract{Accreting compact objects are crucial to understand several important astrophysical phenomena such as Type Ia supernovae, gravitational waves, or X-ray and $\gamma$-ray bursts. In addition, they are natural laboratories to infer fundamental properties of stars, to investigate high-energy phenomena and accretion processes, to test theories of stellar and binary evolution, to explore interactions between high-density plasma and very strong magnetic fields, to examine the interplay between binary evolution and dynamical interactions (in the case they belong to dense star clusters), and they can even be used as a probe for the assembling process of galaxies over cosmic time-scales. Despite the fundamental importance of accreting compact objects for astrophysics and recent progress with the comprehension of these fascinating objects, we still do not fully understand how they form and evolve. In this chapter, we will review the current theoretical status of our knowledge on these objects, and will discuss standing problems and potential solutions to them.}

\section*{Keywords} 
Cataclysmic Variables, 
AM\,CVn Stars, 
Symbiotic Stars,
Super-soft X-ray Binaries,
Low-mass X-ray Binaries,
Ultra-compact X-ray Binaries,
High-mass X-ray Binaries

\tableofcontents

\newpage

\section{Introduction}

Compact objects are remnants of stellar evolution and they are mainly divided into three categories, which strongly depend on the mass and metallicity of their main-sequence (MS) star progenitor. 
Low-mass stars (${\lesssim2}$~\Msun) and intermediate-mass stars (${\sim2-8}$~\Msun) end their lives as white dwarfs (WDs), while more massive stars evolve into either neutron stars (NSs) or black holes (BHs).
Understanding stellar remnants represents a crucial part of understanding stellar evolution and stellar structure.
Compact objects, especially as part of a binary, offer the potential to study physical phenomena in the most extreme environments and produce some of the most fascinating events in the Universe.

Compact objects as members of binaries are frequently found to accrete material from their companion stars.
Accreting compact objects are then binaries in which a compact object accretes matter from a companion, which can be a star, another compact object or even a planet in some cases.
They are excellent laboratories for studying accretion disk physics and the generation/evolution of extremely strong magnetic fields.
These objects are also among the most luminous high-energy sources in the Universe, emitting detectable radiation from X-rays to $\gamma$-rays, thereby permitting detailed studies of plasma physics and the related emission processes under extreme conditions.

Not only the secular evolution of accreting compact objects is intrinsically related to mass transfer, also their formation is in many cases associated with at least one episode of mass transfer between the stars composing the progenitor binary.
Unlike single stars, the stars within a binary may interact and in turn may be affected by one another depending on their nature and the orbital separation.
Mass transfer in a binary through the inner Lagrangian point occurs when at least one star fills its Roche lobe, which is triggered by either nuclear evolution of this star, for example the expansion of a giant star, or by orbital shrinkage due to orbital angular momentum loss (AML).
In the former case, mass transfer can be maintained because the star expands at least at the same rate as its Roche lobe radius.
The latter occurs in all binaries because two masses orbiting their common centre of mass always loose angular momentum due to gravitational wave radiation.
Depending on the nature of the mass losing star, orbital AML mechanisms that are stronger than gravitational wave radiation can be present.
The loss of angular momentum reduces the separation between the two stars, and therefore, the radius of the companion can continuously fill its Roche lobe.

Regardless of what is causing mass transfer, the Roche lobe filling star has to lose mass because it tries to keep itself inside its Roche lobe radius, which implies further mass loss than that originally expected from single stellar evolution.
Mass transfer in binaries affects the evolution of the stellar components.
For example, the mass growth of the helium core of a giant star can be halted before helium burning occurs if the giant star is losing its entire envelope to the accreting companion and this is why most helium-core WDs are found to be members of binaries.
In other cases, the accreting star can rejuvenate by becoming more massive than it was at the onset of mass transfer, which alters its natural evolution.
Such a mass gain could allow, for instance, a star that would otherwise evolve into a WD (or NS) to evolve into a NS (or BH), or more generally a more massive compact object than in the case of single star evolution. 
Another example corresponds to the non-expected chemical abundances in some stars such as barium stars, which can only be explained if their companion were mass-losing stars that managed to synthesise those elements.

In this chapter we will discuss the state-of-the-art of our understanding of the formation and evolution of accreting compact objects.
We start with a very brief review of the different types of accreting compact objects and the main processes that produce their X-ray emission. 
Their formation, the current state, and the future evolution depend fundamentally on the mechanisms that are driving mass transfer in binaries.
We therefore subsequently review the present understanding of the formation and evolution of these systems with a discussion of the modes and the stability of mass transfer, before we discuss their possible formation channels and their secular evolution.
We will then end this review with a discussion of the major current problems in our understanding of the formation and evolution of accreting compact objects and briefly describe potential avenues towards solving these issues.
More details on the physics of binary star evolution can be found in the book by \citet{Tauris_2023BOOK}.

\section{The Accreting Compact Object Zoo}


{\it Cataclysmic variables} (CVs) are close binaries composed of WDs accreting from low-mass Roche lobe filling MS stars and are definitely the best-studied and most numerous type of interacting compact binaries.
While several thousand CVs are known, the most important samples currently available are volume limited \citep{Pala_2020,Inight_2021}. 
These representative samples provide key observational constraints on CV formation and evolution theories, but at the same time we have to admit that they are still rather small and to some degree subject to low-number statistics. 
The majority of CVs are non-magnetic \cite[e.g.][]{Pala_2020} and most non-magnetic CVs are dwarf novae, which exhibit repetitive outbursts due to thermal-viscous instabilities in their accretion disks \cite[e.g.][]{Hameury_1998,Lasota_2001,Lasota_2016}.
All types of CVs are X-ray sources.
In the case of non-magnetic CVs, the X-ray emission is produced in the transition region between the accretion disk and the WD, the boundary layer, from where approximately one-half of the accretion energy is expected to be released \citep{Patterson_1985,Patterson_1985b}.
In magnetic CVs, matter is accreted onto the WD in a field-channeled accretion flow, starting at the region where the magnetic field captures the mass flow from the secondary star, and extending down to the surface of the WD.
This accretion flow is supersonic when it reaches regions close to the surface of the white dwarf and a shock is formed.
The matter in the post-shock region, which is the region between the shock and the WD surface, is compressed and heated to temperatures up to a few tens of keVs.
Bremsstrahlung from this post-shock region is responsible for most of the X-ray emission of magnetic CVs \citep{Mukai_2017,Belloni_ApJSS_2021}.
When a certain amount of hydrogen-rich matter is accumulated on the surface of an accreting WD, hydrogen shell burning is ignited. Non-magnetic and magnetic CVs experience unstable hydrogen burning that produces nova eruptions \cite[e.g.][]{Chomiuk_2021} and because of the associated mass loss, the WD masses in CVs are supposed to remain approximately constant over a nova cycle \cite[e.g.][]{Yaron_2005}.


While mass transfer in CVs is driven by orbital AML, binaries consisting of a WD and a slightly more massive Roche lobe filling MS star experience thermal time-scale mass transfer. 
In such systems, accretion rates exceeding values of ${\sim10^{-7}-10^{-6}}$~\Msun\,yr$^{-1}$, depending on the WD mass, can be reached and the hydrogen-rich material accreted by the WD can stably burn on the surface of the WD \cite[e.g.][]{Nomoto_2007}. 
Nova eruptions and the corresponding erosion of the WD mass are avoided in these systems, and therefore, they are considered to be potential Type Ia supernova progenitors \citep{Distefano_2010}. 
Hydrogen burning on the surface of a WD is producing effective temperatures somewhat below $100$~eV, producing X-ray emission in the $0.01–2$~keV range.
Stably hydrogen-burning WDs in close binaries are therefore called {\it super-soft X-ray binaries} (SSXBs).
The first two SSXBs were discovered in the eighties 
\citep{Long_1981} and about a decade later one additional system was found \citep{Truemper_1991}.
Today the discovery of more than 100 systems has been reported in external galaxies, the Magellanic clouds, and the Milky Way, but this number and the total emission in X-rays seem to be too low to account for the full rate of Type Ia supernovae \citep{Gilfanov_2010,Gilfanov_2021}.
The relatively low number of SSXBs and the corresponding low soft X-ray emission are potentially related to the fact that circumstellar matter can suppress the soft X-ray flux \citep{Nielsen_2013}.


A binary where a compact object is accreting material at a relatively high rate (greater than several $10^{-9}$~\Msun\,yr$^{-1}$) from a low-mass-evolved red giant is called a {{\it symbiotic star}} (SySt), in case the compact object is a WD \citep{Kenyon_1986}, or a {{\it symbiotic X-ray binary}} (SyXB), in case the accretor is a NS \citep{Yungelson_2019}.
In most cases (${\sim80}$\%), the donor is a normal evolved giant, typically M-type red giant, but in some cases it can be a G-/K-type yellow giant. Their near-infrared emission shows the presence of a stellar photosphere, and for this reason these systems are called S-type SySt \citep{Mikolajewska_2003,Mikolajewska_2007,Mikolajewska_2010,Mikolajewska_2012}.
In the remaining systems, the donor is a Mira-type variable and their near-infrared emission is consistent with being originated from a combination of a reddened Mira-type variable and a warm optically thick dust shell, which makes these systems to be called D-type SySts \citep{Gromadzki_2009}.
In SySts, the accretion rates are typically large enough to cause stable nuclear burning on the surface of the WD, which leads to the emission of soft X-rays similar to those of SSXBs.
In most SySts and SyXBs, the mass transfer from the red giant to the compact object is due to gravitationally focused wind accretion and only in some cases due to atmospheric Roche lobe overflow.
SySts are important for understanding mass transfer in wide binaries and the formation of astrophysical jets. Most importantly, as SSXBs, SySts are promising candidates for being Type Ia supernova progenitors \citep{Distefano_2010b,Liu_2019_SySt,Ikiewicz_2019}.


{{\it AM\,CVn}} stars are compact binaries with orbital periods in the range of ${\sim5-65}$~min \citep{Ramsay_2018}.
These binaries are so close that a MS star cannot be part of the system as it would just not fit.
In AM\,CVns, a WD accretes helium-dominated material from a Roche lobe filling companion, usually a lower-mass WD or a semi-degenerate helium-rich star.
Mass transfer in AM\,CVns is driven by orbital AML through gravitational wave radiation which is expected to be detectable by the Laser Interferometer Space Antenna (LISA) mission \citep{Nelemans_2013,Korol_2017}.
Observationally, AM CVns are characterized by their hydrogen-deficient optical spectra and their blue colour.
More than 50 AM\,CVns are known by the time of writing \cite[e.g.][]{Ramsay_2018} and their number is constantly increasing thanks to dedicated spectroscopic and photometric transit surveys \citep{Carter_2013,Breedt_2014}. 
X-ray emission can be triggered in two different ways in AM\,CVns and the details depend to some degree on the formation channel and the nature of the donor.
In case the donor is a WD, the onset of mass transfer occurs as a direct impact during which a small region of the accreting WD is heated and emits a fraction of the released accretion luminosity at high energy wavelength \citep{Postnov_2014}.
In the presence of an accretion disk around the accreting WD, X-ray emission is produced in the boundary layer in full analogy to the X-ray emission of non-magnetic CVs.


{{\it Low-mass X-ray binaries}} (LMXBs) consist of a NS or a BH that is accreting from a Roche lobe filling low-mass star \citep{Tauris_2006}.
As in CVs, mass transfer in LMXBs is driven by orbital AML due to magnetic braking and/or gravitational wave radiation.
In analogy to dwarf novae, LMXBs also show outbursts that are most likely triggered by thermal-viscous instabilities in 
their accretion disks \citep{Hameury_2020}.
An important difference between LMXBs and CVs is that in the former irradiation of the accretion disk is more important and may fully suppress disk instabilities.
In CVs, disk irradiation is dominating most likely only in post-novae systems \citep{Schreiber_2000,Schreiber_2001}.
LMXBs also show thermonuclear X-ray bursts originating from unstable hydrogen and helium burning on the surface of the accreting NS \citep{StrohmayerBildsten_2006}, similarly to the nova eruptions occurring in accreting WDs.
A typical LMXB emits almost all of its radiation in X-rays and these systems are therefore among the brightest X-ray sources.
Since the seventies, an over-abundance of LMXBs in globular cluster in comparison to the Milky Way field has been observationally established, which naturally suggests that dynamical interactions, which are very common in the dense central parts of globular clusters, facilitate their formation \cite[e.g.][]{vandenBerg_2020}.


{{\it Ultra-compact X-ray binaries}} (UCXBs) are binaries with periods shorter than about an hour.
UCXBs are very similar to AM\,CVns with the main difference being that the accreting compact object is either a NS or a BH.
As in the case of AM\,CVns, such short orbital periods exclude MS stars to form part of the binary, and therefore, the donors are either WDs or semi-degenerate helium-rich stars \citep{Nelemans_2010}. 
UCXBs are mainly observed at X-rays and some of these objects show thermonuclear X-ray bursts \citep{Koliopanos_2021}.
As in the case of LMXBs, there is currently a clear consensus that the relative number of UCXBs is much larger in globular clusters than in the Milky Way field \citep{vandenBerg_2020}, and dynamics therefore seem to provide an important formation channel also for these systems.


If a compact object is accreting from a high-mass star (${\gtrsim10}$~\Msun), the system is known as a {{\it high-mass X-ray binary}} \citep[HMXB, e.g.][]{Kretschmar_2019}. 
As in the cases above, the potential energy of the accreted material in HMXBs is converted into X-ray emission.
Although HMXBs are therefore strong X-ray sources, they are also bright at optical wavelengths given that their optical emission is dominated by the massive companion. 
In some cases, the donor is an O- or early B-type supergiant that is filling or close to filling its Roche lobe and these systems are consequently called supergiant HMXBs \cite[sg-HMXBs, e.g.][]{M2017}. 
Despite hosting a supergiant, the orbital periods of some sg-HMXBs are shorter than ${\sim10}$~days, and they are persistent X-ray sources, powered by the capture of matter from the strong stellar wind of the supergiant or, in a few cases, through Roche lobe overflow \citep{Chaty_2012}. 
In contrast, supergiant fast X-ray transients (SFXTs) have orbital periods typically longer than ${\sim10}$~days and exhibit peculiar characteristics compared to the remaining sg-HMXBs, such as bright short time-scale flares.
A second sub-type of HMXBs are B-emission HMXBs (Be-HMXBs), which were discovered in the 1970s \citep{Maraschietal_1976}.
In virtually all cases these systems contain accreting NSs, but there is one confirmed Be-HMXB hosting a BH \citep{Casares_2014}, and a few Be-HMXBs have been proposed to contain WDs \cite[e.g.][]{Kennea_2021,Coe_2020}.
The donors in these systems are rapidly rotating B stars on or close to the MS, and unlike the donors in sg-HMXBs, given their typical orbital period (${\sim10-400}$~days), the donors in Be-HMXBs are deep inside their Roche lobes \cite[e.g.][]{Reig_2011}.  
Be-HMXBs can be in slightly eccentric orbits \citep{Pfahl_2002}, or highly eccentric orbits, and their X-ray emission is originated when the compact object accretes from the Be star circumstellar decretion disk either during periastron passages, resulting in orbital-phase-dependent X-ray outbursts lasting several days, or due to enhanced activity of the Be star, which leads to X-ray outbursts lasting several weeks not correlated with any particular orbital phase \citep{Ziolkowski_2002}.
As a final sub-class, we mention Wolf--Rayet HMXBs (WR-HMXBs), in which the donor is a massive helium star.
Only few such systems are known and they are likely descendants from sg-HMXBs and Be-HMXBs \citep{van_den_Heuvel_2017}.
Among HMXBs, Be-HMXBs correspond to the most numerous class \citep[${\sim60}$\%,][]{Liu_2006}, followed by sg-HMXBs \citep[${\sim32}$\%,][]{Liu_2006}.
This is a direct consequence of selection effects favouring the discovery of Be-HMXBs in comparison with sg-HMXBs.
As previously mentioned, Be-HMXBs are discovered during outbursts, which faint persistent sg-HMXBs would not have.
For this reason, it is very likely that the contribution of sg-HMXBs to the intrinsic population of HMXBs is much higher.

\section{Modes of Mass Transfer}

The Roche lobe is the last equipotential surface in a binary that keeps all the mass attached to one star. Any expansion of a star beyond its Roche lobe or any orbital shrinkage of the Roche lobe due to orbital AML may cause a star to fill its Roche lobe thereby forcing mass to flow over to the other star.
Determining the rate at which mass is transferred through the inner Lagrangian point in binaries represents in general a difficult hydro-dynamical problem.   
One of the most frequently used and reasonable prescriptions properly takes into account the finite scale height of the stellar atmosphere of the Roche lobe filling donor.

For stars with extended atmosphere Roche lobe overflow can occur through the inner Lagrangian point even when the donor is slightly under-filling its Roche lobe, i.e. ${R_{\rm d}<R_{\rm RL}}$, where $R_{\rm d}$ is the donor radius and $R_{\rm RL}$ is the Roche lobe radius of the donor.
Taking this into account, the mass transfer rate $\dot{M}_{\rm d}$ can be estimated as \citep{Ritter_1988}

\begin{equation}
\dot{M}_{\rm d} \ = \ -\,\dot{M}_0\,
\exp\left(\,-\,
   \frac{R_{\rm RL}\,-\,R_{\rm d}}{H_{\rm P}\,/\,\gamma(q)}\,
\right) \ ,
\label{eq:Ritter}
\end{equation}

\noindent
where $H_{\rm P}$ is the pressure scale height at the photosphere of the donor.

The quantity $\gamma(q)$ is a function of the mass ratio $q=M_{\rm d}/M_{\rm a}$, where $M_{\rm d}$ is the donor mass and $M_{\rm a}$ is the accretor mass, given by

\begin{eqnarray}
\gamma(q) \ = \ 
\begin{cases}
0.954 \ + \ 0.025\,\log_{10}q \ - \ 0.038\,\left(\log_{10}q\right)^2, & \ {\rm if} \ \  0.04\lesssim q \le 1\\
0.954 \ + \ 0.039\,\log_{10}q \ + \  0.114\,\left(\log_{10}q\right)^2, &  \ {\rm if} \ \  1\le q \lesssim 20 
\end{cases}
\end{eqnarray}

Finally, $\dot{M}_{0}$ is the mass transfer rate when the donor exactly fills its Roche lobe given by

\begin{equation}
\dot{M}_0 \ = \ 
F(q) \, \rho_{\rm ph} \,
\left(\,
   \frac{2\pi}{\sqrt{2.71828}}\,
\right)\,   
\left(\,
   \frac{R_{\rm RL}^3}{G\,M_{\rm d}}\,
\right)\,   
\left(\,
   \frac{k_{\rm B}\,T_{\rm eff}}{m_{\rm p} \  \mu_{\rm ph}}
\right)^{3/2} \ ,
\end{equation}

\noindent
where $G$ is the gravitational constant, $k_{\rm B}$ is the Boltzmann constant, $m_{\rm p}$ is the proton mass, $M_{\rm d}$ is the donor mass, $T_{\mathrm{eff}}$ is the donor effective temperature, $\mu_{\rm ph}$ and $\rho_{\rm ph}$ are the mean molecular weight and density at its photosphere, respectively and $F(q)\,=\,1.23\,+\,0.5\log_{10}q$, valid for $0.5\lesssim q \lesssim 10$.
Outside the ranges of validity, $F(q)$ and $\gamma(q)$ are usually evaluated using the value of $q$ at the edge of their respective ranges \citep{Paxton2015}.

In case the donor is over-filling its Roche lobe, i.e. $R_{\rm d}>R_{\rm RL}$, the mass transfer rate can be expressed as \citep{Kolb_1990}

\begin{equation}
\dot{M}_{\mathrm{d}} \ = \ 
-\dot{M}_0 \, - \, 2 \, \pi \, F(q) \,
\frac{R_{\rm RL}^3}{G\,M_{\rm d}} \, 
\int_{P_{\mathrm{ph}}}^{P_{\mathrm{RL}}} \Gamma_1^{1/2}\left(\frac{2}{\Gamma_1+1}\right)^{\frac{\Gamma_1+1}{2\Gamma_1-2}}
\left( \, 
   \frac{k_{\rm B} \, T}{m_{\rm p} \, \mu}
\right)^{1/2} dP \ ,
\label{eq:Kolb}   
\end{equation}

\noindent
where $T$ is the temperature, $\Gamma_1$ is the first adiabatic exponent, and $P_\mathrm{ph}$ and $P_\mathrm{RL}$ are, respectively, the pressures at the photosphere and at the radius corresponding to $R_{\rm RL}$.

We shall mention that, in case donor is strongly under-filling its Roche lobe, the mass transfer rates provided by Equation~\ref{eq:Ritter} are negligible.
However, mass can still be transferred from the donor, if the stellar winds from the donor are sufficiently strong and dense and a part of the this wind is accreted by the companion.
In what follows, we discuss in more detail these modes of mass transfer, starting with the stability of mass transfer when the donor is filling its Roche lobe, i.e. ${R_{\rm d} \sim R_{\rm RL}}$.

\section{Stability of Mass Transfer Through Roche Lobe Filling}

To understand the evolution of binaries containing a Roche lobe filling star, it is of fundamental importance to understand the reaction of the donor and the Roche geometry to mass loss.  
Despite the fact that the constraints on a Roche lobe filling star are modified due to mass loss and the star consequently can no longer be considered as evolving in isolation, under certain conditions the star may adjust its internal structure in response to mass loss because most of its mass is concentrated towards its centre.
Whether hydrostatic and/or thermal equilibrium can be maintained in a Roche lobe filling star and whether mass transfer is stable or unstable depend entirely on the response of the donor radius and its Roche lobe radius to mass loss on the corresponding time-scales \citep{Webbink_1985}.

\subsection{Dynamical Time-scale Mass Transfer}

Changes of the radius of the donor star as a response to variations of its mass and through time can be expressed as
\begin{equation}
 {\rm d} \ln R_{\rm d} \ = \ 
 \left( \, 
        \frac{\partial \ln R_{\rm d}}{\partial t}
 \right)
 \, 
 {\rm d} t
 \ + \
 \left( \, 
        \frac{\partial \ln R_{\rm d}}{\partial \ln M_{\rm d}} \, 
 \right) 
 \, 
 {\rm d} \ln M_{\rm d} \ ,
\label{dlnRd}
\end{equation}
\

\noindent
where $t$ is the time.
While the first term on the right side is connected with stellar evolution and relaxation processes connected with thermal readjustment, the second term corresponds to the adiabatic response of the donor to mass loss, i.e. the hydrodynamic response of the donor radius to variations in its mass.

The changes in the donor Roche lobe radius as a response to variations in mass, time, and angular momentum can be written as

\begin{equation}
 {\rm d} \ln R_{\rm RL} \ = \ 
 \left( \, 
        \frac{\partial \ln R_{\rm RL}}{\partial t}
 \right)
 \, 
 {\rm d} t
 \ + \
 \left( \, 
  \frac{\partial \ln R_{\rm RL}}{\partial \ln M_{\rm d}} \, 
 \right) \, 
 {\rm d} \ln M_{\rm d} \ .
\
\label{dlnRrl}
\end{equation}
\

\noindent
The first term on the right side represents the changes in the Roche lobe radius due to orbital AML in the absence of mass transfer, and the second term is associated with the response of the Roche lobe radius to mass loss.

Whether mass transfer from a Roche lobe filling star is stable or not can be evaluated by comparing two logarithmic derivatives associated with the donor.
These derivatives are that of the donor radius with respect to its mass, for a fixed initial profile in composition and specific entropy, and that of its Roche lobe radius with respect to its mass.
The former is usually called \textit{adiabatic radius--mass exponent} $\zeta_{\rm ad}$, defined as

\begin{equation}
\zeta_{\rm ad}  \equiv 
\left. 
\frac{\partial \ln\,R_{\rm d}}{\partial \ln\,M_{\rm d}} 
\right|_{\rm ad} \ ,
\label{zetaAD}
\end{equation}

\noindent
while the latter is termed the \textit{Roche lobe radius--mass exponent} $\zeta_{\rm RL}$, defined as

\begin{equation}
\zeta_{\rm RL} \equiv  \frac{{\rm d}\ln\,R_{\rm RL}}{{\rm d}\ln\,M_{\rm d}} \ .
\label{zetaRL}
\end{equation}

Combining Equations~\ref{dlnRd} and~\ref{dlnRrl}, and using the definitions~\ref{zetaAD} and~\ref{zetaRL}, we can express the change in the donor radius relative to its Roche lobe radius as

\begin{equation}
{\rm d} \ln \left( \, \frac{ R_{\rm d} }{ R_{\rm RL} } \, \right) 
\ = \ 
\left( \,
    \frac{\partial \ln R_{\rm d}}{\partial t}  
    \ - \ 
    \frac{\partial \ln R_{\rm RL}}{\partial t} \,
\right) \, {\rm d}t 
\ + \ 
\Bigl( \,
    \zeta_{\rm ad} - \zeta_{\rm RL} \,
\Bigr) \, {\rm d} \ln M_{\rm d} \ .
\label{dlnDeltaR}
\end{equation}

The condition for the donor to remain within its Roche lobe in hydrostatic equilibrium during an episode of nearly stationary mass transfer is obviously ${\rm d} \ln (R_{\rm d}/R_{\rm L}) \sim 0$.
Given that mass transfer is occurring, the difference in the time derivatives in Equation~\ref{dlnDeltaR} must be positive because the donor radius needs to change more strongly than its Roche lobe radius to guarantee that the donor keeps filling its Roche lobe.
In addition, as the donor is losing mass, ${\rm d} \ln M_{\rm d}$ is necessarily negative.
Therefore, the difference between the adiabatic and Roche lobe radius--mass exponents needs to be positive to keep the donor in hydrostatic equilibrium.
In other words, by comparing these two derivatives we are able to determine whether the response to mass transfer of the donor inner layers can occur adiabatically, i.e. whether the entropy profile in the stellar interior remains unchanged despite the mass loss.
If, on the other hand, $\zeta_{\rm ad}<\zeta_{\rm RL}$, the mass transfer is unstable and occurs on the \textit{dynamical time-scale}, because the mass transfer rate is only limited by the sonic expansion of the envelope through the inner Lagrangian point.

Another way of reaching the same conclusion is by analyzing perturbations of the mass transfer rate \cite[e.g.][]{Ritter_1988}.
During nearly steady mass transfer, i.e. ${\rm d} \ln (R_{\rm d}/R_{\rm L}) \sim 0$, the well-known mass transfer rate equation as a function of $\zeta_{\rm ad}$ and $\zeta_{\rm RL}$ can be derived from Equation~\ref{dlnDeltaR} as

\begin{equation}
\dot{M}_{\rm d} \ = \ - \
\left( 
   \frac{M_{\rm d}}{\zeta_{\rm ad}-\zeta_{\rm RL}} 
\right) \, 
\left( \,
   \frac{\partial \ln R_{\rm d}}{\partial t}  
   \ - \ 
   \frac{\partial \ln R_{\rm RL}}{\partial t} \,
\right) \ .
\label{Mdot}
\end{equation}

Considering a system in an unperturbed phase of stationary mass transfer, i.e. $\ddot{M}_{\rm d}\sim0$, we can evaluate the dynamical stability of mass transfer after a sudden non-null perturbation $\delta\dot{M}_{\rm d}$.
After the perturbation, 
$\dot{M}_{\rm d}=\dot{M}_{\rm d,0} + \delta\dot{M}_{\rm d}$, where $\dot{M}_{\rm d,0}$ is the unperturbed mass transfer rate,
and 
$\ddot{M}_{\rm d}\propto\delta\dot{M}_{\rm d}\,\dot{M}_{\rm d}\,(\zeta_{\rm ad}-\zeta_{\rm RL})$.
Given that $\dot{M}_{\rm d}<0$, the mass transfer will be dynamically stable only if the response to a sudden increase in $|\dot{M}_{\rm d}|$ is a decrease in $|\dot{M}_{\rm d}|$.
This happens only if $\zeta_{\rm ad}-\zeta_{\rm RL}\,>\,0$.
Otherwise, the sudden increase in $|\dot{M}_{\rm d}|$ will result in further increase of $|\dot{M}_{\rm d}|$, which leads then to dynamically unstable mass transfer.

Even though realistic and detailed binary calculations must be performed to investigate the stability of mass transfer \cite[e.g.][]{CH08,W12,PI15,GeI,GeII,GeIII}, polytropes can shed some light on this process in a very simple way and provide a reasonable first-order approach to the problem in some cases \cite[e.g.][]{HW1987,SPV97}.
More specifically, composite polytropes, which are supposed to reproduce rather well the structure of fully convective low-mass MS stars, provide $\zeta_{\rm ad}$ in very good agreement with those computed from detailed stellar models \citep{GeIII}.
On the other hand, condensed polytropes, which are approximations of red giants, are unable to reproduce the non-ideal gas effect and inefficient convection occurring in detailed stellar models, which implies that condensed polytropes likely underestimated $\zeta_{\rm ad}$ of red giants \citep{GeIII}. 
If simple stellar models do not provide realistic estimates, the adiabatic radius--mass exponent needs to be calculated by solving the differential equations for stellar structure.

Once $\zeta_{\rm ad}$ has been obtained, it needs to be compared to $\zeta_{\rm RL}$ in order to evaluate whether mass transfer is dynamically stable or not.
The Roche lobe radius--mass exponent $\zeta_{\rm RL}$ can be expressed as

\begin{equation}
\zeta_{\rm RL} 
\ = \ 
\frac{\partial \ln a}{\partial \ln M_\mathrm{d}} 
\ + \
\left( \,
        \frac{\partial \ln (R_\mathrm{RL}/a)}{\partial \ln q} \,
\right) \,
\left( \,
        \frac{\partial \ln q}{\partial \ln M_\mathrm{d}} \,
\right),
\label{zetaRLq}
\end{equation}

\noindent
where $a$ the semimajor axis.
The three terms in Equation~\ref{zetaRLq} can be written as \citep{Eggleton_1983,SPV97,W12}

\begin{equation}
\frac{\partial \ln a}{\partial \ln M_\mathrm{d}} 
\ = \ 
(\,1\,-\,\alpha_{\rm ml}\,)\,
\left[ \,
       \frac{2\,q^2\,-2\,+\,q\,(1\,-\,\beta_{\rm ml})}{q\,+\,1}  \,
\right]
\ - \ 
\frac{q}{q\,+\,1}
\ ,
\label{eq:dlna_dlnm}
\end{equation}

\begin{equation}
\frac{\partial \ln (R_{\rm RL}/a)}{\partial \ln q} 
\ = \ 
\frac{2}{3} \, - \, 
\left( 
   \frac{q^{1/3}}{3} 
\right)
\left[
   \frac{1.2q^{1/3} \, + \, 1/(1 \, + \, q^{1/3}) }{0.6q^{2/3} \, + \, \ln (1 \, + \, q^{1/3})}
\right] \ ,
\label{EQRLAQ}
\end{equation}

\noindent
and

\begin{equation}
\frac{\partial \ln q}{\partial \ln M_{\rm d}} 
\ = \ 
1 \, + \, \left(\,1\,-\,\alpha_{\rm ml}\,-\,\beta_{\rm ml}\,\right) \, q \ ,
\label{EQQMD}
\end{equation}

\noindent
where $0\le\alpha_{\rm ml}\le1$ corresponds to the fraction of mass lost from the vicinity of the donor in the form of fast isotropic winds, and $0\le\beta_{\rm ml}\le1$ corresponds to the fraction of mass lost from the vicinity of the accretor due to isotropic re-emission.
Given that the fraction of mass leaving the donor that is effectively accreted is $1\,-\,\alpha_{\rm ml}\,-\,\beta_{\rm ml}$, the fully conservative case corresponds to $\alpha_{\rm ml}=\beta_{\rm ml}=0$, while $\alpha_{\rm ml}+\beta_{\rm ml}=1$ represents the fully non-conservative case.
In a more general situation, $\zeta_{\rm RL}$ depends on several other processes, which are related to orbital AML and mass loss from the system \citep{SPV97}, such as emission of gravitational waves, magnetic wind braking, tidal interactions, and torques exerted by circumbinary disks.
The most important dependencies of $\zeta_{\rm RL}$ with respect to the stability of mass transfer are those on the mass ratio as well as on the fraction of mass lost from the binary.

\begin{figure*}[htb!]
\centering
\includegraphics[width=0.49\linewidth]{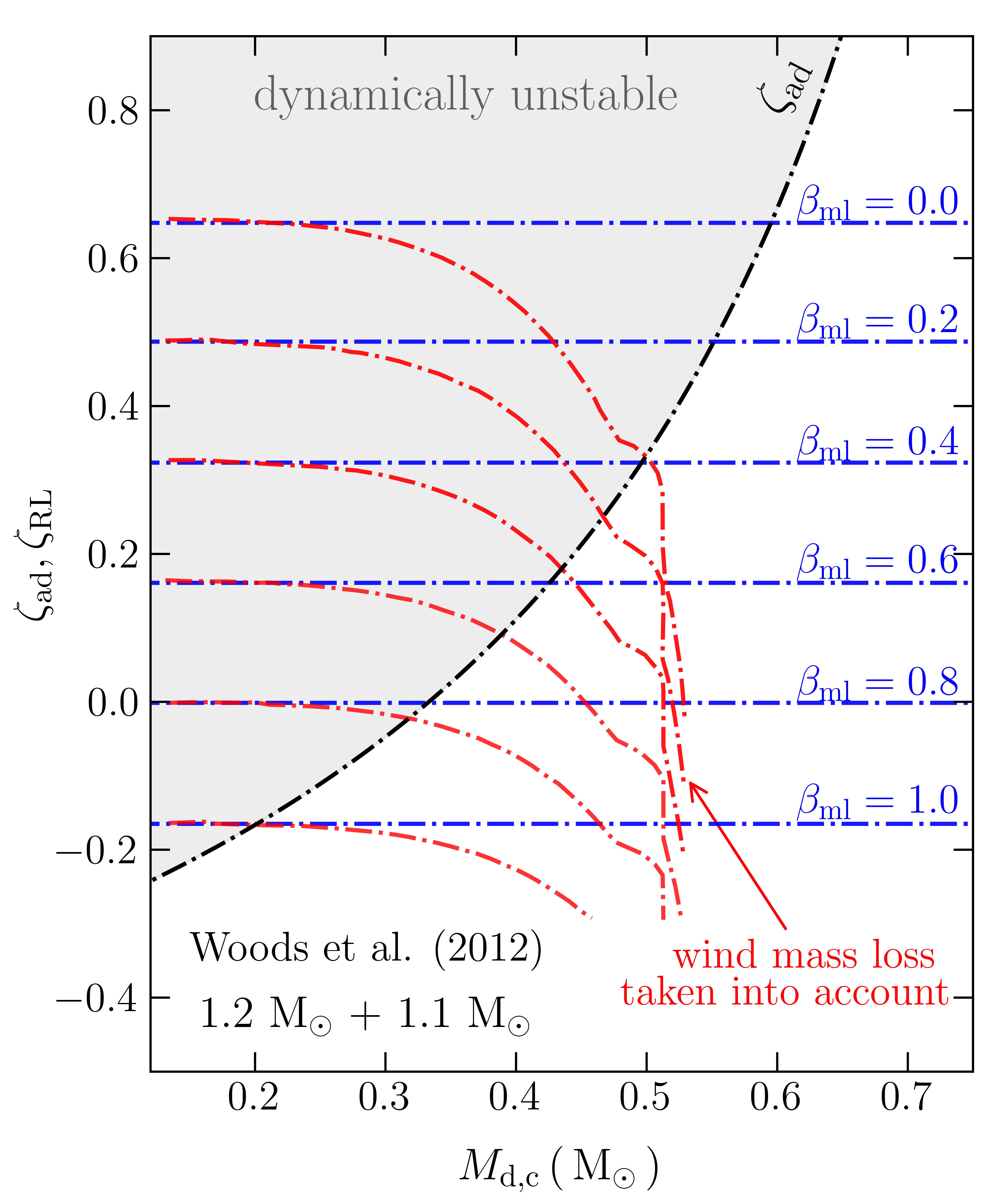}
\includegraphics[width=0.49\linewidth]{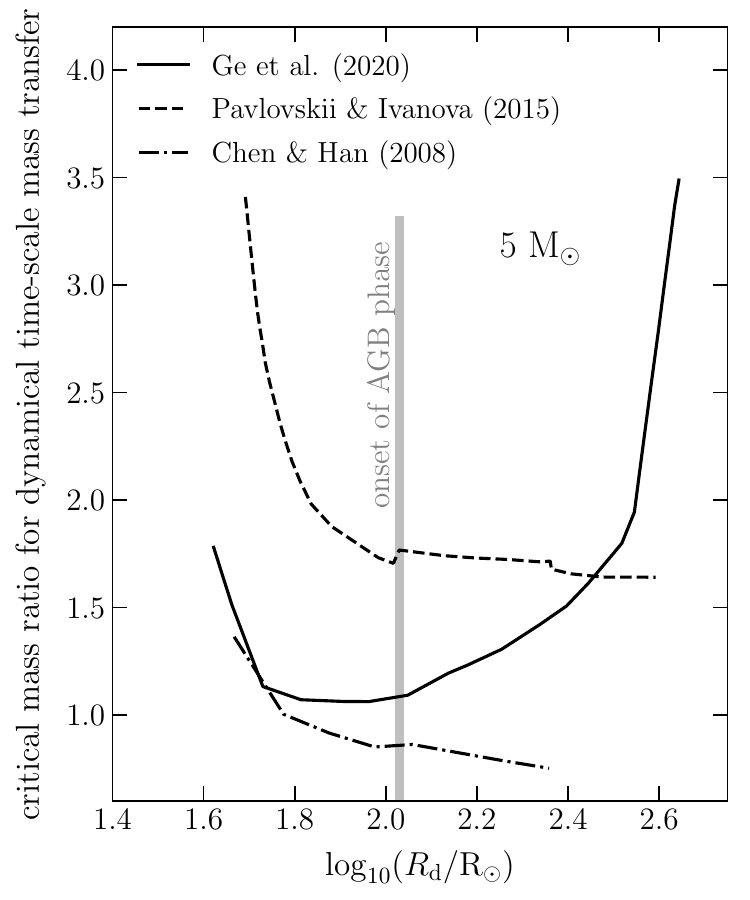}
\caption{Adiabatic ($\zeta_{\rm ad}$) and Roche lobe ($\zeta_{\rm RL}$) mass–radius exponents for a red giant donor of initial mass $1.2$~\Msun~with a companion of mass $1.1$~\Msun, as a function of the red giant core mass ($M_{\rm d,c}$), taken from a binary model \citep[left-hand panel,][]{W12}, and the critical mass ratio separating dynamically stable from unstable mass transfer for a donor of mass $5$~\Msun, as function of its radius ($R_{\rm d}$) during its evolution, taken from three different binary models \citep[right-hand panel,][]{CH08,PI15,GeIII}.
In the computation of $\zeta_{\rm RL}$ in the left-hand panel, for six different fractions of mass lost due to isotropic re-emission from the vicinity of the accretor ($\beta_{\rm ml}$), mass loss through stellar winds is considered (red lines) or not (blue horizontal lines).
In the right-hand panel, wind mass loss is not considered in either binary model and fully conservative mass transfer is assumed.
It is clear from the left-hand panel that non-conservative ($\beta_{\rm ml}>0$) mass transfer is more easily dynamically stable in comparison with the fully conservative case ($\beta_{\rm ml}=0$), regardless of whether wind mass loss is taken into account or not. 
This happens because $\zeta_{\rm RL}$ becomes smaller as $\beta_{\rm ml}$ increases, which allows mass transfer to be dynamically stable for smaller core mass, i.e. the more mass loss through isotropic re-emission is allowed, the more extended the range of stability.
The inclusion of wind mass loss enhances this effect, by decreasing even further $\zeta_{\rm RL}$, for core masses greater than ${\sim0.3}$~\Msun, extending in turn even further the range of stability.
The right-hand panel, on the other hand, clearly illustrates how the assumptions made in binary models can drastically change the boundary separating dynamically stable from unstable mass transfer.
For instance, the solid critical mass ratio strongly increases  during the AGB evolution, which is not seen in the other two models.
In addition, the dashed critical mass ratio is on average much greater than the other two models.
}
\label{FigCriterion}
\end{figure*}

The relation between $\zeta_{\rm ad}$ and $\zeta_{\rm RL}$ for a binary composed of a red giant donor with initial mass $1.2$~\Msun~and a companion of mass $1.1$~\Msun~is illustrated in the left-hand panel of Figure~\ref{FigCriterion} \citep[taken from a binary model,][]{W12}. 
As mentioned before, the limit for dynamical stability strongly depends on $\beta_{\rm ml}$.
In particular, the greater $\beta_{\rm ml}$, the smaller $\zeta_{\rm RL}$.
This means that the greater $\beta_{\rm ml}$, the better the chances for the red giant donor to remain in hydrostatic equilibrium.
If wind mass loss is included in the simulations, $\zeta_{\rm RL}$ decreases even further for more evolved red giants (i.e. red giants with larger core masses), for a given $\beta_{\rm ml}$.
For instance, when mass loss through winds is not considered, fully conservative mass transfer is dynamically stable only for core masses ${\gtrsim0.63}$~\Msun, but this limit becomes ${\gtrsim0.46}$~\Msun, if wind mass loss is included.

Even though the fundamental principles related to dynamical stability of mass transfer are well-understood, outcomes of binary models are strongly dependent on the physical assumptions.
As mentioned earlier, for a given binary at the onset of mass transfer, $\zeta_{\rm RL}$ strongly depends on the mass ratio and on the assumed mass loss and orbital AML prescriptions.
In addition, $\zeta_{\rm ad}$ depends entirely on the donor mass and its evolutionary status.
Thus, for a given binary, it is possible to derive a critical mass ratio separating dynamically stable from unstable mass transfer, as a function of the evolutionary status of the donor.
This critical mass ratio is plotted as a function of the donor radius on the right-hand side of Figure~\ref{FigCriterion}, for a donor with mass $5$~\Msun, taken from three different binary models \citep{CH08,PI15,GeIII}. 
Even though all binary models show the initial decrease in the critical mass ratio as the donor evolves along the first giant branch (FGB), it is clear from the figure that these models provide qualitative and quantitative different results.
In particular, one of them delivers on average a much higher critical mass ratio, in comparison with the other two.
In addition, while in two models the critical mass ratio slowly decreases as the donor climbs the asymptotic giant branch (AGB), the other model predicts a rapid increase of the critical mass ratio.
The main differences in the three works are connected with the mass transfer rate equation (whether it is limited by the local sound speed or not), the radius used to quantify $\zeta_{\rm ad}$ (surface or inner) and the Lagrangian radius adopted to quantify  $\zeta_{\rm RL}$ (inner or outer).

Therefore, despite significant progress during the last decade, the limiting mass ratio for the onset of dynamically unstable mass transfer remains quite uncertain.
This has significant implications for our general understanding of the formation of close binaries hosting compact objects, as dynamically unstable mass transfer is considered a crucial ingredient in most formation scenarios proposed for these objects. 
If mass transfer is dynamically unstable, there are mainly two outcomes.
If the donor is either an MS star or a degenerate object, then most likely the binary will merge on the dynamical time-scale of the donor, with a non-null fraction of the total mass being eventually lost in such an energetic event.
On the other hand, if the donor has evolved beyond the MS, common-envelope (CE) evolution will be triggered, resulting in a strong decrease of the orbital period.
In fact, CE evolution triggered by dynamically unstable mass transfer represents the standard scenario for the formation of close binaries containing a compact object.
Without a proper understanding of the conditions for dynamically unstable mass transfer, it will be impossible to reliably predict the relative numbers of objects as important as SN\,Ia progenitors and binaries that are main gravitational wave sources.

\subsection{Thermal Time-scale Mass Transfer}

If the Roche lobe filling donor in a binary can maintain hydrostatic equilibrium, the natural follow-up question is whether it must sacrifice its own thermal equilibrium or not to restrain its surface within its Roche lobe.
The thermal equilibrium of a star, i.e. the balance between the nuclear energy production and the atmospheric radiation losses, is disturbed as a response to mass loss.
However, as in the case of hydrostatic equilibrium, the donor may be able to re-establish thermal equilibrium, if mass transfer proceeds on a time-scale slower than its thermal time-scale, i.e. if the donor interior has enough time to relax to thermal equilibrium. 
Otherwise, mass transfer will be thermally unstable, i.e. it will proceed on the thermal time-scale of the donor \citep{Ge_2020}.
In order to verify whether this happens or not, we use the thermal \textit{equilibrium radius--mass exponent} $\zeta_{\rm eq}$, defined by

\begin{equation}
\zeta_{\rm eq}  \equiv 
\left. 
\frac{{\rm d}\ln\,R_{\rm d}}{{\rm d}\ln\,M_{\rm d}} 
\right|_\mathrm{th} \ .
\end{equation}

In order to calculate $\zeta_{\rm eq}$, the mass-radius relation for stars in thermal equilibrium must be known.
If mass transfer is dynamically stable, it will also be thermally stable if $\zeta_\mathrm{eq}>\zeta_\mathrm{RL}$.
In this case, mass transfer needs to be driven by an external process and proceeds on a time-scale different from the dynamical and thermal time-scales. 
On the other hand, if $\zeta_{\rm ad}>\zeta_{\rm RL} > \zeta_{\rm eq}$, mass transfer will be driven by thermal readjustment and proceed on the donor thermal time-scale.

\subsection{Nuclear or Orbital Angular Momentum Loss Time-scale Mass Transfer}

If mass transfer is dynamically and thermally stable, continuous mass transfer can still be generated by either stellar or binary evolutionary processes.
In this case, mass transfer is not self-stimulated as in the previous cases, since mass transfer does not proceed by the virtue of the inability of the donor to remain in either dynamical or thermal equilibrium.
Dynamically and thermally stable mass transfer is driven by either nuclear expansion of the Roche lobe filling star or orbital AML.
In the former case, the donor keeps filling its expanding Roche lobe, while in the latter case, the Roche lobe and the donor shrink as a response to mass loss.  
In case mass transfer is driven by nuclear expansion of the donor, mass transfer proceeds at a rate such that $R_{\rm d} \sim R_{\rm RL}$.
On the other hand, if mass transfer is driven by orbital AML, then the mass transfer rate is entirely dependent on the strength of orbital AML.
The main orbital AML mechanisms driving the evolution of binaries are gravitational wave radiation \cite[GR, e.g.][]{Paczynski_1967} and magnetic wind braking \cite[MB, e.g.][]{Huang_1966,Mestel_1968}.

Using the weak-field approximation of general relativity, it is possible to derive the orbital angular momentum changes due to emission of gravitational waves from two point masses as \citep{Hurley_2002}

\begin{equation}
\frac{{\dot{J}}_{\rm GR}}{J_{\rm orb}} \ = \ 
- 8.315 \times 10^{-10} \,
\left[
  \frac{M_1 \, M_2 \, (\,M_1\,+\,M_2\,)}{a^4} \,
\right] \,
\left[ \,
\frac{1 + \frac{7}{8} e^2}
{\left( 1 - e^2 \right)^{5/2}} \,
\right]
 \ \ {\rm yr}^{-1} \ ,
 \label{Eq:GR}
\end{equation}

\noindent
where $a$ is the semimajor axis in $R_\odot$, $M_1$ and $M_2$ are the binary component masses in \Msun, $e$ is the eccentricity, and $J_{\rm orb}$ is the orbital angular momentum given by

\begin{equation}
J_{\rm orb} \ = \  
a^2 \, 
\left( \frac{M_1 \, M_2}{M_1+M_2}  \right) \,
\left( \frac{2\,\pi}{P_{\rm orb}} \right) \, 
\sqrt{1-e^2} \ ,
\end{equation}

\noindent
where $P_{\rm orb}$ is the orbital period in years.
Emission of gravitational waves acts all the time in every binary.
However, from the strong dependence on the orbital separation, only for very close binaries it will have any impact on the orbital angular momentum evolution. Of course, how close a given binary must be so that GR is non-negligible depends on the component masses and on the eccentricity.

While Equation~\ref{Eq:GR} is well-established and gravitational wave radiation is well-understood, the same can unfortunately not be said about MB.
The main physical processes that drive MB are quite easy to understand but as usual the devil is in the details. 
It is clear that a highly ionized wind leaving the star is forced to corotate with the magnetic field of the star out to the Alfv\'en radius, thereby taking away more angular momentum than the specific angular momentum of the star that consequently spins down.
If the star is tidally locked to the orbit, i.e. if the star is synchronized with the orbital motion, the mass loss through winds removes significant amounts of orbital angular momentum, even if the wind mass-loss rate is relatively small.
The torques produced by the magnetized winds depend on the strength of the magnetic field, the Alfv\'en radius, the wind mass-loss rate, and the stiffness of the magnetic field at large distances from the star.
However, the details of these dependencies are very uncertain and neither observationally nor theoretically well-constrained.
Consequently, in the last decades several attempts to formulate empirical laws or recipes that are applicable to particular cases have been developed and a large number of MB prescriptions is available on the market \citep{Knigge_2011_OK}.
Unfortunately, as we will see, for a given fixed star mass, these MB prescriptions provide dramatically different orbital AML rates spanning easily several orders of magnitude. 

In what follows, we briefly mention some important MB prescriptions but discuss later in more detail their impact on the evolution of accreting compact objects.
A MB prescription that is widely used for stars in open star clusters and based on several simplifying assumptions for the winds, magnetic field geometry, and flow acceleration profile was developed by \citet{Kawaler_1988} as

\begin{equation}
\dot{J}_{\rm MB,Kaw} \ = \ - \, K_{W} \,
\left(
   \frac{\Omega}{{\rm s^{-1}}}
\right)^{1\,+\,4a_Kn/3} \, 
\left(
   \frac{\dot{M}_{\rm wind}}{-\,10^{-14}\,\,{\rm M}_\odot\,{\rm yr}^{-1}}
\right)^{1\,-\,2n/3} \, 
\left(
   \frac{R}{{\rm R}_{\odot}}
\right)^{2\,-\,n} \, 
\left(
   \frac{M}{{\rm M}_{\odot}}
\right)^{-n/3} \ \ {\rm dyn~cm} \ ,
\label{Eq:MB-KAW}
\end{equation}
\

\noindent
where $a_K$ depends on the scaling between the secondary magnetic field and spin, $n$ depends on the secondary magnetic field geometry, $\dot{M}_{\rm wind}<0$ is the star mass-loss rate through winds, and $\Omega$, $R$, and $M$ are the star spin, radius, and mass, respectively.
Even though the parameter $K_W$ depends on the ratio of the wind speed to the escape speed at the Alfv\'en radius and how the star magnetic field scales with the star radius and spin, it is usually approximated by ${\sim2.035\times10^{33}(24.93^n)(6.63\times10^5)^{4n/3}}$.
With ${a_K=1}$ and ${n=1.5}$, Equation~\ref{Eq:MB-KAW} describes the standard Skumanich law \citep{VZ}, as orbital AML due to MB becomes proportional to $\Omega^3$.

On the other hand, in the context of accreting compact objects driven by orbital AML, usually the so-called RVJ prescription \citep{RVJ} is adopted, which is given by

\begin{equation}
\dot{J}_{\rm MB,RVJ} \ = \ - \, 3.8\times10^{-30} \,
\left(\,\frac{M}{\rm g}\,\right) \,
\left(\,\frac{R_\odot}{{\rm cm}}\,\right)^4 \, 
\left(\,\frac{\Omega}{\rm s^{-1}}\,\right)^3\, 
\left(\,\frac{R}{R_\odot}\,\right)^{\gamma_{\rm MB}} \ \ {\rm dyn~cm} \ ,
\label{Eq:MB-RVJ}
\end{equation}

\noindent
where $0\leq\gamma_{\rm MB}\leq4$ is a free parameter that largely affects how $\dot{J}_{\rm MB}$ varies during the binary evolution.
With $\gamma_{\rm MB}=4$,  Equation~\ref{Eq:MB-RVJ} describes the standard Skumanich law.

Despite the fact that the RVJ prescription can explain several observations of accreting objects, it has serious problems in different contexts, mainly connected with the mass transfer rates it provides and the corresponding evolutionary time-scales during CV and LMXB evolution \cite[e.g.][]{Podsiadlowski_2002,GH2017,Belloni_2020a,Pala_2022}.
In order to overcome part of these difficulties, recently a new convection and rotation boosted (CARB) prescription that enhances orbital AML due to MB was proposed \citep{Van_2019,Van2019CARB}.
The CARB prescription can be expressed as \citep{Van2019CARB}

\begin{equation}
\Dot{J}_{\rm MB,CARB} \ = \ 
-\,\frac{2}{3}\,\left(-\dot{M}_{\rm wind}\right)^{-1/3}\,R^{14/3}\,
\Omega_\odot\,B_{\odot}^{8/3}\,
\left(\,
   \frac{\Omega }{\Omega_\odot}
\right)^{11/3}\,
\left(
   \frac{\tau_{\rm conv} }{\tau_{\odot, \rm conv}}
\right)^{8/3}\,
\left(\,
    v_{\rm esc}^2\,+\,\frac{2\,\Omega^2\,R^2}{K_2^2}\,
\right )^{-2/3} \ ,
\label{Eq:MB-CARB}
\end{equation}

\noindent 
where $v_{\rm esc}$ is the surface escape velocity, $\tau_{\rm conv}$ is the convective turnover time, $B$ is the surface magnetic field strength, and $K_2=0.07$.
The main difference between the RVJ and the CARB prescription is that the former is the result of an empirical fitting scheme, while the latter was obtained in a self-consistent physical way, considering wind mass loss, the dependence of the magnetic field strength on the outer convective zone and the dependence of the Alfv\'en radius on the donor spin.

\begin{figure*}[htb!]
\centering
\includegraphics[width=0.95\linewidth]{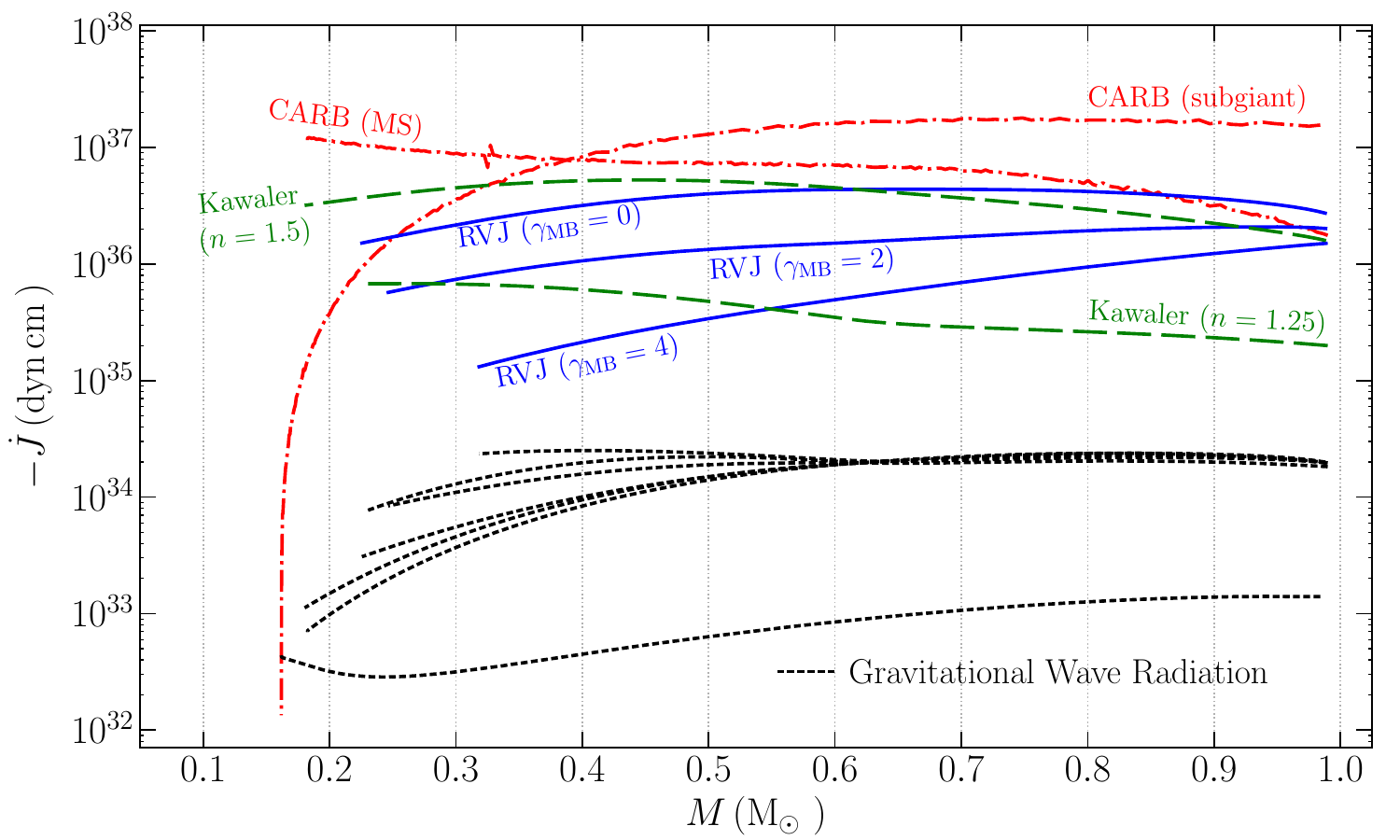}
\caption{Comparison of the orbital AML rates ($-\dot{J}$) due to GR and due to some MB prescriptions, during the evolution of a low-mass Roche lobe filling star, as a function of its mass ($M$).
All tracks have been computed by us with the \mesa~code, assuming solar metallicity, wind mass loss (Reimers parameter equals to 1), $\beta_{\rm ml}=0.8$, $\alpha_{\rm ml}=0$, a point-mass accretor with initial mass of $1.4$~\Msun, and the initial mass of the donor is $1.0$~\Msun.
Each track represents the evolution from the onset of Roche lobe filling to the moment when the star becomes fully convective (when MB is supposed to cease or at least significantly decrease). 
The dot-dashed red lines correspond to the CARB prescription \citep{Van2019CARB}, for which we chose two different initial orbital periods, namely $12$~h to illustrate the case of an unevolved MS star (helium-core mass equals to zero), and $4$~days to illustrate the subgiant case (helium-core mass ${\sim0.1}$~\Msun).
The solid blue lines show the RVJ prescription \citep{RVJ}, for three different values of $\gamma_{\rm MB}$, and in all these cases we chose an initial orbital period of $12$~h, corresponding in turn to unevolved MS stars.
The \citet{Kawaler_1988} prescription is marked by the long-dashed green lines, for three different values of $n$, assuming in all three sequences $a_K=1$, for which we also chose an initial orbital period of $12$~h, corresponding to the unevolved MS star case.
Finally, the short-dashed black lines correspond to the strength of GR in each of the evolutionary sequences, which can be easily associated with them by comparing the mass at which the MB becomes inefficient, i.e. the lowest mass in each track.
It is clear that the CARB recipe provides on average higher orbital AML rates than the RVJ and the Kawaler prescriptions, the latter for $n\lesssim1.5$, especially for masses ${\lesssim0.4}$~\Msun.
On the other hand, the CARB, RVJ and Kawaler prescriptions provide comparable orbital AML rates for larger masses, if $n_K=1.5$ and $\gamma_{\rm MB}=0$.
In addition, the Kawaler prescription is much more sensitive to the chosen parameters, in comparison with the RVJ prescription, as it provides much more different orbital AML rates, for small variations of $n_K$.
Moreover, except for the Kawaler prescription with $n\lesssim1$, GR in all evolutionary sequences contributes only negligibly to the total orbital AML.
Finally, regarding the type of the star in the CARB tracks, we can see that the orbital AML rate is initially higher for a more evolved star but decreases as the star mass decreases, resulting in a weaker MB at masses ${\lesssim0.4}$~\Msun, in comparison with an unevolved MS star.
}
\label{FigMB}
\end{figure*}

We illustrate in Figure~\ref{FigMB} the orbital AML rates provided by the three MB prescriptions mentioned above.
The evolutionary sequences have been computed by us using the Modules for Experiments in Stellar Astrophysics (\mesa) code \citep{Paxton2011,Paxton2013,Paxton2015,Paxton2018, Paxton2019}, in which we evolved a Roche lobe filling star with initial mass of $1$~\Msun, orbiting a point-mass companion of mass $1.4$~\Msun.
In the simulations, we take into account stellar wind mass loss, adopting the \citet{Reimers_1975} prescription with the scaling factor equal to $1$, and assume ${\beta_{\rm ml}=0.8}$ and ${\alpha_{\rm ml}=0}$.
For MS masses ${\lesssim0.4}$~\Msun, the CARB prescription clearly provides higher orbital AML rates than the RVJ and the Kawaler prescriptions.
For MS masses greater than that, the CARB, RVJ and Kawaler prescriptions provide comparable orbital AML rates if $n_K=1.5$ and $\gamma_{\rm MB}=0$.
Finally, GR is typically much weaker than any of these prescriptions, expect for the Kawaler recipe with $n\lesssim1$.

In addition to GR and MB, there are other sources of orbital AML acting on interacting binaries that arise as a consequence of mass transfer. For this reason this type of orbital AML is called consequential AML.
Among the several mechanisms proposed for consequential AML, the most important are a fast isotropic wind from the vicinity of the donor, isotropic re-emission from the vicinity of the accretor \citep{King_1995}, a circumbinary disk \citep{SpruitTaam_2001} and winds from the accretion disk \citep{Cannizzo_1988}.
Consequential AML is by definition an orbital AML mechanism that is mass-transfer-dependent, i.e. it is null in the absence of mass transfer.
It is usually written in a simple parameterized form as

\begin{equation}
\frac{\dot{J}_{\rm CAML}}{J_{\rm orb}} \ = \ \nu \, \frac{\dot{M_d}}{M_d}  \ ,
\label{Eq:CAML}
\end{equation}

\noindent
where $\nu$ depends on the binary properties, including perhaps the mass transfer rate itself, and is the parameter characterizing the strength of consequential AML. 
Its form depends entirely on the mechanism driving orbital AML.

In the case of a circumbinary disk, $\nu$ is given by \citep{SpruitTaam_2001}

\begin{equation}
\nu_{\rm CB} \ = \ \gamma \, \delta_{\rm ml} \, ( \, 1 \, + \, q \,)  \ ,
\label{Eq:nuCB}
\end{equation}

\noindent
where $\delta_{\rm ml}$ is the fraction of the mass lost from the binary components that feeds the circumbinary disk, and
the parameter $\gamma$ takes into account the characteristics of the circumbinary disk as a function of time and depends on the inner edge of the circumbinary disk, the binary semimajor axis, the viscous time-scale at the inner edge of the circumbinary disk and the mass of the circumbinary disk.

In the case of mass loss from the binary, $\nu$ may correspond to the specific angular momentum of the accretor (isotropic re-emission) or the donor (fast isotropic wind), given by \citep{King_1995} 

\begin{equation}
\nu_{\rm ml,a} \ = \ \beta_{\rm ml}  \, \left( \, \frac{q^2}{1 \, + \, q \,} \, \right) \ ,
\hspace{1.5cm}
\nu_{\rm ml,d} \ = \ \alpha_{\rm ml} \, \left( \, \frac{1}{1 \, + \, q \,} \, \right) \ .
\hspace{1.5cm}
\label{Eq:nuML}
\end{equation}

\noindent
Fast isotropic winds and isotropic re-emission may also generate additional orbital AML due to friction arising from the interaction of the ejected material and the accretor/donor.
This frictional AML then strongly depends on the velocity of the ejected material and on the orbital velocity of the accretor/donor as well as on the geometrical cross-section of the accretor/donor \citep{Shara_1986,Schenker_1998,Schreiber_2016,Sparks_2021}.

We have already illustrated in Figure~\ref{FigMB} that MB typically provides orbital AML rates several orders of magnitude higher than those expected from GR.
The rates predicted by consequential AML may be higher or lower in comparison with MB, depending on the form of the two types of orbital AML.
For instance, frictional AML due to isotropic winds and/or isotropic re-emission is typically negligible, unless the expansion velocity of the ejected material is very small \citep{Liu_2019}.
The same also occurs for consequential AML due to mass loss (isotropic winds or isotropic re-emission), which typically provides rates smaller than those related to either MB or GR alone.
On the other hand, consequential AML due to a circumbinary disk can be the dominant source of orbital AML depending on the parameters of the disk and the rate at which it is fed \citep{Willems_2005,Ma_2009}.

So far we have briefly described the current understanding of the different processes that can drive mass transfer in binaries.
This included the stability of a given binary against mass transfer on the dynamical and thermal time-scales of the donor as well as mass transfer driven by the nuclear evolution of the donor and orbital AML.
The largest uncertainties in current models are probably related to the conditions separating dynamically stable from dynamically unstable mass transfer, the strength and parameter dependencies of orbital AML through MB, and the processes that drive consequential AML. 
On the basis of the above-described knowledge of mass transfer in binaries, formation channels for the different types of accreting compact binaries have been developed in the last decades which we will review in the following section.

\section{Formation Channels}

Most accreting compact objects form from the evolution of zero-age MS--MS binaries.
Some may form from stellar mergers in hierarchical systems, i.e. from triples, quadruples and, eventually, higher-order systems.
Dynamical interactions, which are common in high-density environments such as the central parts of dense star clusters, could also lead to the formation of a sufficiently important fraction of accreting compact objects, especially for those originating from high-mass stars. 
While for some specific systems, and some particular types of accreting compact objects, the triple and/or the dynamical scenario is required to explain their properties, in most cases the contribution from these two scenarios is expected to be much smaller than that of binary evolution.
This is simply because of the lower frequency of triples and higher-order hierarchical systems in comparison to binaries \cite[e.g.][]{MD_2017} and because strong dynamical interactions only occur frequently in dense star clusters.
In what follows, we therefore mainly focus on accreting compact objects formed through binary evolution but add a few comments on additional formation channels through dynamical interactions at the end of this section.

The current small orbital separations of close binaries in which a compact object accretes from a non-degenerate low-mass companion, i.e. CVs and LMXBs, imply that these systems are currently much smaller than the progenitor of the compact object was as a giant. 
Consequently, the orbit of the binary must have been significantly reduced during the formation of the compact object, most likely through CE evolution generated by dynamically unstable mass transfer \citep{Webbink_1975,Paczynski_1976}.

The current orbital periods of most wide binaries in which a WD accretes from the winds of an evolved low-mass red giant, i.e. most S-type SySts, are significantly longer than those of post-CE binaries but short enough that mass transfer must have occurred when the WD was formed. 
Therefore, in these cases, the most likely scenario is that the orbit expanded during the formation of the WD, which happens when the mass transfer is dynamically stable and most likely non-conservative \citep{Webbink_1988}.

SyXBs are similar to SySts, in the sense that they are also wide-orbit binaries in which the compact object accretes from the wind of an evolved low-mass red giant.
However, their compact objects are NSs, which implies that stable non-conservative mass transfer cannot be the formation channel, because the required mass ratio of the zero-age MS--MS binary implies that the non-degenerate companions should be high-mass stars, which is not the case.  
A possible formation scenario for these systems could be the formation of a long-period \pceb, hosting a helium star paired with a low-mass star, followed by either electron-capture or core-collapse supernova leading to the NS \citep{Lu_2012,Yungelson_2019}.

Regarding compact objects accreting from high-mass non-degenerate stars, i.e. HMXBs, those in wide orbits (e.g. Be-HMXBs) are most likely formed through an episode of dynamically stable non-conservative mass transfer, while those having orbital periods of only a few days (e.g. sg-HMXBs) seem to have formed through CE evolution \citep{Tauris_2017}.
On the other hand, NSs/BHs accreting from the strong and dense winds of Wolf--Rayet stars, i.e. WR-HMXBs, require two episodes of mass transfer, one to form the accreting compact object and the other one to form the Wolf--Rayet star.
Even though the first episode of mass transfer is most likely dynamically stable non-conservative mass transfer \citep{Qin_2019}, there is evidence that in some WR-HMXBs the second episode of mass transfer was CE evolution, while others can be more easily explained when the second episode was dynamically stable non-conservative mass transfer \citep{van_den_Heuvel_2017}.

Finally, in those very close binaries in which the compact object is accreting from a Roche lobe filling WD or semi-degenerate helium-rich star, i.e. AM\,CVns and UCXBs, two episodes of mass transfer are needed, and at least one of these episodes must have been CE evolution.
More specifically, these systems may form either through two CE phases, or through dynamically stable non-conservative mass transfer followed by CE evolution, or finally through CE evolution followed by dynamically stable non-conservative mass transfer \citep{Solheim_2010,Nelemans_2010REF}.

\begin{figure*}[htb!]
\centering
\includegraphics[width=0.88\linewidth]{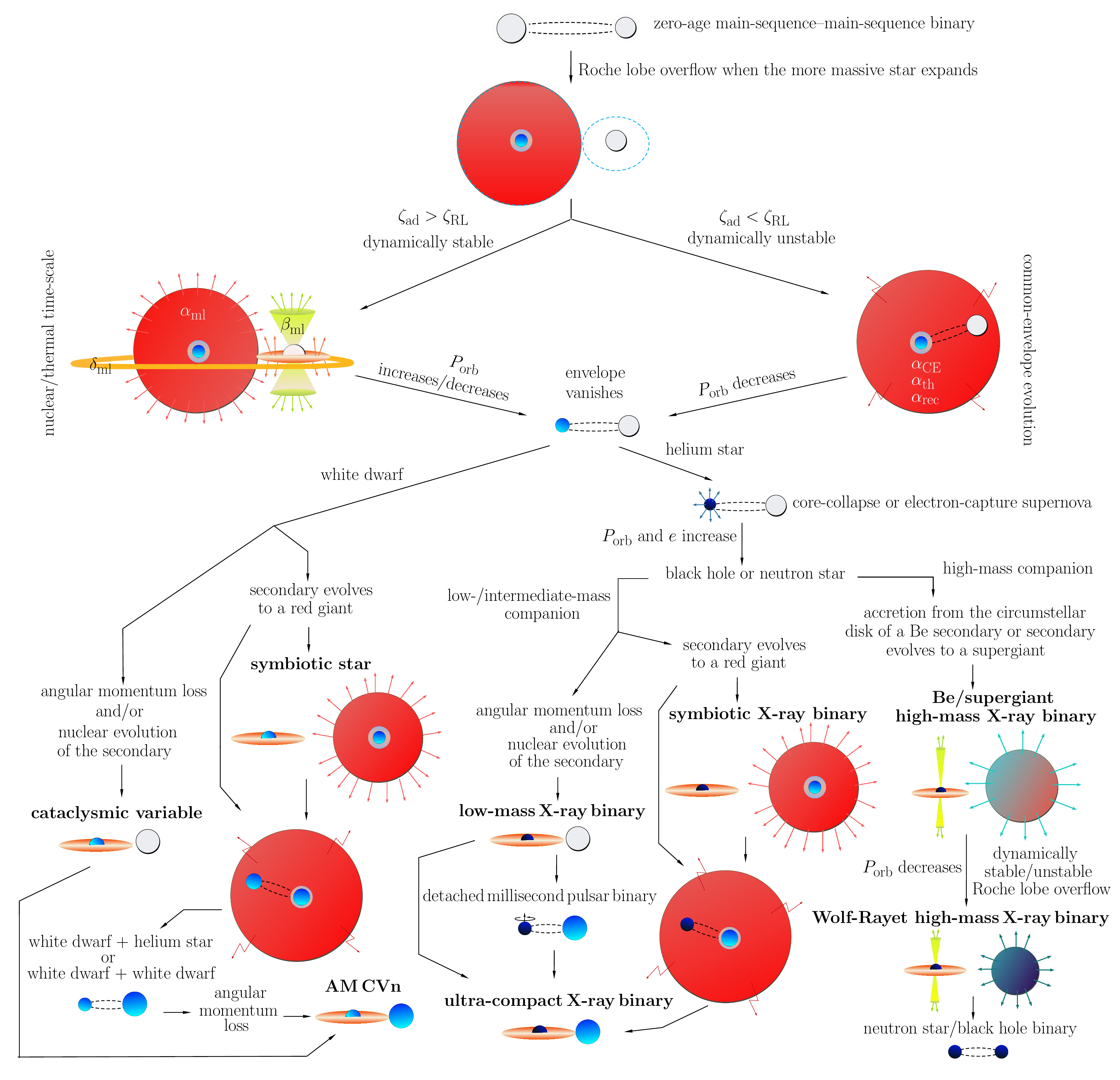}
\caption{Main formation channels leading to accreting compact objects. Starting from a zero-age MS--MS binary, when the more massive MS star evolves and fills its Roche lobe, mass transfer stars. Mass transfer is dynamically stable if the adiabatic radius--mass exponent ($\zeta_{\rm ad}$) is greater than the Roche lobe radius--mass exponent ($\zeta_{\rm ad}$), and unstable otherwise, leading to CE evolution. During dynamically stable mass transfer, the orbital period increases or decreases as a response to mass transfer, and the orbital period at the end of mass transfer depends on the binary properties at the onset of Roche lobe filling and on the fraction of the transferred mass that is lost ($\alpha_{\rm ml} + \beta_{\rm ml} + \delta_{\rm ml}$) during the event. On the other hand, if CE evolution happens, drag forces remove orbital energy, which is used with a certain efficiency ($\alpha_{\rm CE}$) to eject the envelope, potentially together with a fraction of other sources of energy, such as recombination energy ($\alpha_{\rm rec}$) and/or thermal energy ($\alpha_{\rm th}$). The outcome of CE evolution is a binary with a much shorter orbital period. Compact objects in close orbits accreting from a Roche lobe filling non-degenerate low-mass companion (i.e. CVs and LMXBs) are believed to form during one episode of CE evolution. NSs accreting from the winds of low-mass red giants in wide orbits (i.e. SyXBs) are also expected to form in an episode of CE evolution. On the other hand, WDs in wide orbits accreting from the winds of low-mass red giants (i.e. SySt) are expected to form through an episode of dynamically stable non-conservative mass transfer. Regarding NSs or BHs accreting from high-mass companions (i.e. HMXBs), they can be formed through dynamically stable non-conservative mass transfer or CE evolution. Finally, compact objects accreting from Roche lobe filling WDs or semi-degenerate helium-rich stars (i.e. AM\,CVns and UCXBs) are formed through two episodes of mass transfer, at least one of them being CE evolution.}
\label{FigChannels}
\end{figure*}

In what follows, we will in more detail address the main formation channels of accreting compact objects, which are illustrated in Figure~\ref{FigChannels}.

\subsection{Common-Envelope Evolution}

When the more massive star of a zero-age MS--MS binary evolves off the MS and fills its Roche lobe, it will give rise to dynamically unstable mass transfer, if $\zeta_{\rm ad}<\zeta_{\rm RL}$.
In this case, the mass-loss time-scale becomes so short that the donor cannot remain within its Roche lobe, which leads to the formation of a CE around the dense giant core and the MS star.
Even though a CE event in many situations is triggered by dynamically unstable mass transfer, there are also other mechanisms able to do that, such as the Darwin instability \cite[e.g.][]{Hut_1980}, which occurs if the future donor angular momentum at synchronism exceeds one-third of that of the orbit.
The Darwin instability corresponds to a tidal instability. It occurs when not enough orbital angular momentum is extracted from the orbit to maintain the star synchronized as it evolves.
Tidal forces will spin up the star by removing
orbital angular momentum, resulting in a binary with smaller orbital separation and in turn even less orbital angular momentum.
This implies that the star will need even more orbital angular momentum to stay synchronized with the orbital motion.
This leads to a runaway process of orbital decay and ultimately a CE event.

Because of drag forces during CE evolution, the MS star and the core spiral in towards their common centre of mass and the CE is expelled. 
In this process, a large fraction of the initial orbital energy and orbital angular momentum is lost with the envelope which leads to a post-CE binary with an orbital period orders of magnitude shorter than that of the initial MS--MS binary.
In case dynamically unstable mass transfer was generated from a massive star to its companion, the post-CE binary hosts a helium star, which is the progenitor of the compact object, and the binary separation may increase due to a natal kick during the collapse of the helium star.

Despite a lot of effort and (computational) resources that are spent towards a better understanding of CE evolution, we still struggle to identify and describe some of the main physical processes involved in this crucial evolutionary phase. 
The main reason for this failure lies in the fact that modelling CE evolution is very demanding from a computational point of view, since it involves a complex mix of physical processes operating over a huge range of scales \citep{Ivanova_REVIEW}.
In the absence of reliable simulations that cover the full complexity of CE evolution, a very simplistic approach is frequently used to overcome this problem and to make predictions for populations of post-CE binary characteristics. 
These simple equations based on energy and angular momentum conservation provide useful constraints on the outcome of CE evolution.

\subsubsection{The Energy Budget of Common-Envelope Evolution}

In the so-called energy formalism, the outcome of CE evolution is usually approximated by the balance between the change in the orbital energy and the envelope binding energy.
This energy conservation equation is parameterized with a parameter $0\leq\alpha_{\rm CE}\leq1$, 
which corresponds to the efficiency with which orbital energy is used to eject the envelope.
In other words, $\alpha_{\rm CE}$ represents the fraction of the difference in orbital energy (before and after the CE phase) that unbinds the envelope, i.e.

\begin{equation}
E_{\rm bind} \ = \ 
\alpha_{\rm CE} \ \Delta E_{\rm orb} \ = \ - \ 
\alpha_{\rm CE} \
 \left( \, \frac{G\,M_{\rm d,c}\,M_{\rm a}}{2\,a_f}  \ - \ 
           \frac{G\,M_{\rm d}\,M_{\rm a}}{2\,a_i} \, \right) \ ,
\label{EQALPHACE}
\end{equation}

\noindent
where $E_{\rm bind}$ is the envelope binding energy, $E_{\rm orb}$ is the orbital energy, $M_{\rm d,c}$ is the core mass of the donor, $a_i$ is the semimajor axis at onset of CE evolution, and $a_f$ is the semimajor axis after CE ejection.

The binding energy is usually approximated by

\begin{equation}
E_{\rm bind} \ = \ - \ \frac{G\,M_{\rm d}\,(M_{\rm d}\,-\,M_{\rm d,c})}{\lambda \, R_{\rm d}} \ ,
\label{EQLAMBDA}
\end{equation}

\noindent
where $\lambda$ is the binding energy parameter, which depends on the structure of the donor \citep{Dewi_2000,Claeys_2014}.

In case thermal and/or recombination energies stored inside the envelope are assumed to help in ejecting the envelope, the binding energy equation is usually written as

\begin{equation}
E_{\rm bind} \ = \ 
- \int_{M_{\rm d,c}}^{M_{\rm d}} \frac{G \; m}{r(m)} \; {\rm d}m 
 \ + \  
\alpha_{\rm th} \int_{M_{\rm d,c}}^{M_{\rm d}} \varepsilon_{\rm th}(m) \; {\rm d}m 
\ + \  
\alpha_{\rm rec} \int_{M_{\rm d,c}}^{M_{\rm d}} \varepsilon_{\rm rec}(m) \; {\rm d}m 
\label{EQALPHAINT}
\end{equation}

\noindent
where $\varepsilon_{\rm th}$ and $\varepsilon_{\rm rec}$ are the specific thermal and recombination energies and $\alpha_{\rm th}$ and $\alpha_{\rm rec}$ are the fractions of the thermal and recombination energies that are assumed to contribute to unbinding the envelope. 
Thermal energy is the thermodynamic internal energy of the CE, which is constrained by the virial theorem, and can be released and partially used to help in the CE ejection.
Recombination energy is stored in ionized helium or hydrogen within the CE.
These ions can recombine in the expanding layers of the gaseous CE,
once it has sufficiently cooled and expanded, thereby releasing energy \citep{Lucy_1967} that can contribute to unbinding the CE.
The binding energy parameter $\lambda$ can be computed by equating Equations~\ref{EQLAMBDA} and \ref{EQALPHAINT}.
When $\alpha_{\rm th}>0$ and/or $\alpha_{\rm rec}>0$, additional sources of energy contribute to the ejection of the CE, which implies that less orbital energy is needed to unbind the envelope.

Deriving constraints on the values of $\alpha_{\rm CE}$, $\alpha_{\rm th}$, and $\alpha_{\rm rec}$ has been a focus of close compact binary research in the last decades.
Unfortunately, several slightly different CE prescriptions are available, which needs to be taken into account when comparing the results of different studies.
In what follows, we briefly describe the constraints obtained so far for binaries hosting WDs, for which the progenitors are intermediate-mass MS stars, and for those hosting NSs or BHs, which originate from the evolution of high-mass MS stars.

Regarding low-/intermediate-mass stars, the best way for constraining the CE evolution is by using samples of detached WD--MS \pcebs~because of the large number of known systems and because it is relatively easy to characterize them \citep{Nebot_2011}. 
Reconstruction of the evolution of observationally characterized \pcebs~consisting of WDs with M-type companions suggests that CE ejection is rather inefficient, with $\alpha_{\rm CE}\sim0.2-0.3$ \citep{Zorotovic_2010}.
Comparisons between predictions from binary population synthesis with observations reached similar conclusions for this type of \pceb~\citep{Toonen_2013,Camacho_2014,Cojocaru_2017}.
Recently, observational surveys have become capable of identifying close WD binaries with MS companions more massive than M-type MS stars (spectral types A, F, G, and K), and early results indicate the existence of a population of \pcebs~that can be reproduced with the same small CE efficiency \citep{parsonsetal15-1,hernandezetal21-1}.  
However, for some of these systems, a small fraction of recombination energy seems to be required to contribute to CE ejection \citep{Zorotovic_2014_KOI}.

Close binaries hosting hot subdwarf B stars (i.e. helium-core-burning stars with very thin hydrogen-rich envelopes), mostly paired with WDs but some also with M-type MS stars, are also reasonably good targets for estimating the efficiency of CE ejection, as CE evolution is certainly involved in their formation.
Contrary to WD--MS \pcebs, results from binary population synthesis of hot subdwarf B stars indicate that the observations are better reproduced with $\alpha_{\rm CE}\sim0.75$ and a large fraction (${\sim0.75}$) of the internal energy (thermal plus recombination) contributing to expelling the envelope \citep{Han_2002,Han_2003} which indicates that both CE ejection and the conversion of thermal and recombination energies into kinetic energy are efficient processes.
However, given that the sample of WD--MS binaries is cleaner in the sense that only one phase of mass transfer is involved, the observational constraints indicating a small value for the CE efficiency should be considered as being stronger.
In addition, we have to be aware that the energy prescription for CE evolution is too simplistic and several unknown physical processes are not properly taken into account which might affect the constraints derived from observations of different systems on the CE efficiency.

With respect to progenitors of accreting compact objects hosting NSs and BHs, the situation is more complicated because we know less about CE evolution in this mass range \citep{Ivanova_REVIEW} and the situation might be quite different from the previous case.
The main reason for these expected differences is that CE evolution leading to WD--MS \pcebs~is generated by dynamically unstable mass transfer from red giant donors with a convective envelope \citep{TS00}, while for high-mass stars with solar metallicity the onset of Roche lobe filling usually happens when the donor is in the \hergap~and has a radiative envelope. 
In addition, metallicity has a strong impact on the evolution of high-mass stars prior to the onset of Roche lobe overflow, which is independent of the effect of metallicity-dependent winds, thereby altering the evolutionary stage at which Roche lobe filling occurs \citep{Klencki_2020}.
In most cases, this causes $\lambda$ to be far lower for high-mass stars than for intermediate-mass stars \citep{Kruckow_2016}, implying that in several situations, there would be apparently not enough energy available to eject the CE.

For a long time, this has been believed to be the case for BH-LMXBs, which would require an abnormally high efficiency for the envelope ejection in binary models \cite[e.g.][]{YL08}, since it is very difficult for the binary hosting a Roche lobe filling high-mass star paired with a much less massive companion to eject the CE at the expense of only orbital energy. 
However, more recent calculations have shown that this is not an actual issue in the formation of BH-LMXBs, since they could be produced without invoking unrealistically high CE efficiencies \citep{Wiktorowicz_2014}.
The real problem arises when trying to reproduce the donor mass distribution in this population.
Despite the fact that it is possible to reproduce individual BH-LMXBs in binary models, we cannot yet reproduce the properties of the entire population with just one model \citep{Wiktorowicz_2014}.
We will come back to this issue later.

Regarding SyXBs, in the absence of other sources of energies, in population synthesis aiming at explaining their observational properties, $\alpha_{\rm CE}$ needs to be as high as $4$ \citep{Yungelson_2019}, which is unrealistically high.
This result could suggest that there are other sources of energy that are required to help to expel the envelope or that these X-ray binaries do not form through standard CE evolution .
For instance, they could instead be preferentially formed from accretion-induced collapses of a massive oxygen--neon--magnesium WD \cite[e.g.][]{Hinkle_2006}.
Another possibility, perhaps more realistic, is that the uncertainties and simplifications involved in stellar and binary evolution calculations, for example mixing in stellar interiors, core-collapse/electron-capture supernova and CE evolution, are the main reason for the problems in reproducing these systems, as well as BH-LMXBs \cite[e.g.][]{Belczynski_2021_uncer}.

\subsubsection{Common-Envelope Evolution from Hydro-dynamical simulations}

Despite the above-described attempts to constrain $\alpha_{\rm CE}$, $\alpha_{\rm th}$ and $\alpha_{\rm rec}$, it seems unlikely that there is a universal recipe within the simplistic energy formalism able to account for all types of accreting compact objects.
In the case of low-/intermediate-mass stars, evidence is growing evidence that the efficiency of expelling common envelopes is rather low, and overall the energy formalism delivers reasonable results.    
In contrast, it is very difficult to explain the properties of accreting compact objects coming from high-mass stars based on the energy balance of CE evolution, regardless of the assumed values for the CE evolution parameters.
This difficulty must to some degree arise from our inability to understand the role played by other sources of energies acting together with orbital energy, as well as correctly computing the CE binding energy in an unbiased way.

CE evolution can be divided into several subsequent phases, which helps to better understand why this event is so complicated to be modelled.
Most of the problems arise from the fact that each of these phases occurs at a particular time-scale, involving a particular physical process \cite[e.g.][]{Podsiadlowski_2001}.
The first phase is the loss of co-rotation, which can last hundreds of years, and corresponds to very small changes in the orbital separation in comparison with the orbital period.
During this phase, the accretor orbits either outside the future CE or inside the CE outer layers \cite[e.g.][]{IN16}.
The second phase is called plunge-in, which corresponds to the rapid spiral-in with huge changes in the orbital separation.
During this phase, the accretor plunges inside the CE, and the orbital energy is transferred to the CE, driving its expansion, in case there is enough energy available for that.
At the end of this phase, most of the CE mass is outside the orbit of accretor and the core of the donor \cite[e.g.][]{IN16}.
During the third phase, when the CE is well-extended, a self-regulating spiral-in occurs on the thermal time-scale of the CE, during which energy released due to the spiral-in is transported to the CE surface and subsequently radiated away.
This phase corresponds to small changes in the orbital energy \cite[e.g.][]{IN16}.
CE evolution terminates at the end of the self-regulating spiral-in phase with either the ejection of the CE or the accretor or the core of the donor filling its Roche lobe.
The latter most likely result is a merger but could also provide another pathway for CE ejection \cite[e.g.][]{Podsiadlowski_2010}.

\begin{figure*}[htb!]
\centering
\includegraphics[width=0.49\linewidth]{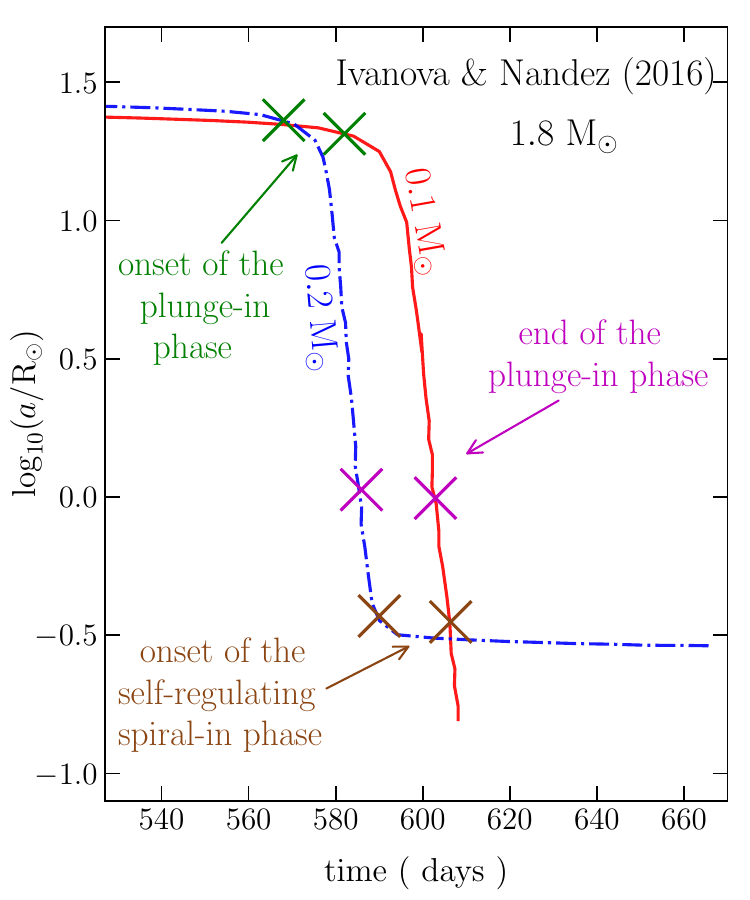}
\includegraphics[width=0.49\linewidth]{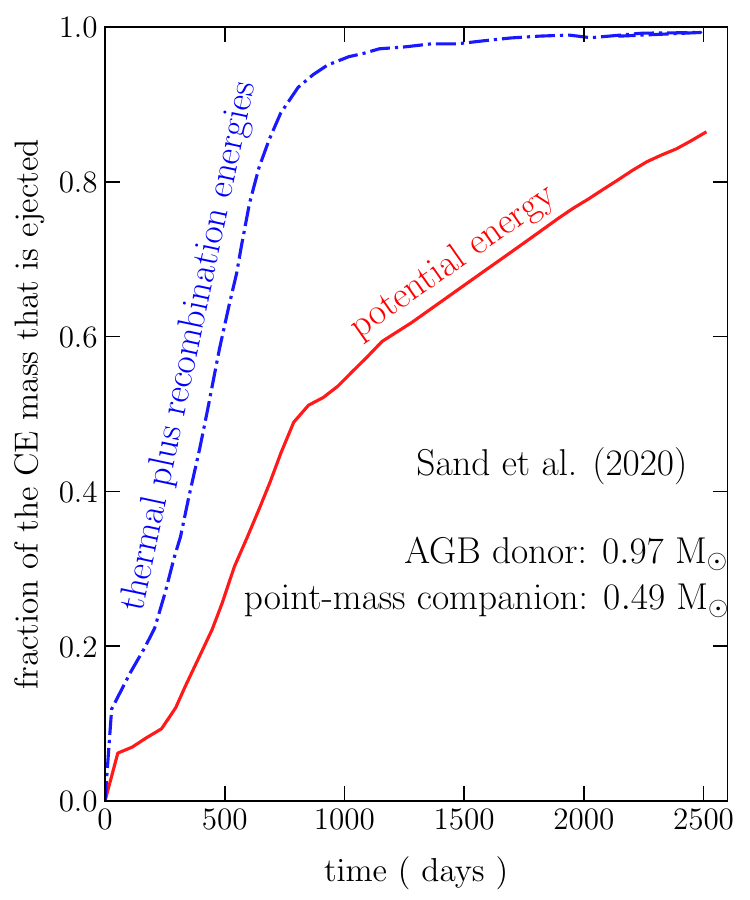}
\caption{Change in the orbital separation \citep[left-hand panel,][]{IN16} and envelope mass \citep[right-hand panel,][]{Sand_2020} through time during CE evolution, taken from two different CE evolution models.
In the left-hand panel, the onset of the plunge-in phase (when $(\,|\,\dot{a}\,|\,P_{\rm  orb}\,)\,/\,a$ becomes greater than 0.1), the end of the plunge-in phase (when $(\,|\,\dot{a}\,|\,P_{\rm  orb}\,)\,/\,a$ becomes smaller than 0.1), and the onset of the self-regulating spiral-in (when $|\, (\,\dot{E}_{\rm orb}\,P_{\rm orb}\,)\,/\,E_{\rm orb}\,|$ becomes smaller than 0.01)  are indicated with crosses.
The transition between the end of the plunge-in phase and the onset of the self-regulating spiral-in is not sudden in these simulations, although it lasts for only a few days.
In the model with companion mass of $0.1$~\Msun, there is not enough orbital energy available to expel the CE, and the binary is expected to merge during the self-regulating spiral-in phase.
However, in the model with companion mass of $0.2$~\Msun, even though orbital energy is transferred to the CE, it is not enough to entirely eject the CE during the self-regulating spiral-in phase.
Despite the fact that a strong reduction in the orbital separation occurs, like in the $0.2$~\Msun~model, in most numerical simulations, the CE is not entirely ejected, which indicates that other sources should contribute to the CE ejection together with orbital energy. 
In the right-hand panel, there is a comparison between the case without internal energy and the case with internal energy (thermal plus recombination), which has been widely proposed as the key source of energy usually missing in the simulations.
It is clear in this simulation that the CE can be ejected with the help from internal energy.
However, there are strong physical arguments suggesting that this energy would be simply radiated way, not actually contributing to the CE ejection.
If confirmed beyond doubts, other sources of energy or transfer mechanisms must be playing a role and should be further explored as alternatives to recombination energy.
}
\label{FigCE}
\end{figure*}

We illustrate CE evolution in the left-hand panel of Figure~\ref{FigCE}, where we show the evolution of the orbital separation from just before the plunge-in phase until just after the self-regulating spiral-in, for two models taken from a study based on a three-dimensional smoothed particle hydrodynamics code \citep{IN16}.
In these simulations, the authors chose a low-mass red giant donor with a mass of $1.8~M_\odot$, a core mass of $0.318~M_\odot$ and radius of $16.3~R\odot$, and several companion masses.
The figure clearly illustrates that there is a huge orbital shrinkage during the plunge-in phase until the onset of the self-regulating spiral-in phase, which is precisely what the energy formalism tries to reproduce.
However, in those cases that avoid merger, the CE is not always effectively ejected, which represents a long-standing problem in numerical simulations of CE evolution \cite[e.g.][]{Ohlmann_2016,Soker_2017}.

Given the above-mentioned inefficient ejection of the CE predicted in simulations, there should exist at least one energy source, perhaps more, in addition to orbital energy, that would be able to expel the remaining bound matter \cite[e.g.][]{Soker_2017}.
It has been argued for a long time that thermal energy and mainly hydrogen and helium recombination energy could solve this problem \cite[e.g.][]{Webbink_2008,NIL2015,Ivanova_2018}.
The impact of including thermal and recombination energies is illustrated in the right-hand panel of Figure~\ref{FigCE}, which shows the evolution of the unbound mass fraction through time, taken from a numerical simulation with a moving-mesh hydrodynamics code \citep{Sand_2020}.
In this simulation, the donor is on the AGB undergoing helium-shell burning and no thermal pulses, with a mass of $0.97$~\Msun~and a core mass of $0.545$~\Msun.
Almost the entire CE mass is unbound when the stored internal (thermal plus recombination) energy is used, most of which is only released at rather late times, when the expanding CE sufficiently cools down.
This is not the case when only the potential energy contributes to the binding energy, since ${\sim10}$\% of the CE mass is still inside the orbit of the donor core and the companion, after the plunge-in phase.
No further dynamical spiral-in is expected, because of the low density of the remaining envelope and the co-rotation of the material, avoiding in turn any drag force.

The example above, together with other simulations, provides support to the claim that internal (mainly recombination) energy might be a key ingredient in CE evolution.
On the other hand, most of the available recombination energy could be simply radiated away \cite[e.g.][]{Soker_2018,Grichener_2018}, since the optical depth is expected to be low when recombination occurs, allowing radiation to more easily escape.
In addition, despite the fact that the amount of available recombination energy depends on the core definition, it is expected to be smaller in high-mass stars than in intermediate-mass stars.
Therefore, some simulations suggest that recombination energy is unlikely to help in the removal of the CE in this mass range \citep{Ricker_2019}.

In any event, it is widely accepted by now that only dynamical spiral-in is most likely not enough to completely remove the CE.
In case recombination energy is not significantly helping, other sources of energy would be needed \cite[e.g.][]{Soker_2017}. These additional contributions could come from nuclear energy \citep{Podsiadlowski_2010}, accretion energy \citep{VT2003}, dust-driven winds \citep{GP2018}, interaction of the binary (core of the donor + companion) with a circumbinary disk \citep{KS2011}, large-amplitude pulsations \citep{Clayton_2017},
or jets launched by the companion as it accretes mass from the circumbinary envelope when it is about to exit the CE from inside \citep{Soker_2017,Diego_2017,Diego_2019,Diego_2020,Diego_2022}.
Even though there are several mechanisms available, they have been poorly explored in hydro-dynamical simulations, and it is still not clear which one is dominant, if any at all, and under which conditions.
This means that CE evolution is a subject far from being well-understood, and, hopefully, in the next years, further hydro-dynamical simulations will shed more light on this extremely important phase for the formation of accreting compact objects.

\subsection{Dynamically Stable Non-conservative Mass Transfer}

Even though the energy formalism is successful to some extent in avoiding the messy physics involved in the several phases of CE evolution, it fails to reproduce some observed systems, in particular double-helium WD binaries.
Explaining these systems with two consecutive CE events requires that energy is generated during CE (which implies that the energy conservation law is violated), or at best, that an unknown source of energy, much greater than the orbital energy, is contributing to expelling the envelope (which can be considered unrealistic).

To solve this problem, an alternative but similarly simplistic formalism was invented based on conservation of angular momentum, which is usually called angular momentum (or $\gamma$) formalism \citep{Nelemans_2000,Nelemans_2005}.
Although this new approach could somehow work mathematically, it does not provide a physical explanation for the required CE phases without spiral-in \cite[e.g.][]{Webbink_2008,Ivanova_REVIEW}.

We believe that most systems for which no reasonable solutions in the framework of CE energy balance exist are actually not formed through CE evolution, but rather dynamically stable non-conservative mass transfer. 
Indeed, there is strong theoretical evidence that an increase in the orbital separation can be reached during an episode of dynamically stable non-conservative mass transfer on a thermal and/or nuclear time-scale, which can potentially reproduce otherwise inexplicable systems.
In what follows, we discuss in more detail systems that may form through this dynamically stable non-conservative mass transfer channel.

\subsubsection{Low-/Intermediate-Mass Stars}

There are several types of binaries hosting (proto) WDs with orbital periods longer than ${\sim100}$~days.
In some of them, it is clear that some sort of interaction in their history must have occurred, because of observed chemical anomalies.
For instance, barium stars, which are paired with WDs, are F-/G-type MS stars or G-/K-type red giants exhibiting strong absorption lines of ionized barium in their spectrum as well as of other s-process elements \cite[e.g.][]{Jorissen_2019,Escorza_2019}, which could not have been synthesized by them.
In a similar way, carbon-enhanced metal-poor stars show abundances of s-process elements \cite[e.g.][]{Jorissen_2016,Hansen_2016} that cannot be explained in the context of single star evolution.
S-process elements are produced during the thermally-pulsing AGB phase \cite[e.g.][]{Kappeler_2011}, and it is therefore widely accepted that these elements were transferred to the barium stars and carbon-enhanced metal-poor stars via stellar winds or dynamically stable Roche lobe overflow, when the WD progenitor was at this evolutionary stage.

Other types of binaries having comparable orbital periods include post-AGB binaries, with orbital periods in the range of ${\sim100-2500}$~days \citep{Oomen_2018}.
In addition, some blue stragglers, which are typically found in old star clusters and correspond to MS stars bluer, brighter, and more massive than those around the cluster turnoff point, are in binaries with orbital periods between ${\sim100}$ and ${\sim3000}$~days.
While some hot subdwarf B stars are members of close binaries, a certain fraction of these stars are members of binaries with rather wide orbits. Hot subdwarf B stars are post-FGB binaries that must have formed when their progenitors filled their Roche lobe just before reaching the tip of the FGB. The typical orbital periods for wide binaries containing hot subdwarf B stars range from ${\sim700}$ to ${\sim1300}$~days.
Moreover, most S-type SySts have orbital periods between ${\sim200}$ and ${\sim1500}$~days.
In some cases, there is evidence for negligible mass accretion during these episodes of mass transfer, such as for hot subdwarf B stars \citep{Vos_2018}, while in other cases, such as for barium stars and carbon-enhanced metal-poor stars, only a non-negligible amount of mass accretion can explain their chemical properties \cite[e.g.][]{Miszalski_2013,Abate_2015a,Abate_2015b}.

\begin{figure*}[htb!]
\centering
\includegraphics[width=0.9\linewidth]{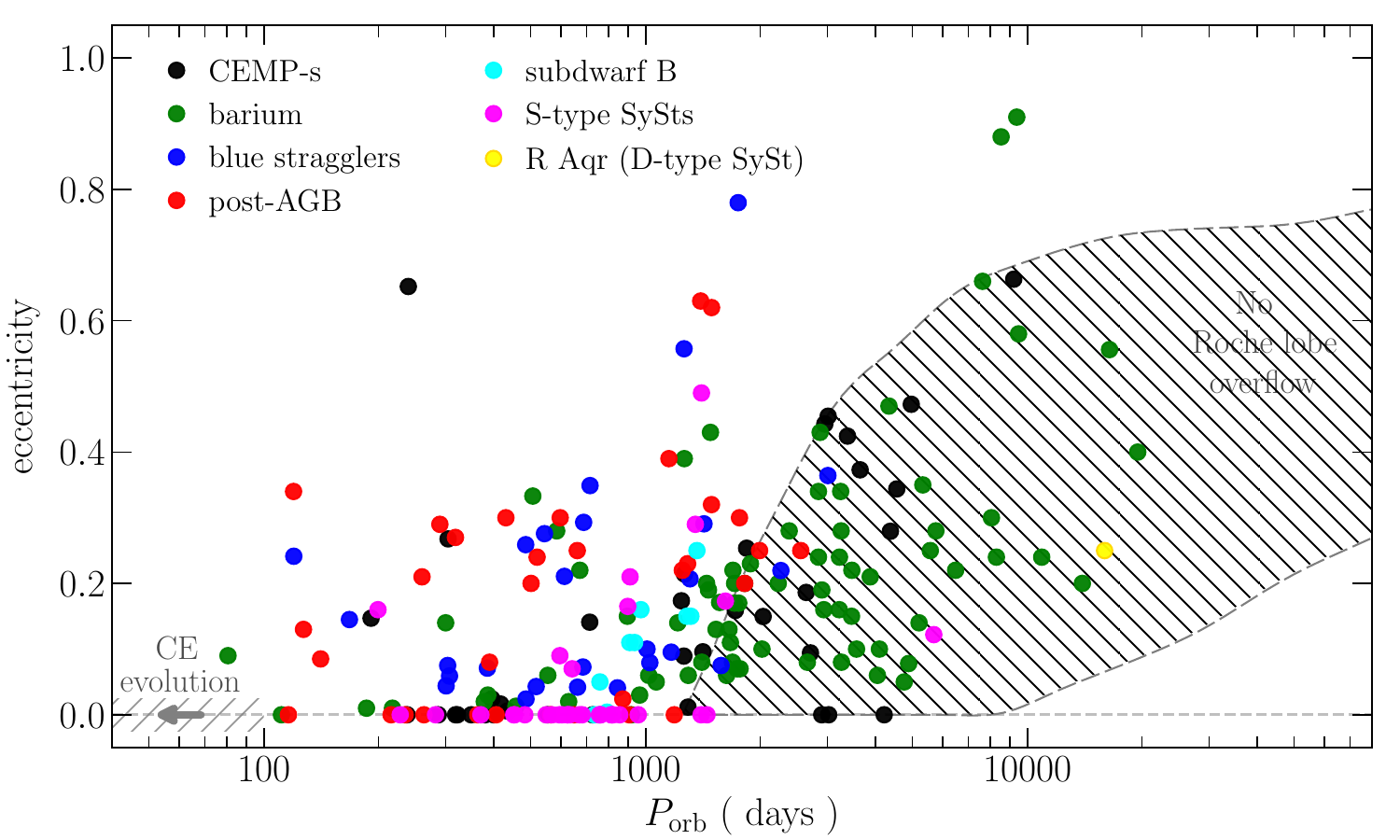}
\caption{Diagram depicting the orbital period ($P_{\rm orb}$) versus eccentricity for several types of low-/intermediate-mass post-mass-transfer binaries with orbital period longer than ${\sim100}$~day, namely barium star binaries \citep{Swaelmen_2017}, blue straggler binaries \citep{Carney_2001,Mathieu_2009}, post-AGB binaries \citep{Oomen_2018}, s-process-polluted carbon-enhanced metal-poor star (CEMP-s) binaries  \citep{Jorissen_2016,Hansen_2016}, wide-orbit hot subdwarf B binaries \citep{Vos_2015,Otani_2018}, S-type SySts \citep{Mikolajewska_2003,Brandi_2009,Fekel_2007,Fekel_2008,Hinkle_2009,Fekel_2010}, and R~Aqr, which is the only D-type SySt with radial velocity data suitable for orbital determination \citep{Gromadzki_2009}.
The hatched regions correspond to the location of \pcebs~($P_{\rm orb}\lesssim100$~days) and of systems formed without Roche lobe filling ($P_{\rm orb}\gtrsim2000$~days), taken from a binary model \citep{Belloni_2020b}.
It is clear from the diagram that there is a significant fraction of systems in eccentric binaries with orbital periods between ${\sim100}$ and ${\sim2000}$~days, which is precisely between the region occupied by \pcebs~and the region occupied by systems formed without Roche lobe filling.
Therefore, most of these systems apparently cannot be explained by CE evolution, nor by the lack of Roche lobe filling.
Indeed, the overwhelming majority of these systems are most likely formed during dynamically stable non-conservative mass transfer, accompanied by an eccentricity pumping mechanism that would efficiently work against tidal circularization for some system, but not for all of them, as it must also account for those systems in circular orbits.
}
\label{FigPorbEcc}
\end{figure*}

All these binaries, in addition to having orbital periods usually longer than ${\sim100}$~days, have in many cases non-null eccentricities.
We show the orbital period versus eccentricity diagram for them in Figure~\ref{FigPorbEcc}, where we also added a hatched region indicating the location of the post-FGB and post-AGB binaries formed without Roche lobe overflow \citep[taken from a binary model,][]{Belloni_2020b}.
We can see from the figure that those systems with orbital periods longer than ${\sim2000}$~days 
fit reasonably well within the expected region for WD binaries where the WD progenitor never filled its Roche lobe. 
The same certainly holds for most D-type SySts, which have long orbital periods (${\gtrsim40}$~years), although only for one system the orbital period has been accurately measured.
The mass exchange that is required to explain the formation of barium stars and s-process-polluted carbon-enhanced metal-poor stars with periods exceeding ${\sim2000}$~days has likely been driven by efficient wind accretion, instead of Roche lobe overflow.

In contrast, systems with orbital periods shorter than ${\sim2000}$~days must have experienced Roche lobe overflow.
Given their current orbital period, 
the progenitors of the WD (or hot subdwarf B stars) 
in these systems must have filled their Roche lobes at some point along their evolution.
It is difficult to explain the periods of these systems as a result of CE evolution, since orbital periods of \pcebs~are usually shorter than ${\sim100}$~days \citep{Nie_2012,Zorotovic_2014} and often just a few h \citep{Nebot_2011}.
In addition, even though some of these binaries are on circular orbits, many others have eccentric orbits, which is not expected for post-CE binaries, since tidal forces during CE evolution would clearly have managed to circularize their orbits.
Consequently, Figure~\ref{FigPorbEcc} indicates that a different mode of mass transfer has to be considered to understand these systems. 
For orbital periods longer than the typical outcome of CE evolution, but shorter than those that might avoid Roche lobe overflow, i.e. ${\gtrsim100}$ and ${\lesssim2000}$~days, dynamically stable and non-conservative mass transfer represents a reasonable alternative.

Indeed, several theoretical studies show that dynamically stable non-conservative mass transfer can explain these systems \cite[e.g.][]{Webbink_1988,CH08,Chen_2013,Gosnell_2019,Vos_2020}.
Depending on the combination of initial masses, metallicity, and orbital period of the low-/intermediate-mass zero-age MS--MS binaries, the more massive star could fill its Roche lobe while on the FGB (having or not a degenerate core) or on the AGB. 
The properties of post-stable-mass-transfer binaries then depend on the above-mentioned combination of initial masses, initial orbital period, and metallicity as well as on the fraction of mass and orbital angular momentum lost by the binary during the process.

\begin{figure*}[htb!]
\centering
\includegraphics[width=0.49\linewidth]{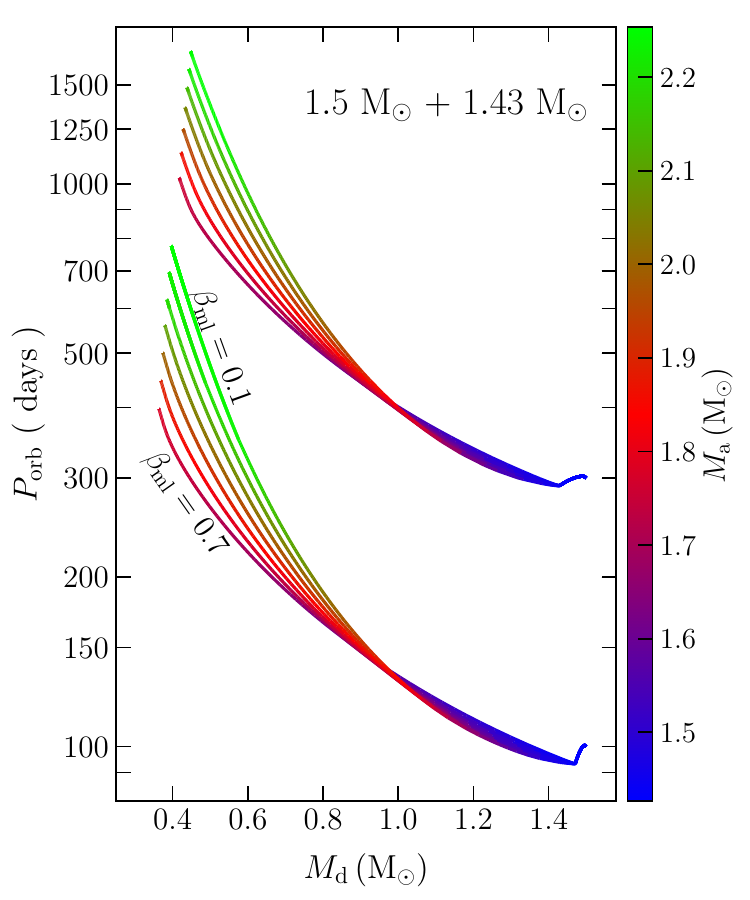}
\includegraphics[width=0.49\linewidth]{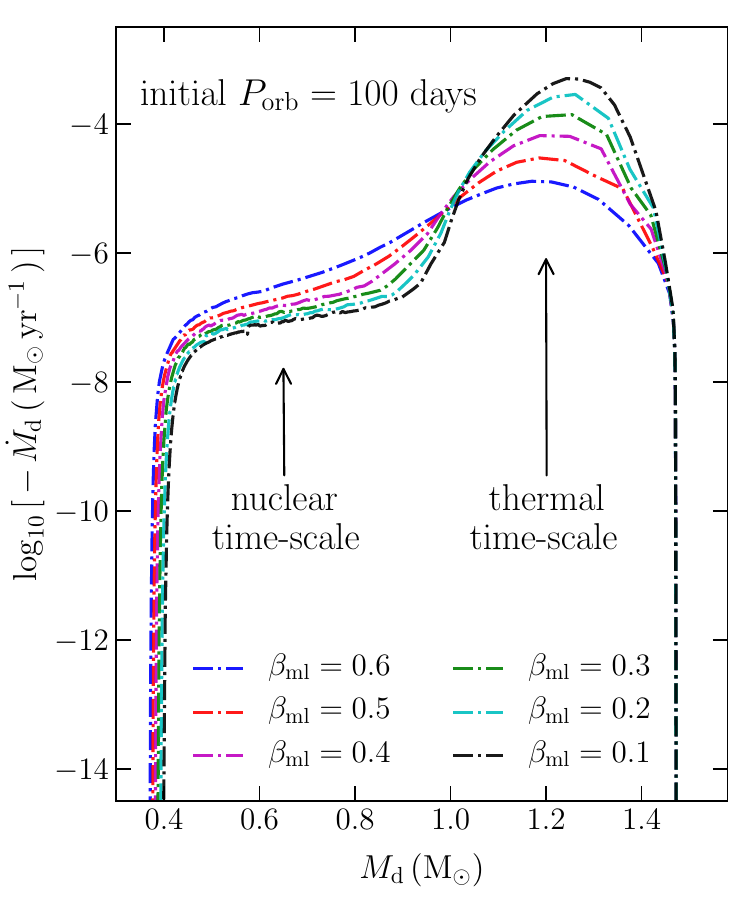}
\caption{Evolution towards a post-FGB binary with the donor mass ($M_{\rm d}$) of the orbital period ($P_{\rm orb}$), in the left-hand panel, and the mass transfer rate ($\dot{M}_{\rm d}$), in the right-hand panel, during dynamically stable non-conservative mass transfer using the \mesa~code.
We chose two different initial orbital periods ($100$ and $300$~days) and seven different values for the transferred mass-loss fraction from the vicinity of the accretor due to isotropic re-emission ($\beta_{\rm ml}$), from $0.1$ to $0.7$ in steps of $0.1$, which are indicated in the figure (the upper curve in each set of tracks in the left-hand panel corresponds to $\beta_{\rm ml}=0.1$).
The simulations start with a zero-age MS--MS circular binary, with stars of masses $1.50$ and $1.43$~\Msun, with negligible rotation.
In addition, we considered solar metallicity, changes in the rotation of the stars, tidal synchronization as well as wind mass loss (Reimers parameter equals to 0.5).
Furthermore, except for stellar winds, we assumed that any mass loss in the form of fast isotropic wind from the donor is negligible ($\alpha_{\rm ml}=0$), and ignored mass loss and orbital AML due to a circumbinary disk ($\delta_{\rm ml}=0$).
We can see some nice correlations in the left-hand panel of this figure, in which the colours indicate the evolution of the accretor mass ($M_{\rm a}$) along each track.
In particular, the smaller $\beta_{\rm ml}$, the longer $P_{\rm orb}$ at the end of mass transfer, which occurs because the increase in the orbital period due to mass transfer is much stronger than the increase due to mass loss from the binary.
In addition, as expected, the smaller $\beta_{\rm ml}$, the larger the fraction of mass effectively accreted by accretor, and in turn, the greater $M_{\rm a}$ at the end of the mass transfer episode.
Finally, the longer the initial $P_{\rm orb}$, the longer $P_{\rm orb}$ at the end of mass transfer.
The right-hand panel of the figure illustrates the typical behaviour in such episodes of dynamically stable mass transfer \cite[e.g.][]{W12}, namely the transition from a thermally unstable to a thermally stable phase of mass transfer, the latter being driven then by the nuclear evolution.
The mass transfer starts thermally unstable because the equilibrium radius--mass exponent ($\zeta_{\rm eq}$) is smaller than the Roche lobe radius--mass exponent ($\zeta_{\rm RL}$), which means that it is driven by the thermal readjustment of the donor, and proceeds initially on the donor thermal time-scale.
However, as the donor loses mass and changes its structure, $\zeta_{\rm eq}$ eventually becomes larger than $\zeta_{\rm RL}$, and mass transfer proceeds on the donor nuclear time-scale.
We can also see a nice correlation in this panel, the smaller $\beta_{\rm ml}$, the greater the peak of $\dot{M}_{\rm d}$ during thermally unstable mass transfer.
This happens because the donor is increasingly driven out of thermal equilibrium, as mass transfer tends to be more conservative.
On the other hand, the smaller $\beta_{\rm ml}$, the lower the mass transfer rate during thermally stable mass transfer, i.e. the slower the mass transfer, which has an impact on the post-FGB mass.
In particular, the smaller $\beta_{\rm ml}$, the larger $M_{\rm d}$ at the end of mass transfer, which can also be seen in the left-hand panel and happens because the overall evolution is slower when mass transfer tends to be more conservative, allowing in turn the core mass to grow more as $\beta_{\rm ml}$ decreases.
}
\label{FigStable}
\end{figure*}

We show evolutionary tracks for dynamically stable non-conservative mass transfer in Figure~\ref{FigStable}. 
It is clear that orbital periods between a few hundreds and a few thousands days can be achieved during dynamically stable non-conservative mass transfer. Extrapolating the results shown in the figure towards smaller initial periods one can easily imagine that stable mass transfer may also lead to orbital periods ${\lesssim100}$~days.

According to Figures~\ref{FigPorbEcc} and \ref{FigStable}, formation through dynamically stable but non-conservative mass transfer therefore represents a reasonable scenario for most blue stragglers, post-AGB binaries, long-period binaries hosting subdwarf B stars, and a significant fraction of barium stars and s-process-polluted carbon-enhanced metal-poor stars. 
Also most S-type SySts, which correspond to ${\sim80}$\% of known SySts \citep{Mikolajewska_2003,Mikolajewska_2007,Mikolajewska_2010,Mikolajewska_2012}, should be formed through an episode of dynamically stable non-conservative mass transfer.
Their WD and red giant masses are mostly in the range ${\sim0.4-0.6}$~\Msun~and ${\sim1.2-2.2}$~\Msun~\citep{Mikolajewska_2003}, which does not seem difficult to obtain with dynamically stable non-conservative mass transfer, especially when taking into account a potential WD mass growth during SySt evolution.
Following the episode of mass transfer, the WD companion will eventually evolve beyond the MS and become an evolved red giant, either close to the tip of the FGB, or on the AGB.
In most cases, the mass transfer from the red giant to the WD is enhanced due to gravitationally focused wind accretion \citep{Mohamed_2007,Mohamed_2012,Abate_2013,ValBorro_2009,Skopal_2015,Borro_2017}.
If the accretion rate onto the WD is sufficiently high, the binary will become a SySt.

Despite the general agreement between predictions and observed systems, there are three S-type SySts that have very different WD and red giant masses, and their red giants are nearly filling their Roche lobe, namely T~CrB, \citep{Stanishev_2004}, RS~Oph \citep{Brandi_2009,Mikolajewska_2017}, and V3890~Sgr \citep{Mikolajewska_2021}.
In these three systems, the compact objects are massive carbon--oxygen WDs with masses in the range ${\sim1.2-1.4}$~\Msun, and their Roche lobe filling red giants have relatively low masses (${\sim0.7-1.2}$~\Msun). These characteristics exclude that these systems formed through stable mass transfer as the mass of the donor star (i.e. WD progenitor) in the progenitor system was far more massive than the accretor (i.e. red giant progenitor). 
It is therefore very likely that these three SySts have formed through CE evolution \citep{Liu_2019_SySt}, and their WD/red giant masses have substantially increased/decreased during their evolution.
Why CE evolution resulted in exceptionally long orbital periods in these cases (${\sim200-600}$~days) remains unclear.
However, if additional energy (e.g. recombination energy) is efficiently contributing to expelling the envelope, such long post-CE periods seem possible \citep{RebassaMansergas_2012,Zorotovic_2014}.
Following the CE evolution, the WD companion becomes an evolved red giant, its extended atmosphere eventually fills its Roche lobe, and mass transfer proceeds similarly to what is illustrated in Figure~\ref{FigStable}.
We will discuss in more detail the secular evolution of the different types of SySts later.

Even though dynamically stable non-conservative mass transfer is a very good candidate to explain the orbital periods of a large fraction of systems with orbital periods ranging from ${\sim100}$ to ${\sim2000}$~days, the observed eccentricities remain to be explained.
In order to have orbital periods consistent with observations at the end of mass transfer, the initial orbital periods cannot be too short, nor too long.
This would imply then that, even for an initially eccentric orbit, tidal interaction should be strong enough to circularize the orbit before the onset of mass transfer.
However, it is clear from Figure~\ref{FigPorbEcc} that many systems have eccentric orbits at the end of mass transfer.
This leads then to three possibilities, either we still do not fully understand the tidal interaction phenomenon, specially the circularization time-scales, or there is an eccentricity pumping mechanism operating during mass transfer, or both effects are acting together.

Concerning the first possibility, i.e. tidal forces, recent observational efforts on ellipsoidal red giant binaries in the Large Magellanic Cloud suggest that tidal interaction should be much weaker than expected from theoretical considerations \citep{Nie_2017}.
Among the 81 binaries that were investigated, $20$\% have eccentric orbits, which should not exist according to our understanding on tidal dissipation in convective stars.
In order to explain such systems, the circularization efficiency needs to be two orders of magnitude smaller than predicted. 

Regarding the possibility that the large eccentricities are generated by some kind of eccentricity pumping, we note that there are currently several mechanisms proposed that could at least help to explain the non-circular orbits of the post-stable-mass-transfer binaries.
First, the eccentricity can increase if the mass-loss rate from the donor is higher near periastron than near apastron.
The required enhanced mass loss occurs during periastron passages can be caused by either wind accretion or Roche lobe overflow or even a combination of both, in which there is a smooth transition between wind accretion and Roche lobe filling \citep{Bonacic_2008,Soker_2000,Vos_2015}.

Second, in the circumbinary disk scenario, eccentricity pumping occurs due to the resonances in the interaction of the binary with a circumbinary disk \citep{Dermine_2013,Vos_2015}.
Since circumbinary disks are most likely created only from mass lost from the outer Lagrangian point, the eccentricity pumping due to circumbinary disks should actually work together with the phase-dependent Roche lobe overflow scenario, by enhancing even further the eccentricity.
A third possibility is that the proto WD receives a natal kick after the mass transfer episode, similarly to natal kicks in the formation of NSs, although not that strong.

The final possibility is the triple scenario, in which dynamical and stellar evolution with a tertiary companion would be responsible for creating eccentric binaries hosting compact objects, via chaotic orbital evolution of the stars, which can trigger close encounters, collisions, and exchanges between the stellar components \cite[e.g.][]{PK12}.
Even though all these scenarios are promising and provide reasonable results for particular systems, they all fail to some extent to explain the overall properties of the systems, and further investigations are needed \cite[e.g.][]{Vos_2015,Rafikov_2016,Oomen_2020,Escorza_2020}.

There is an important question to be answered regarding the picture we described above.
If the accretion efficiency during dynamically stable non-conservative mass transfer is high, what happens with the accretor spin?
This issue has been addressed in binary models \citep{Matrozis_2017}, which suggest that only ${\sim0.05}$~\Msun~of accreted material is usually enough to drive the accretor to critical rotation. This is a major problem to explain barium stars and s-process-polluted carbon-enhanced metal-poor stars unless there exists a mechanism allowing the accretor to transfer its spin angular momentum back to the orbit with the help of tidal torques from the donor \citep{Matrozis_2017}.
However, it is not yet clear what kind of mechanism could do this job.

\subsubsection{High-Mass Stars}

After discussing dynamically stable non-conservative mass transfer among low-/intermediate-mass stars, we now turn to the implication of this mass transfer mode for the formation of HMXBs.
We start by discussing the eclipsing sg-HMXB M33~X--7, which hosts a rapidly spinning, $15.65$~\Msun~BH orbiting a close to Roche lobe filling underluminous, ${\sim70}$~\Msun~O-type supergiant in a slightly eccentric orbit with a short orbital period of $3.45$~days \citep{Orosz_2007}. 
The existence of this HMXB can neither be explained by CE evolution nor rotational mixing, but dynamically stable non-conservative mass transfer offers a solution \citep{Valsecchi_2010}.
Starting with a zero-age MS--MS binary with stellar masses of ${\sim90}$~\Msun~and ${\sim30}$~\Msun and an orbital period of ${\sim3}$~days, the more massive MS star fills its Roche lobe and dynamically stable mass transfer occurs. 
A fraction of the mass leaving the donor escapes the binary as strong isotropic winds and part is effectively accreted by the accretor.
After mass transfer finishes, due to the strong winds when the initially more massive star becomes a Wolf--Rayet star,
the binary hosts a Wolf--Rayet star synchronized with the orbital motion, due to tidal interaction, and a massive O-type MS star. 
The Wolf--Rayet star with an iron--nickel core later collapses to a BH, leading to rapidly spinning BH in a slightly eccentric short-period orbit with a high-mass O-type MS star.
When the O-type MS star becomes an O-type supergiant almost filling its Roche lobe, the binary becomes a BH-HMXB and wind accretion explains its X-ray luminosity.
This formation picture is also consistent with other sg-HMXBs hosting rapidly rotating BHs and O-type supergiants almost filling their Roche lobes, such as Cyg~X--1 \citep{MillerJones_2021} and LMC~X--1 \citep{Orosz_2009}, in which dynamically stable non-conservative mass transfer can not only explain their masses and orbital periods, but also their BH spins \citep{Qin_2019}.

Even though the initial orbital periods in such models are very short (only a few days), they are neither odd nor rare, when compared with observations.
The orbital period distribution of binaries hosting O-type MS stars, the so-called Sana distribution \citep{Sana_2012},  although extending until ${\sim10^6}$~days, is dominated by systems with orbital periods shorter than ${\sim10}$~days.
According to the described formation channel, the orbital period does not significantly increase after the onset of mass transfer, which explain why sg-HMXBs can be formed through dynamically stable non-conservative mass transfer and still have current orbital periods shorter than ${\sim10}$~days.

In most known HMXBs, a NS accretes from the circumstellar decretion disk of a rapidly rotating Be star \cite[e.g.][]{Reig_2011}, and their orbital periods are predominantly in range from ${\sim10}$ to ${\sim400}$~days \cite[e.g.][]{Karino_2021}.
According to the wind compression disk model \citep{BC1993}, an equatorial disk is formed around a rapidly rotating star due to ram pressure confinement by the stellar wind.
The disk is formed because the supersonic wind leaving the star surface at high latitudes travels along paths that move it down to the equatorial plane, where the material passes through a standing oblique shock on the top of the disk.
The ram pressure of the polar wind thus confines and compresses the disk.
The emission lines typically observed in an active Be star are attributed to this equatorial disk \citep{PR2003}.
As mass from this outflowing disk is transferred to the compact object via the accretion gate mechanism \citep{Ziolkowski_2002}, preferentially during periastron passages or enhanced activity of the Be star, the system becomes bright in X-rays and observable as a Be-HMXB.

The typical age of a Be-HMXB is ${\sim40-50}$~Myr \cite[e.g.][]{Williams_2013,Garofali_2018}, which is believed to be a direct consequence of the NS formation and B star mass-loss activity time-scales.
Given their properties, specially their relatively long orbital periods and their rapidly rotating Be stars, they have most likely formed through an episode of stable non-conservative mass transfer \cite[e.g.][]{RH1982,Tauris_2017}, during which the B star spins up.
Despite the fact that there is only one known Be-HMXB in which the compact object is a BH \citep[MWC~656,][]{Casares_2014}, it seems more likely that Be-HMXBs hosting BHs are formed during CE evolution \citep{Grudzinska_2015}, although given the uncertainties in the stability criterion \citep{Olejak_2021,Belczynski_2021_uncer}, dynamically stable non-conservative mass transfer is not necessarily ruled out.
The main problem with having Be-HMXBs hosting BHs formed through CE evolution is that the existence of the Be star cannot be easily explained, since a mechanism different from Roche lobe overflow would be needed to spin up the B star.
Gravitationally focused wind accretion prior CE evolution seems to be a good candidate for that \citep{Grudzinska_2015,Mellah_2019b}.

In the Corbet's diagram \citep{Corbet_1986}, i.e. the orbital period versus NS spin period diagram, Be-HMXBs exhibit a clear correlation, which is explained  in terms of the equilibrium period, defined as the period at which the rotation velocity at the magnetospheric radius equals the Keplerian velocity \cite[e.g.][]{Waters_1989}.
If the NS spin period is shorter than the equilibrium period, then matter is efficiently ejected from the vicinity of the NS due to the propeller mechanism \citep{Illarionov_1975}, and the NS spin decreases.
On the other hand, if the NS spin period is longer than the equilibrium period, then mass and angular momentum is transferred onto the NS thereby increasing its rotational velocity. 
The correlation comes from the fact that the equilibrium period depends on the accretion rate, and in turn, on the orbital period.
In addition, Be-HMXBs can be separated into two groups in the orbital period versus eccentricity diagram, one being those in low-eccentricity orbits (${\lesssim0.2}$) and the other one composed of high-eccentricity orbits (${\gtrsim0.2}$) \citep{Pfahl_2002,Townsend_2011}.
This could then indicate two distinct formation channels, with (core-collapse supernova) and without (electron-capture supernova) strong natal kicks during the helium star collapse into the NS.

\subsection{Combination of Dynamically Stable Non-conservative Mass Transfer and Common-Envelope Evolution}

The orbits of AM\,CVns and UCXBs are so close that two episodes of mass transfer must occur to explain their orbital periods, at least one of them being CE evolution.
Similarly, WR-HMXBs, which are compact objects accreting part of the winds from massive helium stars, are also expected to be formed through two episodes of mass transfer, given their short orbital periods.
However, unlike the previous case, these systems can be formed in two episodes of dynamically stable non-conservative mass transfer.
There are then three possible combinations for AM\,CVns and UCXBs: (i) double CE evolution, (ii) CE evolution followed by dynamically stable non-conservative mass transfer, and (iii) dynamically stable non-conservative mass transfer followed by CE evolution; and there is another possible combination for WR-HMXBs, in addition to stable non-conservative mass transfer followed by CE evolution, which is (iv) double dynamically stable non-conservative mass transfer.
In what follows, we will discuss in more detail these formation channels, based on results from binary models.

\subsubsection{Evolution Through Two Episodes of Common-Envelope Evolution}

Starting with a zero-age MS--MS binary, if the initial orbital period is not too short nor too long, the initially more massive star will fill its Roche lobe after evolving beyond the MS.
If the initially more massive star is much more massive than its companion, then mass transfer will be dynamically unstable and CE evolution commences.
The result of such an interaction will be a helium star or a WD, depending on the mass and core mass of the donor at the onset of CE evolution, orbiting the initially less massive MS star with an orbital period orders of magnitude shorter than initially.
In case of a helium star, depending on its mass and composition, it will quickly collapse into a WD, NS or BH, and in the latter cases, this collapse will produce a core-collapse or electron-capture supernova.

If the orbital period of the \pceb~is not too short, the initially less massive MS will have room to grow and evolve off the MS before filling its Roche lobe.
If the compact object is much less massive than its companion, then another CE will be triggered and the orbital period is shortened even further.
In case the companion of the compact object is initially a low-/intermediate-mass MS star and if it fills its Roche lobe on the FGB, then it will become a helium WD after CE evolution, if its mass is initially smaller than ${\sim2}$~\Msun, or a helium star if its mass is initially larger than that.
On the other hand, if the low or intermediate-mass star fills its Roche lobe on the AGB, then the compact object companion will likely become a carbon--oxygen WD.
After the second CE evolution, the \pceb~will be brought into contact due to emission of gravitational waves, becoming in turn an AM\,CVn or an UCXB, depending on the nature of the compact object, accreting from either a helium/carbon--oxygen WD or a helium star, or even a hybrid WD, in case the helium star becomes a WD before filling its Roche lobe.

This is a standard formation channel for AM\,CVns and UCXBs \cite[e.g.][]{Solheim_2010} in which the accreting compact object is formed first during the binary evolution.
However, even if this scenario ends up having a non-negligible contribution to these populations, it cannot account for the existence of some particular types of binaries related to AM\,CVns and UCXBs, such as close detached double-helium WDs in which the younger WD is more massive than the older WD \cite[e.g.][]{Nelemans_2000}, or millisecond pulsars in close detached binaries with helium WDs \cite[e.g.][]{Tauris_2011}.
Therefore, alternative scenarios involving dynamically stable non-conservative mass transfer are needed.

\subsubsection{Dynamically Stable Non-conservative Mass Transfer Followed by Common-Envelope Evolution}

Starting from a zero-age MS--MS binary, if the orbital period is sufficiently short, the initially more massive star will fill its Roche lobe on the \hergap, or when it is an unevolved giant.
If the initial masses of the stars are comparable, or at least the initially more massive star is not much more massive than its companion, then mass transfer will be dynamically stable.
The orbital period in this case may increase, as illustrated in Figure~\ref{FigStable}, leading then to a binary with an orbital period longer than the orbital period the zero-age binary had initially, hosting a (proto) WD and a companion that is more massive than the zero-age MS stars.
This companion can still be on the MS or already slightly evolved, since it could have started its evolution off the MS, in case the initial masses are comparable.
The companion of the (proto) WD then keeps evolving until it fills its Roche lobe, leading to dynamically unstable mass transfer, since it is much more massive than the WD.
The \pceb~will have an orbital period orders of magnitude shorter than at the onset of CE evolution, hosting a WD paired with a WD or helium star, which will later collapse into a WD, NS or BH, depending on its mass.
If the \pceb~orbital period is sufficiently short, the (proto) WD will fill its Roche lobe, giving rise to an AM\,CVn or UCXBs, depending on the nature of the accreting compact object.

Unlike the double CE evolution scenario, in the scenario in which dynamically stable non-conservative mass transfer is followed by CE evolution, the accreting compact object is formed second.
This scenario has been invoked to explain the existence of double-helium WDs in which the younger WD is more massive than the older WD \citep{W12}, which is a natural consequence of this scenario.
Most importantly, this scenario can solve the energy budget problem we mentioned earlier \cite[e.g.][]{Webbink_2008,Ivanova_REVIEW}, suggesting that the importance of dynamically stable non-conservative mass transfer for binary evolution has been underestimated in several contexts.
However, the existence of millisecond pulsars in close detached binaries with helium WDs cannot be explained within this formation channel, which means that another channel most likely plays a key role in formation of UCXBs.

Regarding WR-HMXBs, they are believed to be the direct progenitors of merging binary BHs, which means that these merger events can also be used to further constrain WR-HMXBs \cite[e.g.][]{Kalogera_2007}.
We have already discussed that the properties of sg-HMXBs hosting BHs can be reasonably well explained by an episode of dynamically stable non-conservative mass transfer.
This is also the most common outcome for the first episode of mass transfer in merging BH binary models \cite[e.g.][]{B2016,B2020}.
In these binaries, the BH is accreting from the winds of an O-type supergiant that is almost filling its Roche lobe.
When it effectively fills its Roche lobe, depending on the binary properties, mass transfer can be dynamically unstable or not.
A CE evolution is the most common scenario for the formation of merging binary BHs, specially those mergers with massive stellar-mass BHs, such as GW190521 \cite[e.g.][]{Belczynski_2020}.

The sg-HMXB M33~X--7 will apparently undergo CE evolution, given that its O-type supergiant is much more massive than its BH companion \citep{Orosz_2007}.
On the other hand, in the sg-HMXB Cyg~X--1, the O-type supergiant is more massive than its BH companion only by a factor of two \citep{MillerJones_2021}, which implies that mass transfer will be most likely dynamically stable when the O-type supergiant fills its Roche lobe.
Finally, the mass of the O-type supergiant in sg-HMXB LMC~X--1 is around three times its BH companion mass \citep{Orosz_2009}, and it is much less clear whether mass transfer could be dynamically stable or unstable in this case.
Now regarding the current population of confirmed WR-HMXBs, there are two systems that have very short orbital periods, namely Cyg~X--3, in the Milky Way \citep[$4.8$~h,][]{Zdziarski_2013}, and CG~X--1, in the  Circinus galaxy \citep[$7.8$~h,][]{Q2019}, which mostly implies that the BHs orbit within the Wolf--Rayet photospheres.
These extremely short orbital periods suggest that there was a strong orbital shrinkage, mostly likely only possible during CE evolution.
For the remaining confirmed WR-HMXBs, the orbital periods are longer than a day and these systems could be explained by a weak spiral-in during dynamically stable non-conservative mass transfer \citep{van_den_Heuvel_2017}, as we will discuss later.

\subsubsection{Common-Envelope Evolution Followed by Dynamically Stable Non-conservative Mass Transfer}

The situation here is initially similar to the double CE evolution scenario, i.e. from a zero-age MS--MS binary, the initially more massive star fills its Roche lobe when it evolves off the MS, leading to dynamically unstable mass transfer, because it is much more massive than its companion.
However, unlike in the double CE evolution scenario, the \pceb~orbital period is shorter, and the compact object companion is not allowed to significantly grow before filling its Roche lobe.
Therefore, in this case, the zero-age MS--MS binary orbital period must be shorter than that in double CE scenario.

When the companion of the compact object fills its Roche lobe, depending on its mass and the mass of the compact object, mass transfer can be either dynamically unstable or dynamically stable.
If the donor is much more massive than the compact object, mass transfer will be dynamically unstable, which implies that the compact object and its companion will most likely merge on the dynamical time-scale of the donor.
In case the donor is an MS star, this is the natural outcome of dynamically unstable mass transfer.
On the other hand, if the donor is a \hergap~star, or an unevolved FGB star, then a merger is also expected, since there would not be enough orbital energy available to eject the CE, due to the short orbital period of the binary.

If the donor is a low-mass star, then mass transfer will most likely be dynamically stable and non-conservative.
The beginning of the evolution can be thermally stable or not, depending on the masses of the accreting compact object and the donor, and on the properties of the donor.
In any event, the binary will become a CV or a LMXB, depending on the nature of the compact object.
If the donor is an M-/K-type MS star, i.e. an MS star with mass ${\lesssim0.8}$~\Msun, then it will eventually become a hydrogen-rich degenerate star, i.e. a brown dwarf, when enough mass is lost.
However, if the donor is an evolved MS star of an earlier type or a \hergap~star, then it has already or will develop a growing degenerate helium core.
In these cases, when the donor hydrogen envelope is entirely consumed, it becomes a semi-degenerate helium-rich star.
At this moment, the CV/LMXB becomes an AM\,CVn/UCXB.

The donor transition from a non-generate to a semi-degenerate configuration may be preceded by a short detached phase, depending on the \pceb~properties \cite[e.g.][]{Chen_2021,Soethe_2021}.
Alternatively, if the \pceb~properties are such that this detached phase happens at longer orbital periods (${\gtrsim1}$~day), then emission of gravitational waves will not be strong enough to bring the binary close enough so that the compact object companion could fill again its Roche lobe within a Hubble time.
The existence of millisecond pulsars in close detached binaries is essentially explained in such a way, since the NS becomes a millisecond pulsar during the LMXB evolution \citep[the so-called recycling scenario, e.g.][]{BH1991,SS20}, and the close detached population are those systems observed in the transition from LMXB to UCXB.
Those systems that directly become UCXBs, or effectively manage to become UCXBs after the detached phase, are then UCXBs hosting accreting millisecond pulsars.

\subsubsection{Evolution Through Two Episodes of Dynamically Stable Non-conservative Mass Transfer}

We have argued before that WR-HMXBs with extremely short orbital periods 
can most likely only be explained with CE evolution.
However, the remaining confirmed WR-HMXBs have orbital periods longer than ${\sim1}$~day, namely X--1 in the IC~10 galaxy \citep[$34.9$~h,][]{Silverman_2008}, X--1 in the NGC~300 galaxy \citep[$32.8$~h,][]{Crowther_2010}, and ULX--1 in the galaxy M101 \citep[$196.8$~h,][]{Liu_2013}.
Despite their short orbital periods, they could have formed through dynamically stable non-conservative mass transfer, since it is possible in such cases to avoid CE evolution during the formation of the Wolf--Rayet stars, due to the high masses of the Wolf--Rayet progenitors.

A slow spiral-in during dynamically stable non-conservative mass transfer is possible if the Wolf--Rayet progenitor has a radiative envelope and if it is at most ${\sim3-4}$~times the BH mass \citep{van_den_Heuvel_2017,Neijssel_2019}, although the exact critical mass ratio depends on the several model assumptions \cite[e.g.][]{Olejak_2021,Belczynski_2021_uncer}.
This seems to be the fate of several WR-HMXB progenitors.
Based on an analysis of the small sample of Wolf--Rayet $+$ O-type MS binaries, with orbital periods clustering at around one week \citep{Hucht_2001}, it has been recently suggested that, after the Wolf--Rayet star becomes a BH and its O-type companion fills its Roche lobe, they are expected to become WR-HMXBs with orbital periods as short as ${\sim1}$~day  \citep{van_den_Heuvel_2017}.
These WR-HMXBs will then evolve to binary BHs, when the Wolf--Rayet donors collapse to BHs.
This then suggests that not only the formation of WR-HMXBs can avoid CE, but also the formation of merging BHs \citep{Olejak_2021,Belczynski_2021_uncer}.

\subsubsection{Further Considerations on the Formation of Ultra-Compact X-ray Binaries}

The first three scenarios highlighted above correspond to the most traditional scenarios for the formation of AM\,CVns and UCXBs.
Regarding UCXBs, there is an additional channel, via accretion-induced collapse of an oxygen--neon--magnesium WD \cite[e.g.][]{BT2004}.
If such a WD triggers electron-capture reactions at its centre upon reaching the Chandrasekhar mass limit, then the burning that propagates outward makes central temperatures and pressures high enough to cause a collapse of the WD into a NS \cite[e.g.][]{Nomoto_1991}.
Binary models, in which different types of donors have been considered, mainly MS stars and helium star donors, have shown that this is indeed possible \cite[e.g.][]{Tauris_2013}, leading to the formation of millisecond pulsar binaries.

In the case of MS stars, after the accretion-induced collapse, the binary  temporarily detaches (for $10^3-10^5$~years), due to the small kicks, but, due to orbital AML, it soon becomes an LMXB.
And the subsequent formation of the UCXB is similar to that already discussed.
The UCXB progenitors in this case are SSXBs with MS donors with masses of ${\sim2.0-2.6}$~\Msun, for a metallicity of 0.02, and of ${\sim1.4-2.2}$~\Msun, for a metallicity of 0.001.
It is worth mentioning that this formation channel may also help to explain the existence of millisecond pulsars in eccentric detached binaries \cite[e.g.][]{FT2014}.
In the case of helium stars, the binary also temporarily detaches and soon becomes an UCXB, due to orbital AML through emission of gravitational waves.
Unlike the previous case, the helium stars quickly (${\lesssim1}$~Myr) become carbon--oxygen WDs with masses of ${\sim0.6-0.9}$, causing the binary to detach.
Further orbital AML could eventually make the binary semi-detached again, and the UCXB would host a millisecond pulsar accreting from a carbon--oxygen WD donor.
This happens for AM\,CVns with helium star donors with masses of ${\sim1.1-1.5}$~\Msun

Even though possible, the contribution from this channel is most likely small.
This is because of the narrow range in the parameter space of the accreting WD that allows it to eventually exceed the Chandrasekhar mass limit.
In particular, during the accreting WD evolution, there is a very narrow range of mass transfer rates such that the WD mass can effectively increase \cite[e.g.][]{Nomoto_2007,Shen_2007,Wolf_2013}.
In addition, given that a typical mass of an oxygen--neon--magnesium WD is ${\sim1.2}$~\Msun, the donor cannot be much more massive than that, since otherwise mass transfer becomes dynamically unstable, which restricts even more the parameter space of the UCXB progenitors.

\subsection{Additional Channels Through Dynamical Interactions in High-Density Environments}

In addition to the formation channels highlighted above, dynamical interactions in crowded regions of star clusters, specially globular clusters, can create conditions to form accreting compact objects from systems that would otherwise not evolve into them.
A way of addressing whether a type of X-ray sources is predominantly dynamically formed in globular clusters is by means of the cluster stellar encounter rate, which depends on the central density, the core radius and the central velocity dispersion \citep{Verbunt_1987,Pooley_2003}.
If dynamical interactions are supposed to play any significant role in the formation of X-ray sources, a correlation between the number of sources of a particular type within a cluster and its stellar encounter rate could indicate that dynamical interactions played a significant role in their formation, despite the several potential problems with this interpretation \citep{BR21}.
For a long time, this has been believed to happen for LMXBs and UCXBs, which correspond to the brightest X-ray sources found in Milky Way globular clusters, as there is continuously growing strong evidence for such a correlation \citep{Pooley_2006,Ivanova_2008}.
There are currently 21 bright X-ray sources in 15 Milky Way globular clusters, being 8 persistent and 13 transients \citep{vandenBerg_2020}, and they are still consistent with being dynamically formed, especially considering their mass density, which is much higher than that of systems in the Milky Way disk.

Given the problem with reproducing the properties of BH-LMXBs in binary models \citep{Wiktorowicz_2014} and the above-mentioned over-abundance of LMXBs in the Milky Way globular cluster population in comparison with the Milky Way disk population, a natural solution to the problem could be a dynamical origin for the field population (i.e. those LMXBs not belonging to stellar aggregates).
Recent Monte Carlo numerical simulations of evolving globular clusters \citep{Kremer_2018} show that progenitors of LMXBs are ejected from their host clusters over a wide range of ejection times, contributing to ${\sim300}$ NS-LMXBs and ${\sim180}$ BH-LMXBs to the Milky Way field population, not all of them being indeed observable due to observational selection effects.
More than 20 galaxies within ${\sim25}$~Mpc have been deeply investigated with Chandra and Hubble Space Telescope \citep{Lehmer_2020}.
These observations have revealed that, in order to properly describe the X-ray luminosity function of the field population of LMXB, a component that scales with the globular cluster specific frequency is needed.
In addition, it has a shape consistent with that found for the globular cluster population of LMXBs.
This then indicates that most, if not all, LMXBs in the fields of these galaxies are formed similarly to those in globular clusters and in turn could have origin in globular clusters.
These binaries could be ejected from the globular clusters due to energetic dynamical interaction leading to their formation, or the cluster itself could dissolve due to strong tidal interaction.
However, for galaxies with low globular cluster specific frequencies, the X-ray luminosity function has a component that scales with stellar mass, suggesting then that a substantial population of LMXBs in these galaxies are formed through binary evolution, without invoking dynamics.

These findings may alleviate quite a bit the problems of reproducing the proprieties of BH-LMXBs in the Milky Way field.
However, it is not clear whether enough systems are predicted to form through this channel, nor whether their positions in the Milky Way as well as their kinematic properties, nor their numbers could be explained as ejected from globular clusters, or from dissolving star clusters.
For instance, while the observed systems are almost entirely concentrated in the Milky Way disk \citep{Corral_2016}, those predicted to form in globular clusters and latter ejected to the field are spherically distributed around the Milky Way disk, reflecting the observed distribution of globular clusters in the Milky Way.
In addition, the number of systems predicted by this channel seems too low to account for the large number of BH-LMXBs inferred from observations \citep[${\sim10^4-10^8}$,][]{Tetarenko_2016}.
Therefore, the current theoretical and observational evidence indicates that a dynamical origin in star cluster is unlikely to be the dominant formation channel for these systems.
A more promising alternative to bring into agreement predicted and observed properties seems to be the standard CE evolution, coupled with a revision of the strength of orbital AML during their evolution, as we will see later.

Regarding the fainter X-ray sources such as CVs, the situation is much more difficult to be addressed, due to the several biases and selection effects \citep{BR21}, e.g. the photometric incompleteness in the more crowded central parts of the clusters is most likely playing a key role in the characterization of these sources \cite[e.g.][]{Cohn_2021}.
Therefore, any conclusions drawn from a comparison between the numerical simulations and observations of faint X-ray sources should be taken with a grain of salt.
Despite that, it is still possible to infer the impact of dynamics on this population, if observational selection effects are somehow addressed.
Recent numerical simulations suggest that fewer CVs should be expected in dense globular clusters relative to the Milky Way field, due to the fact that dynamical destruction of CV progenitors is more important in globular clusters than dynamical formation of CVs \citep{Belloni_2019}.
This is consistent with observations \citep{Cheng_2018}, because the faint X-ray populations, primarily composed of CVs and chromospherically active binaries, are under-abundant in globular clusters with respect to the Solar neighbourhood and Local Group dwarf elliptical galaxies.
Regarding the dynamical formation of CVs in globular clusters, these simulations show that the detectable CV population is predominantly composed of CVs formed via standard CE evolution (${\gtrsim70}$\%).
This happens because the main dynamical scenarios have a very low probability of occurring, which resulted in (or very weak, if at all) correlation between the predicted number of detectable CVs and the predicted globular cluster stellar encounter rate.
Despite the fact that these results are consistent with recent observations \citep{Cheng_2018}, we shall mention that, even though it is not expected to be dominant, dynamics should play a sufficiently important role to explain the relatively larger fraction of bright CVs (i.e. those close to the MS in the colour-magnitude diagram) observed in denser clusters.

Regarding SySts and SyXBs, despite a few reasonable candidates have been proposed over the years, none so far has been confirmed in the entire Milky Way globular cluster population.
This issue has been partially addressed in Monte Carlo numerical simulations of evolving globular clusters \citep{Belloni_2020b}, in which SySts with WDs formed without Roche lobe overflow have been investigated.
These simulations show that their orbital periods are typically sufficiently long (${\gtrsim10^3}$~days) so that dynamical disruption of their progenitors is virtually unavoidable in very dense clusters.
On the other hand, in less dense clusters, some of these SySts are still predicted to exist, although their identification should be rather difficult in observational campaigns, mainly due to their locations within the clusters and rareness. 
Despite these simulations aimed at explaining the lack of SySts in globular clusters, they do not necessarily correspond to definite answers to the issue, since most known SySts have orbital period between ${\sim200}$ and ${\sim10^3}$~days, which means that most SySts were not addressed in the simulations.
Further simulations are still missing, in which the formation of SySts through dynamically stable non-conservative mass transfer is consistently taken into account and the impact of dynamics on their progenitors is evaluated.

So far, we have discussed the population of accreting compact objects in old star clusters.
These star clusters cannot harbour HMXBs, since at their present-day ages, the entire population of HMXBs have already become single NSs/BHs or binaries hosting either NSs and/or BHs.
However, in very young starburst galaxies (only a few Myr old), HMXBs are expected to be the dominant source of X-rays.
Given that high-mass stars, and potentially all stars \citep{Kroupa_1995a,Lada_2003}, are most likely formed in binaries \cite[e.g.][]{Sana_2012} in embedded clusters \cite[e.g.][]{Marks_2012}, it is not surprising that observations suggest that the HMXBs in starburst galaxies are consistent with being formed in star clusters \cite[e.g.][]{Kaaret_2004}.
Even though young star clusters are most likely HMXB factories, these HMXBs are not necessarily formed dynamically.
The analysis of the X-ray sources in the merging Antennae galaxies and in the dwarf starburst NGC 4449 suggests that high stellar density and in turn dynamical interactions are not a primary driver of HMXB formation \citep{Mulia_2019}, which is consistent with star cluster direct $N$-body simulations \citep{Garofali_2012}.
Therefore, despite being rather frequent in starburst galaxies and likely formed in compact star clusters, there is evidence supporting that HMXBs form directly from primordial high-mass binaries, without any significant influence of dynamics.

\section{Secular Evolution}

The mass transfer in accreting compact objects is dynamically stable and can be separated into three main modes: 
(i) Roche lobe overflow (if ${R_{\rm d} \sim R_{\rm RL}}$), 
(ii) atmospheric Roche lobe overflow (if ${0.9\,R_{\rm RL} \lesssim R_{\rm d} \lesssim \,R_{\rm RL}}$), 
and 
(iii) wind accretion (if ${R_{\rm d} \lesssim 0.9\,R_{\rm RL}}$).
In CVs, LMXBs, AM\,CVns, and UCXBs, the donor is filling its Roche lobe.
Therefore, mass transfer in these systems can be driven by orbital AML or by nuclear expansion of the donor, with or without an initial phase of thermally unstable mass transfer.
On the other hand, despite the fact that in most SySts, SyXBs, sg-HMXBs, and WR-HMXBs, the donor is under-filling its Roche lobe, mass transfer still occurs, mainly when the extended atmosphere of the donor fills its Roche lobe, or due to accretion of a significant part of the winds from the donor.

\begin{figure*}[htb!]
\centering
\includegraphics[width=0.99\linewidth]{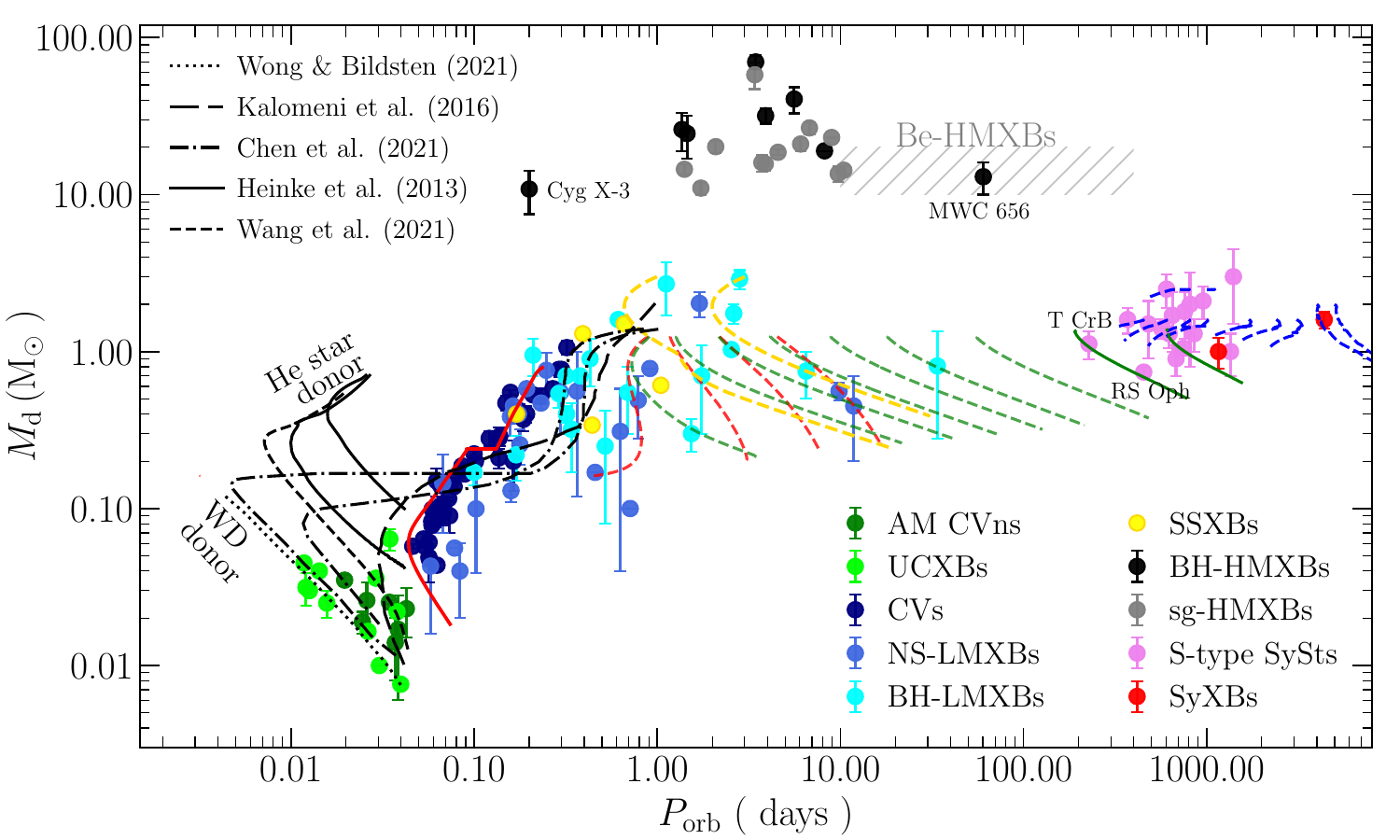}
\caption{Comparison between observed and predicted properties of accreting compact objects in the plane donor mass ($M_{\rm d}$) versus orbital period ($P_{\rm orb}$).
The observational data correspond to BH-LMXBs \citep{Wiktorowicz_2014,SL2020}, BH-HMXBs \citep{Crowther_2010,Orosz_2009,Orosz_2007,Casares_2014,Zdziarski_2013,Silverman_2008,Liu_2013,MillerJones_2021}, NS-LMXBs \citep{Bassa_2009,Van_2019}, UCXBs \citep{Van_2019}, eclipsing sg-HMXBs \citep{Clark_2002,Quaintrell_2003,vanderMeer_2007,Mason_2010,Mason_2011,Mason_2012,Bhalerao_2012,Falanga_2015,Pearlman_2019}, eclipsing AM CVns \citep{Copperwheat_2011,Green_2018,vanRoestel_2021}, eclipsing CVs \citep{McAllister_2019}, S-type SySts \citep[most of which are eclipsing,][]{Mikolajewska_2003,Stanishev_2004,Brandi_2009,Mikolajewska_2021,Hinkle_2009}, SyXBs \citep{Hinkle_2006,Hinkle_2019}, and SSXBs \citep{Kalomeni_2016}.
In particular, we labeled the only known Be-HMXB hosting a BH \citep[MWC~656,][]{Casares_2014}, a WR-HMXB with an extremely short orbital period \citep[Cyg~X--3,][]{Zdziarski_2013}, and the two SySts T~CrB \citep{Stanishev_2004} and RS~Oph \citep{Brandi_2009}, which are among the few SySts in which the red giant is nearly filling its Roche lobe, and in addition, given their high carbon--oxygen WD masses, they will most likely evolve towards Type Ia supernovae \citep{Liu_2019_SySt}. 
The theoretical evolutionary tracks correspond to the lines, of which all coloured tracks were computed by us using the \mesa~code, adopting solar metallicity, assuming the RVJ prescription \citep{RVJ} (Equation~\ref{Eq:MB-RVJ}, with $\gamma_{\rm MB}=3$), for the MB, and considering tidal interaction \citep{Hurley_2002} and stellar wind mass loss, with the Reimers parameter equals to 0.5 for FGB stars and the Bl\"ocker parameters equals to 0.02 for AGB stars.
In all these coloured tracks, we chose several different initial orbital periods and initial masses for the point-mass compact object and its zero-age MS star companion.
The black tracks correspond to previous binary evolution models, in which the donor is initially a helium WD \citep{Wong_2021}, a helium star \citep{Heinke_2013,Wang_2021}, or a MS star \citep{Kalomeni_2016,Chen_2021}.
Given that the evolutionary tracks pass through most observational points, we could naively conclude that our current knowledge on how accreting compact objects evolve is not bad.
However, major problems arise when we take into account the entire populations of different types of accreting compact objects, and confront results from population synthesis with observations.
These problems are most likely connected with our poor understanding of the mechanisms driving the evolution of accreting compact objects.
Despite that, there has been important progress in the last couple of years, which can definitely guide us towards a solution to these problems.
}
\label{FigEvolMdonor}
\end{figure*}

\begin{figure*}[htb!]
\centering
\includegraphics[width=0.99\linewidth]{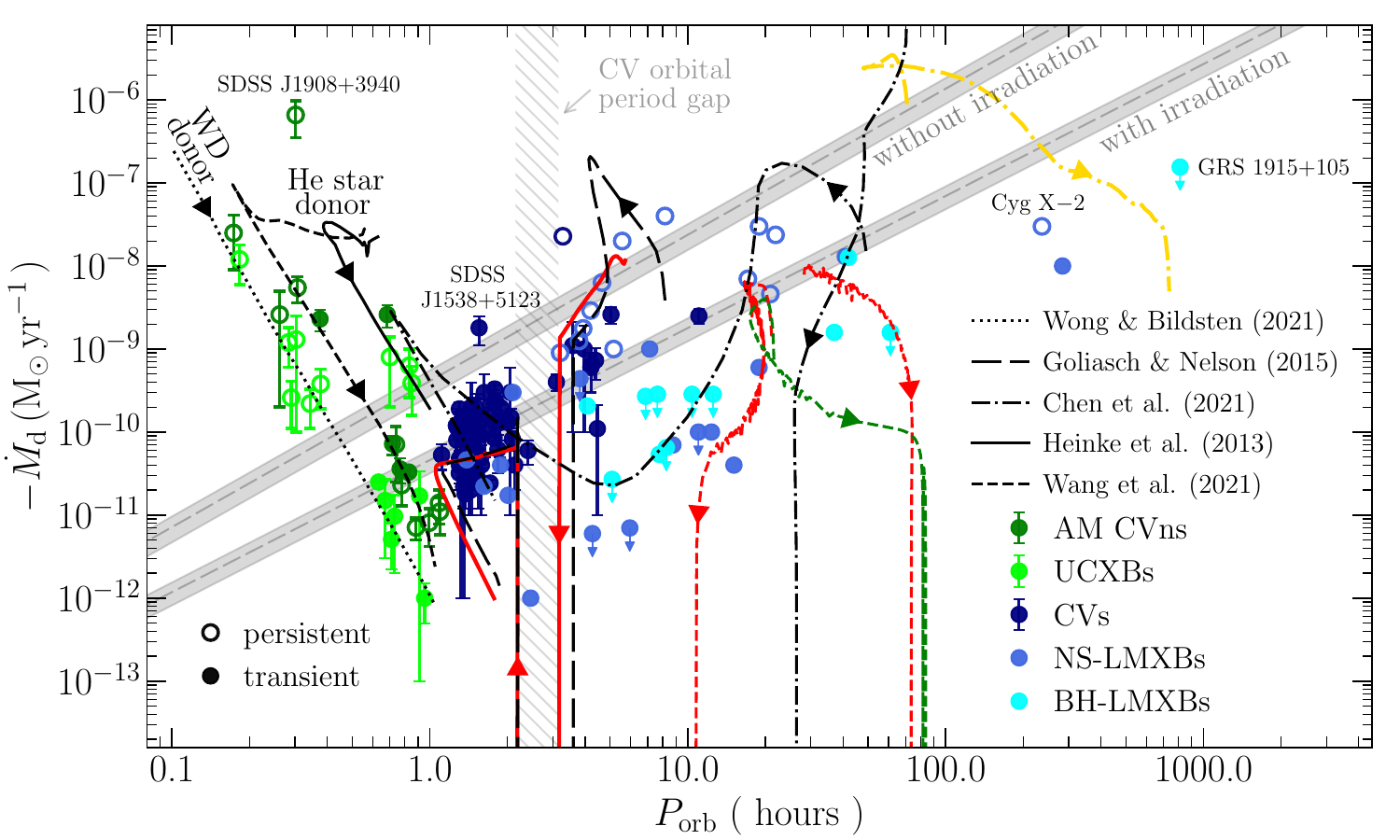}
\caption{Comparison between observed and predicted properties of accreting compact objects in the plane mass transfer rate ($\dot{M}_{\rm d}$) versus orbital period ($P_{\rm orb}$).
We only considered in the plot those systems in which the donor is a Roche lobe filling low-mass star, i.e. BH-LMXBs \citep{Coriat_2012}, NS-LMXBs \citep{Van_2019}, UCXBs \citep{Van_2019}, AM CVns \citep{Ramsay_2018}, and CVs \citep{Pala_2022}.
Some of the coloured evolutionary tracks in Figure~\ref{FigEvolMdonor} are shown, which were computed by us with the \mesa~code.
The black lines correspond to evolutionary tracks from previous binary models, in which the donor in initially either a WD \citep{Wong_2021}, or a helium star \citep{Heinke_2013,Wang_2021}, or a MS star \citep{Goliasch_2015,Chen_2021}, the latter having an initial phase of thermally unstable mass transfer.
We also included in the figure, as grey thick lines, the critical mass transfer rates \citep{Coriat_2012}, above which the accretion disks are stable and the systems are persistent sources, and transient otherwise, considering irradiated accretion disks or not.
By comparing the observational data with these critical mass transfer rates, we can see that the disk instability model reasonably well explains persistent and transient sources, except perhaps in the case of Cyg~X--2 \citep[but see][]{Coriat_2012} and some AM\,CVns.
There are two representative tracks for CVs, the solid red line computed by us and the long-dashed black line \citep{Goliasch_2015}, which were both computed with the RVJ prescription \citep{RVJ} (Equation~\ref{Eq:MB-RVJ}, with $\gamma_{\rm MB}=3$).
We can clearly see that this MB prescription is not adequate to explain the persistent and transient sources with orbital periods longer than ${\sim3}$~h. 
In addition, CVs hosting fully convective M-type MS stars, at orbital periods shorter than ${\sim2}$~h, exhibit a spread in the mass transfer rates that cannot be explained by only gravitational wave radiation.
With respect to NS-LMXBs, it is also clear that the RVJ prescription (dashed red lines) is not able to account for the persistent sources, which has been a long-standing problem of LMXB evolution.
However, when a MB law \citep{Van_2019} similar to the recently developed CARB prescription \citep{Van2019CARB} is adopted \citep[dot-dashed black lines,][]{Chen_2021}, mass transfer rates are great enough so that they can be explained.
The persistent UCXBs with long orbital periods (${\gtrsim40}$~min) can apparently only be explained if the donor is initially a helium star, which is consistent with observations \cite[e.g.][]{Heinke_2013}.
There are two particular systems, one AM\,CVn (SDSS~J1908$+$3940) and one CV (SDSS~J1538$+$5123), that are very interesting, since they apparently cannot be explained by any of the tracks we included in the figure.
While it has been proposed that this CV might be either a newly born system or a post-nova system, it is not clear what might be happening with this AM\,CVn.
}
\label{FigEvolMdot}
\end{figure*}

Among the observational properties, the donor mass and the mass transfer rate deserve special attention, as these two quantities are somewhat tied to the secular evolution of accreting compact objects.
We show these properties of accreting compact objects in Figures~\ref{FigEvolMdonor} and \ref{FigEvolMdot}, against their orbital periods, together with evolutionary tracks computed by us with the \mesa~code and other tracks from previous binary models.
In what follows, we will describe the dynamically stable evolution of accreting compact objects as well as the up-to-date observational information that can be used to constrain evolutionary models, which includes their orbital periods, their mass transfer rates, and their donor and compact object masses, but without being exclusively limited to them.

\subsection{Cataclysmic Variables and Low-Mass X-ray Binaries}

As mentioned earlier, CVs and LMXBs are formed through an episode of CE evolution.
More specifically, after CE evolution, a \pceb~is formed in a detached binary hosting a compact object and a non-degenerate star.
Because of orbital AML and/or nuclear evolution of the future donor, the orbital separation decreases up to the formation of the accreting compact binary, when the non-degenerate companion starts filling its Roche lobe and becomes a donor.
The donors in CVs and LMXBs are usually MS stars, subgiants or unevolved red giants.

\subsubsection{Low-Mass Unevolved M-/K-Type Main-Sequence Star Donors}

Let us start with the case in which the donor is initially a low-mass unevolved MS star, corresponding to \pcebs~hosting MS stars with types M and K (${\lesssim0.8}$~\Msun), which do not significantly evolve within the Hubble time.
This case is illustrated in Figures~\ref{FigEvolMdonor} and \ref{FigEvolMdot} by the solid red track, for which the point-mass compact object and zero-age MS star have both the same initial mass of $0.8$~\Msun, and the initial orbital period is $12$~h.
In this case, the MS donor has a radiative core, which implies that MB is the main orbital AML mechanism driving the binary evolution.
MB causes mass transfer rates high enough so that the MS donor is only able to maintain thermal equilibrium if it is slightly bloated, which means that MS donors are somewhat oversized in comparison with isolated MS stars \citep{Knigge_2011_OK}. 
When the MS donor becomes fully convective, MB is expected to cease or, at least, become much less efficient (disrupted MB scenario), most likely as a consequence of a rise in the secondary star magnetic complexity \citep{TaamSpruit_1989,Garraffo_2018}, rather than the originally proposed demise of the dynamo \citep{RVJ,SpruitRitter_1983}.
From this point on, the main driving mechanism becomes GR.
This leads to a significant drop in the mass transfer rate and a slowdown of the evolution, as AML through GR is in general orders of magnitude weaker than MB.
Such a drop in the mass transfer rate allows the MS donor to re-establish thermal equilibrium at its normal size (i.e. decrease in size), and the system becomes a detached binary since the companion of the compact object is no longer filling its Roche lobe.
This detachment occurs at the upper edge of the so-called orbital period gap (a paucity of observed accreting systems in the orbital period range of ${\sim2-3}$~h).

Even though the system is now detached, it keeps losing angular momentum due to GR and continues to evolve towards shorter orbital periods.
When the orbital period is sufficiently short, the MS donor fills its Roche lobe again, and mass transfer restarts, i.e. the system is again a CV evolving now with lower mass transfer rates towards shorter orbital periods, which happens at the lower edge of the orbital period gap.
When the donor mass becomes too low to sustain hydrogen burning, it becomes a hydrogen-rich degenerate object, i.e. a brown dwarf.
From this point on, the donor expands in response to the mass loss, leading to an increase in the orbital period.
The so-called orbital period minimum is then a natural consequence of the transition from a fully convective MS star to a brown dwarf.

The edges of the orbital period gap, and consequently its width, as well as the orbital period minimum depend on the metallicity \citep{Stehle_1997,Kalomeni_2016}, the accreting compact object mass, and the orbital AML rate \citep{Knigge_2011_OK,Belloni_2020a}.
For instance, regarding the impact of the metallicity, the higher the central helium fraction, the shorter the orbital period minimum, and the orbital period minimum can be as short as ${\sim10-40}$~min, when the central helium fraction is nearly unity \citep{Kalomeni_2016}.
The orbital period minimum is also affected by orbital AML in the following way.
For systems having fully convective donors at a given orbital period, the stronger/weaker the orbital AML, the higher/lower the mass transfer rate, and in turn the smaller/greater the donor mass, which allows the donor to reach the mass limit for hydrogen burning at longer/shorter orbital periods.
Orbital AML is also expected to have a strong impact on the orbital period gap, due to variations in the MB strength for systems with donors having convective cores.
The stronger/weaker MB, the more/less bloated the donor is, which means that it becomes fully convective at a lower/higher mass (see Figure~\ref{FigMB}), having in turn a smaller/greater radius.
This implies that system will become semi-detached again at a shorter/longer orbital period, i.e. the lower edge of the orbital period gap is located at shorter/longer orbital periods.
For a sufficiently weak MB , such as the RVJ prescription with $\gamma_{\rm MB}\geq5$, the donor bloating is negligible, especially near fully convective boundary, and the system evolves virtually without detaching.
On the other hand, for a very strong MB, such as that provided by the CARB prescription or by the Kawaler prescription with $n\gtrsim1.5$, the donor becomes fully convective at a mass as small as ${\sim0.18}$~\Msun, and the orbital period gap can start at an orbital period as long as ${\sim7-8}$~h.
This happens because, as the donor approaches the fully convective boundary, it is increasingly driven out of thermal equilibrium, leading to an increase in its radius by a factor of ${\gtrsim2}$ in comparison with isolated M-type MS stars with similar masses.
Then, as a response of the donor to mass loss, the orbit expands.

The above outlined evolution successfully describes several features of the evolution of LMXBs and CVs. However, it is clearly not complete.
For instance, it does not include the impact of the magnetic field of the compact object on the evolutionary picture for CV and LMXB, which can be significant as we will explain in what follows.  
If the compact object magnetic moment, which is proportional to its magnetic field strength and to the cube of its radius, is sufficiently large (${\gtrsim10^{33}}$~G~cm$^{3}$), then a coupling of the donor magnetic field lines and the compact object magnetic field lines may lead to an increase of the portion of the donor winds that remains trapped to the binary \citep{LWW94,WW02}.
This effect would make winds to carry away less angular momentum, resulting in reduced orbital AML via MB and, in turn, diminished mass transfer rates.
While NS magnetic fields seem to considerably decay during LMXB evolution \citep{Zhang_2006}, the magnetic field strength in most accreting WDs are expected to change only negligibly, which is a consequence of mass loss due to nova eruptions \citep{Zhang_2009}.
Given the small sizes of NSs (only ${\sim10-20}$~kilometers) and their relatively weak magnetic fields (${\sim10^8-10^9}$~G) due to accretion-induced field decay, their magnetic moments are usually not strong enough to cause an important impact on LMXB evolution.

The situation is different for WDs, as they are much bigger (${\sim5000-10000}$~kilometers), which implies that, if they host sufficiently strong magnetic fields (${\gtrsim10^7}$~G), the CV evolution is most likely different from the remaining CVs hosting WDs with weaker magnetic fields \citep{Belloni_2020a}.
In case the WD magnetic moment is higher than ${\sim10^{34}-10^{35}}$~G~cm$^{3}$, MB is completely suppressed and the CV evolution is driven by only GR, i.e. at a much lower orbital AML rate, without detaching.
Recent binary models provide strong support to this magnetic evolutionary picture for such CVs, as the predicted mass transfer rates are in agreement with the observations and the orbital period gap is absent in this population, which reasonably well agrees with observations \citep{Belloni_2020a}.
It even seems that CV evolution in general cannot be understood ignoring the origin of WD magnetic fields and its impact on orbital AML and exchange between the stellar components and the orbit.
Including magnetic field generation due to a rotation- and crystallization-driven dynamo \citep{Isern_2017,SchreiberNATURE} and the impact of the generated field on the evolution of CVs can reproduce several observational facts that remain otherwise inexplicable, such as the existence of AR\,Sco, which hosts a radio-pulsing WD \citep{Marsh_2016}, the relative numbers of magnetic and non-magnetic CVs \citep{Pala_2020}, and the absence of strong magnetic WDs among young detached \pcebs~\citep{Liebert_2005}.

\subsubsection{Subgiant or A-/F-/G-Type Main-Sequence Star Donors}

Let us now consider the case in which the donor is a nuclear-evolved MS star (i.e. near the end of the MS) or a subgiant at the onset of mass transfer to the compact object.
In other words, we are now looking at  \pcebs~hosting MS stars of spectral type G or earlier (${\gtrsim1}$~\Msun).
If the orbital period of the \pceb~is sufficiently long, the future donor star will have enough time to evolve on the MS, or even become a subgiant, before filling its Roche lobe.
For stars having convective envelopes, the onset of mass transfer occurs due to a combination of orbital AML due to MB and nuclear evolution of the future donor, while for stars having radiative envelopes, it is triggered only by nuclear evolution of the future donor.
The evolution for these two types of donors is illustrated in Figures~\ref{FigEvolMdonor} and \ref{FigEvolMdot} by the long-dashed and dot-dashed black tracks, which were taken from different binary models \citep{Goliasch_2015,Kalomeni_2016,Chen_2021}, by the two dashed red and green tracks with shortest initial orbital periods ($1$~day), and by the two yellow tracks (initial orbital periods set as $1$ and $3$~days).
The difference between the three sets of coloured tracks is the initial mass of the compact object, which is $0.8$~\Msun~in the red tracks, $1.2$~\Msun~in the green tracks, $1.4$~\Msun~in the yellow tracks; and the zero-age MS mass, which is $1.2$~\Msun~for the green and red tracks, and $3.0$~\Msun~in the yellow tracks.

Depending on the mass and structure of the donor as well as on the mass of the compact object, the mass transfer can be thermally unstable at the very beginning of the evolution.
Thermally unstable mass transfer typically occurs when the nuclear-evolved MS or subgiant donor is significantly more massive than the compact object at the onset of Roche lobe overflow, as in the case of the yellow tracks in Figures~\ref{FigEvolMdonor} and \ref{FigEvolMdot}.
More generally, for a given donor evolutionary state, the higher the mass ratio at the onset of Roche lobe overflow, the greater the chances that the mass transfer will be thermally unstable.
The mass transfer rates during this phase are typically larger than those provided by thermally stable mass transfer and can exceed ${\sim10^{-7}-10^{-6}}$~\Msun~yr$^{-1}$, which is high enough to completely drive the donor out of thermal equilibrium.
In case the compact object is a WD, systems observed at this phase appear typically as SSXBs.
The orbital period during this thermally unstable phase decreases as a consequence of the donor response to thermal time-scale mass loss.
However, when mass transfer stabilizes, the donor re-establishes thermal equilibrium and the orbital period can further decrease or reverse and increase.

Let us now discuss the evolution following the onset of thermally stable mass transfer.
For stars with radiative envelopes, orbital AML due to MB does not take place, and the orbital period should increase as a response to mass transfer.
For stars with convective envelopes, there are two possibilities.
In case the nuclear time-scale of the donor is shorter than its MB-driven mass-loss time-scale, then the binary evolution will be driven by the donor nuclear evolution.
Mass transfer then proceeds on the nuclear time-scale of the donor, the orbital period increases, and the evolution is usually called \textit{divergent}.
On the other hand, in case MB is sufficiently strong, such that the MB-driven mass-loss time-scale becomes shorter than nuclear time-scale, the evolution will be driven by orbital AML and the orbital period will decrease.
In this case, the evolution is called \textit{convergent}, and the system will evolve to either a CV or LMXB with a hydrogen-rich degenerate donor (in case the MS donor did not have time to start developing a degenerate helium core) or an AM\,CVn or UCXB with helium-rich degenerate donors (in case the donor is initially a subgiant or in case the MS donor managed to develop a degenerate helium core).

The evolution towards a hydrogen-rich degenerate donor is similar to the case in which the donor is an unevolved low-mass MS star (compare long-dashed black and solid red tracks in Figure~\ref{FigEvolMdot}).
However, the evolution towards AM\,CVns and UCXBs is rather different from that of the previous cases, which can be seen by comparing the long-dashed and dot-dashed black tracks with the solid red track in Figure~\ref{FigEvolMdonor}, and the dot-dashed black track with the long-dashed black and solid red tracks in Figure~\ref{FigEvolMdot}.
For instance, the orbital period minimum becomes much shorter for donors that were more evolved at the onset of Roche lobe overflow.
This happens because the donors have helium cores that are partially degenerate, with masses that are too low to trigger helium burning.
This allows their radii to considerably shrink as the binary evolves, so that the binary can reach orbital periods as short as ${\sim6}$~min, before the donor becomes degenerate and the orbital period subsequently increases as a response to mass transfer.

More generally, whether convergent or divergent evolution takes place depends on the strength of orbital AML due to MB, on the fraction of mass loss from the binary, on the metallicity, and on the binary properties at onset of Roche lobe overflow, most importantly the orbital period, donor structure, and compact object and donor masses \cite[e.g.][]{Nelson_2004,Deng_2021}.
For instance, the impact of the compact object mass can be seen Figures~\ref{FigEvolMdonor} and \ref{FigEvolMdot} by comparing the dashed red and green tracks with the shortest initial orbital period, in which the RVJ prescription with $\gamma_{\rm MB}=3$ was adopted.
We can see that the red track in which the point-mass compact object mass is $0.8$~\Msun is convergent, while the green track in which the point-mass compact object mass is $1.2$~\Msun is divergent.
In addition, when the RVJ prescription is compared with the CARB prescription, the bifurcation period obtained with the latter is much longer \citep{Istrate_2014,Deng_2021,Chen_2021,Soethe_2021}.
This is a natural consequence of the orbital AML rates provided by the CARB prescription being much larger than those obtained with the RVJ prescription.

\subsubsection{Comparison with Observations}


Let us start our comparison between theory and observations discussing how good the disk instability model \cite[e.g.][]{Lasota_2001} is to explain persistent and transient systems.
By inspecting Figure~\ref{FigEvolMdot}, we can see that in general this model successfully explains why some CVs and LMXBs exhibit outbursts, while others do not.
Irradiation heating of the accretion disk is usually negligible for the stability of disks in CVs \citep{Dubus_2018} while in the case of LMXBs X-ray irradiation needs to be taken into account \citep{Coriat_2012}.
The net effect of including irradiation is that a system at a given orbital period can be persistent for lower mass transfer rates in comparison with non-irradiated accretion disks which is in general the case (compare CVs and LMXBs in Figure~\ref{FigEvolMdot}).
However, the disk instability model apparently fails to explain two systems in Figure~\ref{FigEvolMdot}, namely the CV SDSS~J1538$+$5123 and the NS-LMXB Cyg~X--2.

%
%

The time-averaged mass transfer rates in CVs can be estimated with the disk instability model itself \citep{Dubus_2018} or by analyzing the ultraviolet emission from the WD during quiescence \citep{Townsley_2009}, the latter being used for SDSS~J1538$+$5123.
From the ultraviolet spectra, the WD effective temperature can be estimated, and this is considered a good indicator of the mass transfer rate, since the WD is compressionally heated due to the accretion process, and the higher the mass transfer rate, the higher the WD effective temperature \citep{Townsley_2003,Townsley_2004}.
SDSS~J1538$+$5123 is a dwarf nova (i.e. transient) located below the orbital period gap and therefore should according to the disk instability model have mass transfer rates much lower than the ${\sim1.8\pm0.7\times10^{-9}}$~\Msun~yr$^{-1}$ inferred from its high WD effective temperature.
However, there is no problem with the disk instability model here and this high WD effective temperature is most likely related to a recent nova eruption \citep{palaetal_2017}, and therefore, the temperature is not a good indicator of the time-averaged mass transfer rate in this particular case.

%
%
The persistent system Cyg~X--2 is the only NS-LMXB that does not agree with the predictions of the disk instability model, which clearly indicates that this system should be transient.
One possibility for this disagreement would be that Cyg~X--2 is in fact a transient source undergoing a very long outburst \citep{Coriat_2012}, similarly to what happens with the well-known BH-LMXB GRS~1915$+$105, which is a transient source and has the longest orbital period ($33.85$~days) of this population.
Given that the longer the orbital period, the larger the accretion disk, the disks in these wider systems can accumulate enough mass to fuel an outburst for decades.
The accretion disk in Cyg~X--2 is estimated to have a mass of ${\sim2.5\times10^{28}}$~g, which is massive enough to sustain outbursts lasting up to ${\gtrsim80}$~years \citep{Coriat_2012}.


Let us move now to compare expected secular evolution of CVs and LMXBs with observations.
When we compare the tracks with the observational data in these figures, we can see that most CVs are consistent with having initially unevolved low-mass MS donors (solid red track).
However, some CVs clearly reveal abnormal (non-solar) abundances in their spectra, indicating that the material has undergone thermonuclear processing by the CNO cycle \citep[see][for list of systems]{Sparks_2021}.
These CVs contribute to ${\sim5-15}$\% of the entire CV population \citep{Gaensicke2003,Pala_2020}, and given that their donors are not expected to possess this material, two scenarios have been proposed to explain them.
As a first possibility, they are descendants of SSXBs, i.e. they are systems hosting initially nuclear-evolved MS stars that underwent a short phase of thermally unstable mass transfer before reaching their present-day configurations \citep{schenkeretal02-1,Podsiadlowski_2003_CV}.
In this way, the donor has its outer layers stripped during this phase, revealing CNO-processed material.
The other possibility would be donor contamination by the accretion of CNO-processed material from ejected by the WD during nova eruptions \citep{StehleRitter_1999}.

Even though the first scenario over predicts the number of such CVs by a factor of ${2-6}$ \citep{schenkeretal02-1}, it has been for a long time accepted as most likely.
This is because it is apparently rather difficulty for the donor to capture enough nova-processed material \citep{Sparks_2021}, e.g. because its geometrical cross-section is small and the high-velocity material that the secondary geometrically intercepts is not highly nuclear-processed. 
However, it has been recently suggested that these problems can be overcome if a significant part of the ejected material leaves the WD with velocities smaller than the escape velocity \citep{Sparks_2021}.
This is possible because the thermonuclear runaway on the WD causes a strong propagating shock wave, which leaves behind a large velocity gradient in the expanding shell.
This implies that part of this shell will reach escape velocity but the remaining material will be gravitationally bound to the binary, leading to a non-rotating CE-like structure surrounding the CV.
This nova-induced CE-like structure will then contain a large amount of non-solar metallicity material, which can be efficiently accreted by the donor, creating a significant non-solar convective region.
However, given that virtually all CVs are supposed to undergo nova cycles and that only ${\sim10-15}$\% of them exhibit peculiar abundances, it remains unclear under which conditions this sort of pollution can efficiently work, and whether it can explain why most CV donors fail in to efficiently accrete.
These questions could and should be addressed in future detailed binary models, since 
their answers are potentially related to major problems connected with CV secular evolution, as we will see later.

Before moving to LMXBs, we just mention a few interesting long period (${\sim14-48}$~h) CVs that host subgiant donors, i.e. BV~Cen \citep{Gilliland_1982} and GK~Per \citep{Alvarez_2021} in the Milky Way field, and AKO~9 in the globular cluster 47~Tuc \citep{Knigge_2003}.
Interestingly, the masses of the WDs in BV~Cen \citep[${\sim1.02-1.44}$~\Msun,][]{Watson_2007} and GK~Per \citep[${\sim0.92-1.19}$~\Msun,][]{Alvarez_2021} are significantly higher than the average among CVs \citep[${\sim0.8\pm0.2}$~\Msun,][]{Zorotovic_2011,McAllister_2019,Pala_2022}, which is consistent with the evolutionary picture in which the WD mass has grown due to the high mass transfer rates expected when the donor is initially an evolved MS star or a subgiant. 
This makes this type of CVs very good candidates for being Type Ia supernova progenitors, under certain conditions, although they correspond to a negligible fraction of the entire CV population.


When we compare the distribution of LMXBs and CVs in Figure~\ref{FigEvolMdonor}, together with the evolutionary tracks, it is very likely that the contribution of systems with initially evolved donors is much larger for LMXBs in comparison to CVs, as they have on average significantly longer orbital periods for the same range of donor masses.
While the evolutionary sequence given by solid red line in this figure seems to nicely explain at least around half of NS-LMXBs, a lot of them can only be explained if the donor is initially an evolved MS star or a subgiant.
The situation is even more dramatic for BH-LMXBs, since only two systems host M-type MS donors, and the majority of systems with orbital periods shorter than ${\sim1}$~day have K-type MS donors \citep{SL2020}.
In addition, it seems that a significant fraction of LMXBs start their evolution having initially intermediate-mass donors \citep{Podsiadlowski_2000}, especially in the case of BH-LMXBs.
Similarly to CVs, the LMXBs with longest orbital periods (${\gtrsim1}$~days) most likely have subgiant or red giant donors, such as 
Cyg~X--2 \citep{Casares_1998}, GRO~J1744--28 \citep{Doroshenko_2020}, GX~13$+$1 \citep{Bandyopadhyay_1999},
XTE~J1550--564 \citep{Orosz_2011_LMXB}, GX~339--4 \citep{MD2008}, GS~1354--64 \citep{Casares_2009}, GRO~J1655--40 \citep{Israelian_1999}, V4641~Sgr \citep{MacDonald_2014}, GS~2023$+$338 \citep{Khargharia_2010}, and GRS~1915$+$105~\citep{Harlaftis_2004}.
These systems are then evolving towards longer orbital periods, as mass transfer is driven by the nuclear expansion of the donor.

Even though the evolutionary picture that we have discussed reasonably well explains the main observational features of individual CVs and LMXBs, as illustrated in Figures~\ref{FigEvolMdonor} and \ref{FigEvolMdot}, there are still major problems when we take into account the entire populations and compare observed properties with those predicted by population synthesis that are most likely intrinsically connected with the orbital AML mechanisms driving the binary evolution.
In particular, the standard RVJ prescription (Equation~\ref{Eq:MB-RVJ}) for MB usually assumed for models of CVs and LMXBs is probably the origin of most of the problems we are currently facing.
For instance, the predicted fraction of CVs hosting brown dwarfs is much higher than observed \citep{Belloni_2020a,Pala_2020}, which suggests that the overall evolutionary time-scale should be much longer than predicted by models using the RVJ prescription.
In addition, the predicted mass transfer rates of CVs above the orbital period gap are typically much higher than those inferred from observations, except for systems near the upper edge of the orbital period gap, for which predicted rates are much smaller than those inferred from observations \citep{Belloni_2020a,Pala_2022}.
This suggests that the dependence of MB with the orbital period should be the opposite of what is predicted by the RVJ prescription. While this recipe leads to a decreasing orbital AML rate as the CV evolves (see Figure~\ref{FigMB}), observations suggest that the overall rate should increase as the CV evolves.
Moreover, the predicted mass transfer rates for NS-LMXB are so low that they cannot explain the persistent sources \citep{Podsiadlowski_2002}, which suggests that MB should be much stronger in these systems than predicted by the RVJ prescription.
Furthermore, the predicted donor mass distribution of BH-LMXBs peaks at a much higher value than the observed distribution \citep{Wiktorowicz_2014}, which suggests that MB should also be stronger in these systems.
Finally, when the RVJ prescription is adopted, a well-known fine-tuning problem arises when trying to reproduce millisecond pulsars in close detached binaries (orbital periods in the range ${\sim2-9}$~h) with helium WD companions of mass ${\lesssim0.2}$~\Msun, \citep{Istrate_2014,Chen_2021}, which represents a big problem in the formation of these binaries.

Regarding LMXBs and their descendants, recent binary models \citep{Van_2019,Van2019CARB}, in which the CARB prescription has been proposed, indicate that the problem with the persistent NS-LMXBs can be solved with this recipe that provides much higher orbital AML rates (see Figure~\ref{FigMB}) and therefore higher mass transfer rates, especially for persistent systems with shorter orbital periods.
For this reason, the CARB prescription might also solve the problem with BH-LMXBs, since it could shift the predicted donor mass distribution towards lower values and hopefully reproduce the observed distribution.
In addition, this new prescription might also be consistent with the fast orbital decay observed in some BH-LMXBs, such as XTE~J1118$+$480 \citep{GH2012}, A0620--00 \citep{GH2014}, and GRS~1124--68 \citep{GH2017}, which cannot be explained with the RVJ prescription.
When applied to close detached millisecond pulsar binaries, the CARB prescription also provides a solution to the fine-tuning problem \citep{Soethe_2021}.
In addition, not only compact binaries are well-reproduced by this prescription but also wide-orbit binaries can be explained, and in general, the relation between the helium WD mass and the orbital period predicted by this prescription is in agreement with observations \citep{Soethe_2021}.

Regarding CVs, it has been proposed that not only the problem with the mass transfer rates but also the problem with the fraction of post-orbital period minimum could be solved with another MB prescription \citep{Belloni_2020a,Pala_2020}.
A reasonable candidate to solve both problems is the Kawaler prescription with $n\sim1.25$, which is shown in Figure~\ref{FigMB}.
This prescription predicts an overall evolutionary time-scale substantially longer than that obtained with the RVJ prescription, which may perhaps contribute significantly to reduce the predicted number of CVs that manage to reach the orbital period minimum.
Moreover, given that the orbital AML rates provided by this prescription are around one order of magnitude higher than GR, the constraints provided by detached WD--MS \pcebs~\citep{Schreiber_2010} are not expected to be violated, which strengthens even more the potential of solving CV problems with this prescription.

In addition to the problems most likely connected with the standard RVJ prescription for MB, there are other problems that are probably associated with another orbital AML mechanism, i.e. consequential AML, which is not included in the standard evolutionary picture.
For instance, the observed CV WD mass distribution has a peak at ${\sim0.8}$~\Msun~\citep{Zorotovic_2011,McAllister_2019,Pala_2022}, while most predicted CV WD masses are on average smaller than that \citep{Zorotovic_2011,McAllister_2019,Pala_2022,ZS2020}.
In addition, for CVs hosting fully convective MS donors, there is a clear spread in their mass transfer rates inferred from observation that is definitely not predicted when GR is taken as the only driver of their evolution \citep{Pala_2022}.
Moreover, still when only GR is considered, the predicted mass transfer rates among post-orbital-period-minimum CVs are much lower than those inferred from observations \citep{Pala_2022}, and the predicted orbital period minimum is significantly shorter than observed \citep{Howell_2001,Gansicke_2009,Knigge_2011_OK}.
These problems might be solved if the total AML for these systems is stronger than predicted by GR  \citep{Knigge_2011_OK,Schreiber_2016,Nelemans_2016}.
In case there is another orbital AML mechanism driving the evolution together with GR for CVs with fully convective MS donors and brown dwarf donors, then the predicted mass transfer rates might be significantly enhanced, which would cause the CVs to evolve faster so that their donors become degenerate at longer orbital periods.
The problem with the WD masses can be solved if the strength of this additional orbital AML is inversely proportional to the WD mass, such that most CVs with low-mass WDs would become dynamically unstable at the onset of mass transfer \citep{Schreiber_2016}.
Similarly, the spread in the mass transfer rates for CVs with fully convective MS donors might also be solved with this dependence on the WD mass, if perhaps the additional orbital AML also depends on the age of the CV.

The most natural choice of additional orbital AML to enhance the mass transfer rates would be consequential AML, and there are very good candidates for the above-highlighted dependence on the WD mass, and eventually on the time since the onset of mass transfer
\citep{Schenker_1998,Schreiber_2016,Nelemans_2016}.
The frictional orbital AML produced by novae depends strongly on the expansion velocity of the shell \citep{Schenker_1998}.
For low-mass WDs (${\lesssim0.55}$~\Msun), the expansion velocity is small \citep{Yaron_2005}, and this may lead to strong orbital AML by friction that makes most CVs with low-mass WDs dynamically unstable.
In particular, given the velocity gradient in the nova shell we mentioned earlier, it is reasonable to expect that a nova-induced CE-like structure surrounding the binary will be formed after the eruption.
This idea has been tested in binary models \citep{Nelemans_2016}, which confirms that such a configuration drastically affects the stability of mass transfer in CVs with low-mass WDs.
As we mentioned earlier, this form of consequential AML could also explain the existence of CVs with non-solar abundance, as an alternative to the post-SSXB scenario, which predicts more such CVs (a factor of ${2-6}$) than observed.
Other promising mechanisms are circumbinary disks \citep{SpruitTaam_2001,Willems_2005}, or a combination of both, e.g. if the circumbinary disk is fed by a fraction of the accreted matter that is re-emitted by WD due to the nova eruptions.

\subsection{AM\,CVns and Ultra-Compact X-ray Binaries}

As we mentioned earlier, AM\,CVns and UCXBs can be formed in three ways, either through two CE evolution, or through dynamically stable non-conservative mass transfer followed by CE evolution, or through CE evolution followed by dynamically stable non-conservative mass transfer.
Since the later channel from CVs/LMXBs has just been addressed, we focus on the other two, which have always a CE evolution for the second episode of mass transfer.
Similarly to CVs and LMXBs, after CE evolution, a detached \pceb~is formed and evolves towards shorter orbital periods due to orbital AML, leading to the onset of mass transfer, when the donor fills its Roche lobe.
However, unlike these systems, the main mechanism driving the evolution of AM\,CVns and UCXBs is always GR, since typically their donors are initially helium WDs or helium stars.

\subsubsection{Helium White Dwarf or Helium Star Donors}

Let us start with the case in which the donor is initially a helium WD.
This case is illustrated in Figures~\ref{FigEvolMdonor} and \ref{FigEvolMdot} by the dotted black track in each figure, which were taken from a binary model \citep{Wong_2021}.
Even though the orbital period corresponding to the onset of mass transfer depends on the mass ratio and on helium WD properties, mainly its radius, and in turn its mass, it is of order of a few minutes.
After that, the further binary evolution depends on the initial entropy and the subsequent thermal evolution of the donor, and is usually separated into three phases \cite[e.g.][]{Deloye_2007,Solheim_2010,Kaplan_2012,Wong_2021}.

In the very beginning, the donor contracts in response to mass loss as the outermost radiative layer is stripped off, leading to an increase in the mass transfer rate.
The mass transfer rate then grows from zero to its maximum, which can be up to ${\sim10^{-8}-10^{-6}}$~\Msun~yr$^{-1}$, depending the donor properties, while the donor radius decreases to a minimum value.
During this phase, the mass-loss time-scale becomes much shorter than the donor thermal time-scale, and the orbital period decreases as a response to mass transfer.
After this quick phase, which lasts ${\lesssim10^6}$~years, the donor radius evolution reverses and the donor expands in response to mass loss.
This happens because, during the first phase, the entropy profile eventually becomes sufficiently shallow that the expansion of the underlying layers starts to dominate.
The binary then evolves towards longer orbital periods driven by GR, with decreasing mass transfer rates.
During this second phase, which can last ${\sim1}$~Gyr, the mass-loss time-scale is much shorter than the donor time-scale and the donor responds adiabatically to the mass loss.
The third phase starts when the mass loss time-scale becomes comparable to or shorter than the donor thermal time-scale in the underlying layers.
This allows the donor to cool down and contract, although the contraction ends when the donor has shed sufficient entropy to reach its fully degenerate configuration.
This phase starts at an orbital period of ${\sim40}$~min.


Let us now address the case in which the companion of the compact object was very close to the tip of the FGB just before CE evolution, resulting in turn in a \pceb~hosting the compact object and a helium star.
In case the \pceb~orbital period is sufficiently short, the compact object companion will still be a helium star, i.e. still burning helium, at the onset of mass transfer.
The binary evolution for this case is illustrated in Figures~\ref{FigEvolMdonor} and \ref{FigEvolMdot} by the solid and short-dashed black tracks taken from two binary models \citep{Heinke_2013,Wang_2021}.
Depending on the mass, the lifetime of a helium star is typically ${\sim10-10^3}$~Myr, more massive stars having shorter lifetimes, and its radius can increase up to ${\sim30}$\% during its life \citep{Yungelson_2008}, and how large it can become strongly depends on the amount of hydrogen left after CE evolution \citep{BK2021}.
Thus, unlike the previous case, the onset of Roche lobe filling is driven by a combination of orbital AML due to GR and nuclear expansion of the donor.
Therefore, the orbital period at which the onset of mass transfer occurs depends on the masses of the compact object and helium star as well as on the evolutionary status of the helium star.
In particular, the longer the \pceb~orbital period, the more time the helium star has to evolve, becoming larger, and in turn the longer the orbital period at the onset of mass transfer.
Given the orbital AML rates provided by GR and the helium star lifetimes, to have a helium star at onset of mass transfer, the \pceb~orbital must be typically shorter than ${\sim2}$~h \citep{Yungelson_2008}.
After mass transfer starts the further evolution for systems hosting helium star donors is different from those having helium WDs \citep{Savonije_1986,Yungelson_2008,Solheim_2010,Wang_2021}.

During the Roche lobe overflow, the mass-loss time-scale is typically shorter than the thermal time-scale of the donor, which makes the donor slightly bloated as a response to mass loss.
This causes a rapid increase in the mass transfer rate, reaching values of a few $10^{-8}$~\Msun~yr$^{-1}$, after the residual hydrogen is entirely consumed.
In a more realistic situation, a non-negligible residual outer hydrogen envelope is left in the helium star after CE evolution, which plays a significant role during the early evolution \citep{BK2021}.
The donors can spend tens of Myr transferring this small amount of hydrogen-rich material at low rates (${\sim10^{-10}-\sim10^{-9}}$~\Msun~yr$^{-1}$), before finally transitioning to a phase where helium is transferred at much higher rates (${\gtrsim10^{-8}}$~\Msun~yr$^{-1}$).
Since the donor should be modestly evolving and expanding, in the absence of orbital AML, the orbital period should increase in response to mass loss.
However, since GR is sufficiently strong, the donor nuclear evolution time-scale is longer than the GR-driven mass-loss time-scale, which leads to convergent evolution, and the orbital period decreases.
A significant fraction of the energy released through nuclear burning is absorbed in the donor envelope, which leads to a decrease in its luminosity and an increase of its thermal time-scale.
As the binary evolves and the donor mass drops, nuclear burning quickly becomes less important, which makes the convective core to be replaced by an outer convection zone that penetrates inwards.
At the same time, as the orbital period decreases, the mass-loss time-scale also decreases, causing the mass transfer rate to slightly increase.
Towards the end of this initial phase, the donor thermal time-scale becomes much longer than the mass-loss time-scale, and the thermal structure of the donor is severely perturbed.
The mass transfer rate then increases faster, as the donor tries to remain within its Roche lobe. 
At some point, helium burning in the centre of the donor virtually vanishes and the donor is driven far out of thermal equilibrium.
The mass transfer rate then reaches its maximum and the orbital period its minimum.
This typically occurs when the donor mass is ${\sim0.20-0.26}$~\Msun, at an orbital period of ${\sim8-11}$~min.
Since the star becomes more degenerate, its radius starts to increase and the further evolution resembles that of systems hosting helium WD donors, in which the donor is evolving adiabatically in response to mass loss.
The mass transfer rate drops while the binary evolves towards longer orbital periods.

In case the compact object is a WD, some interesting features are expected in the first phases of AM\,CVn evolution hosting initially helium star donors.
It has been shown by binary models \citep{Fink_2007,Brooks_2015} that, if the initial mass of the helium star is ${\gtrsim0.4}$~\Msun~and the accreting WD mass is initially ${\gtrsim0.8}$~\Msun, enough mass is accumulated onto the accreting WD, leading to a first thermonuclear flash that is likely vigorous enough to trigger a detonation in the helium layer.
This thermonuclear runaway originated from the detonation in the helium layer of the accreting WD may create converging shock waves that reach the core.
If the carbon in the core is also detonated, the thermonuclear runaway is seen as a Type Ia supernova.
Those that survive the first flash and eject mass will have a temporary increase in orbital separation, but orbital AML through GR drives the donor back into contact, resuming mass transfer and triggering several subsequent
weaker flashes.

Despite the fact that the three evolutionary channels leading to AM\,CVns and UCXBs are in principle viable, there are theoretical arguments against two of them.
It has been argued that most, if not all, WDs accreting initially from helium WDs merge at the beginning of the evolution \citep{Shen_2015}.
This would happen due to nova eruptions during the initial phases of accretion that would trigger dynamical friction within the expanding nova shell, shrinking the orbit and causing the mass transfer rate to become dynamically unstable.
If that is true,  most, if not all, AM\,CVns would come from WDs initially accreting from helium stars or from CVs with subgiant or nuclear-evolved MS donors.
It has been argued, based on the RVJ prescription for orbital AML due to MB, that only a negligible fraction of AM\,CVns could have originated from CVs \citep{Goliasch_2015,Liu_2021}.
The reason for that is the narrow range of CV properties from which AM\,CVns originate.
However, the situation here is very similar to the fine-tuning problem related to the formation of close detached millisecond pulsar binaries.
As the CARB prescription, which predicts higher orbital AML rates than the standard RVJ prescription, can solve the millisecond pulsar fine-tune problem, it would not be surprising if it can also increase the probability of forming AM\,CVns from CVs \citep{BelloniSchreiber_2023}.

\subsubsection{Comparison with Observations}

After describing the main features of AM\,CVn and UCXB evolution, we can confront prediction with observations.
Let us start with a brief discussion of the persistent and transient sources with respect to what the disk instability model predicts.
By inspecting Figure~\ref{FigEvolMdot}, we can see that this model provides quite convincing results, similarly to the case of CVs and LMXBs, especially for UCXBs \citep{Heinke_2013}.
However, some AM\,CVns deserve attention, like ASASSN-14cc and ASASSN-14mv.
The disk instability model predicts that these two systems should be persistent, but they have undergone outbursts \citep{Ramsay_2018}.
In addition, SDSS~J1525$+$3600 and GP~Com have never shown outbursts \citep{Ramsay_2018}, although the model predicts that they should be transient.
Finally, despite their very low mass transfer rates (${\lesssim10^{-11}}$~\Msun~yr$^{-1}$), SDSS~J1208$+$3550, SDSS~J1137$+$4054, V396~Hya and SDSS~J1319$+$5915 are marked in the figure as persistent, since they have never been observed in outburst.
Even though it is very likely that the lack of outbursts related to the last four systems is due to the fact that their accretion disks are stable and cold, it is not clear what might be happening with the other systems.

When we compare the evolutionary tracks for the three channels leading to AM\,CVns and UCXBs in Figures~\ref{FigEvolMdonor} and \ref{FigEvolMdot}, we can see that they can reasonably well explain observations, and some particular systems can be better described by particular channels.
For instance, as noted in previous studies \citep{Heinke_2013}, UCXBs can be separated into three groups: (i) transients with orbital periods longer than ${\sim40}$~min, (ii) persistent with orbital periods shorter than ${\sim25}$~min, and (iii) transient with orbital period longer than that.
Since the mass transfer rate drops as an UCXB evolves with a WD donor, sources in the groups (i) and (ii) could in principle be explained by any scenario.
However, those in the group (iii), i.e. the long-period persistent UCXBs 4U~1626--67, 4U~1916--053, and 4U~0614$+$091, can apparently only be explained if the donor is initially a helium star or perhaps if they are descendants of LMXBs.
This is because if the donor is initially a helium WD, mass transfer rates in this orbital period range are expected to be similar to those in group (i), i.e. orders of magnitude lower than those inferred for the group (iii).

This could have important implications when comparing the population of UCXBs in the Milky Way field with those belonging to globular clusters.
As already noted before \citep{Zurek_2009,Heinke_2013}, the persistent UCXBs in globular clusters have orbital periods much shorter than those of 4U~1626--67, 4U~1916--053, and 4U~0614$+$091 \citep[the longest orbital period is $22.58$~min for M15~X--2,][]{Dieball_2005}.
According to \citet{Heinke_2013}, this fact could be naturally explained if long-period persistent UCXBs have initially helium star donors that ignite helium under non-degenerate conditions in the centre.
This is because UCXBs in globular clusters cannot form in such a way, and the reason for that is twofold.
First, the typical evolutionary time-scale for an UCXB to be seen as persistent is only a few tens of Myr \citep{Wang_2021}.
Second, for low metallicities, only MS stars with masses ${\gtrsim1.85}$~\Msun~will evolve without developing a degenerate helium core on the FGB \citep{Han_2002}, which is a much higher mass than that of \msto~in globular clusters \citep[${\sim0.8-0.9}$~\Msun, e.g.][]{Cohn_2010,Lugger_2017,Rivera_2018,Cohn_2021}.
This would then explain the lack of long-period persistent UCXBs in globular clusters.
However, helium stars can also originate from less massive stars, which undergo the helium core flash to ignite helium off-centre \citep[e.g.][]{Han_2002,BK2021}, and it is not clear why such helium stars should behave in a different way than those formed under non-degenerate conditions.
In addition, there is an over-abundance of bright X-ray sources in globular clusters in comparison to the Milky Way field, which indicates that dynamics are playing a significant role in their formation.
Furthermore, it has been shown that strong MB can lead to mass transfer rates high enough to explain long-period persistent UCXBs \citep{Van_2019,Van2019CARB}.
Therefore, the difference in the orbital period distribution of persistent UCXBs in the two environments most likely arises from a more complex interplay between dynamics and binary evolution than previously thought by \citet{Heinke_2013}.

There is one AM\,CVn (SDSS~J1908$+$3940) that seems challenging to be explained by any evolutionary channel. 
This system has an orbital period of ${\sim18}$~min and an inferred mass transfer rate of ${\sim4\times10^{-7}-10^{-6}}$~\Msun~yr$^{-1}$ \citep{Ramsay_2018}.
For its orbital period, this mass transfer rate is orders of magnitude higher than expected from any type of initial donor, which is ${\lesssim10^{-8}}$~\Msun~yr$^{-1}$.
Future observational efforts and binary models with focus on this system could help to understand whether any of these formation channels could explain its properties.
If not, perhaps something atypical is occurring with this system that pushed its inferred mass transfer rate to such a high value.

Unfortunately, except for a few particular systems, we still do not know which of these three formation channels is the dominant among AM\,CVns and UCXBs, if any is dominating at all, which compromises studies of their intrinsic populations.
It seems clear as well that, for the overwhelming majority of systems, having only the orbital period, donor mass and mass transfer rate are not enough to distinguish between the different evolutionary scenarios, and more information is needed.
There are a few ways to better evaluate the contribution of each channel, by gathering more information about these systems.

For instance, by identifying and studying the properties of potential progenitors, such as the system LAMOST~J0140355$+$392651 \citep{ElBadry_2021}.
This system has an orbital period of $3.81$~h and hosts a bloated, relatively cool, low-mass (${\sim0.15}$~\Msun) proto WD and a massive (${\sim0.95}$~\Msun) WD companion.
Its properties can be well-explained as a CV entering the detached phase, when the donor envelope has been entirely consumed.
In this case, the proto WD was a nuclear-evolved MS star before the onset of mass transfer.
The system will evolve through this detached phase due to GR and mass transfer will eventually resume, after a few Gyr.
This sort of investigation provides key information to understand binary evolution leading to AM\,CVns and UCXBs.
However, the number of these systems is far too low to provide conclusive evidences for a particular channel being dominant or not.

Another possibility is by constraining the properties of their donors with detailed analyses of their chemical composition, which allows us to investigate in more detail their evolutionary history \citep{Nelemans_2010REF}.
Those formed through CV/LMXB evolution are expected to retain a greater fraction of their hydrogen in comparison with systems formed in the other two channels.
This would imply that the presence of hydrogen in their spectra is needed to rule out the other channels as well as systems not exhibiting hydrogen in their spectra are most likely not formed in such a way.
In both helium WD and helium star channels, the transferred material is expected to be mainly composed of helium with CNO-processed material, mainly nitrogen, and abundance ratios of these elements are believed to be an effective discriminant between these two channels.
A third possibility is by means of the donor mass-radius measurements obtained from precise analyses of eclipsing systems, since each evolutionary channel predicts on average different radius for the same mass \citep{Deloye_2007,Yungelson_2008,Goliasch_2015}, although some overlap is also expected depending on the initial properties of the donor as well as how the WD donor cooling is treated.
In general, systems having initially helium WD donors evolve with smaller radii than systems having initially helium star donors, and these evolve with smaller radii than those originated from CVs/LMXBs.

The last option we discuss is applicable to UCXBs and consists of investigating the existence/lack of thermonuclear explosions that occur on the surface of the NS as well as their features.
It has been proposed that there is a strong correlation between the iron K$\alpha$ line strength in their X-ray spectra and the abundance of carbon--oxygen in the accretion disc \citep{Koliopanos_2013}.
This line occurs at ${\sim6.4-6.9}$~keV and is a result of the reflection of the radiation by the accretion disk and the surface of the WD facing the NS, and for this reason it can provide clues on the disk chemical composition.
This correlation coupled with the fact that UCXBs with prominent and persistent iron K$\alpha$ emission also show bursting activity, it has been proposed that UCXBs with persistent iron emission have helium-rich donors, while those that do not, likely have carbon--oxygen or oxygen--neon--magnesium-rich donors \citep{Koliopanos_2021}.
This possibility can be further explored, but more elaborated X-ray reflection models will be eventually needed.


Having in mind these possibilities, we illustrate with eclipsing AM\,CVns how they may help to disentangle the formation channels.
Thus far, there are seven eclipsing AM\,CVns with accurate measurements of the masses and radii of their donors \citep{Copperwheat_2011,Green_2018,vanRoestel_2021}.
When considering only the radii and masses, YZ~LMi can be easily explained by either the helium WD or the helium star channels, ZTF~J0407--0007 by the helium WD channel, and ZTF~J2252--0519 by the helium star channel.
The remaining (Gaia14aae, ZTF~J1637$+$4917, ZTF~J0003$+$1404 and ZTF~J0220$+$2141) have very large radii, which could be only explained by the CV channel.
However, when the chemical properties of these seven systems are taken into account, the situation becomes much more complicated.
Regarding those systems apparently formed through the CV channel, since no hydrogen was detected in the spectra of any of them, it seems more likely that they have formed in a different way.
Therefore, the situation for these systems apparently seems rather challenging.
A possibility to bring into agreement predicted and observed WD donor radii and chemical properties is by avoiding the phase in which the donor can contract and cool down.
This phase is predicted to exist because, at some point, the donor thermal time-scale becomes comparable or shorter than the mass-loss time-scale, since the orbital AML rate is continuously dropping as the AM\,CVn evolves.
If there is an additional source of orbital AML, e.g. consequential AML, such that the donor cooling is significantly delayed, then the donors could keep their large sizes, even when the orbital period is longer than ${\sim40}$~min.
However, this possibility at the moment is very speculative and future detailed binary models could verify whether consequential AML is able to do that, and if so, for which mechanisms and under which conditions.

After describing compact objects accreting from Roche lobe filling donors, we can turn to a discussion of other modes of transfer, in which the donor is not filling the Roche lobe and is either a low-/intermediate-mass-evolved red giant, or a supergiant, or a Wolf--Rayet star.

\subsection{Symbiotic Stars and Symbiotic X-ray Binaries}

We have argued before that, in order to explain their orbital periods (${\sim200-10^3}$~days), most S-type SySts should form through an episode of dynamically stable non-conservative mass transfer.
After this event, the resulting post-stable-mass-transfer binary, with an orbital period longer than ${\sim100}$~days, hosts a WD orbiting a low-mass MS star or perhaps a subgiant or red giant, depending on the mass ratio of the zero-age MS--MS binary.
On the other hand, those S-type SySts hosting massive WDs as well as SyXBs are most likely formed through CE evolution, leading to long-period \pcebs, since dynamically stable non-conservative mass transfer cannot explain their properties.
Finally, given the very long orbital periods of D-type SySts (${\gtrsim40}$~years), their WD progenitors are not expected to form during an episode of Roche lobe filling, i.e. their WDs are formed similarly to single WDs.
In the three situations, given the relatively long orbital period of the binary, the WD companion has enough room to evolve and expand, and the binary will become a SySt or a SyXB under certain conditions.
Even though in what follows we discuss the evolution of these systems with focus on SySts, the case of SyXBs is very similar, except from the fact that the compact object is a NS.

The donors in SySts are either located near the tip of the FGB or on the AGB \cite[e.g.][]{Gromadzki_2013,Mikolajewska_2007}.
The typical luminosity of accreting WDs in SySts is ${\sim10^2-10^4}$~\Lsun~\citep{Mikolajewska_2010}, and in most cases these are powered by nuclear burning due to the very high accretion rates, which are typically ${\sim10^{-8}-10^{-7}}$~\Msun~yr$^{-1}$.
In addition, in most SySts, the red giant is under-filling its Roche lobe filling \citep{Mikolajewska_2012,Boffin_2014}, but in some cases it is nearly filling its Roche lobe, as evidenced by the clear signature of ellipsoidal variations in their light curve \citep{Gromadzki_2013}.
Therefore, in what follows, we discuss two promising modes of mass transfer for these systems, which are (i) atmospheric Roche lobe overflow \citep{Ritter_1988}, if ${0.9\,R_{\rm RL} \lesssim R_{\rm d} \lesssim R_{\rm RL}}$, or (ii)  gravitationally focused wind accretion \citep{Mohamed_2007,Mohamed_2012,Abate_2013,ValBorro_2009,Skopal_2015,Borro_2017}, if ${R_{\rm d} \lesssim 0.9\,R_{\rm RL}}$.

\subsubsection{Atmospheric Roche Lobe Overflow}

As the red giant evolves, its radius approaches its Roche lobe, and at some point the mass transfer rate provided by Equation~\ref{eq:Ritter} will become relevant.
If the red giant mass is not much higher than the compact object mass, $\zeta_{\rm ad}>\zeta_{\rm RL}$, and mass transfer is dynamically stable.
We show in Figure~\ref{FigEvolMdonor}, as dashed green lines, several evolutionary tracks in which the initial point-mass compact object mass is $1.2$~\Msun, and the zero-age MS mass is $1.25$.
For those with initial orbital periods longer than ${\sim5}$~days but shorter than ${\sim100}$~days, the onset of mass transfer occurs when the donor is an unevolved red giant on the FGB.
The mass transfer in this case is similar to what is depicted in Figure~\ref{FigStable}, and it is characterized by an initial phase of thermally unstable mass transfer, until the donor mass is sufficiently low such that the mass transfer stabilizes.
After that, mass transfer proceeds on the nuclear time-scale of the donor.
The mass transfer rates nicely correlate with the evolutionary status of the donor.
First, the longer the initial orbital period, the more evolved the donor at the onset of mass transfer.
Consequently, the more evolved the donor, the higher the mass transfer rates.
For instance, during thermally unstable mass transfer, the rates for the binaries with shortest and longest initial orbital period are ${\sim3\times10^{-8}}$ and ${\sim6\times10^{-6}}$~\Msun~yr$^{-1}$, respectively.
During nuclear time-scale mass transfer, the rates for the same binaries are ${\sim5\times10^{-9}}$ and ${\sim7\times10^{-8}}$~\Msun~yr$^{-1}$, respectively.

There are at least three SySts in which their red giant donors are very close to filling their Roche lobes and their WDs are massive enough so that mass transfer can be regarded as dynamically stable.
T~CrB is a well-studied SySt with a relatively short orbital period ($227$~days), hosting a massive carbon--oxygen WD ($1.37\pm0.13$~\Msun) and a red giant with mass $1.12\pm0.23$~\Msun~\citep{Stanishev_2004}.
RS~Oph is another well-known SySt, which also hosts a massive carbon--oxygen WD ($1.2-1.4$~\Msun).
However, its orbital period ($453$~days) is longer than that of T~CrB and it hosts a less massive red giant donor \citep[${0.68-0.80}$~\Msun,][]{Brandi_2009}.
This likely indicates that both SySts could share similar evolutionary sequences, in which the orbital period is expanding as the red giant is consumed.
The last SySt we know in which the red giant donor is nearly filling its Roche lobe is V3890~Sgr \citep{Mikolajewska_2021}.
Similarly to the other two systems, this SySt also hosts a massive carbon--oxygen WD ($1.35\pm0.13$~\Msun).
However, its longer orbital period ($747$~days) combined with the mass of its red giant donor ($1.05\pm0.11$~\Msun) suggests that the evolutionary pathway of V3890~Sgr is slightly different from those of T~CrB and RS~Oph, starting at a longer orbital period.


The properties of T~CrB and RS~Oph can be nicely reproduced if the \pceb~hosts a WD with mass ${\sim1.15}$~\Msun, and a MS star companion with mass ${\sim1.4}$~\Msun, as illustrated in Figure~\ref{FigEvolMdonor} by the solid green track passing through these two systems.
As mentioned earlier, T~CrB and RS~Oph could indeed be explained with the same evolutionary sequence, if the WD mass in the former is smaller, which would imply that T~CrB may evolve to something similar to RS~Oph.
In order to reproduce the properties of these two systems with the same evolutionary sequence, the \pceb~orbital period needs to be as long as ${\sim200}$~days, and the fraction of mass lost from the vicinity of the WD due to isotropic re-emission needs to be $\beta_{\rm ml}\sim0.6$.
The predicted masses of the WD and red giant during the T~CrB stage are ${\sim1.25}$ and ${\sim1.07}$~\Msun, and during the RS~Oph stage, they are ${\sim1.38}$ and ${\sim0.67}$~\Msun.
The WD will exceed the Chandrasekhar limit and explode in a Type Ia supernova when the orbital period is ${\sim500-800}$~days, depending on the assumed Chandrasekhar limit.
After the onset of mass transfer, the binary reaches the T~CrB stage after ${\sim0.2}$~Myr.
The transition from the T~CrB stage to the RS~Oph stage takes ${\sim1}$~Myr, and from RS~Oph stage to the Type Ia supernova ${\lesssim1-4}$~Myr, depending on the assumed Chandrasekhar limit.

The properties of V3890~Sgr can also be reproduced with similar WD and companion masses after CE evolution, as well as the same mass-loss fraction during SySt evolution.
However, the \pceb~orbital period must be much longer (${\sim600}$~days) than in the previous case.
An example of evolutionary sequence (rightmost solid green track) is shown in Figure~\ref{FigEvolMdonor}.
The predicted masses of the WD and red giant at the orbital period of V3890~Sgr are ${\sim1.25}$ and ${\sim0.96}$~\Msun.
Similarly to the previous case, the evolution will terminate when the WD exceeds the Chandrasekhar limit.

\begin{figure*}[htb!]
\centering
\includegraphics[width=0.99\linewidth]{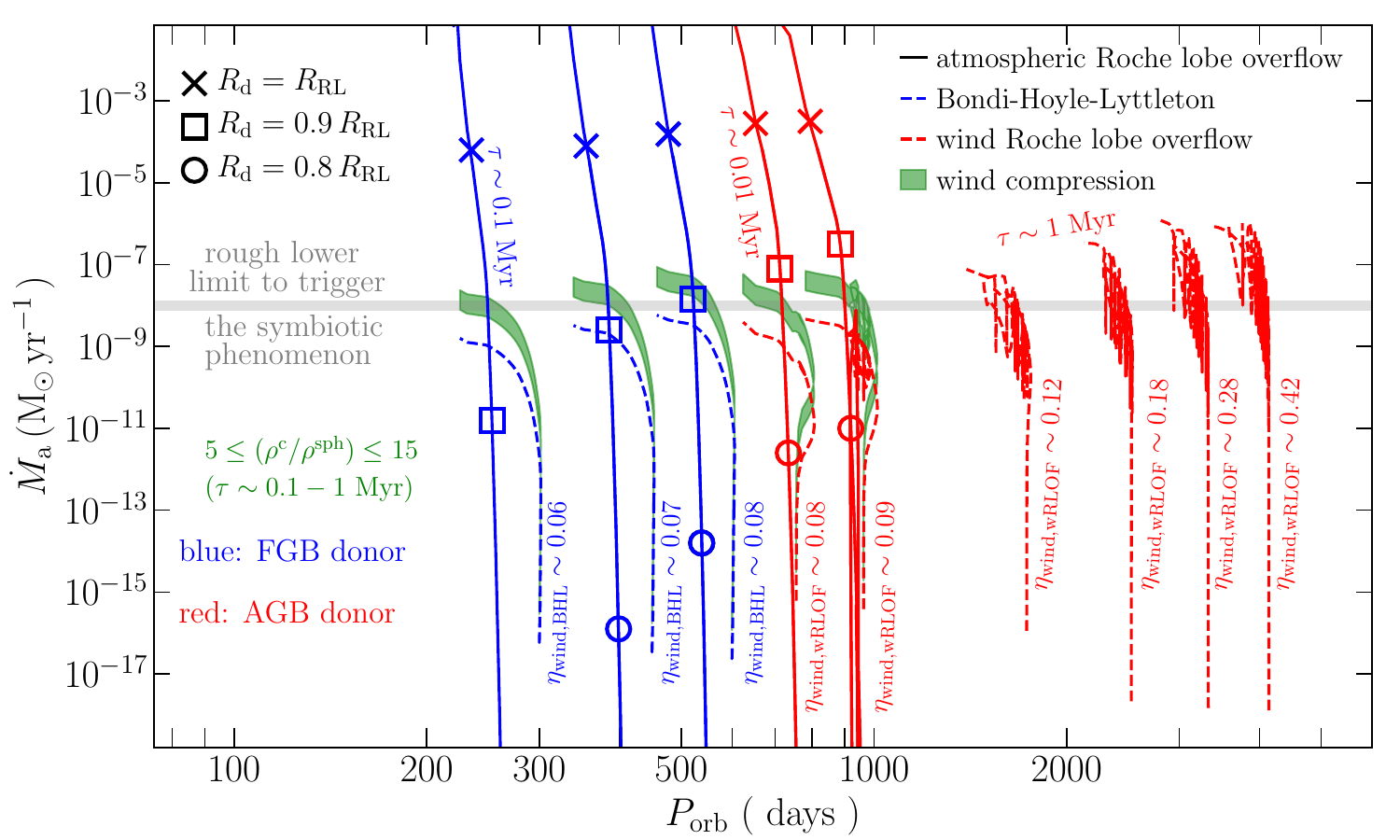}
\caption{Predicted accretion rate ($\dot{M}_{\rm a}$) onto the WD during pre-SySt and SySt evolution, as a function of the orbital period $P_{\rm orb}$, according to the atmospheric Roche lobe overflow model (Equation~\ref{eq:Ritter}), the \bhl~model (Equations~\ref{eq:MdotAcc} and \ref{eq:BH}, with $\alpha_{\rm wind}=1.5$), the wind Roche lobe overflow model (Equations~\ref{eq:MdotAcc}, \ref{eq:wRLOF}, and \ref{eq:Rdust}, with {${T_{\rm dust}=1\,500}$\,K} and {${p=1}$}), and the wind compression model (Equation~\ref{eq:WC}, assuming ${5\leq\rho^{\rm c}/\rho^{\rm sph}\leq15}$).
The grey thick horizontal line marks the minimum accretion rate such that the system can be considered a SySt, which is ${\sim10^{-8}}$~\Msun~yr$^{-1}$.
The evolutionary tracks were computed by us using the \mesa~code and correspond to some of those shown in Figure~\ref{FigEvolMdonor}.
For these selected tracks, we start the simulations with a zero-age MS star of mass $1.6$~\Msun~orbiting a point-mass WD with mass $0.53$~\Msun, which are consistent with the average values for the red giant and WD in SySts \citep{Mikolajewska_2003}.
We chose nine different initial orbital periods, from left to right: $300$, $450$, $600$, $800$, $1000$, $1700$, $2400$, $3100$, and $3800$~days, and evolve each one of them until the onset of CE evolution.
In all tracks, the orbital period increases/decreases as the SySt evolves mainly as a response to red giant donor spin evolution, since it quickly becomes synchronized with the orbit while ascending the FGB.
For the three binaries with shortest initial orbital period, Roche lobe filling occurs when the donor is on the FGB, while for the remaining binaries it takes place when the donor is on the AGB.
In case the orbital period during the SySt phase is shorter than $10^3$~days, we computed the accretion rates provided by the atmospheric Roche lobe overflow model (solid lines) and those obtained with gravitationally focused wind accretion, for which we assumed either the \bhl~model, if the donor is on the FGB (dashed blue lines), or the wind Roche lobe overflow model, if the donor is on the AGB (dashed red lines), or the wind compression model (green areas).
Furthermore, we only show the accretion rates provided by the wind Roche lobe overflow model when the orbital periods are longer than $10^3$~days.
The average \bhl~and wind Roche lobe overflow accretion rate efficiencies ($\eta_{\rm wind,BHL}$ and $\eta_{\rm wind,wRLOF}$) are indicated in the figure, while those for the compression wind model are given by $(5-15)\times\eta_{\rm wind,BHL}$.
Also indicated is the typical SySt lifetime ($\tau$), based on the atmospheric Roche lobe overflow and wind compression model, for orbital periods shorter than $10^3$~days, and based on the wind Roche lobe overflow model, for orbital periods longer than that.
SySts with orbital periods shorter than $10^3$~days cannot be easily explained by the \bhl~model, nor the wind Roche lobe overflow model, as these mass transfer modes do not provide accretion rates high enough to trigger the symbiotic phenomenon.
However, the required rates in this period range can be obtained with the wind compression model, if the accretion efficiency is ${\gtrsim5}$ times higher than that from the \bhl~model.
On the other hand, the wind Roche lobe overflow model is clearly able to explain SySts with AGB donors at orbital periods longer than $10^3$~days.
Finally, the atmospheric Roche lobe overflow model always provides sufficiently high accretion rates, when $R_{\rm d}\gtrsim0.85-0.9\,R_{\rm RL}$.
}
\label{FigEvolMdotSySt}
\end{figure*}

As mentioned earlier, mass transfer will be dynamically stable only if the red giant mass is comparable to or only slightly more massive than the compact object mass.
This condition is usually not satisfied in SySts \citep{Mikolajewska_2003}, since their WD masses are typically ${\lesssim0.6}$~\Msun, while their red giant donors usually have masses of ${\sim1.2-2.2}$~\Msun, implying mass ratios ${\gtrsim2.5}$.
Such mass ratios are likely high enough to cause mass transfer to be dynamically unstable.
In this case, as the red giant radius approaches the Roche lobe radius, the mass transfer rates provided by Equation~\ref{eq:Ritter} always increase, leading to CE evolution.
This behaviour is illustrated in Figure~\ref{FigEvolMdotSySt} by the solid lines.
The evolutionary tracks in this figure correspond to some of the blue tracks in Figure~\ref{FigEvolMdonor}, and they all start with a zero-age MS star with mass $1.6$~\Msun~orbiting a point-mass WD with mass $0.53$~\Msun, but each one of them starts at a different orbital period. 
Before the onset of CE evolution, the atmospheric Roche lobe overflow model provides accretion rates that can only explain SySts that are close filling their Roche lobe, i.e. ${R_{\rm d}\gtrsim0.90\,R_{\rm RL}}$.

Moreover, the SySt lifetime predicted by this model is ${\sim0.01-0.1}$~Myr.
This is orders of magnitude shorter than those typically assumed in studies that try to estimate the expected number of SySts in the Milky Way \cite[e.g.][]{Kenyon_1993}.
In general, regardless of the assumptions, these investigations predict much more Milky Way SySts than observed.
This problem could be partially/completely overcome with the atmospheric Roche lobe overflow model, if the intrinsic population of S-type SySts in the Milky Way hosts red giants close to Roche lobe filling.
However, if only a small fraction in the intrinsic population shows evidence for ellipsoidal variability, like currently observed \citep{Gromadzki_2013}, then this model most likely does not correspond to the dominant mode of mass transfer for these systems.
In what follows, we discuss the case in which the red giant is strongly under-filling its Roche lobe.

\subsubsection{Gravitationally Focused Wind Accretion}

In the case of low-mass red giants, the mass transfer rates provided by Equation~\ref{eq:Ritter} in the atmospheric Roche lobe overflow model are only relevant if ${R_{\rm d} \gtrsim 0.85-0.9\,R_{\rm RL}}$.
If this condition is not satisfied, then the accretion rates onto the WD are negligible, and in turn this model cannot explain the observed WD luminosities in SySts.
This implies that mass transfer must be associated with accretion of part of the stellar winds from the red giant, i.e.

\begin{equation}
\dot{M}_{\rm a} \ = \ -\,\eta_{\rm wind}\,\dot{M}_{\rm wind} \ ,
\label{eq:MdotAcc}
\end{equation}

\noindent
where $\eta_{\rm wind}$ is the orbit-averaged wind accretion rate efficiency.
Observations suggest that the wind mass-loss rate from a typical red giant donor in SySts is ${\gtrsim10^{-7}}$~\Msun~yr$^{-1}$ \cite[e.g.][]{Skopal_2005}, implying that ${\eta_{\rm wind}\gtrsim0.1}$ to reach the accretion rates of ${\gtrsim10^{-8}-10^{-7}}$~\Msun~yr$^{-1}$ needed to explain the hydrogen nuclear burning.

The most simple case of wind accretion occurs when we assume that stellar winds are supersonic and spherically symmetric, which is the so-called \bhl~model \citep{HL39,BH44,Edgar_2004}.
In this case, the WD can accrete part of the stellar wind as it orbit through it and the accretion rate efficiency, i.e. the fraction of the stellar winds that is effectively accreted by the WD, can be estimated as

\begin{equation}
\eta_{\rm wind,BHL} \ = \ 
\frac{\alpha_{\rm wind}}{2\sqrt{1\,-\,e^2}} \,
\left(
      \frac{G\,M_{\rm a}}{a\,v_{\rm wind}^2}\,
\right)^2 \,
\left[1\,+\,
    \left(
      \frac{v_{\rm orb}}{v_{\rm wind}}
    \right)^2\,
\right]^{-3/2} \ \leq 1 \ ,
\label{eq:BH}
\end{equation} 
\

\noindent
where $v_{\rm wind}$ and $v_{\rm orb}$ are the wind and orbital velocities, respectively, and $\alpha_{\rm wind}$ is a parameter to be adjusted based on observations.

Depending on the properties of the red giant and the orbital separation, the winds can be gravitationally focused towards the orbital plane and more easily accreted by the WD.
Even though there are a few models available to estimate gravitationally focused wind accretion rates \cite[e.g.][]{Abate_2013,Skopal_2015,Saladino_2019}, we will only consider two models to illustrate how this mechanism can lead to accretion efficiencies much higher than those obtained with the \bhl~model.
These two models are (i) the wind Roche lobe overflow model \citep{Mohamed_2007,Mohamed_2012,Abate_2013} and (ii) the wind compression model \citep{BC1993,Skopal_2015}, and they look very promising in the context of SySts.

Let us start with the wind Roche lobe overflow model.
During the AGB phase, winds are most likely driven by a combination of pulsation-induced shock waves and radiation pressure on dust grains \citep{Hofner_2018}.
Stellar pulsation and convection induce strong shock waves in the extended atmosphere, pushing the gas outwards, which reaches cooler regions.
Since most of the gas in a cool AGB star is in molecular form, some species may condense into dust grains, giving rise to a dusty shell around the star.
These dust grains are then accelerated outwards by radiation pressure and the gas is dragged along.
In the presence of a companion, the gas can remain within the AGB Roche lobe, if the wind acceleration radius is larger than the AGB Roche lobe radius \citep{Abate_2013}.
When this happens, instead of having the red giant effectively filling its Roche lobe due to nuclear evolutionary processes, the slow winds fill its Roche lobe, enhancing in turn the wind accretion onto the WD \citep{Mohamed_2007,Mohamed_2012,ValBorro_2009,Borro_2017,Saladino_2019}.
This implies that the accretion rates predicted by the wind Roche lobe overflow model are expected to be higher than those provided by the \bhl~model, but smaller than those obtained with the atmospheric Roche lobe overflow model.

The accretion rate efficiency in the wind Roche lobe overflow model can be expressed as \citep{Abate_2013}

\begin{equation}
\eta_{\rm wind,wRLOF} \ = \
q^2\,\left(\,\frac{25}{9}\,\right)\,
\left[\,
-\,0.284\,
  \left(\,
\frac{R_{\rm dust}}{R_{\rm RL}}\,
  \right)^2\,
+\,0.918\,
\left(\,
  \frac{R_{\rm dust}}{R_{\rm RL}}\,
\right)\,
-\,0.234\,
\right] \ \leq 0.5 \ ,
\label{eq:wRLOF}
\end{equation}
\

\noindent
where the maximum value $0.5$ comes from the results of hydro-dynamical simulations \citep{Mohamed_2007,Mohamed_2012}, and $R_{\rm dust}$ is the dust condensation radius, corresponding the radius of the wind acceleration zone, given by \citep{Hofner_2007}

\begin{equation}
R_{\rm dust} \ = \ 
\left(  \,
   \frac{R_{\rm d}}{2} \,
\right) \,
\left( \,
\frac{T_{\rm dust}}{T_{\rm eff}} \,
\right)^{-(4+p)/2}
\label{eq:Rdust}
\end{equation}
\

\noindent
where $T_{\rm dust}$  is the dust condensation temperature and $p$ is a parameter characterizing wavelength dependence of the dust opacity.


Another way to enhance wind accretion is by means of asymmetric radiation-driven winds from rotating red giants.
Based on the wind compression disk model \citep{BC1993}, which explains the equatorial circumstellar disks in rapidly rotating Be stars, it has been proposed that gravitational focusing in SySts can be achieved if the radiation-driven wind from the red giant donor is not spherically symmetric and is partially confined to the equatorial plane around the donor \citep{Skopal_2015}.
Compared to Be stars, normal red giants are slow rotators.
Despite that, red giants in SySts rotate faster than isolated stars with comparable spectral type \citep{Zamanov_2007,Zamanov_2008}, and this mechanism can still work, leading to the formation of a circumstellar gaseous disk-like component around the red giant donor.
Therefore, part of the wind leaving the red giant can be compressed to its equatorial plane, which increases the mass loss around the equatorial plane, proving then accretion rate efficiencies that are enhanced in comparison with the spherically symmetric case, i.e. the \bhl~model.

The required assumption that winds are asymmetric seems to be consistent with observations.
For instance, based on the velocity profile of the wind from the giant near the orbital plane of the eclipsing SySts EG~And and SY~Mus, it was possible to conclude that their winds are not spherically symmetric \citep{Skopal_2016}.
In addition, the properties of the nebular [\Ion{O}{iii}]~$\lambda$5007 line in EG~And provides further support for the wind being focusing towards the orbital plane \citep{Skopal_2021}.

The accretion rate efficiency in the wind compression model can be expressed as \citep{Skopal_2015}

\begin{equation}
\eta_{\rm wind,WC} \ = \ \eta_{\rm wind,BHL} \, 
\left( \, 
    \frac{\rho^{\rm c}(r)}{\rho^{\rm sph}(r)} \,
\right)
\ ,
\label{eq:WC}
\end{equation}
\

\noindent
where $r$ is the distance between the center of the red giant and the WD, $\rho^{\rm sph}(r)$ is the density of the spherically symmetric wind at a distance $r$, and $\rho^{\rm c}(r)$ is the local density of the compressed wind at a distance $r$ in the equatorial plane.
The quantity $\rho^{\rm c}(r)/\rho^{\rm sph}(r)$ also depends on the radius and rotational velocity of the red giant, on the initial and terminal wind velocities, and on how the wind is accelerated.
For typical rotational velocities of red giants in SySts \citep[${\sim6-10}$~km~s$^{-1}$,][]{Zamanov_2008} and wind terminal velocities (${\sim20-50}$~km~s$^{-1}$), the accretion rate efficiencies obtained with the wind compressional model, in comparison with those from the \bhl~model, can be enhanced by a factor of ${\sim5-15}$ \citep{Skopal_2015}.

We compare in Figure~\ref{FigEvolMdotSySt} the accretion rates onto the WDs in SySts predicted by the atmospheric Roche lobe overflow model, the \bhl~model, the wind Roche lobe overflow model, and the wind compression model.
It is clear from the figure that the \bhl~model cannot account for the accretion rates in SySts.
On the other hand, the wind Roche lobe overflow model can reasonably well explain the SySts with orbital periods longer than a few thousand days.
However, for those systems with orbital periods shorter than ${\sim10^3}$~days, this model is most likely unable to provide the required accretion rates, since in this orbital period range it provides accretion rate efficiencies comparable to the \bhl~model \citep{Abate_2013}.
The wind compression model is a good candidate to explain SySts in this range, as it provides accretion rates ${\gtrsim10^{-8}}$~\Msun~yr$^{-1}$.
As final comment, the SySt lifetimes predicted by these gravitationally focused wind accretion models (${\sim0.1-1}$~Myr) may also help to bring into agreement predicted and observed numbers of Milky Way SySts, although further and more detailed calculations are needed to better address this issue.

\subsection{Supergiant and Wolf--Rayet High-Mass X-ray Binaries}

We have just described the modes of mass transfer in SySts and SyXBs.
Given that sg-HMXBs and WR-HMXBs are the massive counterparts of these systems, the same modes could be in principle directly applicable to sg-HMXBs and WR-HMXBs.

Let us start with sg-HMXBs hosting NSs in which the supergiants are nearly filling their Roche lobes, such as Cen~X--3, SMC~X--1, LMC~X--4 \citep{vanderMeer_2007}, and XMMU~J013236.7$+$303228 \citep{Bhalerao_2012}.
The mass transfer in these systems can be described by the atmospheric Roche lobe overflow mode in which the extended photosphere of the supergiant is in contact with the Roche radius.
In this case, matter flows to the compact object through the inner Lagrangian point and an accretion disk around the NS is formed.
Given that the donor masses in these systems are several times higher than the masses of their NS companions, mass transfer will quickly become dynamically unstable mass transfer, triggering CE evolution.
The most likely outcome of this situation is a complete coalescence of the system, leading to the formation of a Thorne–\.Zytkow object \citep{TZ1975}, which would be a very cool red supergiant with a degenerate neutron core, although these objects remain thus far hypothetical.

In other sg-HMXBs and WR-HMXBs, their donors are under-filling their Roche lobes, and the most likely mode of mass transfer for them is gravitationally focused wind accretion.
An important implication of this mass transfer mode is that it can naturally explain wind-fed ultraluminous X-ray sources \citep[ULXs,][]{Kaaret_2017}, such as the WR-HMXB M101~ULX--1 \citep{Mellah_2019b}.
In addition, it also offers an explanation for the existence of accretion disks in the WR-HMXBs IC~10~X--1 and NGC~300~X--1, since this model is consistent with the formation of a wind-captured disk \citep{Mellah_2019a}.
Moreover, results from recent binary models indicate that the contribution of wind-fed ULXs to the overall ULX population is up to ${\sim75-96}$\% for young (${<100}$~Myr) star-forming environments and up to ${\sim49-87}$\% for environments like the Milky Way disk \citep{Wiktorowicz_2021}.
All this implies that Roche lobe filling is not the only mode of mass transfer able to provide large accretion rates onto the NS/BH in ULXs.

As a final comment, there is potential important application of gravitationally focused wind accretion in the context of HMXBs, which is related to the formation of Be-HMXBs hosting BHs.
Having the uncertainties in the stability criterion for dynamical mass transfer in mind \citep{Olejak_2021,Belczynski_2021_uncer}, binary models suggest that these systems should form through an episode of CE evolution \citep{Grudzinska_2015}.
In case there is no efficient mass and angular momentum transfer from the BH progenitor to the B star, the \pceb~will host a Wolf--Rayet star orbiting a slowly rotating B star, which will not exhibit the typical characteristics of Be-HMXBs.
However, this problem may be overcome in case of efficient wind accretion prior the CE evolution, such that the B star would enter this event already as a rapidly rotating star.
Further binary models, with consistent spin-up prescription, are still needed to better evaluate this possibility.

\section{Conclusion}

Multi-wavelength observations of several types of binaries as well as large systematic surveys, such as the Sloan Digital Sky Survey, have provided new observational constraints on theories of the formation and evolution of accreting compact objects.
Together with improved numerical modelling, these constraints have led to substantial progress in our understanding of the formation of these fascinating objects that have been outlined in the previews sections of this review.  
Despite these improvements, there are still important problems to be solved.
In what follows, we will list the most critical open questions and potential avenues towards obtaining answers.

For the formation of accreting compact objects, mass transfer from a giant star to its companion star is fundamental. We clearly need to better understand the conditions that lead to dynamically stable and unstable mass transfer.
This will only be possible if we find ways to test which assumptions for the mass transfer equation, the radius of the donor and the Lagrangian radius provide more realistic results. 
Comparison of detailed model predictions with observed samples of stars that evolve through dynamically stable mass transfer or common-envelope evolution may perhaps allow us to reach that goal. 
Post-common-envelope binaries consisting of WDs and MS star companions may be numerous enough and easy enough to characterize to provide the required observational constraints.
The increasing number of merging binary black holes/neutron stars detected by the LIGO and Virgo interferometers as well as combined gravitational wave and electromagnetic observations of compact binary mergers will also provide useful and strong multi-messenger constraints in the high-mass regime.

Very much related to the point above, we need to better understand common-envelope evolution.
If we ever want to understand the relative numbers of accreting compact objects and different binaries hosting compact objects, we need to be able to reliably predict the outcome of common-envelope evolution for a given system, which apparently is currently not the case.
To reach this goal, on one hand we need to aim at developing modelling tools that cover the different physical processes and time-scales involved in common-envelope evolution.
On the other hand, observational surveys that provide clean and ideally volume limited samples of close binaries can provide key information for theoretically unconstrained parameters.
First steps in this direction have been performed for white dwarf binaries with low-mass star companions, but large, complete, and unbiased samples are still not available.

Dynamically stable non-conservative mass transfer is potentially the dominant (but most likely not the only) formation mechanism for systems hosting white dwarfs with orbital periods between ${\sim100}$ and ${\sim1000}$~days.
These systems are often found in eccentric binaries, which should not be the case if our understanding of tidal evolution is correct.
Despite the several proposed mechanisms to enhance the eccentricity during dynamically stable non-conservative mass transfer, they all fail to some extent.
Either the theory of tidal interaction and/or the eccentricity pumping mechanisms need serious revision, or another more promising, still unknown, mechanism is needed.
On the top of that, we have virtually no idea how much mass and angular momentum are lost during dynamically stable non-conservative mass transfer, which makes predictions rather uncertain. 
Given that mass loss is a huge problem already in single star evolution, it is unlikely that we will soon find the answer by performing theoretical simulations. 
Decisive progress therefore again depends to a significant degree on observational constraints from samples of post-mass-transfer systems.

A problem equally serious as those listed above is that we do not properly understand orbital angular momentum loss in close binaries with both accreting and non-accreting compact objects. 
Problems related very likely to our ignorance of magnetic braking and consequential angular momentum loss are numerous when we confront standard prescriptions with observations, e.g. the average white dwarf mass in cataclysmic variables is much larger than expected, mass transfer rates in cataclysmic variables and low-mass X-ray binaries inferred from observations are very different from those predicted by our models, the predicted donor masses in black hole low-mass X-ray binaries are significantly higher than observed, the observed orbital period minimum in cataclysmic variables is significantly longer than predicted, the predicted close millisecond pulsar binaries come from a range of the parameter space so narrow that their formation becomes compromised.

Promising revisions in our understanding of the orbital angular momentum loss due to magnetic braking have been proposed in the past years, such as the convection and rotation boosted prescription, which provides stronger orbital angular momentum loss rates and can potentially solve the major problems related to the mass transfer rates in low-mass X-ray binaries and consequently the formation of their millisecond pulsar binary descendants. 
In addition, given that a dynamical origin of black hole low-mass X-ray binaries appears unlikely, this new prescription perhaps in combination with consequential angular momentum loss, e.g. due to circumbinary disks, seems a fruitful path to explore in binary models aiming at reproducing their characteristics.
However, this prescription cannot be the solution for cataclysmic variables, since observations suggest lower orbital angular momentum loss rates than those predicted by the standard prescription, which implies a weaker magnetic braking.

Any important question arising when all these systems are taken into account is whether there is a universal magnetic braking prescription able to simultaneously explain all of them.
Given the intrinsic different donor types in cataclysmic variables and low-mass X-ray binaries, a possible universal prescription could be achieved if the donor structure is also taken into account in the formulation of magnetic braking.
The overwhelming majority of donors in cataclysmic variables are M-type MS stars, while most donors in low-mass X-ray binaries seem to be MS stars of earlier spectral types or subgiants.
If a magnetic braking prescription derived from first principles could provide weaker orbital angular momentum loss rates for MS stars with initial masses smaller than ${\sim0.8}$~\Msun~and stronger for MS stars initially more massive than that, then such a recipe would be a reasonable candidate for a universal magnetic braking prescription.

All that said, despite the significant progress achieved during decades of investigation, formation and evolution of accreting compact objects is a remarkably interesting wide open field.
Future observational and theoretical efforts will hopefully improve our physical understanding of these systems.







\bibliographystyle{mnras.bst}
\bibliography{references}

\begin{thebibliography}{}
\makeatletter
\relax
\def\mn@urlcharsother{\let\do\@makeother \do\$\do\&\do\#\do\^\do\_\do\%\do\~}
\def\mn@doi{\begingroup\mn@urlcharsother \@ifnextchar [ {\mn@doi@}
  {\mn@doi@[]}}
\def\mn@doi@[#1]#2{\def\@tempa{#1}\ifx\@tempa\@empty \href
  {http://dx.doi.org/#2} {doi:#2}\else \href {http://dx.doi.org/#2} {#1}\fi
  \endgroup}
\def\mn@eprint#1#2{\mn@eprint@#1:#2::\@nil}
\def\mn@eprint@arXiv#1{\href {http://arxiv.org/abs/#1} {{\tt arXiv:#1}}}
\def\mn@eprint@dblp#1{\href {http://dblp.uni-trier.de/rec/bibtex/#1.xml}
  {dblp:#1}}
\def\mn@eprint@#1:#2:#3:#4\@nil{\def\@tempa {#1}\def\@tempb {#2}\def\@tempc
  {#3}\ifx \@tempc \@empty \let \@tempc \@tempb \let \@tempb \@tempa \fi \ifx
  \@tempb \@empty \def\@tempb {arXiv}\fi \@ifundefined
  {mn@eprint@\@tempb}{\@tempb:\@tempc}{\expandafter \expandafter \csname
  mn@eprint@\@tempb\endcsname \expandafter{\@tempc}}}

\bibitem[\protect\citeauthoryear{{Abate}, {Pols}, {Izzard}, {Mohamed}  \& {de
  Mink}}{{Abate} et~al.}{2013}]{Abate_2013}
{Abate} C.,  {Pols} O.~R.,  {Izzard} R.~G.,  {Mohamed} S.~S.,   {de Mink}
  S.~E.,  2013, \mn@doi [\aap] {10.1051/0004-6361/201220007}, \href
  {https://ui.adsabs.harvard.edu/abs/2013A&A...552A..26A} {552, A26}

\bibitem[\protect\citeauthoryear{{Abate}, {Pols}, {Karakas}  \&
  {Izzard}}{{Abate} et~al.}{2015a}]{Abate_2015a}
{Abate} C.,  {Pols} O.~R.,  {Karakas} A.~I.,   {Izzard} R.~G.,  2015a, \mn@doi
  [\aap] {10.1051/0004-6361/201424739}, \href
  {https://ui.adsabs.harvard.edu/abs/2015A&A...576A.118A} {576, A118}

\bibitem[\protect\citeauthoryear{{Abate}, {Pols}, {Izzard}  \&
  {Karakas}}{{Abate} et~al.}{2015b}]{Abate_2015b}
{Abate} C.,  {Pols} O.~R.,  {Izzard} R.~G.,   {Karakas} A.~I.,  2015b, \mn@doi
  [\aap] {10.1051/0004-6361/201525876}, \href
  {https://ui.adsabs.harvard.edu/abs/2015A&A...581A..22A} {581, A22}

\bibitem[\protect\citeauthoryear{{{\'A}lvarez-Hern{\'a}ndez}
  et~al.,}{{{\'A}lvarez-Hern{\'a}ndez} et~al.}{2021}]{Alvarez_2021}
{{\'A}lvarez-Hern{\'a}ndez} A.,  et~al., 2021, \mn@doi [\mnras]
  {10.1093/mnras/stab2547}, \href
  {https://ui.adsabs.harvard.edu/abs/2021MNRAS.507.5805A} {507, 5805}

\bibitem[\protect\citeauthoryear{{Bandyopadhyay}, {Shahbaz}, {Charles}  \&
  {Naylor}}{{Bandyopadhyay} et~al.}{1999}]{Bandyopadhyay_1999}
{Bandyopadhyay} R.~M.,  {Shahbaz} T.,  {Charles} P.~A.,   {Naylor} T.,  1999,
  \mn@doi [\mnras] {10.1046/j.1365-8711.1999.02547.x}, \href
  {https://ui.adsabs.harvard.edu/abs/1999MNRAS.306..417B} {306, 417}

\bibitem[\protect\citeauthoryear{{Bassa}, {Jonker}, {Steeghs}  \&
  {Torres}}{{Bassa} et~al.}{2009}]{Bassa_2009}
{Bassa} C.~G.,  {Jonker} P.~G.,  {Steeghs} D.,   {Torres} M.~A.~P.,  2009,
  \mn@doi [\mnras] {10.1111/j.1365-2966.2009.15395.x}, \href
  {https://ui.adsabs.harvard.edu/abs/2009MNRAS.399.2055B} {399, 2055}

\bibitem[\protect\citeauthoryear{{Bauer} \& {Kupfer}}{{Bauer} \&
  {Kupfer}}{2021}]{BK2021}
{Bauer} E.~B.,  {Kupfer} T.,  2021, \mn@doi [\apj] {10.3847/1538-4357/ac25f0},
  \href {https://ui.adsabs.harvard.edu/abs/2021ApJ...922..245B} {922, 245}

\bibitem[\protect\citeauthoryear{{Belczynski}}{{Belczynski}}{2020}]{Belczynski_2020}
{Belczynski} K.,  2020, \mn@doi [\apjl] {10.3847/2041-8213/abcbf1}, \href
  {https://ui.adsabs.harvard.edu/abs/2020ApJ...905L..15B} {905, L15}

\bibitem[\protect\citeauthoryear{{Belczynski} \& {Taam}}{{Belczynski} \&
  {Taam}}{2004}]{BT2004}
{Belczynski} K.,  {Taam} R.~E.,  2004, \mn@doi [\apj] {10.1086/381491}, \href
  {https://ui.adsabs.harvard.edu/abs/2004ApJ...603..690B} {603, 690}

\bibitem[\protect\citeauthoryear{{Belczynski}, {Holz}, {Bulik}  \&
  {O'Shaughnessy}}{{Belczynski} et~al.}{2016}]{B2016}
{Belczynski} K.,  {Holz} D.~E.,  {Bulik} T.,   {O'Shaughnessy} R.,  2016,
  \mn@doi [\nat] {10.1038/nature18322}, \href
  {https://ui.adsabs.harvard.edu/abs/2016Natur.534..512B} {534, 512}

\bibitem[\protect\citeauthoryear{{Belczynski} et~al.,}{{Belczynski}
  et~al.}{2020}]{B2020}
{Belczynski} K.,  et~al., 2020, \mn@doi [\aap] {10.1051/0004-6361/201936528},
  \href {https://ui.adsabs.harvard.edu/abs/2020A&A...636A.104B} {636, A104}

\bibitem[\protect\citeauthoryear{{Belczynski} et~al.,}{{Belczynski}
  et~al.}{2021}]{Belczynski_2021_uncer}
{Belczynski} K.,  et~al., 2021, arXiv e-prints, \href
  {https://ui.adsabs.harvard.edu/abs/2021arXiv210810885B} {p. arXiv:2108.10885}

\bibitem[\protect\citeauthoryear{{Belloni} \& {Rivera}}{{Belloni} \&
  {Rivera}}{2021}]{BR21}
{Belloni} D.,  {Rivera} L.,  2021, in The Golden Age of Cataclysmic Variables
  and Related Objects V. p.~13 (\mn@eprint {arXiv} {2008.12772})

\bibitem[\protect\citeauthoryear{{Belloni} \& {Schreiber}}{{Belloni} \&
  {Schreiber}}{2023}]{BelloniSchreiber_2023}
{Belloni} D.,  {Schreiber} M.~R.,  2023, \mn@doi [\aap]
  {10.1051/0004-6361/202347047}, \href
  {https://ui.adsabs.harvard.edu/abs/2023A&A...678A..34B} {678, A34}

\bibitem[\protect\citeauthoryear{{Belloni}, {Giersz}, {Rivera Sandoval},
  {Askar}  \& {Ciecielag}}{{Belloni} et~al.}{2019}]{Belloni_2019}
{Belloni} D.,  {Giersz} M.,  {Rivera Sandoval} L.~E.,  {Askar} A.,
  {Ciecielag} P.,  2019, \mn@doi [\mnras] {10.1093/mnras/sty3097}, \href
  {http://adsabs.harvard.edu/abs/2019MNRAS.483..315B} {483, 315}

\bibitem[\protect\citeauthoryear{{Belloni}, {Schreiber}, {Pala},
  {G{\"a}nsicke}, {Zorotovic}  \& {Rodrigues}}{{Belloni}
  et~al.}{2020a}]{Belloni_2020a}
{Belloni} D.,  {Schreiber} M.~R.,  {Pala} A.~F.,  {G{\"a}nsicke} B.~T.,
  {Zorotovic} M.,   {Rodrigues} C.~V.,  2020a, \mn@doi [\mnras]
  {10.1093/mnras/stz3413}, \href
  {https://ui.adsabs.harvard.edu/abs/2020MNRAS.491.5717B} {491, 5717}

\bibitem[\protect\citeauthoryear{{Belloni}, {Miko{\l}ajewska}, {I{\l}kiewicz},
  {Schreiber}, {Giersz}, {Rivera Sandoval}  \& {Rodrigues}}{{Belloni}
  et~al.}{2020b}]{Belloni_2020b}
{Belloni} D.,  {Miko{\l}ajewska} J.,  {I{\l}kiewicz} K.,  {Schreiber} M.~R.,
  {Giersz} M.,  {Rivera Sandoval} L.~E.,   {Rodrigues} C.~V.,  2020b, \mn@doi
  [\mnras] {10.1093/mnras/staa1714}, \href
  {https://ui.adsabs.harvard.edu/abs/2020MNRAS.496.3436B} {496, 3436}

\bibitem[\protect\citeauthoryear{{Belloni} et~al.,}{{Belloni}
  et~al.}{2021}]{Belloni_ApJSS_2021}
{Belloni} D.,  et~al., 2021, \mn@doi [\apjs] {10.3847/1538-4365/ac141c}, \href
  {https://ui.adsabs.harvard.edu/abs/2021ApJS..256...45B} {256, 45}

\bibitem[\protect\citeauthoryear{{Bhalerao}, {van Kerkwijk}  \&
  {Harrison}}{{Bhalerao} et~al.}{2012}]{Bhalerao_2012}
{Bhalerao} V.~B.,  {van Kerkwijk} M.~H.,   {Harrison} F.~A.,  2012, \mn@doi
  [\apj] {10.1088/0004-637X/757/1/10}, \href
  {https://ui.adsabs.harvard.edu/abs/2012ApJ...757...10B} {757, 10}

\bibitem[\protect\citeauthoryear{{Bhattacharya} \& {van den
  Heuvel}}{{Bhattacharya} \& {van den Heuvel}}{1991}]{BH1991}
{Bhattacharya} D.,  {van den Heuvel} E.~P.~J.,  1991, \mn@doi [\physrep]
  {10.1016/0370-1573(91)90064-S}, \href
  {https://ui.adsabs.harvard.edu/abs/1991PhR...203....1B} {203, 1}

\bibitem[\protect\citeauthoryear{{Bjorkman} \& {Cassinelli}}{{Bjorkman} \&
  {Cassinelli}}{1993}]{BC1993}
{Bjorkman} J.~E.,  {Cassinelli} J.~P.,  1993, \mn@doi [\apj] {10.1086/172676},
  \href {https://ui.adsabs.harvard.edu/abs/1993ApJ...409..429B} {409, 429}

\bibitem[\protect\citeauthoryear{{Boffin}, {Hillen}, {Berger}, {Jorissen},
  {Blind}, {Le Bouquin}, {Miko{\l}ajewska}  \& {Lazareff}}{{Boffin}
  et~al.}{2014}]{Boffin_2014}
{Boffin} H.~M.~J.,  {Hillen} M.,  {Berger} J.~P.,  {Jorissen} A.,  {Blind} N.,
  {Le Bouquin} J.~B.,  {Miko{\l}ajewska} J.,   {Lazareff} B.,  2014, \mn@doi
  [\aap] {10.1051/0004-6361/201323194}, \href
  {https://ui.adsabs.harvard.edu/abs/2014A&A...564A...1B} {564, A1}

\bibitem[\protect\citeauthoryear{{Bona{\v{c}}i{\'c} Marinovi{\'c}}, {Glebbeek}
  \& {Pols}}{{Bona{\v{c}}i{\'c} Marinovi{\'c}} et~al.}{2008}]{Bonacic_2008}
{Bona{\v{c}}i{\'c} Marinovi{\'c}} A.~A.,  {Glebbeek} E.,   {Pols} O.~R.,  2008,
  \mn@doi [\aap] {10.1051/0004-6361:20078297}, \href
  {https://ui.adsabs.harvard.edu/abs/2008A&A...480..797B} {480, 797}

\bibitem[\protect\citeauthoryear{{Bondi} \& {Hoyle}}{{Bondi} \&
  {Hoyle}}{1944}]{BH44}
{Bondi} H.,  {Hoyle} F.,  1944, \mn@doi [\mnras] {10.1093/mnras/104.5.273},
  \href {https://ui.adsabs.harvard.edu/abs/1944MNRAS.104..273B} {104, 273}

\bibitem[\protect\citeauthoryear{{Brandi}, {Quiroga}, {Miko{\l}ajewska},
  {Ferrer}  \& {Garc{\'\i}a}}{{Brandi} et~al.}{2009}]{Brandi_2009}
{Brandi} E.,  {Quiroga} C.,  {Miko{\l}ajewska} J.,  {Ferrer} O.~E.,
  {Garc{\'\i}a} L.~G.,  2009, \mn@doi [\aap] {10.1051/0004-6361/200811417},
  \href {https://ui.adsabs.harvard.edu/abs/2009A&A...497..815B} {497, 815}

\bibitem[\protect\citeauthoryear{{Breedt} et~al.,}{{Breedt}
  et~al.}{2014}]{Breedt_2014}
{Breedt} E.,  et~al., 2014, \mn@doi [\mnras] {10.1093/mnras/stu1377}, \href
  {https://ui.adsabs.harvard.edu/abs/2014MNRAS.443.3174B} {443, 3174}

\bibitem[\protect\citeauthoryear{{Brooks}, {Bildsten}, {Marchant}  \&
  {Paxton}}{{Brooks} et~al.}{2015}]{Brooks_2015}
{Brooks} J.,  {Bildsten} L.,  {Marchant} P.,   {Paxton} B.,  2015, \mn@doi
  [\apj] {10.1088/0004-637X/807/1/74}, \href
  {https://ui.adsabs.harvard.edu/abs/2015ApJ...807...74B} {807, 74}

\bibitem[\protect\citeauthoryear{{Camacho}, {Torres}, {Garc{\'{\i}}a-Berro},
  {Zorotovic}, {Schreiber}, {Rebassa-Mansergas}, {Nebot G{\'o}mez-Mor{\'a}n}
  \& {G{\"a}nsicke}}{{Camacho} et~al.}{2014}]{Camacho_2014}
{Camacho} J.,  {Torres} S.,  {Garc{\'{\i}}a-Berro} E.,  {Zorotovic} M.,
  {Schreiber} M.~R.,  {Rebassa-Mansergas} A.,  {Nebot G{\'o}mez-Mor{\'a}n} A.,
   {G{\"a}nsicke} B.~T.,  2014, \mn@doi [\aap] {10.1051/0004-6361/201323052},
  \href {http://adsabs.harvard.edu/abs/2014A%26A...566A..86C} {566, A86}

\bibitem[\protect\citeauthoryear{{Cannizzo} \& {Pudritz}}{{Cannizzo} \&
  {Pudritz}}{1988}]{Cannizzo_1988}
{Cannizzo} J.~K.,  {Pudritz} R.~E.,  1988, \mn@doi [The Astrophysical Journal]
  {10.1086/166241}, \href {http://adsabs.harvard.edu/abs/1988ApJ...327..840C}
  {327, 840}

\bibitem[\protect\citeauthoryear{{Carney}, {Latham}, {Laird}, {Grant}  \&
  {Morse}}{{Carney} et~al.}{2001}]{Carney_2001}
{Carney} B.~W.,  {Latham} D.~W.,  {Laird} J.~B.,  {Grant} C.~E.,   {Morse}
  J.~A.,  2001, \mn@doi [\aj] {10.1086/324233}, \href
  {https://ui.adsabs.harvard.edu/abs/2001AJ....122.3419C} {122, 3419}

\bibitem[\protect\citeauthoryear{{Carter} et~al.,}{{Carter}
  et~al.}{2013}]{Carter_2013}
{Carter} P.~J.,  et~al., 2013, \mn@doi [\mnras] {10.1093/mnras/sts485}, \href
  {https://ui.adsabs.harvard.edu/abs/2013MNRAS.429.2143C} {429, 2143}

\bibitem[\protect\citeauthoryear{{Casares}, {Charles}  \& {Kuulkers}}{{Casares}
  et~al.}{1998}]{Casares_1998}
{Casares} J.,  {Charles} P.,   {Kuulkers} E.,  1998, \mn@doi [\apjl]
  {10.1086/311124}, \href
  {https://ui.adsabs.harvard.edu/abs/1998ApJ...493L..39C} {493, L39}

\bibitem[\protect\citeauthoryear{{Casares} et~al.,}{{Casares}
  et~al.}{2009}]{Casares_2009}
{Casares} J.,  et~al., 2009, \mn@doi [\apjs] {10.1088/0067-0049/181/1/238},
  \href {https://ui.adsabs.harvard.edu/abs/2009ApJS..181..238C} {181, 238}

\bibitem[\protect\citeauthoryear{{Casares}, {Negueruela}, {Rib{\'o}}, {Ribas},
  {Paredes}, {Herrero}  \& {Sim{\'o}n-D{\'\i}az}}{{Casares}
  et~al.}{2014}]{Casares_2014}
{Casares} J.,  {Negueruela} I.,  {Rib{\'o}} M.,  {Ribas} I.,  {Paredes} J.~M.,
  {Herrero} A.,   {Sim{\'o}n-D{\'\i}az} S.,  2014, \mn@doi [\nat]
  {10.1038/nature12916}, \href
  {https://ui.adsabs.harvard.edu/abs/2014Natur.505..378C} {505, 378}

\bibitem[\protect\citeauthoryear{{Chaty}}{{Chaty}}{2011}]{Chaty_2012}
{Chaty} S.,  2011, in {Schmidtobreick} L.,  {Schreiber} M.~R.,   {Tappert} C.,
  eds,  Astronomical Society of the Pacific Conference Series Vol. 447,
  Evolution of Compact Binaries. p.~29 (\mn@eprint {arXiv} {1107.0231})

\bibitem[\protect\citeauthoryear{{Chen} \& {Han}}{{Chen} \& {Han}}{2008}]{CH08}
{Chen} X.,  {Han} Z.,  2008, \mn@doi [\mnras]
  {10.1111/j.1365-2966.2008.13334.x}, \href
  {https://ui.adsabs.harvard.edu/abs/2008MNRAS.387.1416C} {387, 1416}

\bibitem[\protect\citeauthoryear{{Chen}, {Han}, {Deca}  \&
  {Podsiadlowski}}{{Chen} et~al.}{2013}]{Chen_2013}
{Chen} X.,  {Han} Z.,  {Deca} J.,   {Podsiadlowski} P.,  2013, \mn@doi [\mnras]
  {10.1093/mnras/stt992}, \href
  {https://ui.adsabs.harvard.edu/abs/2013MNRAS.434..186C} {434, 186}

\bibitem[\protect\citeauthoryear{{Chen}, {Tauris}, {Han}  \& {Chen}}{{Chen}
  et~al.}{2021}]{Chen_2021}
{Chen} H.-L.,  {Tauris} T.~M.,  {Han} Z.,   {Chen} X.,  2021, \mn@doi [\mnras]
  {10.1093/mnras/stab670}, \href
  {https://ui.adsabs.harvard.edu/abs/2021MNRAS.503.3540C} {503, 3540}

\bibitem[\protect\citeauthoryear{{Cheng}, {Li}, {Xu}  \& {Li}}{{Cheng}
  et~al.}{2018}]{Cheng_2018}
{Cheng} Z.,  {Li} Z.,  {Xu} X.,   {Li} X.,  2018, \mn@doi [\apj]
  {10.3847/1538-4357/aaba16}, \href
  {http://adsabs.harvard.edu/abs/2018ApJ...858...33C} {858, 33}

\bibitem[\protect\citeauthoryear{{Chomiuk}, {Metzger}  \& {Shen}}{{Chomiuk}
  et~al.}{2020}]{Chomiuk_2021}
{Chomiuk} L.,  {Metzger} B.~D.,   {Shen} K.~J.,  2020, arXiv e-prints, \href
  {https://ui.adsabs.harvard.edu/abs/2020arXiv201108751C} {p. arXiv:2011.08751}

\bibitem[\protect\citeauthoryear{{Claeys}, {Pols}, {Izzard}, {Vink}  \&
  {Verbunt}}{{Claeys} et~al.}{2014}]{Claeys_2014}
{Claeys} J.~S.~W.,  {Pols} O.~R.,  {Izzard} R.~G.,  {Vink} J.,   {Verbunt}
  F.~W.~M.,  2014, \mn@doi [\aap] {10.1051/0004-6361/201322714}, \href
  {http://adsabs.harvard.edu/abs/2014A%26A...563A..83C} {563, A83}

\bibitem[\protect\citeauthoryear{{Clark}, {Goodwin}, {Crowther}, {Kaper},
  {Fairbairn}, {Langer}  \& {Brocksopp}}{{Clark} et~al.}{2002}]{Clark_2002}
{Clark} J.~S.,  {Goodwin} S.~P.,  {Crowther} P.~A.,  {Kaper} L.,  {Fairbairn}
  M.,  {Langer} N.,   {Brocksopp} C.,  2002, \mn@doi [\aap]
  {10.1051/0004-6361:20021184}, \href
  {https://ui.adsabs.harvard.edu/abs/2002A&A...392..909C} {392, 909}

\bibitem[\protect\citeauthoryear{{Clayton}, {Podsiadlowski}, {Ivanova}  \&
  {Justham}}{{Clayton} et~al.}{2017}]{Clayton_2017}
{Clayton} M.,  {Podsiadlowski} P.,  {Ivanova} N.,   {Justham} S.,  2017,
  \mn@doi [\mnras] {10.1093/mnras/stx1290}, \href
  {https://ui.adsabs.harvard.edu/abs/2017MNRAS.470.1788C} {470, 1788}

\bibitem[\protect\citeauthoryear{{Coe}, {Kennea}, {Evans}  \& {Udalski}}{{Coe}
  et~al.}{2020}]{Coe_2020}
{Coe} M.~J.,  {Kennea} J.~A.,  {Evans} P.~A.,   {Udalski} A.,  2020, \mn@doi
  [\mnras] {10.1093/mnrasl/slaa112}, \href
  {https://ui.adsabs.harvard.edu/abs/2020MNRAS.497L..50C} {497, L50}

\bibitem[\protect\citeauthoryear{{Cohn} et~al.,}{{Cohn}
  et~al.}{2010}]{Cohn_2010}
{Cohn} H.~N.,  et~al., 2010, \mn@doi [\apj] {10.1088/0004-637X/722/1/20}, \href
  {http://adsabs.harvard.edu/abs/2010ApJ...722...20C} {722, 20}

\bibitem[\protect\citeauthoryear{{Cohn} et~al.,}{{Cohn}
  et~al.}{2021}]{Cohn_2021}
{Cohn} H.~N.,  et~al., 2021, \mn@doi [\mnras] {10.1093/mnras/stab2636}, \href
  {https://ui.adsabs.harvard.edu/abs/2021MNRAS.508.2823C} {508, 2823}

\bibitem[\protect\citeauthoryear{{Cojocaru}, {Rebassa-Mansergas}, {Torres}  \&
  {Garc{\'{\i}}a-Berro}}{{Cojocaru} et~al.}{2017}]{Cojocaru_2017}
{Cojocaru} R.,  {Rebassa-Mansergas} A.,  {Torres} S.,   {Garc{\'{\i}}a-Berro}
  E.,  2017, \mn@doi [\mnras] {10.1093/mnras/stx1326}, \href
  {http://adsabs.harvard.edu/abs/2017MNRAS.470.1442C} {470, 1442}

\bibitem[\protect\citeauthoryear{{Copperwheat} et~al.,}{{Copperwheat}
  et~al.}{2011}]{Copperwheat_2011}
{Copperwheat} C.~M.,  et~al., 2011, \mn@doi [\mnras]
  {10.1111/j.1365-2966.2010.17508.x}, \href
  {https://ui.adsabs.harvard.edu/abs/2011MNRAS.410.1113C} {410, 1113}

\bibitem[\protect\citeauthoryear{{Corbet}}{{Corbet}}{1986}]{Corbet_1986}
{Corbet} R.~H.~D.,  1986, \mn@doi [\mnras] {10.1093/mnras/220.4.1047}, \href
  {https://ui.adsabs.harvard.edu/abs/1986MNRAS.220.1047C} {220, 1047}

\bibitem[\protect\citeauthoryear{{Coriat}, {Fender}  \& {Dubus}}{{Coriat}
  et~al.}{2012}]{Coriat_2012}
{Coriat} M.,  {Fender} R.~P.,   {Dubus} G.,  2012, \mn@doi [\mnras]
  {10.1111/j.1365-2966.2012.21339.x}, \href
  {https://ui.adsabs.harvard.edu/abs/2012MNRAS.424.1991C} {424, 1991}

\bibitem[\protect\citeauthoryear{{Corral-Santana}, {Casares},
  {Mu{\~n}oz-Darias}, {Bauer}, {Mart{\'\i}nez-Pais}  \&
  {Russell}}{{Corral-Santana} et~al.}{2016}]{Corral_2016}
{Corral-Santana} J.~M.,  {Casares} J.,  {Mu{\~n}oz-Darias} T.,  {Bauer} F.~E.,
  {Mart{\'\i}nez-Pais} I.~G.,   {Russell} D.~M.,  2016, \mn@doi [\aap]
  {10.1051/0004-6361/201527130}, \href
  {https://ui.adsabs.harvard.edu/abs/2016A&A...587A..61C} {587, A61}

\bibitem[\protect\citeauthoryear{{Crowther}, {Barnard}, {Carpano}, {Clark},
  {Dhillon}  \& {Pollock}}{{Crowther} et~al.}{2010}]{Crowther_2010}
{Crowther} P.~A.,  {Barnard} R.,  {Carpano} S.,  {Clark} J.~S.,  {Dhillon}
  V.~S.,   {Pollock} A.~M.~T.,  2010, \mn@doi [\mnras]
  {10.1111/j.1745-3933.2010.00811.x}, \href
  {https://ui.adsabs.harvard.edu/abs/2010MNRAS.403L..41C} {403, L41}

\bibitem[\protect\citeauthoryear{{Deloye}, {Taam}, {Winisdoerffer}  \&
  {Chabrier}}{{Deloye} et~al.}{2007}]{Deloye_2007}
{Deloye} C.~J.,  {Taam} R.~E.,  {Winisdoerffer} C.,   {Chabrier} G.,  2007,
  \mn@doi [\mnras] {10.1111/j.1365-2966.2007.12262.x}, \href
  {https://ui.adsabs.harvard.edu/abs/2007MNRAS.381..525D} {381, 525}

\bibitem[\protect\citeauthoryear{{Deng}, {Li}, {Gao}  \& {Shao}}{{Deng}
  et~al.}{2021}]{Deng_2021}
{Deng} Z.-L.,  {Li} X.-D.,  {Gao} Z.-F.,   {Shao} Y.,  2021, \mn@doi [\apj]
  {10.3847/1538-4357/abe0b2}, \href
  {https://ui.adsabs.harvard.edu/abs/2021ApJ...909..174D} {909, 174}

\bibitem[\protect\citeauthoryear{{Dermine}, {Izzard}, {Jorissen}  \& {Van
  Winckel}}{{Dermine} et~al.}{2013}]{Dermine_2013}
{Dermine} T.,  {Izzard} R.~G.,  {Jorissen} A.,   {Van Winckel} H.,  2013,
  \mn@doi [\aap] {10.1051/0004-6361/201219430}, \href
  {https://ui.adsabs.harvard.edu/abs/2013A&A...551A..50D} {551, A50}

\bibitem[\protect\citeauthoryear{{Dewi} \& {Tauris}}{{Dewi} \&
  {Tauris}}{2000}]{Dewi_2000}
{Dewi} J.~D.~M.,  {Tauris} T.~M.,  2000, \aap, \href
  {https://ui.adsabs.harvard.edu/abs/2000A&A...360.1043D} {360, 1043}

\bibitem[\protect\citeauthoryear{{Di Salvo} \& {Sanna}}{{Di Salvo} \&
  {Sanna}}{2020}]{SS20}
{Di Salvo} T.,  {Sanna} A.,  2020, arXiv e-prints, \href
  {https://ui.adsabs.harvard.edu/abs/2020arXiv201009005D} {p. arXiv:2010.09005}

\bibitem[\protect\citeauthoryear{{Di Stefano}}{{Di
  Stefano}}{2010a}]{Distefano_2010}
{Di Stefano} R.,  2010a, \mn@doi [\apj] {10.1088/0004-637X/712/1/728}, \href
  {https://ui.adsabs.harvard.edu/abs/2010ApJ...712..728D} {712, 728}

\bibitem[\protect\citeauthoryear{{Di Stefano}}{{Di
  Stefano}}{2010b}]{Distefano_2010b}
{Di Stefano} R.,  2010b, \mn@doi [\apj] {10.1088/0004-637X/719/1/474}, \href
  {https://ui.adsabs.harvard.edu/abs/2010ApJ...719..474D} {719, 474}

\bibitem[\protect\citeauthoryear{{Dieball}, {Knigge}, {Zurek}, {Shara}, {Long},
  {Charles}, {Hannikainen}  \& {van Zyl}}{{Dieball}
  et~al.}{2005}]{Dieball_2005}
{Dieball} A.,  {Knigge} C.,  {Zurek} D.~R.,  {Shara} M.~M.,  {Long} K.~S.,
  {Charles} P.~A.,  {Hannikainen} D.~C.,   {van Zyl} L.,  2005, \mn@doi [\apjl]
  {10.1086/498712}, \href
  {https://ui.adsabs.harvard.edu/abs/2005ApJ...634L.105D} {634, L105}

\bibitem[\protect\citeauthoryear{{Doroshenko}, {Suleimanov}, {Tsygankov},
  {M{\"o}nkk{\"o}nen}, {Ji}  \& {Santangelo}}{{Doroshenko}
  et~al.}{2020}]{Doroshenko_2020}
{Doroshenko} V.,  {Suleimanov} V.,  {Tsygankov} S.,  {M{\"o}nkk{\"o}nen} J.,
  {Ji} L.,   {Santangelo} A.,  2020, \mn@doi [\aap]
  {10.1051/0004-6361/202038093}, \href
  {https://ui.adsabs.harvard.edu/abs/2020A&A...643A..62D} {643, A62}

\bibitem[\protect\citeauthoryear{{Dubus}, {Otulakowska-Hypka}  \&
  {Lasota}}{{Dubus} et~al.}{2018}]{Dubus_2018}
{Dubus} G.,  {Otulakowska-Hypka} M.,   {Lasota} J.-P.,  2018, \mn@doi [\aap]
  {10.1051/0004-6361/201833372}, \href
  {https://ui.adsabs.harvard.edu/abs/2018A&A...617A..26D} {617, A26}

\bibitem[\protect\citeauthoryear{{Edgar}}{{Edgar}}{2004}]{Edgar_2004}
{Edgar} R.,  2004, \mn@doi [\nar] {10.1016/j.newar.2004.06.001}, \href
  {https://ui.adsabs.harvard.edu/abs/2004NewAR..48..843E} {48, 843}

\bibitem[\protect\citeauthoryear{{Eggleton}}{{Eggleton}}{1983}]{Eggleton_1983}
{Eggleton} P.~P.,  1983, \mn@doi [\apj] {10.1086/160960}, \href
  {http://adsabs.harvard.edu/abs/1983ApJ...268..368E} {268, 368}

\bibitem[\protect\citeauthoryear{{El Mellah}, {Sundqvist}  \& {Keppens}}{{El
  Mellah} et~al.}{2019a}]{Mellah_2019b}
{El Mellah} I.,  {Sundqvist} J.~O.,   {Keppens} R.,  2019a, \mn@doi [\aap]
  {10.1051/0004-6361/201834543}, \href
  {https://ui.adsabs.harvard.edu/abs/2019A&A...622L...3E} {622, L3}

\bibitem[\protect\citeauthoryear{{El Mellah}, {Sander}, {Sundqvist}  \&
  {Keppens}}{{El Mellah} et~al.}{2019b}]{Mellah_2019a}
{El Mellah} I.,  {Sander} A.~A.~C.,  {Sundqvist} J.~O.,   {Keppens} R.,  2019b,
  \mn@doi [\aap] {10.1051/0004-6361/201834498}, \href
  {https://ui.adsabs.harvard.edu/abs/2019A&A...622A.189E} {622, A189}

\bibitem[\protect\citeauthoryear{{ElBadry} et~al.,}{{ElBadry}
  et~al.}{2021}]{ElBadry_2021}
{ElBadry} K.,  et~al., 2021, \mn@doi [\mnras] {10.1093/mnras/stab1318}, \href
  {https://ui.adsabs.harvard.edu/abs/2021MNRAS.505.2051E} {505, 2051}

\bibitem[\protect\citeauthoryear{{Escorza} et~al.,}{{Escorza}
  et~al.}{2019}]{Escorza_2019}
{Escorza} A.,  et~al., 2019, \mn@doi [\aap] {10.1051/0004-6361/201935390},
  \href {https://ui.adsabs.harvard.edu/abs/2019A&A...626A.128E} {626, A128}

\bibitem[\protect\citeauthoryear{{Escorza}, {Siess}, {Van Winckel}  \&
  {Jorissen}}{{Escorza} et~al.}{2020}]{Escorza_2020}
{Escorza} A.,  {Siess} L.,  {Van Winckel} H.,   {Jorissen} A.,  2020, \mn@doi
  [\aap] {10.1051/0004-6361/202037487}, \href
  {https://ui.adsabs.harvard.edu/abs/2020A&A...639A..24E} {639, A24}

\bibitem[\protect\citeauthoryear{{Falanga}, {Bozzo}, {Lutovinov},
  {Bonnet-Bidaud}, {Fetisova}  \& {Puls}}{{Falanga}
  et~al.}{2015}]{Falanga_2015}
{Falanga} M.,  {Bozzo} E.,  {Lutovinov} A.,  {Bonnet-Bidaud} J.~M.,  {Fetisova}
  Y.,   {Puls} J.,  2015, \mn@doi [\aap] {10.1051/0004-6361/201425191}, \href
  {https://ui.adsabs.harvard.edu/abs/2015A&A...577A.130F} {577, A130}

\bibitem[\protect\citeauthoryear{{Fekel}, {Hinkle}, {Joyce}, {Wood}  \&
  {Lebzelter}}{{Fekel} et~al.}{2007}]{Fekel_2007}
{Fekel} F.~C.,  {Hinkle} K.~H.,  {Joyce} R.~R.,  {Wood} P.~R.,   {Lebzelter}
  T.,  2007, \mn@doi [\aj] {10.1086/509133}, \href
  {https://ui.adsabs.harvard.edu/abs/2007AJ....133...17F} {133, 17}

\bibitem[\protect\citeauthoryear{{Fekel}, {Hinkle}, {Joyce}, {Wood}  \&
  {Howarth}}{{Fekel} et~al.}{2008}]{Fekel_2008}
{Fekel} F.~C.,  {Hinkle} K.~H.,  {Joyce} R.~R.,  {Wood} P.~R.,   {Howarth}
  I.~D.,  2008, \mn@doi [\aj] {10.1088/0004-6256/136/1/146}, \href
  {https://ui.adsabs.harvard.edu/abs/2008AJ....136..146F} {136, 146}

\bibitem[\protect\citeauthoryear{{Fekel}, {Hinkle}, {Joyce}  \& {Wood}}{{Fekel}
  et~al.}{2010}]{Fekel_2010}
{Fekel} F.~C.,  {Hinkle} K.~H.,  {Joyce} R.~R.,   {Wood} P.~R.,  2010, \mn@doi
  [\aj] {10.1088/0004-6256/139/4/1315}, \href
  {https://ui.adsabs.harvard.edu/abs/2010AJ....139.1315F} {139, 1315}

\bibitem[\protect\citeauthoryear{{Fink}, {Hillebrandt}  \& {R{\"o}pke}}{{Fink}
  et~al.}{2007}]{Fink_2007}
{Fink} M.,  {Hillebrandt} W.,   {R{\"o}pke} F.~K.,  2007, \mn@doi [\aap]
  {10.1051/0004-6361:20078438}, \href
  {https://ui.adsabs.harvard.edu/abs/2007A&A...476.1133F} {476, 1133}

\bibitem[\protect\citeauthoryear{{Freire} \& {Tauris}}{{Freire} \&
  {Tauris}}{2014}]{FT2014}
{Freire} P. C.~C.,  {Tauris} T.~M.,  2014, \mn@doi [\mnras]
  {10.1093/mnrasl/slt164}, \href
  {https://ui.adsabs.harvard.edu/abs/2014MNRAS.438L..86F} {438, L86}

\bibitem[\protect\citeauthoryear{{Galiullin} \& {Gilfanov}}{{Galiullin} \&
  {Gilfanov}}{2021}]{Gilfanov_2021}
{Galiullin} I.,  {Gilfanov} M.,  2021, \mn@doi [\aap]
  {10.1051/0004-6361/202039522}, \href
  {https://ui.adsabs.harvard.edu/abs/2021A&A...646A..85G} {646, A85}

\bibitem[\protect\citeauthoryear{{G{\"a}nsicke} et~al.,}{{G{\"a}nsicke}
  et~al.}{2003}]{Gaensicke2003}
{G{\"a}nsicke} B.~T.,  et~al., 2003, \mn@doi [\apj] {10.1086/376902}, \href
  {http://adsabs.harvard.edu/abs/2003ApJ...594..443G} {594, 443}

\bibitem[\protect\citeauthoryear{{G{\"a}nsicke} et~al.,}{{G{\"a}nsicke}
  et~al.}{2009}]{Gansicke_2009}
{G{\"a}nsicke} B.~T.,  et~al., 2009, \mn@doi [\mnras]
  {10.1111/j.1365-2966.2009.15126.x}, \href
  {http://adsabs.harvard.edu/abs/2009MNRAS.397.2170G} {397, 2170}

\bibitem[\protect\citeauthoryear{{Garofali}, {Converse}, {Chandar}  \&
  {Rangelov}}{{Garofali} et~al.}{2012}]{Garofali_2012}
{Garofali} K.,  {Converse} J.~M.,  {Chandar} R.,   {Rangelov} B.,  2012,
  \mn@doi [\apj] {10.1088/0004-637X/755/1/49}, \href
  {https://ui.adsabs.harvard.edu/abs/2012ApJ...755...49G} {755, 49}

\bibitem[\protect\citeauthoryear{{Garofali}, {Williams}, {Hillis}, {Gilbert},
  {Dolphin}, {Eracleous}  \& {Binder}}{{Garofali} et~al.}{2018}]{Garofali_2018}
{Garofali} K.,  {Williams} B.~F.,  {Hillis} T.,  {Gilbert} K.~M.,  {Dolphin}
  A.~E.,  {Eracleous} M.,   {Binder} B.,  2018, \mn@doi [\mnras]
  {10.1093/mnras/sty1612}, \href
  {https://ui.adsabs.harvard.edu/abs/2018MNRAS.479.3526G} {479, 3526}

\bibitem[\protect\citeauthoryear{{Garraffo}, {Drake}, {Alvarado-Gomez},
  {Moschou}  \& {Cohen}}{{Garraffo} et~al.}{2018}]{Garraffo_2018}
{Garraffo} C.,  {Drake} J.~J.,  {Alvarado-Gomez} J.~D.,  {Moschou} S.~P.,
  {Cohen} O.,  2018, \mn@doi [\apj] {10.3847/1538-4357/aae589}, \href
  {https://ui.adsabs.harvard.edu/abs/2018ApJ...868...60G} {868, 60}

\bibitem[\protect\citeauthoryear{{Ge}, {Hjellming}, {Webbink}, {Chen}  \&
  {Han}}{{Ge} et~al.}{2010}]{GeI}
{Ge} H.,  {Hjellming} M.~S.,  {Webbink} R.~F.,  {Chen} X.,   {Han} Z.,  2010,
  \mn@doi [\apj] {10.1088/0004-637X/717/2/724}, \href
  {https://ui.adsabs.harvard.edu/abs/2010ApJ...717..724G} {717, 724}

\bibitem[\protect\citeauthoryear{{Ge}, {Webbink}, {Chen}  \& {Han}}{{Ge}
  et~al.}{2015}]{GeII}
{Ge} H.,  {Webbink} R.~F.,  {Chen} X.,   {Han} Z.,  2015, \mn@doi [\apj]
  {10.1088/0004-637X/812/1/40}, \href
  {https://ui.adsabs.harvard.edu/abs/2015ApJ...812...40G} {812, 40}

\bibitem[\protect\citeauthoryear{{Ge}, {Webbink}  \& {Han}}{{Ge}
  et~al.}{2020a}]{Ge_2020}
{Ge} H.,  {Webbink} R.~F.,   {Han} Z.,  2020a, \mn@doi [\apjs]
  {10.3847/1538-4365/ab98f6}, \href
  {https://ui.adsabs.harvard.edu/abs/2020ApJS..249....9G} {249, 9}

\bibitem[\protect\citeauthoryear{{Ge}, {Webbink}, {Chen}  \& {Han}}{{Ge}
  et~al.}{2020b}]{GeIII}
{Ge} H.,  {Webbink} R.~F.,  {Chen} X.,   {Han} Z.,  2020b, \mn@doi [\apj]
  {10.3847/1538-4357/aba7b7}, \href
  {https://ui.adsabs.harvard.edu/abs/2020ApJ...899..132G} {899, 132}

\bibitem[\protect\citeauthoryear{{Gilfanov} \& {Bogd{\'a}n}}{{Gilfanov} \&
  {Bogd{\'a}n}}{2010}]{Gilfanov_2010}
{Gilfanov} M.,  {Bogd{\'a}n} {\'A}.,  2010, \mn@doi [\nat]
  {10.1038/nature08685}, \href
  {https://ui.adsabs.harvard.edu/abs/2010Natur.463..924G} {463, 924}

\bibitem[\protect\citeauthoryear{{Gilliland}}{{Gilliland}}{1982}]{Gilliland_1982}
{Gilliland} R.~L.,  1982, \mn@doi [\apj] {10.1086/160504}, \href
  {https://ui.adsabs.harvard.edu/abs/1982ApJ...263..302G} {263, 302}

\bibitem[\protect\citeauthoryear{{Glanz} \& {Perets}}{{Glanz} \&
  {Perets}}{2018}]{GP2018}
{Glanz} H.,  {Perets} H.~B.,  2018, \mn@doi [\mnras] {10.1093/mnrasl/sly065},
  \href {https://ui.adsabs.harvard.edu/abs/2018MNRAS.478L..12G} {478, L12}

\bibitem[\protect\citeauthoryear{{Goliasch} \& {Nelson}}{{Goliasch} \&
  {Nelson}}{2015}]{Goliasch_2015}
{Goliasch} J.,  {Nelson} L.,  2015, \mn@doi [\apj]
  {10.1088/0004-637X/809/1/80}, \href
  {https://ui.adsabs.harvard.edu/abs/2015ApJ...809...80G} {809, 80}

\bibitem[\protect\citeauthoryear{{Gonz{\'a}lez Hern{\'a}ndez}, {Rebolo}  \&
  {Casares}}{{Gonz{\'a}lez Hern{\'a}ndez} et~al.}{2012}]{GH2012}
{Gonz{\'a}lez Hern{\'a}ndez} J.~I.,  {Rebolo} R.,   {Casares} J.,  2012,
  \mn@doi [\apjl] {10.1088/2041-8205/744/2/L25}, \href
  {https://ui.adsabs.harvard.edu/abs/2012ApJ...744L..25G} {744, L25}

\bibitem[\protect\citeauthoryear{{Gonz{\'a}lez Hern{\'a}ndez}, {Rebolo}  \&
  {Casares}}{{Gonz{\'a}lez Hern{\'a}ndez} et~al.}{2014}]{GH2014}
{Gonz{\'a}lez Hern{\'a}ndez} J.~I.,  {Rebolo} R.,   {Casares} J.,  2014,
  \mn@doi [\mnras] {10.1093/mnrasl/slt150}, \href
  {https://ui.adsabs.harvard.edu/abs/2014MNRAS.438L..21G} {438, L21}

\bibitem[\protect\citeauthoryear{{Gonz{\'a}lez Hern{\'a}ndez},
  {Su{\'a}rez-Andr{\'e}s}, {Rebolo}  \& {Casares}}{{Gonz{\'a}lez Hern{\'a}ndez}
  et~al.}{2017}]{GH2017}
{Gonz{\'a}lez Hern{\'a}ndez} J.~I.,  {Su{\'a}rez-Andr{\'e}s} L.,  {Rebolo} R.,
   {Casares} J.,  2017, \mn@doi [\mnras] {10.1093/mnrasl/slw182}, \href
  {https://ui.adsabs.harvard.edu/abs/2017MNRAS.465L..15G} {465, L15}

\bibitem[\protect\citeauthoryear{{Gosnell}, {Leiner}, {Mathieu}, {Geller},
  {Knigge}, {Sills}  \& {Leigh}}{{Gosnell} et~al.}{2019}]{Gosnell_2019}
{Gosnell} N.~M.,  {Leiner} E.~M.,  {Mathieu} R.~D.,  {Geller} A.~M.,  {Knigge}
  C.,  {Sills} A.,   {Leigh} N. W.~C.,  2019, \mn@doi [\apj]
  {10.3847/1538-4357/ab4273}, \href
  {https://ui.adsabs.harvard.edu/abs/2019ApJ...885...45G} {885, 45}

\bibitem[\protect\citeauthoryear{{Green} et~al.,}{{Green}
  et~al.}{2018}]{Green_2018}
{Green} M.~J.,  et~al., 2018, \mn@doi [\mnras] {10.1093/mnras/sty299}, \href
  {https://ui.adsabs.harvard.edu/abs/2018MNRAS.476.1663G} {476, 1663}

\bibitem[\protect\citeauthoryear{{Grichener}, {Sabach}  \& {Soker}}{{Grichener}
  et~al.}{2018}]{Grichener_2018}
{Grichener} A.,  {Sabach} E.,   {Soker} N.,  2018, \mn@doi [\mnras]
  {10.1093/mnras/sty1178}, \href
  {https://ui.adsabs.harvard.edu/abs/2018MNRAS.478.1818G} {478, 1818}

\bibitem[\protect\citeauthoryear{{Gromadzki} \& {Miko{\l}ajewska}}{{Gromadzki}
  \& {Miko{\l}ajewska}}{2009}]{Gromadzki_2009}
{Gromadzki} M.,  {Miko{\l}ajewska} J.,  2009, \mn@doi [\aap]
  {10.1051/0004-6361:200810052}, \href
  {https://ui.adsabs.harvard.edu/abs/2009A&A...495..931G} {495, 931}

\bibitem[\protect\citeauthoryear{{Gromadzki}, {Miko{\l}ajewska}  \&
  {Soszy{\'n}ski}}{{Gromadzki} et~al.}{2013}]{Gromadzki_2013}
{Gromadzki} M.,  {Miko{\l}ajewska} J.,   {Soszy{\'n}ski} I.,  2013, \actaa,
  \href {https://ui.adsabs.harvard.edu/abs/2013AcA....63..405G} {63, 405}

\bibitem[\protect\citeauthoryear{{Grudzinska} et~al.,}{{Grudzinska}
  et~al.}{2015}]{Grudzinska_2015}
{Grudzinska} M.,  et~al., 2015, \mn@doi [\mnras] {10.1093/mnras/stv1419}, \href
  {https://ui.adsabs.harvard.edu/abs/2015MNRAS.452.2773G} {452, 2773}

\bibitem[\protect\citeauthoryear{{Hameury}}{{Hameury}}{2020}]{Hameury_2020}
{Hameury} J.~M.,  2020, \mn@doi [Advances in Space Research]
  {10.1016/j.asr.2019.10.022}, \href
  {https://ui.adsabs.harvard.edu/abs/2020AdSpR..66.1004H} {66, 1004}

\bibitem[\protect\citeauthoryear{{Hameury}, {Menou}, {Dubus}, {Lasota}  \&
  {Hure}}{{Hameury} et~al.}{1998}]{Hameury_1998}
{Hameury} J.-M.,  {Menou} K.,  {Dubus} G.,  {Lasota} J.-P.,   {Hure} J.-M.,
  1998, \mn@doi [\mnras] {10.1046/j.1365-8711.1998.01773.x}, \href
  {https://ui.adsabs.harvard.edu/abs/1998MNRAS.298.1048H} {298, 1048}

\bibitem[\protect\citeauthoryear{{Han}, {Podsiadlowski}, {Maxted}, {Marsh}  \&
  {Ivanova}}{{Han} et~al.}{2002}]{Han_2002}
{Han} Z.,  {Podsiadlowski} P.,  {Maxted} P.~F.~L.,  {Marsh} T.~R.,   {Ivanova}
  N.,  2002, \mn@doi [\mnras] {10.1046/j.1365-8711.2002.05752.x}, \href
  {https://ui.adsabs.harvard.edu/abs/2002MNRAS.336..449H} {336, 449}

\bibitem[\protect\citeauthoryear{{Han}, {Podsiadlowski}, {Maxted}  \&
  {Marsh}}{{Han} et~al.}{2003}]{Han_2003}
{Han} Z.,  {Podsiadlowski} P.,  {Maxted} P.~F.~L.,   {Marsh} T.~R.,  2003,
  \mn@doi [\mnras] {10.1046/j.1365-8711.2003.06451.x}, \href
  {https://ui.adsabs.harvard.edu/abs/2003MNRAS.341..669H} {341, 669}

\bibitem[\protect\citeauthoryear{{Hansen}, {Andersen}, {Nordstr{\"o}m},
  {Beers}, {Placco}, {Yoon}  \& {Buchhave}}{{Hansen}
  et~al.}{2016}]{Hansen_2016}
{Hansen} T.~T.,  {Andersen} J.,  {Nordstr{\"o}m} B.,  {Beers} T.~C.,  {Placco}
  V.~M.,  {Yoon} J.,   {Buchhave} L.~A.,  2016, \mn@doi [\aap]
  {10.1051/0004-6361/201527409}, \href
  {https://ui.adsabs.harvard.edu/abs/2016A&A...588A...3H} {588, A3}

\bibitem[\protect\citeauthoryear{{Harlaftis} \& {Greiner}}{{Harlaftis} \&
  {Greiner}}{2004}]{Harlaftis_2004}
{Harlaftis} E.~T.,  {Greiner} J.,  2004, \mn@doi [\aap]
  {10.1051/0004-6361:20031754}, \href
  {https://ui.adsabs.harvard.edu/abs/2004A&A...414L..13H} {414, L13}

\bibitem[\protect\citeauthoryear{{Heinke}, {Ivanova}, {Engel}, {Pavlovskii},
  {Sivakoff}, {Cartwright}  \& {Gladstone}}{{Heinke}
  et~al.}{2013}]{Heinke_2013}
{Heinke} C.~O.,  {Ivanova} N.,  {Engel} M.~C.,  {Pavlovskii} K.,  {Sivakoff}
  G.~R.,  {Cartwright} T.~F.,   {Gladstone} J.~C.,  2013, \mn@doi [\apj]
  {10.1088/0004-637X/768/2/184}, \href
  {https://ui.adsabs.harvard.edu/abs/2013ApJ...768..184H} {768, 184}

\bibitem[\protect\citeauthoryear{{Hernandez} et~al.,}{{Hernandez}
  et~al.}{2021}]{hernandezetal21-1}
{Hernandez} M.~S.,  et~al., 2021, \mn@doi [\mnras] {10.1093/mnras/staa3815},
  \href {https://ui.adsabs.harvard.edu/abs/2021MNRAS.501.1677H} {501, 1677}

\bibitem[\protect\citeauthoryear{{Hinkle}, {Fekel}, {Joyce}, {Wood}, {Smith}
  \& {Lebzelter}}{{Hinkle} et~al.}{2006}]{Hinkle_2006}
{Hinkle} K.~H.,  {Fekel} F.~C.,  {Joyce} R.~R.,  {Wood} P.~R.,  {Smith} V.~V.,
   {Lebzelter} T.,  2006, \mn@doi [\apj] {10.1086/500350}, \href
  {https://ui.adsabs.harvard.edu/abs/2006ApJ...641..479H} {641, 479}

\bibitem[\protect\citeauthoryear{{Hinkle}, {Fekel}  \& {Joyce}}{{Hinkle}
  et~al.}{2009}]{Hinkle_2009}
{Hinkle} K.~H.,  {Fekel} F.~C.,   {Joyce} R.~R.,  2009, \mn@doi [\apj]
  {10.1088/0004-637X/692/2/1360}, \href
  {https://ui.adsabs.harvard.edu/abs/2009ApJ...692.1360H} {692, 1360}

\bibitem[\protect\citeauthoryear{{Hinkle}, {Fekel}, {Joyce}, {Miko{\l}ajewska},
  {Ga{\l}an}  \& {Lebzelter}}{{Hinkle} et~al.}{2019}]{Hinkle_2019}
{Hinkle} K.~H.,  {Fekel} F.~C.,  {Joyce} R.~R.,  {Miko{\l}ajewska} J.,
  {Ga{\l}an} C.,   {Lebzelter} T.,  2019, \mn@doi [\apj]
  {10.3847/1538-4357/aafba5}, \href
  {https://ui.adsabs.harvard.edu/abs/2019ApJ...872...43H} {872, 43}

\bibitem[\protect\citeauthoryear{{Hjellming} \& {Webbink}}{{Hjellming} \&
  {Webbink}}{1987}]{HW1987}
{Hjellming} M.~S.,  {Webbink} R.~F.,  1987, \mn@doi [\apj] {10.1086/165412},
  \href {https://ui.adsabs.harvard.edu/abs/1987ApJ...318..794H} {318, 794}

\bibitem[\protect\citeauthoryear{{H{\"o}fner}}{{H{\"o}fner}}{2007}]{Hofner_2007}
{H{\"o}fner} S.,  2007, in {Kerschbaum} F.,  {Charbonnel} C.,   {Wing} R.~F.,
  eds,  Astronomical Society of the Pacific Conference Series Vol. 378, Why
  Galaxies Care About AGB Stars: Their Importance as Actors and Probes. p.~145
  (\mn@eprint {arXiv} {astro-ph/0702444})

\bibitem[\protect\citeauthoryear{{H{\"o}fner} \& {Olofsson}}{{H{\"o}fner} \&
  {Olofsson}}{2018}]{Hofner_2018}
{H{\"o}fner} S.,  {Olofsson} H.,  2018, \mn@doi [\aapr]
  {10.1007/s00159-017-0106-5}, \href
  {https://ui.adsabs.harvard.edu/abs/2018A&ARv..26....1H} {26, 1}

\bibitem[\protect\citeauthoryear{{Howell}, {Nelson}  \& {Rappaport}}{{Howell}
  et~al.}{2001}]{Howell_2001}
{Howell} S.~B.,  {Nelson} L.~A.,   {Rappaport} S.,  2001, \mn@doi [\apj]
  {10.1086/319776}, \href {http://adsabs.harvard.edu/abs/2001ApJ...550..897H}
  {550, 897}

\bibitem[\protect\citeauthoryear{{Hoyle} \& {Lyttleton}}{{Hoyle} \&
  {Lyttleton}}{1939}]{HL39}
{Hoyle} F.,  {Lyttleton} R.~A.,  1939, \mn@doi [Proceedings of the Cambridge
  Philosophical Society] {10.1017/S0305004100021150}, \href
  {https://ui.adsabs.harvard.edu/abs/1939PCPS...35..405H} {35, 405}

\bibitem[\protect\citeauthoryear{{Huang}}{{Huang}}{1966}]{Huang_1966}
{Huang} S.-S.,  1966, Annales d'Astrophysique, \href
  {https://ui.adsabs.harvard.edu/abs/1966AnAp...29..331H} {29, 331}

\bibitem[\protect\citeauthoryear{{Hurley}, {Tout}  \& {Pols}}{{Hurley}
  et~al.}{2002}]{Hurley_2002}
{Hurley} J.~R.,  {Tout} C.~A.,   {Pols} O.~R.,  2002, \mn@doi [\mnras]
  {10.1046/j.1365-8711.2002.05038.x}, \href
  {http://adsabs.harvard.edu/abs/2002MNRAS.329..897H} {329, 897}

\bibitem[\protect\citeauthoryear{{Hut}}{{Hut}}{1980}]{Hut_1980}
{Hut} P.,  1980, \aap, \href
  {https://ui.adsabs.harvard.edu/abs/1980A&A....92..167H} {92, 167}

\bibitem[\protect\citeauthoryear{{I{\l}kiewicz}, {Miko{\l}ajewska},
  {Belczy{\'n}ski}, {Wiktorowicz}  \& {Karczmarek}}{{I{\l}kiewicz}
  et~al.}{2019}]{Ikiewicz_2019}
{I{\l}kiewicz} K.,  {Miko{\l}ajewska} J.,  {Belczy{\'n}ski} K.,  {Wiktorowicz}
  G.,   {Karczmarek} P.,  2019, \mn@doi [\mnras] {10.1093/mnras/stz760}, \href
  {https://ui.adsabs.harvard.edu/abs/2019MNRAS.485.5468I} {485, 5468}

\bibitem[\protect\citeauthoryear{{Illarionov} \& {Sunyaev}}{{Illarionov} \&
  {Sunyaev}}{1975}]{Illarionov_1975}
{Illarionov} A.~F.,  {Sunyaev} R.~A.,  1975, \aap, \href
  {https://ui.adsabs.harvard.edu/abs/1975A&A....39..185I} {39, 185}

\bibitem[\protect\citeauthoryear{{Inight}, {G{\"a}nsicke}, {Breedt}, {Marsh},
  {Pala}  \& {Raddi}}{{Inight} et~al.}{2021}]{Inight_2021}
{Inight} K.,  {G{\"a}nsicke} B.~T.,  {Breedt} E.,  {Marsh} T.~R.,  {Pala}
  A.~F.,   {Raddi} R.,  2021, \mn@doi [\mnras] {10.1093/mnras/stab753}, \href
  {https://ui.adsabs.harvard.edu/abs/2021MNRAS.504.2420I} {504, 2420}

\bibitem[\protect\citeauthoryear{{Isern}, {Garc{\'{\i}}a-Berro}, {K{\"u}lebi}
  \& {Lor{\'e}n-Aguilar}}{{Isern} et~al.}{2017}]{Isern_2017}
{Isern} J.,  {Garc{\'{\i}}a-Berro} E.,  {K{\"u}lebi} B.,   {Lor{\'e}n-Aguilar}
  P.,  2017, \mn@doi [\apjl] {10.3847/2041-8213/aa5eae}, \href
  {http://adsabs.harvard.edu/abs/2017ApJ...836L..28I} {836, L28}

\bibitem[\protect\citeauthoryear{{Israelian}, {Rebolo}, {Basri}, {Casares}  \&
  {Mart{\'\i}n}}{{Israelian} et~al.}{1999}]{Israelian_1999}
{Israelian} G.,  {Rebolo} R.,  {Basri} G.,  {Casares} J.,   {Mart{\'\i}n}
  E.~L.,  1999, \mn@doi [\nat] {10.1038/43625}, \href
  {https://ui.adsabs.harvard.edu/abs/1999Natur.401..142I} {401, 142}

\bibitem[\protect\citeauthoryear{{Istrate}, {Tauris}  \& {Langer}}{{Istrate}
  et~al.}{2014}]{Istrate_2014}
{Istrate} A.~G.,  {Tauris} T.~M.,   {Langer} N.,  2014, \mn@doi [\aap]
  {10.1051/0004-6361/201424680}, \href
  {https://ui.adsabs.harvard.edu/abs/2014A&A...571A..45I} {571, A45}

\bibitem[\protect\citeauthoryear{{Ivanova}}{{Ivanova}}{2018}]{Ivanova_2018}
{Ivanova} N.,  2018, \mn@doi [\apjl] {10.3847/2041-8213/aac101}, \href
  {https://ui.adsabs.harvard.edu/abs/2018ApJ...858L..24I} {858, L24}

\bibitem[\protect\citeauthoryear{{Ivanova} \& {Nandez}}{{Ivanova} \&
  {Nandez}}{2016}]{IN16}
{Ivanova} N.,  {Nandez} J.~L.~A.,  2016, \mn@doi [\mnras]
  {10.1093/mnras/stw1676}, \href
  {https://ui.adsabs.harvard.edu/abs/2016MNRAS.462..362I} {462, 362}

\bibitem[\protect\citeauthoryear{{Ivanova}, {Heinke}, {Rasio}, {Belczynski}  \&
  {Fregeau}}{{Ivanova} et~al.}{2008}]{Ivanova_2008}
{Ivanova} N.,  {Heinke} C.~O.,  {Rasio} F.~A.,  {Belczynski} K.,   {Fregeau}
  J.~M.,  2008, \mn@doi [\mnras] {10.1111/j.1365-2966.2008.13064.x}, \href
  {http://adsabs.harvard.edu/abs/2008MNRAS.386..553I} {386, 553}

\bibitem[\protect\citeauthoryear{{Ivanova} et~al.,}{{Ivanova}
  et~al.}{2013}]{Ivanova_REVIEW}
{Ivanova} N.,  et~al., 2013, \mn@doi [\aapr] {10.1007/s00159-013-0059-2}, \href
  {http://adsabs.harvard.edu/abs/2013A%26ARv..21...59I} {21, 59}

\bibitem[\protect\citeauthoryear{{Johns Mulia}, {Chandar}  \&
  {Rangelov}}{{Johns Mulia} et~al.}{2019}]{Mulia_2019}
{Johns Mulia} P.,  {Chandar} R.,   {Rangelov} B.,  2019, \mn@doi [\apj]
  {10.3847/1538-4357/aaf56a}, \href
  {https://ui.adsabs.harvard.edu/abs/2019ApJ...871..122J} {871, 122}

\bibitem[\protect\citeauthoryear{{Jorissen} et~al.,}{{Jorissen}
  et~al.}{2016}]{Jorissen_2016}
{Jorissen} A.,  et~al., 2016, \mn@doi [\aap] {10.1051/0004-6361/201526992},
  \href {https://ui.adsabs.harvard.edu/abs/2016A&A...586A.158J} {586, A158}

\bibitem[\protect\citeauthoryear{{Jorissen}, {Boffin}, {Karinkuzhi}, {Van Eck},
  {Escorza}, {Shetye}  \& {Van Winckel}}{{Jorissen}
  et~al.}{2019}]{Jorissen_2019}
{Jorissen} A.,  {Boffin} H.~M.~J.,  {Karinkuzhi} D.,  {Van Eck} S.,  {Escorza}
  A.,  {Shetye} S.,   {Van Winckel} H.,  2019, \mn@doi [\aap]
  {10.1051/0004-6361/201834630}, \href
  {https://ui.adsabs.harvard.edu/abs/2019A&A...626A.127J} {626, A127}

\bibitem[\protect\citeauthoryear{{Kaaret}, {Alonso-Herrero}, {Gallagher},
  {Fabbiano}, {Zezas}  \& {Rieke}}{{Kaaret} et~al.}{2004}]{Kaaret_2004}
{Kaaret} P.,  {Alonso-Herrero} A.,  {Gallagher} J.~S.,  {Fabbiano} G.,  {Zezas}
  A.,   {Rieke} M.~J.,  2004, \mn@doi [\mnras]
  {10.1111/j.1365-2966.2004.07516.x}, \href
  {https://ui.adsabs.harvard.edu/abs/2004MNRAS.348L..28K} {348, L28}

\bibitem[\protect\citeauthoryear{{Kaaret}, {Feng}  \& {Roberts}}{{Kaaret}
  et~al.}{2017}]{Kaaret_2017}
{Kaaret} P.,  {Feng} H.,   {Roberts} T.~P.,  2017, \mn@doi [\araa]
  {10.1146/annurev-astro-091916-055259}, \href
  {https://ui.adsabs.harvard.edu/abs/2017ARA&A..55..303K} {55, 303}

\bibitem[\protect\citeauthoryear{{Kalogera}, {Belczynski}, {Kim},
  {O'Shaughnessy}  \& {Willems}}{{Kalogera} et~al.}{2007}]{Kalogera_2007}
{Kalogera} V.,  {Belczynski} K.,  {Kim} C.,  {O'Shaughnessy} R.,   {Willems}
  B.,  2007, \mn@doi [\physrep] {10.1016/j.physrep.2007.02.008}, \href
  {https://ui.adsabs.harvard.edu/abs/2007PhR...442...75K} {442, 75}

\bibitem[\protect\citeauthoryear{{Kalomeni}, {Nelson}, {Rappaport}, {Molnar},
  {Quintin}  \& {Yakut}}{{Kalomeni} et~al.}{2016}]{Kalomeni_2016}
{Kalomeni} B.,  {Nelson} L.,  {Rappaport} S.,  {Molnar} M.,  {Quintin} J.,
  {Yakut} K.,  2016, \mn@doi [\apj] {10.3847/1538-4357/833/1/83}, \href
  {http://adsabs.harvard.edu/abs/2016ApJ...833...83K} {833, 83}

\bibitem[\protect\citeauthoryear{{Kaplan}, {Bildsten}  \& {Steinfadt}}{{Kaplan}
  et~al.}{2012}]{Kaplan_2012}
{Kaplan} D.~L.,  {Bildsten} L.,   {Steinfadt} J. D.~R.,  2012, \mn@doi [\apj]
  {10.1088/0004-637X/758/1/64}, \href
  {https://ui.adsabs.harvard.edu/abs/2012ApJ...758...64K} {758, 64}

\bibitem[\protect\citeauthoryear{{K{\"a}ppeler}, {Gallino}, {Bisterzo}  \&
  {Aoki}}{{K{\"a}ppeler} et~al.}{2011}]{Kappeler_2011}
{K{\"a}ppeler} F.,  {Gallino} R.,  {Bisterzo} S.,   {Aoki} W.,  2011, \mn@doi
  [Reviews of Modern Physics] {10.1103/RevModPhys.83.157}, \href
  {https://ui.adsabs.harvard.edu/abs/2011RvMP...83..157K} {83, 157}

\bibitem[\protect\citeauthoryear{{Karino}}{{Karino}}{2021}]{Karino_2021}
{Karino} S.,  2021, \mn@doi [\mnras] {10.1093/mnras/stab2076}, \href
  {https://ui.adsabs.harvard.edu/abs/2021MNRAS.507.1002K} {507, 1002}

\bibitem[\protect\citeauthoryear{{Kashi} \& {Soker}}{{Kashi} \&
  {Soker}}{2011}]{KS2011}
{Kashi} A.,  {Soker} N.,  2011, \mn@doi [\mnras]
  {10.1111/j.1365-2966.2011.19361.x}, \href
  {https://ui.adsabs.harvard.edu/abs/2011MNRAS.417.1466K} {417, 1466}

\bibitem[\protect\citeauthoryear{{Kawaler}}{{Kawaler}}{1988}]{Kawaler_1988}
{Kawaler} S.~D.,  1988, \mn@doi [\apj] {10.1086/166740}, \href
  {https://ui.adsabs.harvard.edu/abs/1988ApJ...333..236K} {333, 236}

\bibitem[\protect\citeauthoryear{{Kennea}, {Coe}, {Evans}, {Townsend},
  {Campbell}  \& {Udalski}}{{Kennea} et~al.}{2021}]{Kennea_2021}
{Kennea} J.~A.,  {Coe} M.~J.,  {Evans} P.~A.,  {Townsend} L.~J.,  {Campbell}
  Z.~A.,   {Udalski} A.,  2021, \mn@doi [\mnras] {10.1093/mnras/stab2632},
  \href {https://ui.adsabs.harvard.edu/abs/2021MNRAS.508..781K} {508, 781}

\bibitem[\protect\citeauthoryear{{Kenyon}}{{Kenyon}}{1986}]{Kenyon_1986}
{Kenyon} S.~J.,  1986, The Symbiotic Stars.
Cambridge Astrophysics, Cambridge University Press,
  \mn@doi{10.1017/CBO9780511586071}

\bibitem[\protect\citeauthoryear{{Kenyon}, {Livio}, {Mikolajewska}  \&
  {Tout}}{{Kenyon} et~al.}{1993}]{Kenyon_1993}
{Kenyon} S.~J.,  {Livio} M.,  {Mikolajewska} J.,   {Tout} C.~A.,  1993, \mn@doi
  [\apjl] {10.1086/186811}, \href
  {https://ui.adsabs.harvard.edu/abs/1993ApJ...407L..81K} {407, L81}

\bibitem[\protect\citeauthoryear{{Khargharia}, {Froning}  \&
  {Robinson}}{{Khargharia} et~al.}{2010}]{Khargharia_2010}
{Khargharia} J.,  {Froning} C.~S.,   {Robinson} E.~L.,  2010, \mn@doi [\apj]
  {10.1088/0004-637X/716/2/1105}, \href
  {https://ui.adsabs.harvard.edu/abs/2010ApJ...716.1105K} {716, 1105}

\bibitem[\protect\citeauthoryear{{King} \& {Kolb}}{{King} \&
  {Kolb}}{1995}]{King_1995}
{King} A.~R.,  {Kolb} U.,  1995, \mn@doi [\apj] {10.1086/175176}, \href
  {http://adsabs.harvard.edu/abs/1995ApJ...439..330K} {439, 330}

\bibitem[\protect\citeauthoryear{{Klencki}, {Nelemans}, {Istrate}  \&
  {Pols}}{{Klencki} et~al.}{2020}]{Klencki_2020}
{Klencki} J.,  {Nelemans} G.,  {Istrate} A.~G.,   {Pols} O.,  2020, \mn@doi
  [\aap] {10.1051/0004-6361/202037694}, \href
  {https://ui.adsabs.harvard.edu/abs/2020A&A...638A..55K} {638, A55}

\bibitem[\protect\citeauthoryear{{Knigge}, {Zurek}, {Shara}, {Long}  \&
  {Gilliland}}{{Knigge} et~al.}{2003}]{Knigge_2003}
{Knigge} C.,  {Zurek} D.~R.,  {Shara} M.~M.,  {Long} K.~S.,   {Gilliland}
  R.~L.,  2003, \mn@doi [\apj] {10.1086/379609}, \href
  {https://ui.adsabs.harvard.edu/abs/2003ApJ...599.1320K} {599, 1320}

\bibitem[\protect\citeauthoryear{{Knigge}, {Baraffe}  \& {Patterson}}{{Knigge}
  et~al.}{2011}]{Knigge_2011_OK}
{Knigge} C.,  {Baraffe} I.,   {Patterson} J.,  2011, \mn@doi [\apjs]
  {10.1088/0067-0049/194/2/28}, \href
  {http://adsabs.harvard.edu/abs/2011ApJS..194...28K} {194, 28}

\bibitem[\protect\citeauthoryear{{Kolb} \& {Ritter}}{{Kolb} \&
  {Ritter}}{1990}]{Kolb_1990}
{Kolb} U.,  {Ritter} H.,  1990, \aap, \href
  {https://ui.adsabs.harvard.edu/abs/1990A&A...236..385K} {236, 385}

\bibitem[\protect\citeauthoryear{{Koliopanos}, {Gilfanov}  \&
  {Bildsten}}{{Koliopanos} et~al.}{2013}]{Koliopanos_2013}
{Koliopanos} F.,  {Gilfanov} M.,   {Bildsten} L.,  2013, \mn@doi [\mnras]
  {10.1093/mnras/stt542}, \href
  {https://ui.adsabs.harvard.edu/abs/2013MNRAS.432.1264K} {432, 1264}

\bibitem[\protect\citeauthoryear{{Koliopanos}, {P{\'e}ault}, {Vasilopoulos}  \&
  {Webb}}{{Koliopanos} et~al.}{2021}]{Koliopanos_2021}
{Koliopanos} F.,  {P{\'e}ault} M.,  {Vasilopoulos} G.,   {Webb} N.,  2021,
  \mn@doi [\mnras] {10.1093/mnras/staa3474}, \href
  {https://ui.adsabs.harvard.edu/abs/2021MNRAS.501..548K} {501, 548}

\bibitem[\protect\citeauthoryear{{Korol}, {Rossi}, {Groot}, {Nelemans},
  {Toonen}  \& {Brown}}{{Korol} et~al.}{2017}]{Korol_2017}
{Korol} V.,  {Rossi} E.~M.,  {Groot} P.~J.,  {Nelemans} G.,  {Toonen} S.,
  {Brown} A. G.~A.,  2017, \mn@doi [\mnras] {10.1093/mnras/stx1285}, \href
  {https://ui.adsabs.harvard.edu/abs/2017MNRAS.470.1894K} {470, 1894}

\bibitem[\protect\citeauthoryear{{Kremer}, {Chatterjee}, {Rodriguez}  \&
  {Rasio}}{{Kremer} et~al.}{2018}]{Kremer_2018}
{Kremer} K.,  {Chatterjee} S.,  {Rodriguez} C.~L.,   {Rasio} F.~A.,  2018,
  arXiv e-prints, \href {https://ui.adsabs.harvard.edu/abs/2018arXiv180204895K}
  {p. arXiv:1802.04895}

\bibitem[\protect\citeauthoryear{{Kretschmar} et~al.,}{{Kretschmar}
  et~al.}{2019}]{Kretschmar_2019}
{Kretschmar} P.,  et~al., 2019, \mn@doi [\nar] {10.1016/j.newar.2020.101546},
  \href {https://ui.adsabs.harvard.edu/abs/2019NewAR..8601546K} {86, 101546}

\bibitem[\protect\citeauthoryear{{Kroupa}}{{Kroupa}}{1995}]{Kroupa_1995a}
{Kroupa} P.,  1995, \mn@doi [\mnras] {10.1093/mnras/277.4.1491}, \href
  {http://adsabs.harvard.edu/abs/1995MNRAS.277.1491K} {277, 1491}

\bibitem[\protect\citeauthoryear{{Kruckow}, {Tauris}, {Langer}, {Sz{\'e}csi},
  {Marchant}  \& {Podsiadlowski}}{{Kruckow} et~al.}{2016}]{Kruckow_2016}
{Kruckow} M.~U.,  {Tauris} T.~M.,  {Langer} N.,  {Sz{\'e}csi} D.,  {Marchant}
  P.,   {Podsiadlowski} P.,  2016, \mn@doi [\aap]
  {10.1051/0004-6361/201629420}, \href
  {https://ui.adsabs.harvard.edu/abs/2016A&A...596A..58K} {596, A58}

\bibitem[\protect\citeauthoryear{{Lada} \& {Lada}}{{Lada} \&
  {Lada}}{2003}]{Lada_2003}
{Lada} C.~J.,  {Lada} E.~A.,  2003, \mn@doi [\araa]
  {10.1146/annurev.astro.41.011802.094844}, \href
  {http://adsabs.harvard.edu/abs/2003ARA%26A..41...57L} {41, 57}

\bibitem[\protect\citeauthoryear{{Lasota}}{{Lasota}}{2001}]{Lasota_2001}
{Lasota} J.-P.,  2001, \mn@doi [\nar] {10.1016/S1387-6473(01)00112-9}, \href
  {http://adsabs.harvard.edu/abs/2001NewAR..45..449L} {45, 449}

\bibitem[\protect\citeauthoryear{{Lasota}}{{Lasota}}{2016}]{Lasota_2016}
{Lasota} J.-P.,  2016, in {Bambi} C.,  ed.,  Astrophysics and Space Science
  Library Vol. 440, Astrophysics of Black Holes: From Fundamental Aspects to
  Latest Developments. p.~1 (\mn@eprint {arXiv} {1505.02172}),
  \mn@doi{10.1007/978-3-662-52859-4_1}

\bibitem[\protect\citeauthoryear{{Lehmer} et~al.,}{{Lehmer}
  et~al.}{2020}]{Lehmer_2020}
{Lehmer} B.~D.,  et~al., 2020, \mn@doi [\apjs] {10.3847/1538-4365/ab9175},
  \href {https://ui.adsabs.harvard.edu/abs/2020ApJS..248...31L} {248, 31}

\bibitem[\protect\citeauthoryear{{Li}, {Wu}  \& {Wickramasinghe}}{{Li}
  et~al.}{1994}]{LWW94}
{Li} J.~K.,  {Wu} K.~W.,   {Wickramasinghe} D.~T.,  1994, \mn@doi [\mnras]
  {10.1093/mnras/268.1.61}, \href
  {http://adsabs.harvard.edu/abs/1994MNRAS.268...61L} {268, 61}

\bibitem[\protect\citeauthoryear{{Liebert} et~al.,}{{Liebert}
  et~al.}{2005}]{Liebert_2005}
{Liebert} J.,  et~al., 2005, \mn@doi [\aj] {10.1086/429639}, \href
  {https://ui.adsabs.harvard.edu/abs/2005AJ....129.2376L} {129, 2376}

\bibitem[\protect\citeauthoryear{{Liu} \& {Li}}{{Liu} \& {Li}}{2019}]{Liu_2019}
{Liu} W.-M.,  {Li} X.-D.,  2019, \mn@doi [\apj] {10.3847/1538-4357/aaf19f},
  \href {https://ui.adsabs.harvard.edu/abs/2019ApJ...870...22L} {870, 22}

\bibitem[\protect\citeauthoryear{{Liu}, {van Paradijs}  \& {van den
  Heuvel}}{{Liu} et~al.}{2006}]{Liu_2006}
{Liu} Q.~Z.,  {van Paradijs} J.,   {van den Heuvel} E.~P.~J.,  2006, \mn@doi
  [\aap] {10.1051/0004-6361:20064987}, \href
  {https://ui.adsabs.harvard.edu/abs/2006A&A...455.1165L} {455, 1165}

\bibitem[\protect\citeauthoryear{{Liu}, {Bregman}, {Bai}, {Justham}  \&
  {Crowther}}{{Liu} et~al.}{2013}]{Liu_2013}
{Liu} J.-F.,  {Bregman} J.~N.,  {Bai} Y.,  {Justham} S.,   {Crowther} P.,
  2013, \mn@doi [\nat] {10.1038/nature12762}, \href
  {https://ui.adsabs.harvard.edu/abs/2013Natur.503..500L} {503, 500}

\bibitem[\protect\citeauthoryear{{Liu}, {Wang}, {Ge}, {Chen}  \& {Han}}{{Liu}
  et~al.}{2019}]{Liu_2019_SySt}
{Liu} D.,  {Wang} B.,  {Ge} H.,  {Chen} X.,   {Han} Z.,  2019, \mn@doi [\aap]
  {10.1051/0004-6361/201833010}, \href
  {https://ui.adsabs.harvard.edu/abs/2019A&A...622A..35L} {622, A35}

\bibitem[\protect\citeauthoryear{{Liu}, {Jiang}  \& {Chen}}{{Liu}
  et~al.}{2021}]{Liu_2021}
{Liu} W.-M.,  {Jiang} L.,   {Chen} W.-C.,  2021, \mn@doi [\apj]
  {10.3847/1538-4357/abdfc7}, \href
  {https://ui.adsabs.harvard.edu/abs/2021ApJ...910...22L} {910, 22}

\bibitem[\protect\citeauthoryear{{Long}, {Helfand}  \& {Grabelsky}}{{Long}
  et~al.}{1981}]{Long_1981}
{Long} K.~S.,  {Helfand} D.~J.,   {Grabelsky} D.~A.,  1981, \mn@doi [\apj]
  {10.1086/159222}, \href
  {https://ui.adsabs.harvard.edu/abs/1981ApJ...248..925L} {248, 925}

\bibitem[\protect\citeauthoryear{{L{\'o}pez-C{\'a}mara}, {De Colle}  \& {Moreno
  M{\'e}ndez}}{{L{\'o}pez-C{\'a}mara} et~al.}{2019}]{Diego_2019}
{L{\'o}pez-C{\'a}mara} D.,  {De Colle} F.,   {Moreno M{\'e}ndez} E.,  2019,
  \mn@doi [\mnras] {10.1093/mnras/sty2959}, \href
  {https://ui.adsabs.harvard.edu/abs/2019MNRAS.482.3646L} {482, 3646}

\bibitem[\protect\citeauthoryear{{L{\'o}pez-C{\'a}mara}, {Moreno M{\'e}ndez}
  \& {De Colle}}{{L{\'o}pez-C{\'a}mara} et~al.}{2020}]{Diego_2020}
{L{\'o}pez-C{\'a}mara} D.,  {Moreno M{\'e}ndez} E.,   {De Colle} F.,  2020,
  \mn@doi [\mnras] {10.1093/mnras/staa1983}, \href
  {https://ui.adsabs.harvard.edu/abs/2020MNRAS.497.2057L} {497, 2057}

\bibitem[\protect\citeauthoryear{{L{\'o}pez-C{\'a}mara}, {De Colle}, {Moreno
  M{\'e}ndez}, {Shiber}  \& {Iaconi}}{{L{\'o}pez-C{\'a}mara}
  et~al.}{2022}]{Diego_2022}
{L{\'o}pez-C{\'a}mara} D.,  {De Colle} F.,  {Moreno M{\'e}ndez} E.,  {Shiber}
  S.,   {Iaconi} R.,  2022, \mn@doi [\mnras] {10.1093/mnras/stac932}, \href
  {https://ui.adsabs.harvard.edu/abs/2022MNRAS.513.3634L} {513, 3634}

\bibitem[\protect\citeauthoryear{{L{\"u}}, {Zhu}, {Postnov}, {Yungelson},
  {Kuranov}  \& {Wang}}{{L{\"u}} et~al.}{2012}]{Lu_2012}
{L{\"u}} G.~L.,  {Zhu} C.~H.,  {Postnov} K.~A.,  {Yungelson} L.~R.,  {Kuranov}
  A.~G.,   {Wang} N.,  2012, \mn@doi [\mnras]
  {10.1111/j.1365-2966.2012.21395.x}, \href
  {https://ui.adsabs.harvard.edu/abs/2012MNRAS.424.2265L} {424, 2265}

\bibitem[\protect\citeauthoryear{{Lucy}}{{Lucy}}{1967}]{Lucy_1967}
{Lucy} L.~B.,  1967, \mn@doi [\aj] {10.1086/110452}, \href
  {https://ui.adsabs.harvard.edu/abs/1967AJ.....72Q.813L} {72, 813}

\bibitem[\protect\citeauthoryear{{Lugger}, {Cohn}, {Cool}, {Heinke}  \&
  {Anderson}}{{Lugger} et~al.}{2017}]{Lugger_2017}
{Lugger} P.~M.,  {Cohn} H.~N.,  {Cool} A.~M.,  {Heinke} C.~O.,   {Anderson} J.,
   2017, \mn@doi [\apj] {10.3847/1538-4357/aa6c56}, \href
  {http://adsabs.harvard.edu/abs/2017ApJ...841...53L} {841, 53}

\bibitem[\protect\citeauthoryear{{Ma} \& {Li}}{{Ma} \& {Li}}{2009}]{Ma_2009}
{Ma} B.,  {Li} X.-D.,  2009, \mn@doi [\apj] {10.1088/0004-637X/698/2/1907},
  \href {https://ui.adsabs.harvard.edu/abs/2009ApJ...698.1907M} {698, 1907}

\bibitem[\protect\citeauthoryear{{MacDonald} et~al.,}{{MacDonald}
  et~al.}{2014}]{MacDonald_2014}
{MacDonald} R. K.~D.,  et~al., 2014, \mn@doi [\apj]
  {10.1088/0004-637X/784/1/2}, \href
  {https://ui.adsabs.harvard.edu/abs/2014ApJ...784....2M} {784, 2}

\bibitem[\protect\citeauthoryear{{Maraschi}, {Treves}  \& {van den
  Heuvel}}{{Maraschi} et~al.}{1976}]{Maraschietal_1976}
{Maraschi} L.,  {Treves} A.,   {van den Heuvel} E.~P.~J.,  1976, \mn@doi [\nat]
  {10.1038/259292a0}, \href
  {https://ui.adsabs.harvard.edu/abs/1976Natur.259..292M} {259, 292}

\bibitem[\protect\citeauthoryear{{Marks} \& {Kroupa}}{{Marks} \&
  {Kroupa}}{2012}]{Marks_2012}
{Marks} M.,  {Kroupa} P.,  2012, \mn@doi [\aap] {10.1051/0004-6361/201118231},
  \href {http://adsabs.harvard.edu/abs/2012A%26A...543A...8M} {543, A8}

\bibitem[\protect\citeauthoryear{{Marsh} et~al.,}{{Marsh}
  et~al.}{2016}]{Marsh_2016}
{Marsh} T.~R.,  et~al., 2016, \mn@doi [\nat] {10.1038/nature18620}, \href
  {https://ui.adsabs.harvard.edu/abs/2016Natur.537..374M} {537, 374}

\bibitem[\protect\citeauthoryear{{Mart{\'\i}nez-N{\'u}{\~n}ez}
  et~al.,}{{Mart{\'\i}nez-N{\'u}{\~n}ez} et~al.}{2017}]{M2017}
{Mart{\'\i}nez-N{\'u}{\~n}ez} S.,  et~al., 2017, \mn@doi [\ssr]
  {10.1007/s11214-017-0340-1}, \href
  {https://ui.adsabs.harvard.edu/abs/2017SSRv..212...59M} {212, 59}

\bibitem[\protect\citeauthoryear{{Mason}, {Norton}, {Clark}, {Negueruela}  \&
  {Roche}}{{Mason} et~al.}{2010}]{Mason_2010}
{Mason} A.~B.,  {Norton} A.~J.,  {Clark} J.~S.,  {Negueruela} I.,   {Roche} P.,
   2010, \mn@doi [\aap] {10.1051/0004-6361/200913394}, \href
  {https://ui.adsabs.harvard.edu/abs/2010A&A...509A..79M} {509, A79}

\bibitem[\protect\citeauthoryear{{Mason}, {Norton}, {Clark}, {Negueruela}  \&
  {Roche}}{{Mason} et~al.}{2011}]{Mason_2011}
{Mason} A.~B.,  {Norton} A.~J.,  {Clark} J.~S.,  {Negueruela} I.,   {Roche} P.,
   2011, \mn@doi [\aap] {10.1051/0004-6361/201117392}, \href
  {https://ui.adsabs.harvard.edu/abs/2011A&A...532A.124M} {532, A124}

\bibitem[\protect\citeauthoryear{{Mason}, {Clark}, {Norton}, {Crowther},
  {Tauris}, {Langer}, {Negueruela}  \& {Roche}}{{Mason}
  et~al.}{2012}]{Mason_2012}
{Mason} A.~B.,  {Clark} J.~S.,  {Norton} A.~J.,  {Crowther} P.~A.,  {Tauris}
  T.~M.,  {Langer} N.,  {Negueruela} I.,   {Roche} P.,  2012, \mn@doi [\mnras]
  {10.1111/j.1365-2966.2012.20596.x}, \href
  {https://ui.adsabs.harvard.edu/abs/2012MNRAS.422..199M} {422, 199}

\bibitem[\protect\citeauthoryear{{Mathieu} \& {Geller}}{{Mathieu} \&
  {Geller}}{2009}]{Mathieu_2009}
{Mathieu} R.~D.,  {Geller} A.~M.,  2009, \mn@doi [\nat] {10.1038/nature08568},
  \href {https://ui.adsabs.harvard.edu/abs/2009Natur.462.1032M} {462, 1032}

\bibitem[\protect\citeauthoryear{{Matrozis}, {Abate}  \&
  {Stancliffe}}{{Matrozis} et~al.}{2017}]{Matrozis_2017}
{Matrozis} E.,  {Abate} C.,   {Stancliffe} R.~J.,  2017, \mn@doi [\aap]
  {10.1051/0004-6361/201730746}, \href
  {https://ui.adsabs.harvard.edu/abs/2017A&A...606A.137M} {606, A137}

\bibitem[\protect\citeauthoryear{{McAllister} et~al.,}{{McAllister}
  et~al.}{2019}]{McAllister_2019}
{McAllister} M.,  et~al., 2019, \mn@doi [\mnras] {10.1093/mnras/stz976}, \href
  {https://ui.adsabs.harvard.edu/abs/2019MNRAS.486.5535M} {486, 5535}

\bibitem[\protect\citeauthoryear{{Mestel}}{{Mestel}}{1968}]{Mestel_1968}
{Mestel} L.,  1968, \mn@doi [\mnras] {10.1093/mnras/138.3.359}, \href
  {https://ui.adsabs.harvard.edu/abs/1968MNRAS.138..359M} {138, 359}

\bibitem[\protect\citeauthoryear{{Miko{\l}ajewska}}{{Miko{\l}ajewska}}{2003}]{Mikolajewska_2003}
{Miko{\l}ajewska} J.,  2003, in {Corradi} R.~L.~M.,  {Mikolajewska} J.,
  {Mahoney} T.~J.,  eds,  Astronomical Society of the Pacific Conference Series
  Vol. 303, Symbiotic Stars Probing Stellar Evolution. p.~9 (\mn@eprint {arXiv}
  {astro-ph/0210489})

\bibitem[\protect\citeauthoryear{{Miko{\l}ajewska}}{{Miko{\l}ajewska}}{2007}]{Mikolajewska_2007}
{Miko{\l}ajewska} J.,  2007, Baltic Astronomy, \href
  {https://ui.adsabs.harvard.edu/abs/2007BaltA..16....1M} {16, 1}

\bibitem[\protect\citeauthoryear{{Mikolajewska}}{{Mikolajewska}}{2010}]{Mikolajewska_2010}
{Mikolajewska} J.,  2010, arXiv e-prints, \href
  {https://ui.adsabs.harvard.edu/abs/2010arXiv1011.5657M} {p. arXiv:1011.5657}

\bibitem[\protect\citeauthoryear{{Miko{\l}ajewska}}{{Miko{\l}ajewska}}{2012}]{Mikolajewska_2012}
{Miko{\l}ajewska} J.,  2012, Baltic Astronomy, \href
  {http://adsabs.harvard.edu/abs/2012BaltA..21....5M} {21, 5}

\bibitem[\protect\citeauthoryear{{Miko{\l}ajewska}, {Shara}, {Caldwell},
  {I{\l}kiewicz}  \& {Zurek}}{{Miko{\l}ajewska}
  et~al.}{2017}]{Mikolajewska_2017}
{Miko{\l}ajewska} J.,  {Shara} M.~M.,  {Caldwell} N.,  {I{\l}kiewicz} K.,
  {Zurek} D.,  2017, \mn@doi [\mnras] {10.1093/mnras/stw2937}, \href
  {http://adsabs.harvard.edu/abs/2017MNRAS.465.1699M} {465, 1699}

\bibitem[\protect\citeauthoryear{{Miko{\l}ajewska}, {I{\l}kiewicz}, {Ga{\l}an},
  {Monard}, {Otulakowska-Hypka}, {Shara}  \& {Udalski}}{{Miko{\l}ajewska}
  et~al.}{2021}]{Mikolajewska_2021}
{Miko{\l}ajewska} J.,  {I{\l}kiewicz} K.,  {Ga{\l}an} C.,  {Monard} B.,
  {Otulakowska-Hypka} M.,  {Shara} M.~M.,   {Udalski} A.,  2021, \mn@doi
  [\mnras] {10.1093/mnras/stab1058}, \href
  {https://ui.adsabs.harvard.edu/abs/2021MNRAS.504.2122M} {504, 2122}

\bibitem[\protect\citeauthoryear{{Miller-Jones} et~al.,}{{Miller-Jones}
  et~al.}{2021}]{MillerJones_2021}
{Miller-Jones} J. C.~A.,  et~al., 2021, \mn@doi [Science]
  {10.1126/science.abb3363}, \href
  {https://ui.adsabs.harvard.edu/abs/2021Sci...371.1046M} {371, 1046}

\bibitem[\protect\citeauthoryear{{Miszalski} et~al.,}{{Miszalski}
  et~al.}{2013}]{Miszalski_2013}
{Miszalski} B.,  et~al., 2013, \mn@doi [\mnras] {10.1093/mnras/stt1795}, \href
  {https://ui.adsabs.harvard.edu/abs/2013MNRAS.436.3068M} {436, 3068}

\bibitem[\protect\citeauthoryear{{Moe} \& {Di Stefano}}{{Moe} \& {Di
  Stefano}}{2017}]{MD_2017}
{Moe} M.,  {Di Stefano} R.,  2017, \mn@doi [\apjs] {10.3847/1538-4365/aa6fb6},
  \href {http://adsabs.harvard.edu/abs/2017ApJS..230...15M} {230, 15}

\bibitem[\protect\citeauthoryear{{Mohamed} \& {Podsiadlowski}}{{Mohamed} \&
  {Podsiadlowski}}{2007}]{Mohamed_2007}
{Mohamed} S.,  {Podsiadlowski} P.,  2007, in {Napiwotzki} R.,  {Burleigh}
  M.~R.,  eds,  Astronomical Society of the Pacific Conference Series Vol. 372,
  15th European Workshop on White Dwarfs. p.~397

\bibitem[\protect\citeauthoryear{{Mohamed} \& {Podsiadlowski}}{{Mohamed} \&
  {Podsiadlowski}}{2012}]{Mohamed_2012}
{Mohamed} S.,  {Podsiadlowski} P.,  2012, \mn@doi [Baltic Astronomy]
  {10.1515/astro-2017-0362}, \href
  {https://ui.adsabs.harvard.edu/abs/2012BaltA..21...88M} {21, 88}

\bibitem[\protect\citeauthoryear{{Moreno M{\'e}ndez}, {L{\'o}pez-C{\'a}mara}
  \& {De Colle}}{{Moreno M{\'e}ndez} et~al.}{2017}]{Diego_2017}
{Moreno M{\'e}ndez} E.,  {L{\'o}pez-C{\'a}mara} D.,   {De Colle} F.,  2017,
  \mn@doi [\mnras] {10.1093/mnras/stx1385}, \href
  {https://ui.adsabs.harvard.edu/abs/2017MNRAS.470.2929M} {470, 2929}

\bibitem[\protect\citeauthoryear{{Mu{\~n}oz-Darias}, {Casares}  \&
  {Mart{\'\i}nez-Pais}}{{Mu{\~n}oz-Darias} et~al.}{2008}]{MD2008}
{Mu{\~n}oz-Darias} T.,  {Casares} J.,   {Mart{\'\i}nez-Pais} I.~G.,  2008,
  \mn@doi [\mnras] {10.1111/j.1365-2966.2008.12987.x}, \href
  {https://ui.adsabs.harvard.edu/abs/2008MNRAS.385.2205M} {385, 2205}

\bibitem[\protect\citeauthoryear{{Mukai}}{{Mukai}}{2017}]{Mukai_2017}
{Mukai} K.,  2017, \mn@doi [\pasp] {10.1088/1538-3873/aa6736}, \href
  {https://ui.adsabs.harvard.edu/abs/2017PASP..129f2001M} {129, 062001}

\bibitem[\protect\citeauthoryear{{Nandez}, {Ivanova}  \& {Lombardi}}{{Nandez}
  et~al.}{2015}]{NIL2015}
{Nandez} J.~L.~A.,  {Ivanova} N.,   {Lombardi} J.~C.~J.,  2015, \mn@doi
  [\mnras] {10.1093/mnrasl/slv043}, \href
  {https://ui.adsabs.harvard.edu/abs/2015MNRAS.450L..39N} {450, L39}

\bibitem[\protect\citeauthoryear{{Nebot G{\'o}mez-Mor{\'a}n} et~al.,}{{Nebot
  G{\'o}mez-Mor{\'a}n} et~al.}{2011}]{Nebot_2011}
{Nebot G{\'o}mez-Mor{\'a}n} A.,  et~al., 2011, \mn@doi [\aap]
  {10.1051/0004-6361/201117514}, \href
  {https://ui.adsabs.harvard.edu/abs/2011A&A...536A..43N} {536, A43}

\bibitem[\protect\citeauthoryear{{Neijssel} et~al.,}{{Neijssel}
  et~al.}{2019}]{Neijssel_2019}
{Neijssel} C.~J.,  et~al., 2019, \mn@doi [\mnras] {10.1093/mnras/stz2840},
  \href {https://ui.adsabs.harvard.edu/abs/2019MNRAS.490.3740N} {490, 3740}

\bibitem[\protect\citeauthoryear{{Nelemans}}{{Nelemans}}{2013}]{Nelemans_2013}
{Nelemans} G.,  2013, in {Auger} G.,  {Bin{\'e}truy} P.,   {Plagnol} E.,  eds,
  Astronomical Society of the Pacific Conference Series Vol. 467, 9th LISA
  Symposium. p.~27 (\mn@eprint {arXiv} {1302.0138})

\bibitem[\protect\citeauthoryear{{Nelemans} \& {Jonker}}{{Nelemans} \&
  {Jonker}}{2010}]{Nelemans_2010}
{Nelemans} G.,  {Jonker} P.~G.,  2010, \mn@doi [\nar]
  {10.1016/j.newar.2010.09.021}, \href
  {https://ui.adsabs.harvard.edu/abs/2010NewAR..54...87N} {54, 87}

\bibitem[\protect\citeauthoryear{{Nelemans} \& {Tout}}{{Nelemans} \&
  {Tout}}{2005}]{Nelemans_2005}
{Nelemans} G.,  {Tout} C.~A.,  2005, \mn@doi [\mnras]
  {10.1111/j.1365-2966.2004.08496.x}, \href
  {http://adsabs.harvard.edu/abs/2005MNRAS.356..753N} {356, 753}

\bibitem[\protect\citeauthoryear{{Nelemans}, {Verbunt}, {Yungelson}  \&
  {Portegies Zwart}}{{Nelemans} et~al.}{2000}]{Nelemans_2000}
{Nelemans} G.,  {Verbunt} F.,  {Yungelson} L.~R.,   {Portegies Zwart} S.~F.,
  2000, \aap, \href {http://adsabs.harvard.edu/abs/2000A%26A...360.1011N} {360,
  1011}

\bibitem[\protect\citeauthoryear{{Nelemans}, {Yungelson}, {van der Sluys}  \&
  {Tout}}{{Nelemans} et~al.}{2010}]{Nelemans_2010REF}
{Nelemans} G.,  {Yungelson} L.~R.,  {van der Sluys} M.~V.,   {Tout} C.~A.,
  2010, \mn@doi [\mnras] {10.1111/j.1365-2966.2009.15731.x}, \href
  {https://ui.adsabs.harvard.edu/abs/2010MNRAS.401.1347N} {401, 1347}

\bibitem[\protect\citeauthoryear{{Nelemans}, {Siess}, {Repetto}, {Toonen}  \&
  {Phinney}}{{Nelemans} et~al.}{2016}]{Nelemans_2016}
{Nelemans} G.,  {Siess} L.,  {Repetto} S.,  {Toonen} S.,   {Phinney} E.~S.,
  2016, \mn@doi [\apj] {10.3847/0004-637X/817/1/69}, \href
  {http://adsabs.harvard.edu/abs/2016ApJ...817...69N} {817, 69}

\bibitem[\protect\citeauthoryear{{Nelson}, {Dubeau}  \& {MacCannell}}{{Nelson}
  et~al.}{2004}]{Nelson_2004}
{Nelson} L.~A.,  {Dubeau} E.,   {MacCannell} K.~A.,  2004, \mn@doi [\apj]
  {10.1086/421698}, \href
  {https://ui.adsabs.harvard.edu/abs/2004ApJ...616.1124N} {616, 1124}

\bibitem[\protect\citeauthoryear{{Nie}, {Wood}  \& {Nicholls}}{{Nie}
  et~al.}{2012}]{Nie_2012}
{Nie} J.~D.,  {Wood} P.~R.,   {Nicholls} C.~P.,  2012, \mn@doi [\mnras]
  {10.1111/j.1365-2966.2012.21087.x}, \href
  {https://ui.adsabs.harvard.edu/abs/2012MNRAS.423.2764N} {423, 2764}

\bibitem[\protect\citeauthoryear{{Nie}, {Wood}  \& {Nicholls}}{{Nie}
  et~al.}{2017}]{Nie_2017}
{Nie} J.~D.,  {Wood} P.~R.,   {Nicholls} C.~P.,  2017, \mn@doi [\apj]
  {10.3847/1538-4357/835/2/209}, \href
  {https://ui.adsabs.harvard.edu/abs/2017ApJ...835..209N} {835, 209}

\bibitem[\protect\citeauthoryear{{Nielsen}, {Dominik}, {Nelemans}  \&
  {Voss}}{{Nielsen} et~al.}{2013}]{Nielsen_2013}
{Nielsen} M.~T.~B.,  {Dominik} C.,  {Nelemans} G.,   {Voss} R.,  2013, \mn@doi
  [\aap] {10.1051/0004-6361/201219195}, \href
  {https://ui.adsabs.harvard.edu/abs/2013A&A...549A..32N} {549, A32}

\bibitem[\protect\citeauthoryear{{Nomoto} \& {Kondo}}{{Nomoto} \&
  {Kondo}}{1991}]{Nomoto_1991}
{Nomoto} K.,  {Kondo} Y.,  1991, \mn@doi [\apjl] {10.1086/185922}, \href
  {http://adsabs.harvard.edu/abs/1991ApJ...367L..19N} {367, L19}

\bibitem[\protect\citeauthoryear{{Nomoto}, {Saio}, {Kato}  \&
  {Hachisu}}{{Nomoto} et~al.}{2007}]{Nomoto_2007}
{Nomoto} K.,  {Saio} H.,  {Kato} M.,   {Hachisu} I.,  2007, \mn@doi [\apj]
  {10.1086/518465}, \href
  {https://ui.adsabs.harvard.edu/abs/2007ApJ...663.1269N} {663, 1269}

\bibitem[\protect\citeauthoryear{{Ohlmann}, {R{\"o}pke}, {Pakmor}  \&
  {Springel}}{{Ohlmann} et~al.}{2016}]{Ohlmann_2016}
{Ohlmann} S.~T.,  {R{\"o}pke} F.~K.,  {Pakmor} R.,   {Springel} V.,  2016,
  \mn@doi [\apjl] {10.3847/2041-8205/816/1/L9}, \href
  {https://ui.adsabs.harvard.edu/abs/2016ApJ...816L...9O} {816, L9}

\bibitem[\protect\citeauthoryear{{Olejak}, {Belczynski}  \& {Ivanova}}{{Olejak}
  et~al.}{2021}]{Olejak_2021}
{Olejak} A.,  {Belczynski} K.,   {Ivanova} N.,  2021, \mn@doi [\aap]
  {10.1051/0004-6361/202140520}, \href
  {https://ui.adsabs.harvard.edu/abs/2021A&A...651A.100O} {651, A100}

\bibitem[\protect\citeauthoryear{{Oomen}, {Van Winckel}, {Pols}, {Nelemans},
  {Escorza}, {Manick}, {Kamath}  \& {Waelkens}}{{Oomen}
  et~al.}{2018}]{Oomen_2018}
{Oomen} G.-M.,  {Van Winckel} H.,  {Pols} O.,  {Nelemans} G.,  {Escorza} A.,
  {Manick} R.,  {Kamath} D.,   {Waelkens} C.,  2018, \mn@doi [\aap]
  {10.1051/0004-6361/201833816}, \href
  {https://ui.adsabs.harvard.edu/abs/2018A&A...620A..85O} {620, A85}

\bibitem[\protect\citeauthoryear{{Oomen}, {Pols}, {Van Winckel}  \&
  {Nelemans}}{{Oomen} et~al.}{2020}]{Oomen_2020}
{Oomen} G.-M.,  {Pols} O.,  {Van Winckel} H.,   {Nelemans} G.,  2020, \mn@doi
  [\aap] {10.1051/0004-6361/202038341}, \href
  {https://ui.adsabs.harvard.edu/abs/2020A&A...642A.234O} {642, A234}

\bibitem[\protect\citeauthoryear{{Orosz} et~al.,}{{Orosz}
  et~al.}{2007}]{Orosz_2007}
{Orosz} J.~A.,  et~al., 2007, \mn@doi [\nat] {10.1038/nature06218}, \href
  {https://ui.adsabs.harvard.edu/abs/2007Natur.449..872O} {449, 872}

\bibitem[\protect\citeauthoryear{{Orosz} et~al.,}{{Orosz}
  et~al.}{2009}]{Orosz_2009}
{Orosz} J.~A.,  et~al., 2009, \mn@doi [\apj] {10.1088/0004-637X/697/1/573},
  \href {https://ui.adsabs.harvard.edu/abs/2009ApJ...697..573O} {697, 573}

\bibitem[\protect\citeauthoryear{{Orosz}, {Steiner}, {McClintock}, {Torres},
  {Remillard}, {Bailyn}  \& {Miller}}{{Orosz} et~al.}{2011}]{Orosz_2011_LMXB}
{Orosz} J.~A.,  {Steiner} J.~F.,  {McClintock} J.~E.,  {Torres} M. A.~P.,
  {Remillard} R.~A.,  {Bailyn} C.~D.,   {Miller} J.~M.,  2011, \mn@doi [\apj]
  {10.1088/0004-637X/730/2/75}, \href
  {https://ui.adsabs.harvard.edu/abs/2011ApJ...730...75O} {730, 75}

\bibitem[\protect\citeauthoryear{{Otani}, {Oswalt}, {Lynas-Gray}, {Kilkenny},
  {Koen}, {Amaral}  \& {Jordan}}{{Otani} et~al.}{2018}]{Otani_2018}
{Otani} T.,  {Oswalt} T.~D.,  {Lynas-Gray} A.~E.,  {Kilkenny} D.,  {Koen} C.,
  {Amaral} M.,   {Jordan} R.,  2018, \mn@doi [\apj] {10.3847/1538-4357/aab9bf},
  \href {https://ui.adsabs.harvard.edu/abs/2018ApJ...859..145O} {859, 145}

\bibitem[\protect\citeauthoryear{{Paczy{\'n}ski}}{{Paczy{\'n}ski}}{1967}]{Paczynski_1967}
{Paczy{\'n}ski} B.,  1967, \actaa, \href
  {http://adsabs.harvard.edu/abs/1967AcA....17..287P} {17, 287}

\bibitem[\protect\citeauthoryear{{Paczy{\'n}ski}}{{Paczy{\'n}ski}}{1976}]{Paczynski_1976}
{Paczy{\'n}ski} B.,  1976, in {Eggleton} P.,  {Mitton} S.,   {Whelan} J.,  eds,
   IAU Symposium Vol. 73, Structure and Evolution of Close Binary Systems.
  p.~75

\bibitem[\protect\citeauthoryear{{Pala} et~al.,}{{Pala}
  et~al.}{2017}]{palaetal_2017}
{Pala} A.~F.,  et~al., 2017, \mn@doi [\mnras] {10.1093/mnras/stw3293}, \href
  {http://adsabs.harvard.edu/abs/2017MNRAS.466.2855P} {466, 2855}

\bibitem[\protect\citeauthoryear{{Pala} et~al.,}{{Pala}
  et~al.}{2020}]{Pala_2020}
{Pala} A.~F.,  et~al., 2020, \mn@doi [\mnras] {10.1093/mnras/staa764}, \href
  {https://ui.adsabs.harvard.edu/abs/2020MNRAS.494.3799P} {494, 3799}

\bibitem[\protect\citeauthoryear{{Pala} et~al.,}{{Pala}
  et~al.}{2022}]{Pala_2022}
{Pala} A.~F.,  et~al., 2022, \mn@doi [\mnras] {10.1093/mnras/stab3449}, \href
  {https://ui.adsabs.harvard.edu/abs/2022MNRAS.510.6110P} {510, 6110}

\bibitem[\protect\citeauthoryear{{Parsons} et~al.,}{{Parsons}
  et~al.}{2015}]{parsonsetal15-1}
{Parsons} S.~G.,  et~al., 2015, \mn@doi [\mnras] {10.1093/mnras/stv1395}, \href
  {http://adsabs.harvard.edu/abs/2015MNRAS.452.1754P} {452, 1754}

\bibitem[\protect\citeauthoryear{{Patterson} \& {Raymond}}{{Patterson} \&
  {Raymond}}{1985a}]{Patterson_1985}
{Patterson} J.,  {Raymond} J.~C.,  1985a, \mn@doi [\apj] {10.1086/163187},
  \href {http://adsabs.harvard.edu/abs/1985ApJ...292..535P} {292, 535}

\bibitem[\protect\citeauthoryear{{Patterson} \& {Raymond}}{{Patterson} \&
  {Raymond}}{1985b}]{Patterson_1985b}
{Patterson} J.,  {Raymond} J.~C.,  1985b, \mn@doi [\apj] {10.1086/163188},
  \href {https://ui.adsabs.harvard.edu/abs/1985ApJ...292..550P} {292, 550}

\bibitem[\protect\citeauthoryear{{Pavlovskii} \& {Ivanova}}{{Pavlovskii} \&
  {Ivanova}}{2015}]{PI15}
{Pavlovskii} K.,  {Ivanova} N.,  2015, \mn@doi [\mnras] {10.1093/mnras/stv619},
  \href {https://ui.adsabs.harvard.edu/abs/2015MNRAS.449.4415P} {449, 4415}

\bibitem[\protect\citeauthoryear{{Paxton}, {Bildsten}, {Dotter}, {Herwig},
  {Lesaffre}  \& {Timmes}}{{Paxton} et~al.}{2011}]{Paxton2011}
{Paxton} B.,  {Bildsten} L.,  {Dotter} A.,  {Herwig} F.,  {Lesaffre} P.,
  {Timmes} F.,  2011, \mn@doi [\apjs] {10.1088/0067-0049/192/1/3}, \href
  {https://ui.adsabs.harvard.edu/abs/2011ApJS..192....3P} {192, 3}

\bibitem[\protect\citeauthoryear{{Paxton} et~al.,}{{Paxton}
  et~al.}{2013}]{Paxton2013}
{Paxton} B.,  et~al., 2013, \mn@doi [\apjs] {10.1088/0067-0049/208/1/4}, \href
  {https://ui.adsabs.harvard.edu/abs/2013ApJS..208....4P} {208, 4}

\bibitem[\protect\citeauthoryear{{Paxton} et~al.,}{{Paxton}
  et~al.}{2015}]{Paxton2015}
{Paxton} B.,  et~al., 2015, \mn@doi [\apjs] {10.1088/0067-0049/220/1/15}, \href
  {https://ui.adsabs.harvard.edu/abs/2015ApJS..220...15P} {220, 15}

\bibitem[\protect\citeauthoryear{{Paxton} et~al.,}{{Paxton}
  et~al.}{2018}]{Paxton2018}
{Paxton} B.,  et~al., 2018, \mn@doi [\apjs] {10.3847/1538-4365/aaa5a8}, \href
  {https://ui.adsabs.harvard.edu/abs/2018ApJS..234...34P} {234, 34}

\bibitem[\protect\citeauthoryear{{Paxton} et~al.,}{{Paxton}
  et~al.}{2019}]{Paxton2019}
{Paxton} B.,  et~al., 2019, \mn@doi [\apjs] {10.3847/1538-4365/ab2241}, \href
  {https://ui.adsabs.harvard.edu/abs/2019ApJS..243...10P} {243, 10}

\bibitem[\protect\citeauthoryear{{Pearlman}, {Coley}, {Corbet}  \&
  {Pottschmidt}}{{Pearlman} et~al.}{2019}]{Pearlman_2019}
{Pearlman} A.~B.,  {Coley} J.~B.,  {Corbet} R. H.~D.,   {Pottschmidt} K.,
  2019, \mn@doi [\apj] {10.3847/1538-4357/aaf001}, \href
  {https://ui.adsabs.harvard.edu/abs/2019ApJ...873...86P} {873, 86}

\bibitem[\protect\citeauthoryear{{Perets} \& {Kratter}}{{Perets} \&
  {Kratter}}{2012}]{PK12}
{Perets} H.~B.,  {Kratter} K.~M.,  2012, \mn@doi [\apj]
  {10.1088/0004-637X/760/2/99}, \href
  {https://ui.adsabs.harvard.edu/abs/2012ApJ...760...99P} {760, 99}

\bibitem[\protect\citeauthoryear{{Pfahl}, {Rappaport}, {Podsiadlowski}  \&
  {Spruit}}{{Pfahl} et~al.}{2002}]{Pfahl_2002}
{Pfahl} E.,  {Rappaport} S.,  {Podsiadlowski} P.,   {Spruit} H.,  2002, \mn@doi
  [\apj] {10.1086/340794}, \href
  {https://ui.adsabs.harvard.edu/abs/2002ApJ...574..364P} {574, 364}

\bibitem[\protect\citeauthoryear{{Podsiadlowski}}{{Podsiadlowski}}{2001}]{Podsiadlowski_2001}
{Podsiadlowski} P.,  2001, in {Podsiadlowski} P.,  {Rappaport} S.,  {King}
  A.~R.,  {D'Antona} F.,   {Burderi} L.,  eds,  Astronomical Society of the
  Pacific Conference Series Vol. 229, Evolution of Binary and Multiple Star
  Systems. p.~239

\bibitem[\protect\citeauthoryear{{Podsiadlowski} \&
  {Rappaport}}{{Podsiadlowski} \& {Rappaport}}{2000}]{Podsiadlowski_2000}
{Podsiadlowski} P.,  {Rappaport} S.,  2000, \mn@doi [\apj] {10.1086/308323},
  \href {https://ui.adsabs.harvard.edu/abs/2000ApJ...529..946P} {529, 946}

\bibitem[\protect\citeauthoryear{{Podsiadlowski}, {Rappaport}  \&
  {Pfahl}}{{Podsiadlowski} et~al.}{2002}]{Podsiadlowski_2002}
{Podsiadlowski} P.,  {Rappaport} S.,   {Pfahl} E.~D.,  2002, \mn@doi [\apj]
  {10.1086/324686}, \href
  {https://ui.adsabs.harvard.edu/abs/2002ApJ...565.1107P} {565, 1107}

\bibitem[\protect\citeauthoryear{{Podsiadlowski}, {Han}  \&
  {Rappaport}}{{Podsiadlowski} et~al.}{2003}]{Podsiadlowski_2003_CV}
{Podsiadlowski} P.,  {Han} Z.,   {Rappaport} S.,  2003, \mn@doi [\mnras]
  {10.1046/j.1365-8711.2003.06380.x}, \href
  {https://ui.adsabs.harvard.edu/abs/2003MNRAS.340.1214P} {340, 1214}

\bibitem[\protect\citeauthoryear{{Podsiadlowski}, {Ivanova}, {Justham}  \&
  {Rappaport}}{{Podsiadlowski} et~al.}{2010}]{Podsiadlowski_2010}
{Podsiadlowski} P.,  {Ivanova} N.,  {Justham} S.,   {Rappaport} S.,  2010,
  \mn@doi [\mnras] {10.1111/j.1365-2966.2010.16751.x}, \href
  {https://ui.adsabs.harvard.edu/abs/2010MNRAS.406..840P} {406, 840}

\bibitem[\protect\citeauthoryear{{Pooley} \& {Hut}}{{Pooley} \&
  {Hut}}{2006}]{Pooley_2006}
{Pooley} D.,  {Hut} P.,  2006, \mn@doi [\apjl] {10.1086/507027}, \href
  {http://adsabs.harvard.edu/abs/2006ApJ...646L.143P} {646, L143}

\bibitem[\protect\citeauthoryear{Pooley et~al.,}{Pooley
  et~al.}{2003}]{Pooley_2003}
Pooley D.,  et~al., 2003, The Astrophysical Journal Letters, 591, L131

\bibitem[\protect\citeauthoryear{{Porter} \& {Rivinius}}{{Porter} \&
  {Rivinius}}{2003}]{PR2003}
{Porter} J.~M.,  {Rivinius} T.,  2003, \mn@doi [\pasp] {10.1086/378307}, \href
  {https://ui.adsabs.harvard.edu/abs/2003PASP..115.1153P} {115, 1153}

\bibitem[\protect\citeauthoryear{{Postnov} \& {Yungelson}}{{Postnov} \&
  {Yungelson}}{2014}]{Postnov_2014}
{Postnov} K.~A.,  {Yungelson} L.~R.,  2014, \mn@doi [Living Reviews in
  Relativity] {10.12942/lrr-2014-3}, \href
  {https://ui.adsabs.harvard.edu/abs/2014LRR....17....3P} {17, 3}

\bibitem[\protect\citeauthoryear{{Qin}, {Marchant}, {Fragos}, {Meynet}  \&
  {Kalogera}}{{Qin} et~al.}{2019}]{Qin_2019}
{Qin} Y.,  {Marchant} P.,  {Fragos} T.,  {Meynet} G.,   {Kalogera} V.,  2019,
  \mn@doi [\apjl] {10.3847/2041-8213/aaf97b}, \href
  {https://ui.adsabs.harvard.edu/abs/2019ApJ...870L..18Q} {870, L18}

\bibitem[\protect\citeauthoryear{{Qiu} et~al.,}{{Qiu} et~al.}{2019}]{Q2019}
{Qiu} Y.,  et~al., 2019, \mn@doi [\apj] {10.3847/1538-4357/ab16e7}, \href
  {https://ui.adsabs.harvard.edu/abs/2019ApJ...877...57Q} {877, 57}

\bibitem[\protect\citeauthoryear{{Quaintrell}, {Norton}, {Ash}, {Roche},
  {Willems}, {Bedding}, {Baldry}  \& {Fender}}{{Quaintrell}
  et~al.}{2003}]{Quaintrell_2003}
{Quaintrell} H.,  {Norton} A.~J.,  {Ash} T.~D.~C.,  {Roche} P.,  {Willems} B.,
  {Bedding} T.~R.,  {Baldry} I.~K.,   {Fender} R.~P.,  2003, \mn@doi [\aap]
  {10.1051/0004-6361:20030120}, \href
  {https://ui.adsabs.harvard.edu/abs/2003A&A...401..313Q} {401, 313}

\bibitem[\protect\citeauthoryear{{Rafikov}}{{Rafikov}}{2016}]{Rafikov_2016}
{Rafikov} R.~R.,  2016, \mn@doi [\apj] {10.3847/0004-637X/830/1/8}, \href
  {https://ui.adsabs.harvard.edu/abs/2016ApJ...830....8R} {830, 8}

\bibitem[\protect\citeauthoryear{{Ramsay} et~al.,}{{Ramsay}
  et~al.}{2018}]{Ramsay_2018}
{Ramsay} G.,  et~al., 2018, \mn@doi [\aap] {10.1051/0004-6361/201834261}, \href
  {https://ui.adsabs.harvard.edu/abs/2018A&A...620A.141R} {620, A141}

\bibitem[\protect\citeauthoryear{{Rappaport} \& {van den Heuvel}}{{Rappaport}
  \& {van den Heuvel}}{1982}]{RH1982}
{Rappaport} S.,  {van den Heuvel} E.~P.~J.,  1982, in {Jaschek} M.,  {Groth}
  H.~G.,  eds,  IAU Symposium Vol. 98, Be Stars. pp 327--344

\bibitem[\protect\citeauthoryear{{Rappaport}, {Verbunt}  \& {Joss}}{{Rappaport}
  et~al.}{1983}]{RVJ}
{Rappaport} S.,  {Verbunt} F.,   {Joss} P.~C.,  1983, \mn@doi [\apj]
  {10.1086/161569}, \href {http://adsabs.harvard.edu/abs/1983ApJ...275..713R}
  {275, 713}

\bibitem[\protect\citeauthoryear{{Rebassa-Mansergas}
  et~al.,}{{Rebassa-Mansergas} et~al.}{2012}]{RebassaMansergas_2012}
{Rebassa-Mansergas} A.,  et~al., 2012, \mn@doi [\mnras]
  {10.1111/j.1365-2966.2012.20880.x}, \href
  {https://ui.adsabs.harvard.edu/abs/2012MNRAS.423..320R} {423, 320}

\bibitem[\protect\citeauthoryear{{Reig}}{{Reig}}{2011}]{Reig_2011}
{Reig} P.,  2011, \mn@doi [\apss] {10.1007/s10509-010-0575-8}, \href
  {https://ui.adsabs.harvard.edu/abs/2011Ap&SS.332....1R} {332, 1}

\bibitem[\protect\citeauthoryear{{Reimers}}{{Reimers}}{1975}]{Reimers_1975}
{Reimers} D.,  1975, Memoires of the Societe Royale des Sciences de Liege,
  \href {https://ui.adsabs.harvard.edu/abs/1975MSRSL...8..369R} {8, 369}

\bibitem[\protect\citeauthoryear{{Ricker}, {Timmes}, {Taam}  \&
  {Webbink}}{{Ricker} et~al.}{2019}]{Ricker_2019}
{Ricker} P.~M.,  {Timmes} F.~X.,  {Taam} R.~E.,   {Webbink} R.~F.,  2019,
  \mn@doi [IAU Symposium] {10.1017/S1743921318007433}, \href
  {https://ui.adsabs.harvard.edu/abs/2019IAUS..346..449R} {346, 449}

\bibitem[\protect\citeauthoryear{{Ritter}}{{Ritter}}{1988}]{Ritter_1988}
{Ritter} H.,  1988, \aap, \href
  {https://ui.adsabs.harvard.edu/abs/1988A&A...202...93R} {202, 93}

\bibitem[\protect\citeauthoryear{{Rivera Sandoval} et~al.,}{{Rivera Sandoval}
  et~al.}{2018}]{Rivera_2018}
{Rivera Sandoval} L.~E.,  et~al., 2018, \mn@doi [\mnras]
  {10.1093/mnras/sty058}, \href
  {https://ui.adsabs.harvard.edu/abs/2018MNRAS.475.4841R} {475, 4841}

\bibitem[\protect\citeauthoryear{{Saladino}, {Pols}  \& {Abate}}{{Saladino}
  et~al.}{2019}]{Saladino_2019}
{Saladino} M.~I.,  {Pols} O.~R.,   {Abate} C.,  2019, \mn@doi [\aap]
  {10.1051/0004-6361/201834598}, \href
  {https://ui.adsabs.harvard.edu/abs/2019A&A...626A..68S} {626, A68}

\bibitem[\protect\citeauthoryear{{Sana} et~al.,}{{Sana}
  et~al.}{2012}]{Sana_2012}
{Sana} H.,  et~al., 2012, \mn@doi [Science] {10.1126/science.1223344}, \href
  {http://adsabs.harvard.edu/abs/2012Sci...337..444S} {337, 444}

\bibitem[\protect\citeauthoryear{{Sand}, {Ohlmann}, {Schneider}, {Pakmor}  \&
  {R{\"o}pke}}{{Sand} et~al.}{2020}]{Sand_2020}
{Sand} C.,  {Ohlmann} S.~T.,  {Schneider} F. R.~N.,  {Pakmor} R.,   {R{\"o}pke}
  F.~K.,  2020, \mn@doi [\aap] {10.1051/0004-6361/202038992}, \href
  {https://ui.adsabs.harvard.edu/abs/2020A&A...644A..60S} {644, A60}

\bibitem[\protect\citeauthoryear{{Savonije}, {de Kool}  \& {van den
  Heuvel}}{{Savonije} et~al.}{1986}]{Savonije_1986}
{Savonije} G.~J.,  {de Kool} M.,   {van den Heuvel} E.~P.~J.,  1986, \aap,
  \href {https://ui.adsabs.harvard.edu/abs/1986A&A...155...51S} {155, 51}

\bibitem[\protect\citeauthoryear{{Schenker}, {Kolb}  \& {Ritter}}{{Schenker}
  et~al.}{1998}]{Schenker_1998}
{Schenker} K.,  {Kolb} U.,   {Ritter} H.,  1998, \mn@doi [\mnras]
  {10.1046/j.1365-8711.1998.01529.x}, \href
  {http://adsabs.harvard.edu/abs/1998MNRAS.297..633S} {297, 633}

\bibitem[\protect\citeauthoryear{{Schenker}, {King}, {Kolb}, {Wynn}  \&
  {Zhang}}{{Schenker} et~al.}{2002}]{schenkeretal02-1}
{Schenker} K.,  {King} A.~R.,  {Kolb} U.,  {Wynn} G.~A.,   {Zhang} Z.,  2002,
  \mn@doi [\mnras] {10.1046/j.1365-8711.2002.05999.x}, \href
  {http://adsabs.harvard.edu/abs/2002MNRAS.337.1105S} {337, 1105}

\bibitem[\protect\citeauthoryear{{Schreiber} \& {G{\"a}nsicke}}{{Schreiber} \&
  {G{\"a}nsicke}}{2001}]{Schreiber_2001}
{Schreiber} M.~R.,  {G{\"a}nsicke} B.~T.,  2001, \mn@doi [\aap]
  {10.1051/0004-6361:20010910}, \href
  {https://ui.adsabs.harvard.edu/abs/2001A&A...375..937S} {375, 937}

\bibitem[\protect\citeauthoryear{{Schreiber}, {G{\"a}nsicke}  \&
  {Cannizzo}}{{Schreiber} et~al.}{2000}]{Schreiber_2000}
{Schreiber} M.~R.,  {G{\"a}nsicke} B.~T.,   {Cannizzo} J.~K.,  2000, \aap,
  \href {https://ui.adsabs.harvard.edu/abs/2000A&A...362..268S} {362, 268}

\bibitem[\protect\citeauthoryear{{Schreiber} et~al.,}{{Schreiber}
  et~al.}{2010}]{Schreiber_2010}
{Schreiber} M.~R.,  et~al., 2010, \mn@doi [\aap] {10.1051/0004-6361/201013990},
  \href {http://adsabs.harvard.edu/abs/2010A%26A...513L...7S} {513, L7}

\bibitem[\protect\citeauthoryear{{Schreiber}, {Zorotovic}  \&
  {Wijnen}}{{Schreiber} et~al.}{2016}]{Schreiber_2016}
{Schreiber} M.~R.,  {Zorotovic} M.,   {Wijnen} T.~P.~G.,  2016, \mn@doi
  [\mnras] {10.1093/mnrasl/slv144}, \href
  {http://adsabs.harvard.edu/abs/2016MNRAS.455L..16S} {455, L16}

\bibitem[\protect\citeauthoryear{{Schreiber}, {Belloni}, {G{\"a}nsicke},
  {Parsons}  \& {Zorotovic}}{{Schreiber} et~al.}{2021}]{SchreiberNATURE}
{Schreiber} M.~R.,  {Belloni} D.,  {G{\"a}nsicke} B.~T.,  {Parsons} S.~G.,
  {Zorotovic} M.,  2021, \mn@doi [Nature Astronomy]
  {10.1038/s41550-021-01346-8}, \href
  {https://ui.adsabs.harvard.edu/abs/2021NatAs...5..648S} {5, 648}

\bibitem[\protect\citeauthoryear{{Shagatova}, {Skopal}  \&
  {Carikov{\'a}}}{{Shagatova} et~al.}{2016}]{Skopal_2016}
{Shagatova} N.,  {Skopal} A.,   {Carikov{\'a}} Z.,  2016, \mn@doi [\aap]
  {10.1051/0004-6361/201525645}, \href
  {https://ui.adsabs.harvard.edu/abs/2016A&A...588A..83S} {588, A83}

\bibitem[\protect\citeauthoryear{{Shagatova}, {Skopal}, {Shugarov},
  {Kom{\v{z}}{\'\i}k}, {Kundra}  \& {Teyssier}}{{Shagatova}
  et~al.}{2021}]{Skopal_2021}
{Shagatova} N.,  {Skopal} A.,  {Shugarov} S.~Y.,  {Kom{\v{z}}{\'\i}k} R.,
  {Kundra} E.,   {Teyssier} F.,  2021, \mn@doi [\aap]
  {10.1051/0004-6361/202039103}, \href
  {https://ui.adsabs.harvard.edu/abs/2021A&A...646A.116S} {646, A116}

\bibitem[\protect\citeauthoryear{{Shao} \& {Li}}{{Shao} \& {Li}}{2020}]{SL2020}
{Shao} Y.,  {Li} X.-D.,  2020, \mn@doi [\apj] {10.3847/1538-4357/aba118}, \href
  {https://ui.adsabs.harvard.edu/abs/2020ApJ...898..143S} {898, 143}

\bibitem[\protect\citeauthoryear{{Shara}, {Livio}, {Moffat}  \& {Orio}}{{Shara}
  et~al.}{1986}]{Shara_1986}
{Shara} M.~M.,  {Livio} M.,  {Moffat} A. F.~J.,   {Orio} M.,  1986, \mn@doi
  [\apj] {10.1086/164762}, \href
  {https://ui.adsabs.harvard.edu/abs/1986ApJ...311..163S} {311, 163}

\bibitem[\protect\citeauthoryear{{Shen}}{{Shen}}{2015}]{Shen_2015}
{Shen} K.~J.,  2015, \mn@doi [\apjl] {10.1088/2041-8205/805/1/L6}, \href
  {https://ui.adsabs.harvard.edu/abs/2015ApJ...805L...6S} {805, L6}

\bibitem[\protect\citeauthoryear{{Shen} \& {Bildsten}}{{Shen} \&
  {Bildsten}}{2007}]{Shen_2007}
{Shen} K.~J.,  {Bildsten} L.,  2007, \mn@doi [\apj] {10.1086/513457}, \href
  {https://ui.adsabs.harvard.edu/abs/2007ApJ...660.1444S} {660, 1444}

\bibitem[\protect\citeauthoryear{{Silverman} \& {Filippenko}}{{Silverman} \&
  {Filippenko}}{2008}]{Silverman_2008}
{Silverman} J.~M.,  {Filippenko} A.~V.,  2008, \mn@doi [\apjl]
  {10.1086/588096}, \href
  {https://ui.adsabs.harvard.edu/abs/2008ApJ...678L..17S} {678, L17}

\bibitem[\protect\citeauthoryear{{Skopal}}{{Skopal}}{2005}]{Skopal_2005}
{Skopal} A.,  2005, \mn@doi [\aap] {10.1051/0004-6361:20034262}, \href
  {https://ui.adsabs.harvard.edu/abs/2005A&A...440..995S} {440, 995}

\bibitem[\protect\citeauthoryear{{Skopal} \& {Carikov{\'a}}}{{Skopal} \&
  {Carikov{\'a}}}{2015}]{Skopal_2015}
{Skopal} A.,  {Carikov{\'a}} Z.,  2015, \mn@doi [\aap]
  {10.1051/0004-6361/201424779}, \href
  {https://ui.adsabs.harvard.edu/abs/2015A&A...573A...8S} {573, A8}

\bibitem[\protect\citeauthoryear{{Soberman}, {Phinney}  \& {van den
  Heuvel}}{{Soberman} et~al.}{1997}]{SPV97}
{Soberman} G.~E.,  {Phinney} E.~S.,   {van den Heuvel} E.~P.~J.,  1997, \aap,
  \href {https://ui.adsabs.harvard.edu/abs/1997A&A...327..620S} {327, 620}

\bibitem[\protect\citeauthoryear{{Soethe} \& {Kepler}}{{Soethe} \&
  {Kepler}}{2021}]{Soethe_2021}
{Soethe} L.~T.~T.,  {Kepler} S.~O.,  2021, \mn@doi [\mnras]
  {10.1093/mnras/stab1916}, \href
  {https://ui.adsabs.harvard.edu/abs/2021MNRAS.506.3266S} {506, 3266}

\bibitem[\protect\citeauthoryear{{Soker}}{{Soker}}{2000}]{Soker_2000}
{Soker} N.,  2000, \aap, \href
  {https://ui.adsabs.harvard.edu/abs/2000A&A...357..557S} {357, 557}

\bibitem[\protect\citeauthoryear{{Soker}}{{Soker}}{2017}]{Soker_2017}
{Soker} N.,  2017, \mn@doi [\mnras] {10.1093/mnras/stx1978}, \href
  {https://ui.adsabs.harvard.edu/abs/2017MNRAS.471.4839S} {471, 4839}

\bibitem[\protect\citeauthoryear{{Soker}, {Grichener}  \& {Sabach}}{{Soker}
  et~al.}{2018}]{Soker_2018}
{Soker} N.,  {Grichener} A.,   {Sabach} E.,  2018, \mn@doi [\apjl]
  {10.3847/2041-8213/aad736}, \href
  {https://ui.adsabs.harvard.edu/abs/2018ApJ...863L..14S} {863, L14}

\bibitem[\protect\citeauthoryear{{Solheim}}{{Solheim}}{2010}]{Solheim_2010}
{Solheim} J.~E.,  2010, \mn@doi [\pasp] {10.1086/656680}, \href
  {https://ui.adsabs.harvard.edu/abs/2010PASP..122.1133S} {122, 1133}

\bibitem[\protect\citeauthoryear{{Sparks} \& {Sion}}{{Sparks} \&
  {Sion}}{2021}]{Sparks_2021}
{Sparks} W.~M.,  {Sion} E.~M.,  2021, \mn@doi [\apj]
  {10.3847/1538-4357/abf2bc}, \href
  {https://ui.adsabs.harvard.edu/abs/2021ApJ...914....5S} {914, 5}

\bibitem[\protect\citeauthoryear{{Spruit} \& {Ritter}}{{Spruit} \&
  {Ritter}}{1983}]{SpruitRitter_1983}
{Spruit} H.~C.,  {Ritter} H.,  1983, \aap, \href
  {https://ui.adsabs.harvard.edu/abs/1983A&A...124..267S} {124, 267}

\bibitem[\protect\citeauthoryear{{Spruit} \& {Taam}}{{Spruit} \&
  {Taam}}{2001}]{SpruitTaam_2001}
{Spruit} H.~C.,  {Taam} R.~E.,  2001, \mn@doi [\apj] {10.1086/319030}, \href
  {https://ui.adsabs.harvard.edu/abs/2001ApJ...548..900S} {548, 900}

\bibitem[\protect\citeauthoryear{{Stanishev}, {Zamanov}, {Tomov}  \&
  {Marziani}}{{Stanishev} et~al.}{2004}]{Stanishev_2004}
{Stanishev} V.,  {Zamanov} R.,  {Tomov} N.,   {Marziani} P.,  2004, \mn@doi
  [\aap] {10.1051/0004-6361:20034623}, \href
  {https://ui.adsabs.harvard.edu/abs/2004A&A...415..609S} {415, 609}

\bibitem[\protect\citeauthoryear{{Stehle} \& {Ritter}}{{Stehle} \&
  {Ritter}}{1999}]{StehleRitter_1999}
{Stehle} R.,  {Ritter} H.,  1999, \mn@doi [\mnras]
  {10.1046/j.1365-8711.1999.02838.x}, \href
  {https://ui.adsabs.harvard.edu/abs/1999MNRAS.309..245S} {309, 245}

\bibitem[\protect\citeauthoryear{{Stehle}, {Kolb}  \& {Ritter}}{{Stehle}
  et~al.}{1997}]{Stehle_1997}
{Stehle} R.,  {Kolb} U.,   {Ritter} H.,  1997, \aap, \href
  {http://adsabs.harvard.edu/abs/1997A%26A...320..136S} {320, 136}

\bibitem[\protect\citeauthoryear{Strohmayer \& Bildsten}{Strohmayer \&
  Bildsten}{2006}]{StrohmayerBildsten_2006}
Strohmayer T.,  Bildsten L.,  2006, New views of thermonuclear bursts.
Cambridge University Press, p. 113–156 (\mn@eprint {arXiv}
  {astro-ph/0301544}), \mn@doi{10.1017/CBO9780511536281.004}

\bibitem[\protect\citeauthoryear{{Taam} \& {Sandquist}}{{Taam} \&
  {Sandquist}}{2000}]{TS00}
{Taam} R.~E.,  {Sandquist} E.~L.,  2000, \mn@doi [\araa]
  {10.1146/annurev.astro.38.1.113}, \href
  {https://ui.adsabs.harvard.edu/abs/2000ARA&A..38..113T} {38, 113}

\bibitem[\protect\citeauthoryear{{Taam} \& {Spruit}}{{Taam} \&
  {Spruit}}{1989}]{TaamSpruit_1989}
{Taam} R.~E.,  {Spruit} H.~C.,  1989, \mn@doi [\apj] {10.1086/167966}, \href
  {https://ui.adsabs.harvard.edu/abs/1989ApJ...345..972T} {345, 972}

\bibitem[\protect\citeauthoryear{{Tauris} \& {van den Heuvel}}{{Tauris} \& {van
  den Heuvel}}{2006}]{Tauris_2006}
{Tauris} T.~M.,  {van den Heuvel} E.~P.~J.,  2006, {Formation and evolution of
  compact stellar X-ray sources, {\rm in Walter Lewin and Michiel van der Klis,
  eds, Compact stellar X-ray sources}}.
Cambridge University Press, Cambridge, UK, pp 623--665 (\mn@eprint {arXiv}
  {astro-ph/0303456})

\bibitem[\protect\citeauthoryear{{Tauris} \& {van den Heuvel}}{{Tauris} \& {van
  den Heuvel}}{2023}]{Tauris_2023BOOK}
{Tauris} T.~M.,  {van den Heuvel} E. P.~J.,  2023, {Physics of Binary Star
  Evolution. From Stars to X-ray Binaries and Gravitational Wave Sources}.
Princeton University Press, Princeton, NJ

\bibitem[\protect\citeauthoryear{{Tauris}, {Langer}  \& {Kramer}}{{Tauris}
  et~al.}{2011}]{Tauris_2011}
{Tauris} T.~M.,  {Langer} N.,   {Kramer} M.,  2011, \mn@doi [\mnras]
  {10.1111/j.1365-2966.2011.19189.x}, \href
  {https://ui.adsabs.harvard.edu/abs/2011MNRAS.416.2130T} {416, 2130}

\bibitem[\protect\citeauthoryear{{Tauris}, {Sanyal}, {Yoon}  \&
  {Langer}}{{Tauris} et~al.}{2013}]{Tauris_2013}
{Tauris} T.~M.,  {Sanyal} D.,  {Yoon} S.~C.,   {Langer} N.,  2013, \mn@doi
  [\aap] {10.1051/0004-6361/201321662}, \href
  {https://ui.adsabs.harvard.edu/abs/2013A&A...558A..39T} {558, A39}

\bibitem[\protect\citeauthoryear{{Tauris} et~al.,}{{Tauris}
  et~al.}{2017}]{Tauris_2017}
{Tauris} T.~M.,  et~al., 2017, \mn@doi [\apj] {10.3847/1538-4357/aa7e89}, \href
  {https://ui.adsabs.harvard.edu/abs/2017ApJ...846..170T} {846, 170}

\bibitem[\protect\citeauthoryear{{Tetarenko} et~al.,}{{Tetarenko}
  et~al.}{2016}]{Tetarenko_2016}
{Tetarenko} B.~E.,  et~al., 2016, \mn@doi [\apj] {10.3847/0004-637X/825/1/10},
  \href {https://ui.adsabs.harvard.edu/abs/2016ApJ...825...10T} {825, 10}

\bibitem[\protect\citeauthoryear{{Thorne} \& {Zytkow}}{{Thorne} \&
  {Zytkow}}{1975}]{TZ1975}
{Thorne} K.~S.,  {Zytkow} A.~N.,  1975, \mn@doi [\apjl] {10.1086/181839}, \href
  {https://ui.adsabs.harvard.edu/abs/1975ApJ...199L..19T} {199, L19}

\bibitem[\protect\citeauthoryear{{Toonen} \& {Nelemans}}{{Toonen} \&
  {Nelemans}}{2013}]{Toonen_2013}
{Toonen} S.,  {Nelemans} G.,  2013, \mn@doi [\aap]
  {10.1051/0004-6361/201321753}, \href
  {http://adsabs.harvard.edu/abs/2013A%26A...557A..87T} {557, A87}

\bibitem[\protect\citeauthoryear{{Townsend}, {Coe}, {Corbet}  \&
  {Hill}}{{Townsend} et~al.}{2011}]{Townsend_2011}
{Townsend} L.~J.,  {Coe} M.~J.,  {Corbet} R.~H.~D.,   {Hill} A.~B.,  2011,
  \mn@doi [\mnras] {10.1111/j.1365-2966.2011.19153.x}, \href
  {https://ui.adsabs.harvard.edu/abs/2011MNRAS.416.1556T} {416, 1556}

\bibitem[\protect\citeauthoryear{{Townsley} \& {Bildsten}}{{Townsley} \&
  {Bildsten}}{2003}]{Townsley_2003}
{Townsley} D.~M.,  {Bildsten} L.,  2003, \mn@doi [\apjl] {10.1086/379535},
  \href {http://adsabs.harvard.edu/abs/2003ApJ...596L.227T} {596, L227}

\bibitem[\protect\citeauthoryear{{Townsley} \& {Bildsten}}{{Townsley} \&
  {Bildsten}}{2004}]{Townsley_2004}
{Townsley} D.~M.,  {Bildsten} L.,  2004, \mn@doi [\apj] {10.1086/379647}, \href
  {http://adsabs.harvard.edu/abs/2004ApJ...600..390T} {600, 390}

\bibitem[\protect\citeauthoryear{{Townsley} \& {G{\"a}nsicke}}{{Townsley} \&
  {G{\"a}nsicke}}{2009}]{Townsley_2009}
{Townsley} D.~M.,  {G{\"a}nsicke} B.~T.,  2009, \mn@doi [\apj]
  {10.1088/0004-637X/693/1/1007}, \href
  {http://adsabs.harvard.edu/abs/2009ApJ...693.1007T} {693, 1007}

\bibitem[\protect\citeauthoryear{{Tr{\"u}mper} et~al.,}{{Tr{\"u}mper}
  et~al.}{1991}]{Truemper_1991}
{Tr{\"u}mper} J.,  et~al., 1991, \mn@doi [\nat] {10.1038/349579a0}, \href
  {https://ui.adsabs.harvard.edu/abs/1991Natur.349..579T} {349, 579}

\bibitem[\protect\citeauthoryear{{Valsecchi}, {Glebbeek}, {Farr}, {Fragos},
  {Willems}, {Orosz}, {Liu}  \& {Kalogera}}{{Valsecchi}
  et~al.}{2010}]{Valsecchi_2010}
{Valsecchi} F.,  {Glebbeek} E.,  {Farr} W.~M.,  {Fragos} T.,  {Willems} B.,
  {Orosz} J.~A.,  {Liu} J.,   {Kalogera} V.,  2010, \mn@doi [\nat]
  {10.1038/nature09463}, \href
  {https://ui.adsabs.harvard.edu/abs/2010Natur.468...77V} {468, 77}

\bibitem[\protect\citeauthoryear{{Van} \& {Ivanova}}{{Van} \&
  {Ivanova}}{2019}]{Van2019CARB}
{Van} K.~X.,  {Ivanova} N.,  2019, \mn@doi [\apjl] {10.3847/2041-8213/ab571c},
  \href {https://ui.adsabs.harvard.edu/abs/2019ApJ...886L..31V} {886, L31}

\bibitem[\protect\citeauthoryear{{Van der Swaelmen}, {Boffin}, {Jorissen}  \&
  {Van Eck}}{{Van der Swaelmen} et~al.}{2017}]{Swaelmen_2017}
{Van der Swaelmen} M.,  {Boffin} H.~M.~J.,  {Jorissen} A.,   {Van Eck} S.,
  2017, \mn@doi [\aap] {10.1051/0004-6361/201628867}, \href
  {https://ui.adsabs.harvard.edu/abs/2017A&A...597A..68V} {597, A68}

\bibitem[\protect\citeauthoryear{{Van}, {Ivanova}  \& {Heinke}}{{Van}
  et~al.}{2019}]{Van_2019}
{Van} K.~X.,  {Ivanova} N.,   {Heinke} C.~O.,  2019, \mn@doi [\mnras]
  {10.1093/mnras/sty3489}, \href
  {https://ui.adsabs.harvard.edu/abs/2019MNRAS.483.5595V} {483, 5595}

\bibitem[\protect\citeauthoryear{{Verbunt} \& {Hut}}{{Verbunt} \&
  {Hut}}{1987}]{Verbunt_1987}
{Verbunt} F.,  {Hut} P.,  1987, in {Helfand} D.~J.,  {Huang} J.~H.,  eds,  IAU
  Symposium Vol. 125, The Origin and Evolution of Neutron Stars. p.~187

\bibitem[\protect\citeauthoryear{{Verbunt} \& {Zwaan}}{{Verbunt} \&
  {Zwaan}}{1981}]{VZ}
{Verbunt} F.,  {Zwaan} C.,  1981, \aap, \href
  {https://ui.adsabs.harvard.edu/abs/1981A&A...100L...7V} {100, L7}

\bibitem[\protect\citeauthoryear{{Vos}, {{\O}stensen}, {Marchant}  \& {Van
  Winckel}}{{Vos} et~al.}{2015}]{Vos_2015}
{Vos} J.,  {{\O}stensen} R.~H.,  {Marchant} P.,   {Van Winckel} H.,  2015,
  \mn@doi [\aap] {10.1051/0004-6361/201526019}, \href
  {https://ui.adsabs.harvard.edu/abs/2015A&A...579A..49V} {579, A49}

\bibitem[\protect\citeauthoryear{{Vos}, {N{\'e}meth}, {Vu{\v{c}}kovi{\'c}},
  {{\O}stensen}  \& {Parsons}}{{Vos} et~al.}{2018}]{Vos_2018}
{Vos} J.,  {N{\'e}meth} P.,  {Vu{\v{c}}kovi{\'c}} M.,  {{\O}stensen} R.,
  {Parsons} S.,  2018, \mn@doi [\mnras] {10.1093/mnras/stx2198}, \href
  {https://ui.adsabs.harvard.edu/abs/2018MNRAS.473..693V} {473, 693}

\bibitem[\protect\citeauthoryear{{Vos}, {Bobrick}  \&
  {Vu{\v{c}}kovi{\'c}}}{{Vos} et~al.}{2020}]{Vos_2020}
{Vos} J.,  {Bobrick} A.,   {Vu{\v{c}}kovi{\'c}} M.,  2020, \mn@doi [\aap]
  {10.1051/0004-6361/201937195}, \href
  {https://ui.adsabs.harvard.edu/abs/2020A&A...641A.163V} {641, A163}

\bibitem[\protect\citeauthoryear{{Voss} \& {Tauris}}{{Voss} \&
  {Tauris}}{2003}]{VT2003}
{Voss} R.,  {Tauris} T.~M.,  2003, \mn@doi [\mnras]
  {10.1046/j.1365-8711.2003.06616.x}, \href
  {https://ui.adsabs.harvard.edu/abs/2003MNRAS.342.1169V} {342, 1169}

\bibitem[\protect\citeauthoryear{{Wang}, {Chen}, {Liu}, {Chen}, {Wu}, {Tang},
  {Guo}  \& {Han}}{{Wang} et~al.}{2021}]{Wang_2021}
{Wang} B.,  {Chen} W.-C.,  {Liu} D.-D.,  {Chen} H.-L.,  {Wu} C.-Y.,  {Tang}
  W.-S.,  {Guo} Y.-L.,   {Han} Z.-W.,  2021, \mn@doi [\mnras]
  {10.1093/mnras/stab2032}, \href
  {https://ui.adsabs.harvard.edu/abs/2021MNRAS.506.4654W} {506, 4654}

\bibitem[\protect\citeauthoryear{{Waters} \& {van Kerkwijk}}{{Waters} \& {van
  Kerkwijk}}{1989}]{Waters_1989}
{Waters} L.~B.~F.~M.,  {van Kerkwijk} M.~H.,  1989, \aap, \href
  {https://ui.adsabs.harvard.edu/abs/1989A&A...223..196W} {223, 196}

\bibitem[\protect\citeauthoryear{{Watson}, {Steeghs}, {Shahbaz}  \&
  {Dhillon}}{{Watson} et~al.}{2007}]{Watson_2007}
{Watson} C.~A.,  {Steeghs} D.,  {Shahbaz} T.,   {Dhillon} V.~S.,  2007, \mn@doi
  [\mnras] {10.1111/j.1365-2966.2007.12173.x}, \href
  {https://ui.adsabs.harvard.edu/abs/2007MNRAS.382.1105W} {382, 1105}

\bibitem[\protect\citeauthoryear{{Webbink}}{{Webbink}}{1975}]{Webbink_1975}
{Webbink} R.~F.,  1975, PhD thesis, University of Cambridge

\bibitem[\protect\citeauthoryear{{Webbink}}{{Webbink}}{1985}]{Webbink_1985}
{Webbink} R.~F.,  1985, {Stellar evolution and binaries}.
Cambridge Univ. Press, Cambridge, p.~39

\bibitem[\protect\citeauthoryear{{Webbink}}{{Webbink}}{1988}]{Webbink_1988}
{Webbink} R.~F.,  1988, in {Mikolajewska} J.,  {Friedjung} M.,  {Kenyon} S.~J.,
    {Viotti} R.,  eds,  Astrophysics and Space Science Library Vol. 145, IAU
  Colloq. 103: The Symbiotic Phenomenon. p.~311,
  \mn@doi{10.1007/978-94-009-2969-2\_69}

\bibitem[\protect\citeauthoryear{{Webbink}}{{Webbink}}{2008}]{Webbink_2008}
{Webbink} R.~F.,  2008, in {Milone} E.~F.,  {Leahy} D.~A.,   {Hobill} D.~W.,
  eds,  Astrophysics and Space Science Library Vol. 352, Astrophysics and Space
  Science Library. p.~233 (\mn@eprint {arXiv} {0704.0280}),
  \mn@doi{10.1007/978-1-4020-6544-6_13}

\bibitem[\protect\citeauthoryear{{Webbink} \& {Wickramasinghe}}{{Webbink} \&
  {Wickramasinghe}}{2002}]{WW02}
{Webbink} R.~F.,  {Wickramasinghe} D.~T.,  2002, \mn@doi [\mnras]
  {10.1046/j.1365-8711.2002.05495.x}, \href
  {http://adsabs.harvard.edu/abs/2002MNRAS.335....1W} {335, 1}

\bibitem[\protect\citeauthoryear{{Wiktorowicz}, {Belczynski}  \&
  {Maccarone}}{{Wiktorowicz} et~al.}{2014}]{Wiktorowicz_2014}
{Wiktorowicz} G.,  {Belczynski} K.,   {Maccarone} T.,  2014, in Binary Systems,
  their Evolution and Environments. p.~37 (\mn@eprint {arXiv} {1312.5924})

\bibitem[\protect\citeauthoryear{{Wiktorowicz}, {Lasota}, {Belczynski}, {Lu},
  {Liu}  \& {I{\l}kiewicz}}{{Wiktorowicz} et~al.}{2021}]{Wiktorowicz_2021}
{Wiktorowicz} G.,  {Lasota} J.-P.,  {Belczynski} K.,  {Lu} Y.,  {Liu} J.,
  {I{\l}kiewicz} K.,  2021, \mn@doi [\apj] {10.3847/1538-4357/ac0cf7}, \href
  {https://ui.adsabs.harvard.edu/abs/2021ApJ...918...60W} {918, 60}

\bibitem[\protect\citeauthoryear{{Willems}, {Kolb}, {Sandquist}, {Taam}  \&
  {Dubus}}{{Willems} et~al.}{2005}]{Willems_2005}
{Willems} B.,  {Kolb} U.,  {Sandquist} E.~L.,  {Taam} R.~E.,   {Dubus} G.,
  2005, \mn@doi [\apj] {10.1086/498010}, \href
  {http://adsabs.harvard.edu/abs/2005ApJ...635.1263W} {635, 1263}

\bibitem[\protect\citeauthoryear{{Williams}, {Binder}, {Dalcanton}, {Eracleous}
   \& {Dolphin}}{{Williams} et~al.}{2013}]{Williams_2013}
{Williams} B.~F.,  {Binder} B.~A.,  {Dalcanton} J.~J.,  {Eracleous} M.,
  {Dolphin} A.,  2013, \mn@doi [\apj] {10.1088/0004-637X/772/1/12}, \href
  {https://ui.adsabs.harvard.edu/abs/2013ApJ...772...12W} {772, 12}

\bibitem[\protect\citeauthoryear{{Wolf}, {Bildsten}, {Brooks}  \&
  {Paxton}}{{Wolf} et~al.}{2013}]{Wolf_2013}
{Wolf} W.~M.,  {Bildsten} L.,  {Brooks} J.,   {Paxton} B.,  2013, \mn@doi
  [\apj] {10.1088/0004-637X/777/2/136}, \href
  {https://ui.adsabs.harvard.edu/abs/2013ApJ...777..136W} {777, 136}

\bibitem[\protect\citeauthoryear{{Wong} \& {Bildsten}}{{Wong} \&
  {Bildsten}}{2021}]{Wong_2021}
{Wong} T. L.~S.,  {Bildsten} L.,  2021, arXiv e-prints, \href
  {https://ui.adsabs.harvard.edu/abs/2021arXiv210913403W} {p. arXiv:2109.13403}

\bibitem[\protect\citeauthoryear{{Woods}, {Ivanova}, {van der Sluys}  \&
  {Chaichenets}}{{Woods} et~al.}{2012}]{W12}
{Woods} T.~E.,  {Ivanova} N.,  {van der Sluys} M.~V.,   {Chaichenets} S.,
  2012, \mn@doi [\apj] {10.1088/0004-637X/744/1/12}, \href
  {https://ui.adsabs.harvard.edu/abs/2012ApJ...744...12W} {744, 12}

\bibitem[\protect\citeauthoryear{{Yaron}, {Prialnik}, {Shara}  \&
  {Kovetz}}{{Yaron} et~al.}{2005}]{Yaron_2005}
{Yaron} O.,  {Prialnik} D.,  {Shara} M.~M.,   {Kovetz} A.,  2005, \mn@doi
  [\apj] {10.1086/428435}, \href
  {http://adsabs.harvard.edu/abs/2005ApJ...623..398Y} {623, 398}

\bibitem[\protect\citeauthoryear{{Yungelson}}{{Yungelson}}{2008}]{Yungelson_2008}
{Yungelson} L.~R.,  2008, \mn@doi [Astronomy Letters]
  {10.1134/S1063773708090053}, \href
  {https://ui.adsabs.harvard.edu/abs/2008AstL...34..620Y} {34, 620}

\bibitem[\protect\citeauthoryear{{Yungelson} \& {Lasota}}{{Yungelson} \&
  {Lasota}}{2008}]{YL08}
{Yungelson} L.~R.,  {Lasota} J.~P.,  2008, \mn@doi [\aap]
  {10.1051/0004-6361:200809684}, \href
  {https://ui.adsabs.harvard.edu/abs/2008A&A...488..257Y} {488, 257}

\bibitem[\protect\citeauthoryear{{Yungelson}, {Kuranov}  \&
  {Postnov}}{{Yungelson} et~al.}{2019}]{Yungelson_2019}
{Yungelson} L.~R.,  {Kuranov} A.~G.,   {Postnov} K.~A.,  2019, \mn@doi [\mnras]
  {10.1093/mnras/stz467}, \href
  {https://ui.adsabs.harvard.edu/abs/2019MNRAS.485..851Y} {485, 851}

\bibitem[\protect\citeauthoryear{{Zamanov}, {Bode}, {Melo}, {Bachev}, {Gomboc},
  {Stateva}, {Porter}  \& {Pritchard}}{{Zamanov} et~al.}{2007}]{Zamanov_2007}
{Zamanov} R.~K.,  {Bode} M.~F.,  {Melo} C.~H.~F.,  {Bachev} R.,  {Gomboc} A.,
  {Stateva} I.~K.,  {Porter} J.~M.,   {Pritchard} J.,  2007, \mn@doi [\mnras]
  {10.1111/j.1365-2966.2007.12150.x}, \href
  {https://ui.adsabs.harvard.edu/abs/2007MNRAS.380.1053Z} {380, 1053}

\bibitem[\protect\citeauthoryear{{Zamanov}, {Bode}, {Melo}, {Stateva},
  {Bachev}, {Gomboc}, {Konstantinova-Antova}  \& {Stoyanov}}{{Zamanov}
  et~al.}{2008}]{Zamanov_2008}
{Zamanov} R.~K.,  {Bode} M.~F.,  {Melo} C.~H.~F.,  {Stateva} I.~K.,  {Bachev}
  R.,  {Gomboc} A.,  {Konstantinova-Antova} R.,   {Stoyanov} K.~A.,  2008,
  \mn@doi [\mnras] {10.1111/j.1365-2966.2008.13751.x}, \href
  {https://ui.adsabs.harvard.edu/abs/2008MNRAS.390..377Z} {390, 377}

\bibitem[\protect\citeauthoryear{{Zdziarski}, {Mikolajewska}  \&
  {Belczynski}}{{Zdziarski} et~al.}{2013}]{Zdziarski_2013}
{Zdziarski} A.~A.,  {Mikolajewska} J.,   {Belczynski} K.,  2013, \mn@doi
  [\mnras] {10.1093/mnrasl/sls035}, \href
  {https://ui.adsabs.harvard.edu/abs/2013MNRAS.429L.104Z} {429, L104}

\bibitem[\protect\citeauthoryear{{Zhang} \& {Kojima}}{{Zhang} \&
  {Kojima}}{2006}]{Zhang_2006}
{Zhang} C.~M.,  {Kojima} Y.,  2006, \mn@doi [\mnras]
  {10.1111/j.1365-2966.2005.09802.x}, \href
  {https://ui.adsabs.harvard.edu/abs/2006MNRAS.366..137Z} {366, 137}

\bibitem[\protect\citeauthoryear{{Zhang}, {Wickramasinghe}  \&
  {Ferrario}}{{Zhang} et~al.}{2009}]{Zhang_2009}
{Zhang} C.~M.,  {Wickramasinghe} D.~T.,   {Ferrario} L.,  2009, \mn@doi
  [\mnras] {10.1111/j.1365-2966.2009.15154.x}, \href
  {https://ui.adsabs.harvard.edu/abs/2009MNRAS.397.2208Z} {397, 2208}

\bibitem[\protect\citeauthoryear{{Ziolkowski}}{{Ziolkowski}}{2002}]{Ziolkowski_2002}
{Ziolkowski} J.,  2002, \memsai, \href
  {https://ui.adsabs.harvard.edu/abs/2002MmSAI..73.1038Z} {73, 1038}

\bibitem[\protect\citeauthoryear{{Zorotovic} \& {Schreiber}}{{Zorotovic} \&
  {Schreiber}}{2020}]{ZS2020}
{Zorotovic} M.,  {Schreiber} M.~R.,  2020, \mn@doi [\adspr]
  {10.1016/j.asr.2019.08.044}, \href
  {https://ui.adsabs.harvard.edu/abs/2020AdSpR..66.1080Z} {66, 1080}

\bibitem[\protect\citeauthoryear{{Zorotovic}, {Schreiber}, {G{\"a}nsicke}  \&
  {Nebot G{\'o}mez-Mor{\'a}n}}{{Zorotovic} et~al.}{2010}]{Zorotovic_2010}
{Zorotovic} M.,  {Schreiber} M.~R.,  {G{\"a}nsicke} B.~T.,   {Nebot
  G{\'o}mez-Mor{\'a}n} A.,  2010, \mn@doi [\aap] {10.1051/0004-6361/200913658},
  \href {http://adsabs.harvard.edu/abs/2010A%26A...520A..86Z} {520, A86}

\bibitem[\protect\citeauthoryear{{Zorotovic}, {Schreiber}  \&
  {G{\"a}nsicke}}{{Zorotovic} et~al.}{2011}]{Zorotovic_2011}
{Zorotovic} M.,  {Schreiber} M.~R.,   {G{\"a}nsicke} B.~T.,  2011, \mn@doi
  [\aap] {10.1051/0004-6361/201116626}, \href
  {http://adsabs.harvard.edu/abs/2011A%26A...536A..42Z} {536, A42}

\bibitem[\protect\citeauthoryear{{Zorotovic}, {Schreiber}  \&
  {Parsons}}{{Zorotovic} et~al.}{2014a}]{Zorotovic_2014_KOI}
{Zorotovic} M.,  {Schreiber} M.~R.,   {Parsons} S.~G.,  2014a, \mn@doi [\aap]
  {10.1051/0004-6361/201424430}, \href
  {http://adsabs.harvard.edu/abs/2014A%26A...568L...9Z} {568, L9}

\bibitem[\protect\citeauthoryear{{Zorotovic}, {Schreiber},
  {Garc{\'{\i}}a-Berro}, {Camacho}, {Torres}, {Rebassa-Mansergas}  \&
  {G{\"a}nsicke}}{{Zorotovic} et~al.}{2014b}]{Zorotovic_2014}
{Zorotovic} M.,  {Schreiber} M.~R.,  {Garc{\'{\i}}a-Berro} E.,  {Camacho} J.,
  {Torres} S.,  {Rebassa-Mansergas} A.,   {G{\"a}nsicke} B.~T.,  2014b, \mn@doi
  [\aap] {10.1051/0004-6361/201323039}, \href
  {http://adsabs.harvard.edu/abs/2014A%26A...568A..68Z} {568, A68}

\bibitem[\protect\citeauthoryear{{Zurek}, {Knigge}, {Maccarone}, {Dieball}  \&
  {Long}}{{Zurek} et~al.}{2009}]{Zurek_2009}
{Zurek} D.~R.,  {Knigge} C.,  {Maccarone} T.~J.,  {Dieball} A.,   {Long} K.~S.,
   2009, \mn@doi [\apj] {10.1088/0004-637X/699/2/1113}, \href
  {https://ui.adsabs.harvard.edu/abs/2009ApJ...699.1113Z} {699, 1113}

\bibitem[\protect\citeauthoryear{{de Val-Borro}, {Karovska}  \& {Sasselov}}{{de
  Val-Borro} et~al.}{2009}]{ValBorro_2009}
{de Val-Borro} M.,  {Karovska} M.,   {Sasselov} D.,  2009, \mn@doi [\apj]
  {10.1088/0004-637X/700/2/1148}, \href
  {https://ui.adsabs.harvard.edu/abs/2009ApJ...700.1148D} {700, 1148}

\bibitem[\protect\citeauthoryear{{de Val-Borro}, {Karovska}, {Sasselov}  \&
  {Stone}}{{de Val-Borro} et~al.}{2017}]{Borro_2017}
{de Val-Borro} M.,  {Karovska} M.,  {Sasselov} D.~D.,   {Stone} J.~M.,  2017,
  \mn@doi [\mnras] {10.1093/mnras/stx684}, \href
  {https://ui.adsabs.harvard.edu/abs/2017MNRAS.468.3408D} {468, 3408}

\bibitem[\protect\citeauthoryear{{van Roestel} et~al.,}{{van Roestel}
  et~al.}{2021}]{vanRoestel_2021}
{van Roestel} J.,  et~al., 2021, \mn@doi [\mnras] {10.1093/mnras/stab2421},
  \href {https://ui.adsabs.harvard.edu/abs/2021MNRAS.tmp.2205V} {}

\bibitem[\protect\citeauthoryear{{van den Berg}}{{van den
  Berg}}{2020}]{vandenBerg_2020}
{van den Berg} M.,  2020, in {Bragaglia} A.,  {Davies} M.,  {Sills} A.,
  {Vesperini} E.,  eds,  IAU Symposium Vol. 351, IAU Symposium. pp 367--376
  (\mn@eprint {arXiv} {1910.07595}), \mn@doi{10.1017/S1743921319007981}

\bibitem[\protect\citeauthoryear{{van den Heuvel}, {Portegies Zwart}  \& {de
  Mink}}{{van den Heuvel} et~al.}{2017}]{van_den_Heuvel_2017}
{van den Heuvel} E.~P.~J.,  {Portegies Zwart} S.~F.,   {de Mink} S.~E.,  2017,
  \mn@doi [\mnras] {10.1093/mnras/stx1430}, \href
  {https://ui.adsabs.harvard.edu/abs/2017MNRAS.471.4256V} {471, 4256}

\bibitem[\protect\citeauthoryear{{van der Hucht}}{{van der
  Hucht}}{2001}]{Hucht_2001}
{van der Hucht} K.~A.,  2001, \mn@doi [\nar] {10.1016/S1387-6473(00)00112-3},
  \href {https://ui.adsabs.harvard.edu/abs/2001NewAR..45..135V} {45, 135}

\bibitem[\protect\citeauthoryear{{van der Meer}, {Kaper}, {van Kerkwijk},
  {Heemskerk}  \& {van den Heuvel}}{{van der Meer}
  et~al.}{2007}]{vanderMeer_2007}
{van der Meer} A.,  {Kaper} L.,  {van Kerkwijk} M.~H.,  {Heemskerk} M.~H.~M.,
  {van den Heuvel} E.~P.~J.,  2007, \mn@doi [\aap]
  {10.1051/0004-6361:20066025}, \href
  {https://ui.adsabs.harvard.edu/abs/2007A&A...473..523V} {473, 523}

\makeatother
\end{thebibliography}

\end{document}